\documentclass{memo-l}

\usepackage{graphicx}
\usepackage{verbatim}
\usepackage{amssymb}

\usepackage{modified_makeidx2}

\makeatletter
\def\printnotation{{%
\def\indexname{Index of notation}
\begin{theindex}
\@input{\jobname.ntn}
\end{theindex}
}}
\makeatother

\makeglossary

\makeindex 

\newtheorem{theorem}{Theorem}[chapter]
\newtheorem{lemma}[theorem]{Lemma}
\newtheorem{proposition}[theorem]{Proposition}
\newtheorem{corollary}[theorem]{Corollary}

\theoremstyle{definition}
\newtheorem{definition}[theorem]{Definition}
\newtheorem{notation}[theorem]{Notation}

\theoremstyle{remark}
\newtheorem{remark}[theorem]{Remark}

\numberwithin{section}{chapter}
\numberwithin{equation}{chapter}

\newcommand{\ls}{\bigl\{}
\newcommand{\rs}{\bigr\}}
\newcommand{\wt}{\widetilde}

\newcommand{\lnf}{\lim_{n \to \infty}}

\newcommand{\lv}{\bigl|}

\newcommand{\indic}{\boldsymbol{1}}
\newcommand{\Lip}{\mathrm{Lip}}
\newcommand{\cond}{\,\lv\,}
\newcommand{\Was}{\mathrm{Was}}
\newcommand{\on}{\operatorname}
\newcommand{\was}[2]{\bigl\| {#1} - {#2} \bigr\|_{\Was}}

\newcommand{\rec}{\frac{1}}

\newcommand{\Acom}{A^{\complement}}
\newcommand{\mP}{\mathfrak{P}}
\newcommand{\mS}{\mathfrak{S}}
\newcommand{\mM}{\mathfrak{M}}
\newcommand{\mR}{\mathfrak{R}}

\newcommand{\D}{\mathcal{D}}
\renewcommand{\H}{\mathcal{H}}
\newcommand{\M}{\mathcal{M}}
\newcommand{\G}{\mathcal{G}}

\newcommand{\K}{\mathcal{K}}
\newcommand{\X}{\mathcal{X}}

\newcommand{\bN}{\mathbb{N}}
\newcommand{\bP}{\mathbb{P}}
\newcommand{\bR}{\mathbb{R}}
\newcommand{\bE}{\mathbb{E}}

\newcommand{\bg}{\mathbf{g}}
\newcommand{\bh}{\mathbf{h}}

\newcommand{\mcA}{\mathcal{A}}

\newcommand{\BM}{\mathcal{B}(\M)}

\newcommand{\hhH}{\check{H}}

\newcommand{\bX}{\mathbf{X}}

\begin{document}

\frontmatter

\title[Mutation-selection with recombination]
{A mutation-selection model with recombination for general genotypes}

\author{Steven N. Evans}
\address{Steven N. Evans\\
Department of Statistics\\
367 Evans Hall\\University of California\\ 
Berkeley, CA 94720-3860\\ 
U.S.A.}
\email{evans@stat.berkeley.edu}
\thanks{SNE supported in part by
grants DMS-04-05778 and DMS-09-07630 from the 
National Science Foundation (U.S.A.)}

\author{David Steinsaltz}
\address{David Steinsaltz\\
Department of Statistics\\
1 South Parks Road\\
Oxford, OX1 3TG\\
UNITED KINGDOM}
\email{steinsal@stats.ox.ac.uk}
\thanks{DS supported by grant K12-AG00981 from the National
Institute on Aging (U.S.A.); a Discovery Grant from the National Science and Engineering Research Council (Canada); and a grant from the New Dynamics of Ageing program coordinated
by the Economics and Social Research Council (U.K.) }

\author{Kenneth W. Wachter}
\address{Kenneth W. Wachter\\
Department of Demography\\
2232 Piedmont Avenue\\
University of California\\ 
Berkeley, CA 94720-2120\\ 
U.S.A.}
\email{wachter@demog.berkeley.edu}
\thanks{KWW supported by grant P01-008454 from the 
National Institute on Aging (U.S.A.) and by the Miller Institute 
for Basic Research in Science at U.C. Berkeley.}

\date{\today}



\begin{abstract}
We investigate a continuous time, probability 
measure-valued dynamical system that describes 
the process of mutation-selection balance 
in a context where the population is infinite, 
there may be infinitely many loci, 
and there are weak assumptions on selective costs.
Our model arises when we incorporate very general 
recombination mechanisms into a previous model of
mutation and selection from Steinsaltz, Evans and Wachter (2005)
and take the relative strength of mutation and selection 
to be sufficiently small.  The resulting dynamical system is 
\index{dynamical system}
a flow of measures on the space of loci.
Each such measure is the intensity measure
of a Poisson random measure on the
space of loci: 
\index{intensity measure}
the points of a realization of the random measure 
record the set of loci at which the genotype of a 
uniformly chosen individual differs from
a reference wild type due to an accumulation of
ancestral mutations.  Our motivation for
working in such a general setting is to provide a basis for
understanding mutation-driven changes in 
age-specific demographic schedules that arise from 
the complex interaction of many genes,
and hence to develop a framework for understanding
the evolution of aging.

We establish the existence and uniqueness of the dynamical
system,  provide conditions for the existence and stability
of equilibrium states,  and prove that our continuous-time
dynamical system is the limit of a sequence of discrete-time 
infinite population mutation-selection-recombination models
in the standard asymptotic regime where selection and mutation
are weak relative to recombination and both scale at the same
infinitesimal rate in the limit. 
\end{abstract}

\maketitle

\setcounter{page}{4}

\tableofcontents

\mainmatter

\chapter{Introduction}
\label{Ch:intro}

A principal goal of population genetics is to understand
how the mechanisms of mutation, selection and recombination
determine the manner in which the distribution of genotypes
in a population changes over time.  The distribution of genotypes 
governs, to a large degree,  the distribution of phenotypes,
an object that is both more readily observable and of
more immediate practical interest. Conversely,
the differential survival of genotypes is mediated almost
exclusively by their phenotypic expression.
\index{genotype}

Mathematical models have particular salience for the 
study of senescence.  Progressive physical deterioration
and mortality rates rising across the lifespan are
not,  on the face of it,  characteristics that natural
selection would be expected to favor.  
Peter Medawar, in his inaugural lecture at 
\index{Medawar, Peter}
University College London delivered in 1951 \cite{pM52}
called the origin of senescence ``an unsolved
problem in biology.''   He went on to outline 
a possible solution.   His idea, called ``mutation
accumulation''  presents senescence as a side-effect
of the interplay between natural selection and 
mutation.   Shaping evolutionary thinking for the
last half-century,  Medawar's idea is the centerpiece
of the present work.  
\index{mutation accumulation}

Medawar's proposal was complemented in 1957 by ideas of
George C. Williams \cite{gW57} directing attention to
\index{Williams, George C.}
alleles with multiple, that is, pleiotropic,  effects on
fitness, some enhancing and some impairing the
production and survival of progeny.      
This concept of ``antagonistic pleiotropy'' has taken
\index{antagonistic pleiotropy}
its place beside the concept of ``mutation accumulation''; 
together they are the twin guiding themes for evolutionary 
theories of aging.    The relevance of both concepts to senescence
turns on the truism that deleterious effects that only
take hold later in life sacrifice less of an organism's 
remaining potential for reproduction and nurturing than
effects that act early.   Medawar leads us to think
of an equilibrium between mutation and selection 
in which a population is maintained at less than
optimal fitness.  Williams depicts alleles that could
in constant environments go on to fixation.  
\index{equilibrium}

It is important to emphasize that the mutations under
consideration here are germ-line mutations passed on
from generation to generation  over 
stretches of evolutionary time.  Mutations may also
occur within cells of a single individual during
the individual's lifespan and be passed on and spread
through cell division.   These ``somatic mutations''
may also be important for aging.  Although the models
in \cite{SEW05} might turn out to have some relevance
to somatic mutation,  the focus in this work is
on germ-line mutations. 
\index{somatic mutations}

There is a rich literature on models for mutation,
selection, and recombination, described further in
Section~\ref{SS:literature}.   Our models build
on that tradition but differ from much of it in 
being specially designed to accommodate general 
forms of detailed demographic structure.   
In leading applications,  mutant alleles are
associated with representations of age-specific
effects on demographic schedules.  
One goal is to provide structure from which 
predictions of the shapes of curves of 
probabilities of death as functions of age can
be derived from assumptions about rates and
kinds of mutations.   This enterprise has come
to the fore with the discovery of common features
in the age-specific demographic schedules of
populations from a variety of species 
which seem to call out for evolutionary explanations
and which have stimulated the field of biodemography
(\cite{WF97} \cite{CT03}).
\index{biodemography}
  
An early precedent for the kinds of models we develop
here is found in the well-known work of 
Motoo Kimura and Takeo Maruyama \cite{KiMa66}.   
\index{Kimura, Motoo} \index{Maruyama, Takeo}
Our demographic formulations 
grow out of the work of Brian Charlesworth
\cite{bC94}, \cite{bC01}.  
\index{Charlesworth, Brian}


Recapitulating,   mutation accumulation
models that incorporate 
the countervailing forces of
recurring, slightly deleterious mutations and
persistent selection 
have been staples of evolutionary theory \cite{rB00}.
In order to use these models to
explain phenomena such as  aging that are presumed to
result from the combination of many small mutational
impacts, it is necessary to adopt a multilocus perspective.
However, as Pletcher and Curtsinger \cite{PC98} point out, early 
\index{Pletcher, Scott} \index{Curtsinger, James}
progress in this area relied on simplifying assumptions 
that are severely limiting and possibly unfounded:
equal impacts from all mutations,
additive effects of mutations on age-specific survival, 
and the existence of mutations that impact a specific, 
narrow range of ages. 

Several general multilocus formalisms have been proposed.
Notable among these is that of \cite{KJB02}, 
which is designed 
to allow almost any conceivable regime of selection, 
mating, linkage, mutation, and phenotypic effects.  
Such frameworks have not been exploited for studies of aging,
in part because so much detail is counterproductive when
what concerns us is not the fate of any individual allele 
but rather the mass of overlapping, age-varying phenotypic effects
that are central to standard theories of aging.

Kevin Dawson \cite{Daw99} applied a variant 
\index{Dawson, Kevin}
of the Kimura-Maruyama rare-allele 
approximation \cite{KiMa66,rB00} (see also \cite{Kon82})
to aging.   While this less detailed view of the genome is more
amenable to theoretical analysis, it is also not suited to 
describing the interacting phenotypic contribution of multiple loci.
\index{Kimura, Motoo} 
\index{Maruyama, Takeo} 
\index{B\"{u}rger, Reinhard}
In a previous paper \cite{SEW05}, we proposed a model,
which we also describe briefly here in the development
leading to \eqref{E:mutation_and_selection}, that
overcame many of these limitations.  
That model leads to computable solutions for 
mutation-selection equilibria, the  hazard functions 
which such equilibria imply, and
the time evolution of the population distribution
of genotypes.   
\index{hazard function}

In this work, we define and analyze a parallel model 
that, in essence, incorporates recombination
into the model of \cite{SEW05}.
The key assumptions behind the model in this work are
the following.
\begin{itemize}
\item
The population is infinite.
\item
The genome may consist of infinitely many 
or even a continuum of loci. 
\item
Reproduction is sexual, in that each individual has
two parents.
\item The mechanism of genetic recombination
randomly shuffles together the genomes of the parents
in order to obtain the genome of the offspring.
\index{recombination}
\item
Mating is random and individuals are identical except
for their genotypes, so the population at any time can be
completely described by a probability measure on
the space of possible genotypes --- heuristically,
the empirical distribution of genotypes
in the population or, equivalently,
the probability distribution of the
genotype of a randomly sampled individual.
\item
Individuals are {\em haploid} ---
an individual has only one copy of each gene
rather than copies from each of two parents.
\item
Down a lineage, mutant alleles only 
accumulate --- there is no back-mutation to cancel out
the alleles introduced by earlier mutation events.
\item 
Fitness is calculated for
individuals rather than for mating pairs.
\item
A genotype becomes less fit when it accumulates 
additional mutant alleles, but otherwise selective costs 
may be freely specified. 
\item
Recombination acts on a faster time scale than 
mutation or selection --- in other words, 
the common quasi-linkage equilibrium (QLE) assumption holds.
\end{itemize}
\index{quasi-linkage equilibrium}
\index{Barton, Nicolas}
\index{Turelli, Michael}
\index{selective cost}
\index{genotype}

Our general picture of the genome and the processes 
of mutation, selection and recombination is
similar to that of \cite{BT91,KJB02}.  
Whereas  \cite{BT91,KJB02} invoke the QLE assumption
to justify treating the effects of alleles at different 
loci as nearly independent, 
we present a detailed asymptotic treatment in a
standard, explicit scaling regime.  

More specifically, we establish conditions under which a 
discrete-time dynamical system with the above features
converges to a continuous-time, deterministic  dynamical
system that has as its state space the probability
measures on the set of possible genotypes.
At any point in time, the genotype distribution
is in the complete linkage equilibrium 
represented by the distribution of a Poisson random measure.
The points of a realization of the Poisson random measure 
represent loci at which a randomly sampled individual's
genotype has accumulated  ancestral mutations away from 
an original ``reference'' genotype which we call the 
{\em wild type}.
Because a Poisson random measure is
defined by its intensity measure, which in our case is a finite
measure on the set of loci, the asymptotic model can be 
more simply described
by a deterministic dynamical system that moves about in
the space of such measures.
\index{intensity measure}
\index{dynamical system}

In order to establish such a convergence result
it is, of course, not enough simply to show that the 
distribution of the genotype of a randomly chosen
individual at a fixed time converges to the distribution 
of a Poisson random measure.  Rather, we must keep track 
of the accumulated perturbations that arise over time from 
the effect of mutation, selection and recombination, and 
demonstrate the convergence of the entire time evolution 
of the genotype distribution to a dynamical system of 
Poisson probability measures.

Moreover, we establish conditions under which the 
continuous-time dynamical system has equilibria, 
investigate when the system converges to an equilibrium 
from the pure wild type genotype, 
and obtain results about the stability and attractivity
of that particular equilibrium when it is present.

Our models provide a basis for the rigorous study
of a number of questions in the biodemography 
of longevity, as surveyed in \cite{kW03}: 
\index{biodemography}
\index{longevity}
\index{Charlesworth, Brian}
\begin{itemize}
\item the adequacy  of Charlesworth's \cite{bC01} 
proposed explanation of Gompertz hazards and mortality plateaus 
as a consequence of mutation accumulation,
\item the causes of the hyperexponential hazards 
described by Horiuchi \cite{sH03},
\index{Horiuchi, Shiro}
\item extensions to age-structured settings of 
Haldane's principle that,
in its original form,
equates the mutation rate to population decline of fitness,
\index{Haldane's Principle}
\item contrasts between proportional and additive mutation effects on hazards
as discussed in \cite{aB05, aB08},
\item the possibility of mortality rates diverging to infinity at ages before the end of reproduction, which have been termed ``Walls of Death''.
\index{walls of death}
\end{itemize}
We pursue these
matters in companion papers \cite{walls, vitalrates}.

\section{Informal description of the limit model}
\label{SS:informal_desc}


In this section we describe the asymptotic model and 
explain by way of motivation 
why it is reasonable that such a dynamical system should
arise as the limit of a sequence of discrete-time systems 
with the features listed above.  We present a formal development
of the model in Sections~\ref{SS:wasserstein}
and \ref{SS:moddef}  of Chapter~\ref{Ch:model}.  
The exposition in \cite{CE09} was based on an early 
version of the present text.

Denote by $\M$ the collection of {\em loci}
\glossary{$\M$}
\index{genotype}
in the portion of the genome that is of interest to us.
There is a distinguished reference
wild type genotype, and each locus represents
a ``position'' at which the genotype of an individual
may differ from that of the wild genotype.
We allow the set $\M$ to be quite
general and  do not necessarily think of it as 
a finite collection of physical DNA base positions
or a finite collection of genes.
For example, the proposed explanation for the
Gompertz mortality curve and mortality plateaus
\index{Gompertz hazards}
at extreme ages in Charlesworth \cite{bC01} 
\index{Charlesworth, Brian}
suggest taking $\M$ to be a class of functions from
$\mathbb{R}_+$ to $\mathbb{R}_+$: the value of such a function
at age $x \ge 0$ represents an additional increment
to mortality at age $x$ conferred
by a mutation away from the wild type at this locus.
Some structure on $\M$ is necessary 
to accommodate rigorous probability theory, so we
take $\M$ to be a complete, separable metric space.

The genotype of an individual is specified by 
the set of loci at which there has been a mutation
somewhere along the ancestral lineage leading to that
individual.  More precisely, 
a genotype is an element of the
space $\G$ of
\glossary{$\G$}
\index{genotype}
integer--valued finite Borel measures on $\M$.
An element of $\G$ is a finite sum $\sum_i \delta_{m_{i}}$,
where $\delta_m$ is the unit point mass at
\glossary{$\delta_m$}
the locus $m \in \M$.  The measure
$\sum_i \delta_{m_{i}}$ corresponds to a genotype
that has ancestral mutations at loci $m_1, m_2, \ldots$.    
The wild type genotype is thus the null measure.  
We do not require that the loci $m_{i} \in \M$  be distinct.  
We thus allow several copies of a mutation.  This is
reasonable, since we do not identify mutations
with changes in nucleotide sequences in 
a one-to-one manner.  

For example,
if $\M$ is finite, so we might as well
take $\M = \{1,2,\ldots,N\}$ for some positive
integer $N$, then $\G$ is essentially
the Cartesian product $\bN_{0}^N$ of $N$ copies
of the nonnegative integers. 
\glossary{$\bN_0$}
A genotype is of the form $\sum_{j=1}^N n_j \delta_j$,
indicating that an ancestral mutation is 
present $n_j$ times at locus $j$, and 
we identify such a genotype with the
nonnegative integer vector $(n_1, n_2, \ldots, n_N)$.

Recall that the population 
is infinite and all that matters about
an individual is the individual's genotype, so that
the dynamics of the population are described by
the proportions of individuals with
genotypes that belong to the various subsets of $\G$.
We are thus led to consider a family
of probability measures $P_t$, $t \ge 0$, on $\G$, 
\glossary{$P_t$}
where $P_t(G)$ for some subset $G \subseteq \G$
represents the proportion of individuals
in the population at time $t$ that have genotypes
belonging to $G$.
Note that we may also think of $P_t(G)$  as the
probability that an individual chosen uniformly at random from
the population has a genotype belonging to the set $G$.
In other words, $P_t$ is the distribution of
a random finite integer-valued measure on $\M$.
For example, if $\M = \{1,2,\ldots,N\}$
and we identify $\G$ with the Cartesian product $\bN_{0}^N$ as above, then
$P_t(\{(n_1, n_2, \ldots, n_N)\})$ represents the
probability that an individual chosen uniformly at random
from the population has $n_j$ ancestral mutations
at locus $j$ for $j=1, 2,\ldots,N$.


We next indicate how we model mutation, selection
and recombination to obtain the
evolution dynamics for $P_t$, $t \ge 0$.

\bigskip\noindent
{\bf Mutation alone.}  
Suppose that there is only mutation
and no selection or recombination.  In this case
all individuals present in the population at a 
given time die and reproduce at the same
rate, because differing genotypes do
not confer differing selective costs.
Mutations accumulate down lineages because
they cannot be replaced by recombination. 
\index{selective cost}

We describe the mutation process using a finite 
measure $\nu$ on the space of loci $\M$, where $\nu(B)$ 
for $B \subseteq \M$ gives the rate at which mutations
from the ancestral wild type belonging to the set $B$
accumulate along a given lineage.
\glossary{$\nu$}

Write $P_t [\Phi] = \int_{\G} \Phi(g) \,dP_t(g)$ for some
\glossary{$\Phi$}
\glossary{$P_t [\Phi]$}
\glossary{$\bR$}
bounded Borel test function $\Phi: \G \rightarrow \bR$. 
That is, $P_t [\Phi]$ is the expected value of the 
real-valued random variable obtained by applying 
the function $\Phi$ to the genotype of an individual 
chosen uniformly at random from the population.
The content of our assumptions when
there is only mutation is that $P_t$, $t \ge 0$,
evolves according to
\begin{equation} 
\label{E:mutation_only}
   \frac{d}{dt}P_{t} [\Phi]  
      = P_{t} \left[ \int_\M \left( \Phi(\cdot+\delta_{m}) -
      \Phi(\cdot) \right) \, d\nu(m) \right].
\end{equation}

For example, when $\M = \{1,2,\ldots,N\}$ we have
the system of ordinary differential equations
\begin{equation*}
\frac{d}{dt} P_t(\{\mathbf{n}\}) \\
=
\sum_{j=1}^N
\nu(\{j\}) 
\left[
P_t(\{\mathbf{n} - \mathbf{e_j}\}) {-} P_t(\{\mathbf{n}\})
\right],
\end{equation*}
where $\mathbf{e_j}$ 
is the $j^{\mathrm{th}}$ coordinate vector.  This equation
is, of course, just a special
case of the usual equation (see, for example, Section III.1.2
of \cite{rB00}) describing evolution due to mutation
of type frequencies in a population where the set
of types is $\mathbb{N}^N$ and mutation from
type $\mathbf{n}$ to type 
$\mathbf{n} + \mathbf{e_j}$ occurs at
rate $\nu(\{j\})$.
\index{B\"{u}rger, Reinhard}

The evolution equation \eqref{E:mutation_only}
is of the form
\[
\frac{d}{dt} P_t = A P_t
\]
for a certain linear operator $A$.  We recognize that $A$ is 
the infinitesimal generator of a $\G$-valued
L\'{e}vy process, and hence \eqref{E:mutation_only}
has the following explicit probabilistic solution:
\index{L\'{e}vy process}

Let $\tilde{\Pi}$ denote a Poisson random 
measure on $\M \times \bR_+$
with intensity measure $\nu \otimes \tilde{\lambda}$, 
where $\otimes$ denotes the product of measures 
and $\tilde{\lambda}$ is Lebesgue measure. 
\glossary{$\tilde{\Pi}$}
\glossary{$\tilde{\lambda}$}
\glossary{$\otimes$}  
That is, $\tilde{\Pi}$ is a random integer-valued 
Borel measure such that:
\begin{itemize}
\item[(1)] 
The nonnegative integer-valued random variable 
$\tilde{\Pi}(A)$ is Poisson with expectation
$(\nu \otimes \tilde{\lambda})(A)$ 
for any Borel subset $A$ of $\M \times \bR_+$.
\item[(2)] 
If $A_1, A_2, \ldots, A_n$ are disjoint Borel subsets of 
$\M \times \bR_+$, then the random variables $\tilde{\Pi}(A_k)$ 
are independent.
\end{itemize}
Define a $\G$-valued random variable $Z_t$ 
(that is, $Z_t$ is a random finite
integer-valued measure on $\M$) by 
\glossary{$Z_t$}
\begin{equation}\label{E:ZtPi}
Z_t := \int_{ \M \times [0,t]} \delta_m \, d\tilde{\Pi}((m,u)).
\end{equation}
Then,
\begin{equation*}
    P_t [\Phi]  = {\bE} \left[\Phi(W+Z_t)\right],
\end{equation*}
where $W$ is a random measure on $\M$
that has probability measure $P_{0}$
and is independent of $\tilde{\Pi}$.
In particular, if $P_0$ is itself the distribution
of a Poisson random measure, then $P_t$ is
also the distribution of a Poisson random measure.
If we write $\rho_t$ for the intensity measure 
associated with $P_t$
\glossary{$\rho_t$}
\index{intensity measure}
(that is, $\rho_t$ is the measure on $\M$ 
defined by $\rho_t(A) := \int_\G g(A) \, dP_t(g)$
for $A \subseteq \M$), then $\rho_t$ evolves according to
the simple dynamics
\begin{equation*}
\rho_t(A) = \rho_0(A) + t \, \nu(A).
\end{equation*}

\bigskip\noindent
{\bf Selection alone.} Now suppose there
is only selection and no mutation or recombination.
We specify the fitnesses of different genotypes by a
{\em selective cost function} $S : \G \to \bR_+$.  
\index{selective cost}
\glossary{$S(g)$}
The difference $S(g') - S(g'')$ for $g', g'' \in \G$ is
the difference in the rate of sub-population growth
between the sub-population of individuals with genotype $g''$
and the sub-population of individuals with genotype $g'$.
We make the normalizing assumption
$S(0) = 0$ and suppose that 
\begin{equation}
\label{monotone_cost}
S(g+h)\geq S(h), \quad g,h \in \G,
\end{equation} 
in line with our assumption above that  genotypes with more
accumulated mutations are less fit.

It follows that at time $t \ge 0$ the per individual rate 
of increase of the proportion of the population
of individuals with genotype $g'$
is $P_t [S] - S(g')$.  
More formally, 
\begin{equation}
\label{E:selection_only}
\begin{split} 
   \frac{d}{dt}P_{t}[\Phi]   
      & = - \, P_{t} [\Phi \cdot (S -P_{t}S)]   \\
      & = - \int_\G \Phi(g')
      \left( S(g') - \int_\G S(g'') \, dP_t(g'')\right) \, dP_t(g'). \\
\end{split}
\end{equation}

For example, when $\M = \{1,2,\ldots,N\}$ we have
\begin{equation*}
\frac{d}{dt} P_t(\{\mathbf{n'}\})
=
-\left[ 
S(\mathbf{n'})
-
\sum_{\mathbf{n''}} P_t(\{\mathbf{n''}\}) S(\mathbf{n''})
\right]
P_t(\{\mathbf{n'}\}).
\end{equation*}

If $S$ is {\em non-epistatic}, that is,
if $S$ has the additive property 
\[
S\left(\sum_i \delta_{m_{i}}\right) = \sum_i S(\delta_{m_{i}}),
\]
then the selective effects of different mutations
do not interact.  In particular, if $P_0$ is the
distribution of a Poisson random measure, then
$P_t$ is also the distribution of a Poisson
random measure and, writing $\rho_t$ for the
intensity measure associated with $P_t$ as before, we have
\begin{equation*}
\rho_t(dm') = \rho_0(dm')
- \int_0^t
\left[S(\delta_{m'}) - \int_{\M} S(\delta_{m''}) \rho_s(dm'')\right] \,
\rho_s(dm') \, ds. 
\end{equation*}
However, in the more general case in which $S$ is epistatic 
(that is, non-additive),
\index{epistatic effects}
then $P_t$ is, in general, not the distribution
of a Poisson random measure --- even when $P_0$ is.

\bigskip\noindent
{\bf Combining mutation and selection.}
If there is mutation and selection, but no recombination,
then the appropriate evolution equation for $P_t$
comes from simply combining equations \eqref{E:mutation_only}
and \eqref{E:selection_only}:
\begin{equation}
\label{E:mutation_and_selection}
\begin{split}
\frac{d}{dt}P_{t}[\Phi] 
& = P_{t} \left[ \int_\M \left( \Phi(\cdot+\delta_{m}) -
      \Phi(\cdot) \right) \, d\nu(m) \right] \\
& \quad - \, P_{t}[\Phi \cdot (S -P_{t}S)] . \\
\end{split}
\end{equation}
This is the model introduced and analyzed at length
in \cite{SEW05} using the Feynman-Kac formula.
When $\M = \{1,2,\ldots,N\}$, 
\eqref{E:mutation_and_selection} is a special
case of the classical system of ordinary differential 
equations for mutation and selection in continuous-time.  
See Section III.1.2 of \cite{rB00}
for the derivation of an analytic solution that agrees with the
one that arises from a Feynman-Kac analysis.
\index{B\"{u}rger, Reinhard}

The principal result from \cite{SEW05} is the following:

\begin{proposition}  \label{P:epistatic}
Let $Z_t$ be the random measure distributed according to the
Poisson probability measure $\tilde{\Pi}$ in  \eqref{E:ZtPi}.
Suppose that there is a positive $T$ such that
\begin{equation} \label{E:defTep}
\mathbb{E} \left[\exp \left( - \int_{0}^{t} S(Z_{u})du
     \right) S(Z_{t}) \right]< \infty
 \end{equation}
for all $t\in [0,T)$. Then,
 \eqref{E:mutation_and_selection} 
has a solution on $[0,T)$, given by
\begin{equation}  \label{E:epissol}
    P_t [\Phi]  = \frac
   { \mathbb{E} \left[\exp\left( - \int_0^t S(Z_u) \, du\right)
    \Phi(Z_t)\right] }
   { \mathbb{E} \left[\exp\left( - \int_0^t S(Z_u) \, du\right)
      \right]}.
\end{equation}
\end{proposition}

Notice that the effect of selection may be understood as reweighting
a Poisson measure, rather than explicitly removing mutations from the population.
This will be an important theme in the convergence proof that starts in
Chapter \ref{Ch:discrete_time}.

The model without recombination in \cite{SEW05},
like the model presented here,  is an infinite-population model.
Alleles are not subject to genetic drift.  
The fittest genotypes present in the initial population
are not lost,  no matter how rare they are.
\index{recombination}
A finite-population model without recombination defined along
the same lines would be vulnerable to Mueller's ratchet,
the process in which the fittest classes
of genotypes in an asexual population can be successively
lost through drift,  carrying the population to extinction.
A thorough discussion of the ratchet and the associated
advantages of sex and recombination with references  
is found in \cite{rB00}, pages 303--308. 
In the face of Mueller's ratchet, a finite-population 
version of the model in \cite{SEW05} would not be
viable.    This observation underscores the importance
of incorporating recombination,  which avoids the ratchet
by allowing fittest classes to be reconstituted 
in every generation.
\index{Mueller's ratchet}
\index{B\"{u}rger, Reinhard}

\bigskip\noindent
{\bf Recombination alone.} 
\index{recombination}
The effect of recombination is to choose an individual
uniformly at random from the population at some rate and replace the
individual's genotype $g'$ with a genotype of the form
$g'(\cdot \cap M) + g''(\cdot \cap M^\complement)$, where $g''$ is the
genotype of another randomly chosen individual, $M$ is a subset
of $\M$ chosen according to a suitable random mechanism,
and $M^\complement$ is the complement of $M$.  Thus, recombination
randomly shuffles together two different genotypes drawn
from the population.   In order to specify the recombination mechanism
fully, we would need to specify the recombination rate
and the distribution of the {\em segregating set} $M$.  
Suppose, however, that we are in the following regime.
\index{segregating set}
\begin{itemize}
\item
Recombination acts alone; that is, there is no mutation or selection. 
\item
The initial population $P_0$ has the property that
there does not exist 
an $m \in \M$ with $P_0(\{g \in \G : g(\{m\}) > 0\}) > 0$; 
that is, no single mutation from the ancestral wild type is possessed
by a positive proportion of the initial population (but see
Remark~\ref{R:clumping}  and Section~\ref{SS:atomstart} below). 
\item
The mechanism for choosing the 
{\em segregating} set $M$
is such that, loosely speaking, 
if $m'$ and $m''$ are
two loci, then there is positive probability that
$m' \in M$ and $m'' \in M^\complement$; 
that is, no region of the genome $\M$ is immune 
from the shuffling effect of recombination.
\end{itemize}
\index{recombination}
%
Then, under these conditions,  the probability 
measure $P_t$ converges as $t \rightarrow \infty$ 
to the distribution of a Poisson random measure on
$\M$ with the same intensity measure as $P_0$.  
Moreover, the speed of this convergence increases with the 
recombination rate and so if we take 
the recombination rate to be effectively infinite,
then the probability measure $P_t$ is essentially Poisson for all $t>0$ with
the same intensity measure as $P_0$, irrespective of the details
of the recombination mechanism.
\index{recombination}

\bigskip\noindent
{\bf Combining mutation, selection and recombination.}
\index{recombination}
We have seen that if $P_0$ is the distribution of a Poisson
random measure, then mutation preserves this property.
On the other hand, 
epistatic selection drives the population distribution 
away from Poisson, while increasing rates of recombination 
push it towards Poisson.  Thus, when all three processes operate 
and we consider a limiting regime where recombination acts 
on a much faster time scale than selection and recombination, 
we expect asymptotically that  $P_t$ is also the distribution 
of a Poisson random measure for all $t>0$.   
As before, $\rho_t$ denotes the intensity measure of $P_t$
(so that $\rho_t$ is a finite measure on the space $\M$ of loci).  
\index{intensity measure}
\index{epistatic effects}
In anticipation of Notation~\ref{N:spaces}, 
we write $X^\pi$ for a Poisson random measure 
on $\M$ with intensity measure $\pi$.  
Combining our previous observations, we expect 
that $\rho_t$ should satisfy the evolution equation
\begin{equation} \label{E:model_informal}
\rho_t(dm) = \rho_0(dm) + t\, \nu(dm) - 
\int_0^t  
\bE\left[S(X^{\rho_s} + \delta_m) - S(X^{\rho_s})\right] 
\, \rho_s(dm) \, ds.
\end{equation} 

We define the rigorous counterpart of 
\eqref{E:model_informal}
in Chapter~\ref{Ch:model} and 
establish the existence and uniqueness
of solutions.  Furthermore,
we show in Chapters~\ref{Ch:discrete_time} ff. that our
dynamical equation is indeed a limit of
a sequence of standard discrete generation,
mutation-selection-recombination models.

Given that it involves computing an expected value
for a quite general Poisson random measure, 
\eqref{E:model_informal} may look rather forbidding.  
However, for certain reasonable choices of selective costs 
we can evaluate the integral explicitly, 
leading to a simpler and more intuitive
system.  We consider three such cases in the
following three sections.  First, though, we make
a useful observation about how we may clump loci 
together in \eqref{E:model_informal}.

\begin{remark}
\label{R:clumping}
Consider two instances 
$(\rho_t')_{t \ge 0}$ and $(\rho_t'')_{t \ge 0}$
of the dynamical system \eqref{E:model_informal}
with respective
locus spaces $\M'$ and $\M''$,
associated genotype spaces $\G'$ and $\G''$,
mutation intensity measures $\nu'$ and $\nu''$,
and selective costs $S'$ and $S''$.
Suppose that there is a Borel measurable map
$T$ from  $\M'$ onto $\M''$ such that the following hold.
\index{dynamical system}
\begin{itemize}
\item
The initial measure $\rho''_{0}$ on $\M''$
is the push-forward of the initial measure $\rho'_{0}$ 
on $\M'$ by the map $T$.
\item
The mutation intensity measure $\nu''$ on $\M''$
is the push-forward of the mutation intensity measure $\nu'$ 
on $\M'$ by the map $T$.
\item
The selective cost $S'$ on $\G'$ has the property that
$S'(g') = S'(h')$ whenever the push-forwards of $g', h' \in \G'$
by $T$ are the same.
\item
The selective cost $S''(g'')$ on $\G''$ is given
by the common value of $S'(g')$ for all
$g' \in \G'$ that have push-forward by $T$ equal to $g''$.
\end{itemize}
Then, for each $t > 0$, the measure $\rho_t''$ on $\M''$
is the push-forward of the measure $\rho_t'$ on $\M'$
by the map $T$.  Intuitively, we have a situation in which 
for a given 
$g'' \in \G''$ any two genotypes
in the set $T^{-1}(g'') \subseteq \G'$ are indistinguishable
in terms of their associated selective cost, and so
we may identify any two such genotypes as being the same.
Of course, we cannot recover the finer
description $\rho_t'$ from the coarser one $\rho_t''$
in general, but if we write $P_t'$ and $P_t''$ for the
population distributions of genotypes corresponding
to $\rho_t'$ and $\rho_t''$ (that is,
$P_t'$ and $P_t''$ are the distributions of Poisson random
measures on $\M'$ and $\M''$ with intensity measures
$\rho_t'$ and $\rho_t''$), then the push-forward of
$P_t'$ by $S'$ is the same as the push-forward of
$P_t''$ by $S''$.  That is, the population distributions of
selective costs agree whether we use the
fine or the coarse description of genotypes.
\end{remark}

\section{Example I: Mutation counting}
\label{SS:counting}

The simplest special case of our framework
occurs when there are many loci but the selective
cost of a genotype $g$ only depends on the total number
of loci $g(\M)$ at which there have been 
ancestral mutations away from the wild type.
\index{mutation counting}
\index{selective cost}

For example, suppose that the space $\M$ of loci is the
unit interval $[0,1]$ and the selective cost $S$ is of the
form $S(g) = s(g(\M))$ for some non-decreasing 
function $s: \mathbb{N}_0 \rightarrow \mathbb{R}_+$ with $s(0) = 0$.
\glossary{$\mathbb{N}_0$}

We may apply Remark~\ref{R:clumping} and ``replace''
the locus space $\M$ by a single point.
Let $q := \nu(\M)$
be the total rate at which mutations occur and write
$r_t := \rho_t(\M)$ for the expected number of ancestral 
mutations in the genotype of an individual chosen
at random from the population at time $t \ge 0$.  
It follows from Remark~\ref{R:clumping}
that the function $r$ evolves 
autonomously according to the (non-linear)
ordinary differential equation
\[
\dot r = q - \psi(r),
\]
where
\[
\begin{split}
\psi(x) 
& := x 
\sum_{k=0}^\infty (s(k+1) - s(k)) e^{-x} \frac{x^k}{k!} \\
& = 
e^{-x} \sum_{k=1}^\infty s(k) \frac{x^k}{k!} (k - x). \\
\end{split}
\]

One example of this simplified model, which lines up with models
familiar from earlier literature, 
assumes the cost per mutation to be a constant $\bar s$,
so that $s(k) = \bar s k$ and $\psi(x) = \bar s x$, 
and assumes the starting point to be the null genotype. In this case,
we readily compute that the intensity at time $t$ is
$  r_t = (q / \bar s)\bigl( 1 - \exp\{ -\bar s t \} \bigr)$.
The intensity converges monotonically to the equilibrium
value $ q / \bar s $,  the elementary expression 
for mutation-selection equilibrium going back to J.B.S. Haldane.  
\index{Haldane, John B. S.}
Thus, the limiting distribution for the number of mutations 
from the ancestral wild type is  Poisson  with mean $q / \bar s$.  
Our assumption about recombination leads to 
different,  simpler answers than Kimura and Maruyama's 
treatment of mutation counting without recombination (see \cite{KiMa66}).
\index{equilibrium}

For another example, consider the case 
when costs are multiplicative, in the sense that
$ s(k) := 1 - \exp\{-k \theta\} $ for some constant $\theta > 0 $.
\index{multiplicative costs}
We have $\psi(x) = a x \exp\{ -a x \} $,
where $ a := 1 - \exp\{-\theta\} $.  An equilibrium exists
only if the mutation rate $q$ is below the maximum
of $  r \exp\{-r\} $, namely $ 1/e $,  and in that case $r_t$
converges monotonically to the smallest positive solution
of the equation $q - \psi(x) = 0$.  
The solution 
%
%
is $x = (1/a)  \sum_{n=1}^\infty (-n)^{n-1} \, (-q)^n \, / \, n!$ 
where the summed expression 
is Lambert's W function evaluated at the negative argument $-q$. 
\index{Lambert's W function}
The power series 
expansion is a simple application of the technique known variously
as reversion of series or the Lagrange inversion formula. 
\index{Lagrange inversion formula}
These properties for multiplicative costs are generalized in 
Section~\ref{SS:multiplicative_equilibrium}.  Note also that
when $\theta$ is small the equilibrium is
approximately $q / \theta$, as one would expect from the
observation that in this case $s(k)$ is approximately $\theta k$
for small $k$ and hence this model is approximately the additive
one of the previous paragraph with $\bar s = \theta$.
\index{equilibrium}

If we are interested only in how the population distribution of selective
costs evolves, then we need consider only $(r_t)_{t \ge 0}$, rather
than $(\rho_t)_{t \ge 0}$.  However, we should be somewhat careful
about how we interpret the biological import of this simplification.
We justified the dynamical system \eqref{E:model_informal} as describing
the evolving population distribution of genotypes defined in terms
of the locus space $\M$ in a population
undergoing mutation, selection and recombination.  
\index{dynamical system}
Mathematically,
we see that in this example we can replace the locus space
$[0,1]$ by a single point for the purposes of studying the dynamics
of the distribution of selective costs, but this does not mean
that biologically the multilocus model is identical with a single
locus model.  As we show later, our instance of \eqref{E:model_informal}
with locus space $[0,1]$ arises as a limit of discrete generation models
in which Poisson random measures appear because of the manner
in which recombination breaks up and shuffles together 
genotypes from different individuals.  
Even though our instance of \eqref{E:model_informal}
with locus space a single point is mathematically well-defined, 
it cannot  arise as a limit of such discrete generation models 
because with a single locus there is no way that recombination 
can drive genotype distributions towards Poisson.
Single-point mutation spaces require the special treatment
described in Remark~\ref{R:clumping}.

\section{Example II: Polynomial selective costs}
\label{SS:poly_cost}
\index{selective cost}
\index{polynomial costs}

Recall that when $\M = \{1,2,\ldots,N\}$ we
can encode genotypes as ordered $N-$tuples of nonnegative integers,
where the entry in the $k^{\mathrm{th}}$ coordinate is
the number of ancestral mutations present at locus $k$.
Then, \eqref{E:model_informal} becomes the system of 
ordinary differential equations
\begin{equation}
\label{model_informal_finite}
\frac{d}{dt} \rho_t(\{j\})
=
\nu(\{j\})
-
\rho_t(\{j\}) \sum_{\mathbf{n}} 
\left[
S(\mathbf{n}+\mathbf{e_j}) - S(\mathbf{n})
\right]
\prod_{k=1}^N e^{-\rho_t(\{k\})} \frac{\rho_t(\{k\})^{n_k}}{n_k!}.
\end{equation}

As in the previous example, we should not think of
$\{1,2,\ldots,N\}$ as being the ``real'' locus space.
Rather, we should imagine that there is something like
a continuum of loci which are partitioned into
$N$ sub-classes in such a way that the loci in each
sub-class have indistinguishable selective effects,
and \eqref{model_informal_finite} is a reduced
description that comes from applying the observation
in Remark~\ref{R:clumping}.

A natural family of selective costs 
is given by those of the polynomial form 
\index{polynomial costs}
\begin{equation*}
S(g)=\sum_{I} \alpha_I g^I, 
\end{equation*}
where the sum is taken over all nonempty 
subsets $I \subseteq \{1,\ldots,N\}$
and we adopt the usual multi-index convention that for a vector 
$v$ the notation $v^I$ denotes the product $\prod_{i \in I} v_i$.
\glossary{$v^I$}
The constant $\alpha_{\{i\}}$ for $1 \le i \le N$ measures the selective
cost of mutation $i$ alone, whereas the constant $\alpha_I$ for 
a subset $I \subseteq \{1,\ldots,N\}$ 
of cardinality greater than one measures the 
selective cost attributable to interactions between all of the mutations
in $I$ over and above that attributable to interactions between
mutations in proper subsets of $I$.

The system of ordinary differential 
equations \eqref{model_informal_finite} becomes
\begin{equation}
\label{finitesysdiff}
\dot{\rho_k} = \nu_k - \sum_{I \in \mathcal{I}_k}  \, 
        \alpha_I \rho^I, \quad 1 \le k \le N,
\end{equation}
where we write $\rho_k := \rho(\{k\})$ and $\nu_k := \nu (\{k\})$,
and where $\mathcal{I}_k$ denotes the collection of 
subsets of $\{1,\ldots,N\}$ that contain $k$, see \cite{CE09}.

It is shown in \cite{CE09} that if $\nu_k > 0$ for all $k$ i
(that is, if mutations may occur at all loci), 
if $\alpha_{\{i\}} > 0$ for all $i$ 
(that is, if the individual effect of any mutation is deleterious, 
in keeping with our general assumption on the selective cost), 
and if  $\alpha_I \ge 0$ for all subsets $I$ 
(that is, the synergistic effects of individually deleterious 
mutations are never beneficial), 
then the system of equations \eqref{finitesysdiff} has a 
unique equilibrium point in the positive orthant $\bR_+^N$.  
Moreover, this equilibrium is globally attractive; 
that is, the system converges
to the equilibrium from any initial conditions in $\bR_+^N$.  The condition
$\alpha_I \ge 0$ for all $I$ certainly implies our standing assumption
that $S(g+h) \ge S(g)$ for all $g,h \in \G$, but it is strictly stronger.
It is also shown in \cite{CE09} that the analogue of this result for general
$\M$ holds with a suitable definition of polynomial selective costs
in terms of sums of integrals against products of the measure $g$
with itself.
\index{attractive equilibrium}
\index{equilibrium}

\section{Example III: Demographic selective costs}
\label{SS:mortality}
\index{selective cost}

The following model is discussed in \cite{walls, vitalrates}, 
where there is a more extensive discussion of the demographic assumptions.
\index{demographic costs}

For the moment, suppose that the space of loci $\M$ is general.
Write $\ell_x(g)$ for the  probability that an individual
\glossary{$\ell_x(g)$}
with genotype $g \in \G$ lives beyond age $x \in \mathbb{R}_+$.
At age $x$, the corresponding {\em cumulative hazard} 
is  $-\log \ell_x(g)$  and the {\em hazard function} is 
its derivative with respect to $x$ when it exists. 
\index{cumulative hazard} 
\index{hazard function}.  
Suppose that the infinitesimal rate that
an individual at age $x \in \mathbb{R}_+$ has offspring  is
$f_x$, independently of the individual's genotype.
\glossary{$f_x$}
For individuals with genotype $g$, the 
size of the next generation relative to the current one is
$\int_0^\infty f_x \ell_x(g) \, dx$, the ``Net Reproduction
Ratio'' or ``NRR''.
\index{hazard function}
\index{cumulative hazard}
\index{net reproduction ratio}

Suppose further that there is a background hazard $\lambda$
and that an ancestral mutation at locus $m \in \M$
contributes a bounded increment $\theta (m,x)$ to the
cumulative hazard function at age $x$.  This function $\theta$
with two arguments is a generalization of the constant 
cost $\theta$ in Section~\ref{SS:counting} 
and of a single-argument cost $\theta(m)$ in play in examples
with multiplicative costs in Section~\ref{SS:multiplicative_equilibrium}. 
\glossary{$\theta(m,x)$}
\glossary{$\lambda$}
The probability that an individual with 
genotype $g \in \G$ lives beyond age $x \in \mathbb{R}_+$ is
\begin{equation}\label{E:ellxofg}
\ell_x(g) =
     \exp \left( -\lambda x - \int_{\M} \theta(m ,x) \, dg(m) \right).
\end{equation}
The corresponding selective cost is
\begin{equation}\label{E:demogcosts}
S(g) =   \int_0^{\infty} f_x \, \ell_x(0) \, dx 
       - \int_0^{\infty} f_x \, \ell_x(g) \, dx
\end{equation}
(Recall that selective costs represent relative rates of increase, and
we have adopted the normalizing convention that $S(0) = 0$).

Marginal selective costs are given by the expression
$$
S(g + \delta_m ) - S(g) =  \int_0^{\infty} \left ( 1 - e^{-\theta(m,x)} \right)
                        f_x \, \ell_x(g) \, dx.
$$
Expected marginal costs have the expected value of $\ell_x(g)$ under
the integral on the right-hand side.  When the genotype $g$ 
is a realization of a Poisson random measure $X^{\pi}$ whose intensity 
is the finite measure $\pi$ on $\M$, this expected survivorship
function can be found from the Poisson identity
\index{Poisson identity}
\begin{equation}
\label{E:Pidentity}
\bE\left[   \exp\left\{ - \int_{\M} \phi(m) \, dX^{\pi}(m)\right\}  \right] 
          =  \exp\left\{ - \int_{\M} ( 1 - e^{-\phi(m)}) \, d\pi(m)\right\}. 
\end{equation}
The identity applies to bounded Borel functions $ \phi : \M \to \bR $.  
It is proved, for instance, in \cite[3.15]{Kin93}.  It also
follows from Campbell's Theorem (see Proposition~\ref{P:Campbell}).
\index{Campbell's Theorem}

The identity gives the expression, in terms of the expectation
operator $\mathbb{E}$, 
$$
 \mathbb{E}\left[ \ell_x(X^\pi) \right]  
       =  \exp \left\{ - \lambda x   
      - \int_{\M} \left( 1- e^{-\theta(m, x)}\right) \,  d\pi(m)  \right\}. 
$$
\glossary{$\mathbb{E}$}

%
An important issue for demographic applications  is whether 
the solution $\rho_t$, $t \ge 0$, of \eqref{E:model_informal} 
converges to an equilibrium $\rho_{*}$ as $t \rightarrow \infty$ 
and, if so, what are the features of that equilibrium.
\glossary{$\rho_{*}$}
In particular, what we can say about $\mathbb{E}[\ell_x(X^{\rho_{*}})]$, 
the probability that a randomly chosen individual 
from the equilibrium population lives beyond age $x$?
It is not hard to show that if the limit $\rho_{*}$ exists, then
it must be absolutely continuous with respect to the mutation rate  
measure $\nu$ and have a Radon-Nikodym derivative $r_{*}$ that satisfies
\glossary{$r_{*}$}
\index{equilibrium}
$$
1 = r_{*}(m) \, \int_0^{\infty} \, ( 1 - e^{-\theta(m,x)} )
       \,  f_x \, \mathbb{E} [  \ell_x(X^{\rho_{*}}) ] \,  dx.
$$

We study such equilibria further 
in Section~\ref{SS:equilibrium_demographic}  
and in \cite{walls, vitalrates}.  
In particular, in \cite{walls}, we consider
mutations which each provide a point mass increment to 
the hazard at a specific age.
When $\M = \mathbb{R}_+$ and
the mutation rate measure $\nu$ is absolutely continuous 
with respect to Lebesgue measure,
and $\theta$ is of the form $\theta(m,x) = \eta(m) \indic_{\{x \ge m\}}$, 
(where, as before, $\indic_A $ is the indicator function of the set $A$)
the equilibrium equation turns
out to be equivalent to a second-order, non-linear, 
ordinary differential equation
in one variable that can be solved explicitly for $r_{*}$.

\section{Comments on the literature}
\label{SS:literature}

We make some brief remarks about the substantial literature 
on multilocus deterministic models in population genetics
and its relation to our work.

A very comprehensive reference 
is Reinhard B\"{u}rger's book \cite{rB00} along 
with B\"{u}rger's review paper \cite{Bur98}).  
\index{B\"{u}rger, Reinhard}
As well as giving an overview of the classical models 
for finitely many alleles at each of a finite number of loci, 
these works consider at length deterministic 
haploid {\em continuum-of-alleles} models in which 
individuals have a {\em type} that is envisioned as the 
contribution of a gene to a given quantitative trait.  
The type belongs to a general state space that represents 
something like the trait value (in which case the state 
space is a subset of $\bR$) and is often regarded 
as the combined effect of a multilocus genotype. 
Each type has an associated fitness, which is some fairly 
arbitrary function from the type space to $(0,1]$.  
However, the models do not explicitly incorporate a 
family of loci, the configuration of alleles
present at those loci, or a function describing the fitness 
of a configuration. Rather, everything
is cast in terms of how fit each type is and how likely one 
type is to mutate into another.

Certain classes of mutation-selection models 
without recombination are solved explicitly
in \cite{MR1657856, BaWa01} using ideas from statistical mechanics.  
Such models may be treated either as multilocus systems 
with complete linkage or as structured single locus systems.
We also mention the constellation of papers 
\cite{MR1952324,MR2191729, MR2297255, MR1842838} presenting a deterministic
model of population change due to recombination alone.
In our setting, the Poisson probability measures 
constitute, very roughly speaking, a hypersurface
within the space of all probability measures
and recombination can be viewed as an operation that pushes
an arbitrary probability measure ``towards'' that hypersurface.  
General results about the convergence of discrete-time
dynamical systems to a continuous-time one that is forced
onto a submanifold by a suitable vector field may be relevant here,
but we are not aware of a particular framework of this
kind into which we can fit our results.

Finally, we make two comments about our model in order to
distinguish it from others in the literature that 
superficially might seem to have similar features.
\begin{itemize}
\item
We do {\bf not} present
a Markovian stochastic model such as the one in 
\cite{SaHa92}, 
where the population is described
in terms of an evolving Poisson random measure that keeps
track of the proportion of alleles at each site that are mutant
(with a typical site being purely wild type and only 
exceptional sites having a positive proportion of mutants present).
For us, the evolution of the proportions of different genotypes in
the population is described by a deterministic dynamical
system living on a space of probability measures.
If we sample from the evolving probability measure at some
fixed time,  then
the resulting individual's genotype is a Poisson random measure.  
\item
Linkage arises in our discrete-time approximation models
as the dependence between loci, a natural consequence 
of non-additive selection costs, is only partially 
broken up in any finite number of rounds of recombination.
However, linkage does not appear in the limit model.  
That is, if we sample from the limit population at some time,
then the fact that the resulting individual's genotype is described
by a Poisson random measure means that the presence
of ancestral mutations from wild type in one part of the genotype
is independent of the presence of mutations in another part.  
Our convergence theorem in Chapter \ref{Ch:convergence} 
therefore delineates the
relative strengths of mutation, selection and recombination
that lead asymptotically to a situation in which
the Poissonizing effect of recombination wins out over
the interactions introduced by non-additive selection.
Understanding how these two forces counteract each other
is far from trivial and is the most demanding technical 
task of the present work.
We cannot stress too strongly that we have not {\em a priori} 
assumed that linkage is absent.
\end{itemize}

\section{Overview of the remainder of the work}

We define the measure-valued dynamical system 
\eqref{E:model_informal} rigorously in 
Chapter~\ref{Ch:model} and 
establish the existence and uniqueness of
solutions.

We investigate in Chapter~\ref{Ch:equilibria}
whether the dynamical system has equilibria and whether these
equilibria are stable and attractive.

We devote Chapters \ref{Ch:discrete_time} to \ref{Ch:convergence}
to showing that the dynamical
system is a limit of discrete-generation,
infinite-population models.  We define
the discrete-generation models in
Chapter~\ref{Ch:discrete_time} and we preview the 
convergence theorem and its hypotheses 
in Chapter~\ref{Ch:hypotheses}.  
As a first step towards the proof of
the convergence result, we show in
Chapter~\ref{Ch:complete_Poisson} that an analogous
convergence result holds when the recombination
mechanism is replaced by a complete Poissonization
operation that destroys all linkage between
loci in a single step.  The proof of the
actual convergence result is quite involved and
requires a number of technical estimates that,
loosely speaking, bound the extent to which
selection reintroduces linkage that has been
partially removed by recombination.  We present
these preliminary results in Chapter~\ref{Ch:techlem}.
We state the convergence theorem and complete its proof 
in Chapter~\ref{Ch:convergence}.  

The appendix contains relevant versions of results 
from the literature 
as well as technical results about Poisson
random measures and Radon-Nikodym derivatives that are used
throughout the main text.

\chapter{Definition, existence, and uniqueness of the dynamical system}
\label{Ch:model}

\section{Spaces of measures}
\label{SS:wasserstein}

Our model, as presented in the Introduction,
has four fundamental ingredients: 
\begin{itemize}
\item
a complete, separable metric space $\M$ of loci;
\item
a finite Borel measure $\nu$ on $\M$
called the mutation measure because it describes the 
rate at which mutations occur in regions of the genome;    
\item
the space $\G$ of
integer--valued finite Borel measures on $\M$; 
\item
a selective cost function $S: \G \to \bR_+$ 
with $S(0) = 0 $  and $S(g+h)\geq S(h)$ for $g,h \in \G$.   
\end{itemize}
\index{selective cost}
\index{genotype}

An element of $\G$ represents a ``genotype''
regarded as the set of loci at which
there have been ancestral mutations away
from the reference wild type.  The
null measure represents the wild genotype.

Recall also that the state of the population at
time $t \ge 0$ in our model is a probability 
measure $P_t$ on $\G$ that is the distribution of
a Poisson random measure on $\M$. 
The distribution of such a Poisson random measure is
determined by its intensity measure, 
which is in general a locally-finite Borel measure on $\M$
and which is a finite measure when the mutation 
measure $\nu$ has finite total mass.
For most results,  we do assume that $\nu$ has finite
total mass,  but a brief discussion of infinite-mass
cases is given in Section~\ref{SS:infinitenu}.

\begin{notation}  \label{N:spaces}
\par\noindent
\begin{itemize}
\item
Denote by $\H$ the Banach space of finite signed Borel measures on $\M$
equipped with the norm $ \| \cdot \|_{\Was}  $ defined below. 
\glossary{$\H$}
\item
Let $\H^{+}$ be the subset of $\H$ consisting of nonnegative measures.
\glossary{$\H^{+}$}
\item
For $\pi \in \H$,   write $\pi^{+},\pi^{-}\in \H^{+}$ 
for the positive and negative parts of $\pi$ appearing in 
the Hahn-Jordan decomposition. Thus, $\pi=\pi^{+}-\pi^{-}$.
\glossary{$\pi^{+}$,$\pi^{-}$}
\index{Hahn-Jordan decomposition}
\item
For any measure $P$ on $\G$, let  $\mu P $
be the intensity measure associated with $P$;  
that is, for a nonnegative Borel function $f$ on $\M$, 
\glossary{$\mu P$}
$$
\int_\M f(m) \, d(\mu P)(m) = \int_\G \int_\M f(m) \, dg(m) \, dP(g).
$$
\item
For $ \pi \in H^{+} $,  write $X^\pi$ for a Poisson random 
measure on $\M$ with intensity measure $\pi$ and $\Pi_{\pi}$  
for the distribution of this Poisson random measure. 
\glossary{$X^{\pi}$}
\glossary{$\Pi_{\pi}$}
\end{itemize}
\end{notation}

The norms on our spaces of measures are based on a metric 
from the class of metrics named for Leonid Wasserstein 
(also transliterated as Vasershtein, or otherwise).
In our investigations we have to deal with spaces of measures 
at a hierarchy of different levels,  including $\G$,  $\H$,  
and finite signed measures on $\G$. 
The  Wasserstein metrics provide a unified way of topologizing 
all these various spaces.
\index{Wasserstein metric}
\index{Wasserstein, Leonid}

\begin{notation}  \label{N:Wass}
\par\noindent
\begin{itemize}
\item
Given a metric space $(E,d)$, let $\mathrm{Lip}$ 
\glossary{$\mathrm{Lip}$} 
\glossary{$\|\cdot \|_{\Lip}$}
be the space of functions $f:E \to \bR$ such that
\begin{equation}
\label{E:def_Lip}
\|f\|_{\Lip}:= \sup_{x}|f(x)| + \sup_{x\ne y} \frac{|f(x)-f(y)|}{d(x,y)}<\infty.
\end{equation}
\item
Define a norm $\|\cdot\|_{\Was}$ on the space of finite signed Borel
\glossary{$\|\cdot\|_{\Was}$}
measures on $(E,d)$ by
\begin{equation}
\label{E:def_Was}
\|\pi\|_{\Was}:= \sup \left\{ \, \left|  \int \, f d\pi \right| \, : 
               \|f\|_{\Lip}\le 1 \right\}.
\end{equation}
The Wasserstein metric metrizes the topology of weak convegence
of measures on the finite signed Borel measures on $(E,d)$.
\\
\item
For any measure $\pi$ in such a space of signed measures, write 
\begin{equation}
\label{E:def_functional_notation}
\pi[f]:=\int f d\pi.
\end{equation}
%
\item
Write
\[
\sigma := 
\sup_{g, h\in \G, \, g \ne h}
\frac{\bigl| S(g)-S(h)\bigr|}  
{\bigl\| g-h \bigr\|_{\Was}}
\]
for the (possibly infinite) Lipschitz constant
of the selective cost function $S$. 
\end{itemize}
\end{notation}
\glossary{$\pi[f]$}
\glossary{$\sigma$}
\index{Wasserstein metric}

We take advantage of the versatility of these definitions 
to build Wasserstein metrics on top of Wasserstein metrics.    
We start by taking $\M$ for the metric space $E$.
Then $\G$, as a space of measures on $\M$, has a Wasserstein metric.
Next we take $\G$ with its Wasserstein metric for the metric space $E$,
and obtain a Wasserstein metric on the finite signed measures on $\G$,
including the measures $P_{t}$.    
Also, with $\M$ again playing the role of $E$, we
obtain a Wasserstein metric on $\H$. 

An extensive account of Wasserstein metrics may be found 
in \cite{MR1105086, MR1619170}. 
The properties used here are described in Problem 3.11.2 of \cite{EK86}.
In particular, it is shown that our version  is
indeed a metric and that $\H$ as well as $\G$ and the finite
signed measures on $\G$ are complete in this norm.  
The Wasserstein distance between two probability measures
is bounded above by their total variation distance.

We note that the designation ``Wasserstein metric'' is 
also often applied to the analogous definition 
where the constraining Lipschitz norm $\|\cdot\|_{\Lip}$
does not include the supremum norm term.  
This latter distance is always greater than or equal 
to the Wasserstein metric as we are defining it.  
It is equivalent to the Kantorovich-Rubinstein distance, 
which is a member of the class of Monge-Kantorovich distances, 
a class defined by a single parameter $p$; 
the Kantorovich-Rubinstein distance is the element of this 
class corresponding to $p=1$.  Details may be found 
in \cite[Section 7.1]{cV03}, \cite[Chapter 6]{MR2459454} 
or \cite[Section 7.1]{AGS05}.
\index{Kantorovich-Rubinstein distance}
\index{Monge-Kantorovich distance}

In some places we need to use the total variation norm, 
defined for a finite signed Borel measure $\pi$ as
\begin{equation} \label{E:def_TV}
\|\pi\|_{\mathrm{TV}}:= \sup\left\{\left| \int f \, d\pi\right|\, 
             : \, \sup_{x} |f(x)|\le 1\right\}.
\end{equation}
\glossary{$\|\cdot\|_{\on{TV}}$}  
\glossary{$\Lip_{\on{TV}} $}  
\index{total variation norm}
There is a corresponding Lipschitz constant for a 
function $F:\G\to\bR$, defined as
\begin{equation} \label{E:LipTV}
\Lip_{\on{TV}} F:= \sup_{g\ne g'} \frac{|F(g)-F(g')|}{\|g-g'\|_{TV}}.
\end{equation}
This yields the identity
\begin{equation} \label{E:LipTV2}
\Lip_{\on{TV}} F = \sup_{g\in\G,x\in \M} |F(g+\delta_{x})-F(g)|.
\end{equation}

Lastly, we sometimes compare two equivalent probability measures 
$P$ and $Q$ on $\G$ by considering the quantity
\begin{equation}
\label{E:ThetaPQ}
\Theta(P,Q) := \Lip_{\on{TV}} \log \frac{dQ}{dP} 
= \sup_{g\ne g'} \frac{\left|\log \frac{dQ}{dP}(g) - \log \frac{dQ}{dP}(g')\right|}{\|g-g'\|_{TV}}.
\end{equation}
\glossary{$\Theta$} 
Note that $\Theta(P,P) = 0$. Conversely, if $\Theta(P,Q) = 0$, then
$\log dQ/dP$ is a constant and so $dQ/dP$ is also a constant; however,
this constant must be $1$ since $P$ and $Q$ are probability measures,
and hence $P=Q$.  Note also that $\Theta(P,Q) = \Theta(Q,P)$ because
$\log dP/dQ = \log (dQ/dP)^{-1} = - \log dQ/dP$.  Finally, 
if $R$ is another probability measure equivalent to $P$ and $Q$, then
\[
\log \frac{dR}{dP} 
= \log \frac{dR}{dQ} \cdot \frac{dQ}{dP} 
= \log \frac{dR}{dQ} + \log \frac{dQ}{dP}
\]
and so $\Theta(P,R) \le \Theta(P,Q) + \Theta(Q,R)$.  Therefore,
$\Theta$ restricted to a suitable class of probability measures
is a metric.  

\section{Definition of the dynamical system}
\label{SS:moddef}

The state of the population at time $t \ge 0$ 
is given by a probability measure $P_t$ on $\G$
that is the distribution of a Poisson random measure
with intensity measure $\rho_t$.  
Our informal description of the evolution of $\rho_t$,
and hence of $P_t$, is motivated by \eqref{E:model_informal},
restated here.   
\index{dynamical system}
\begin{equation}
\label{E:model_informal_again}
\rho_t(dm) = \rho_0(dm) + t\, \nu(dm) - 
\int_0^t  
\bE\left[S(X^{\rho_s} + \delta_m) - S(X^{\rho_s})\right] 
\, \rho_s(dm) \, ds,
\end{equation} 

In order to formalize this definition, it is
convenient to introduce the following two objects.
\\

\begin{definition}
\label{D:F}
$ $\\
\begin{itemize}
\item
Define $F:\M\times \H^{+}\to \bR_{+}$ 
by
\begin{equation}  \label{E:Fpi}
F_{\pi}(x):= \bE\bigl[ S(X^{\pi}+\delta_{x})-S(X^{\pi})\bigr] 
\end{equation}
\glossary{$F_{\pi}$}
for  $ x\in\M $   and  $  \pi \in \H^{+} $. 
\item
Define the operator $D:\H^{+}\to\H^{+}$ by
$$
\label{E:defineD}
\frac{d(D\pi)}{d\pi}(x):= F_{\pi}(x).
$$
\glossary{$D \pi$}

That is, for any bounded Borel function $f:\M  \to\bR$ and $\pi$ in $\H^{+}$,
$$
\int_{\M} f(x) \, d(D\pi)(x) = \int_{\M} f(x) F_{\pi}(x) \, d\pi(x).
$$
\end{itemize}
\end{definition}


With this notation, our evolution equation 
\eqref{E:model_informal_again} becomes 
\begin{equation}  
\label{E:dynam}
\rho_{t}= \rho_{0} + t \nu - \int_{0}^{t} D\rho_{s} \, ds
\end{equation}
A solution is an $\H^{+}$-valued function $\rho$
that is continuous with respect to the Wasserstein metric
and therefore with respect to the topology
of weak convergence of measures and satisfies \eqref{E:dynam} 
for all $t \ge 0$.
Equation \eqref{E:dynam} involves the integration of a 
measure-valued function, and such an integral can have a number 
of different meanings.  We require only that 
\index{Wasserstein metric}
if $\eta: \bR_+ \to \H$ is a Borel function, 
then for $t \ge 0$ the  integral $\int_{0}^{t}\eta_{s} \, ds$ 
is the element of $\H$ satisfying, for every Borel $ A \subseteq \M $, 
\begin{equation}  \label{E:msbl}
\left( \int_{0}^{t} \,\eta_{s} \, ds \right) \, (A) \, 
       =  \, \int_{0}^{t} \eta_{s}(A) \, ds
\end{equation}

This integral certainly exists (and is unique) if the
function $\eta$ is continuous with respect to the
topology of weak convergence of measures.
For more information about integration on infinite
dimensional spaces, see Chapter 2 of \cite{MR0453964}.


\section{Existence and uniqueness of solutions}
\label{SS:existence_uniqueness}   

We now prove the existence and uniqueness of solutions
to \eqref{E:dynam} with the assistance of three 
lemmas that are proved below in Section~\ref{SS:lemmas_ex_uniq}.  
The proof is an application of the standard iterative method
of Charles \'{E}mile Picard.  
\index{Picard, Charles E.}
\index{selective cost}

\begin{theorem} 
\label{T:existence}
Fix a mutation measure $\nu\in\H^{+}$
and a selective cost function $S:\G\to\bR_{+}$, 
that satisfies the conditions 
\begin{itemize}
\item
$S(0)=0$,
\item
$S(g)\le S(g+h)$ for all $g,h\in\G$,
\item
the Lipschitz constant $\sigma$ of the selective cost $S$ 
(in the Wasserstein metric) is finite.
\end{itemize}
Then, \eqref{E:dynam} has a unique solution 
for any $\rho_{0}\in \H^{+}$.
\end{theorem}
\index{Wasserstein metric}

\begin{proof}
Fix a time horizon $T>0$ and  let $c>0$ be a constant 
that will be chosen later.  
Temporarily, write $C([0,T],\H)$ for the Banach space of 
continuous $\H$-valued functions on $[0,T]$, equipped with the norm
$$
\|\alpha\|_{c} = \sup_{0\le t\le T} e^{-ct}\|\alpha_{t}\|_{\Was}.
$$
\glossary{$C([0,T],\H)$}
\glossary{$\|\alpha\|_{c}$}

Denote by $\Gamma$ the closed subset of $C([0,T],\H)$ consisting of $\H$-valued functions $\alpha$ with $\alpha_{0}=\rho_{0}$ and 
$$
\alpha_{t}^{+}(\M)\le \rho_{0}(\M)+t\nu(\M)
$$ 
for $0\le t\le T$. (Recall from Notation~\ref{N:spaces}
that the measure $\alpha_t^+$ is the positive part in the
Hahn-Jordan decomposition of the signed measure $\alpha_t$.)
\glossary{$\Gamma$}
\glossary{$\Delta$}
\index{Hahn-Jordan decomposition}

Define a map $\Delta: C([0,T],\H) \to C([0,T],\H) $ by 
$$
(\Delta \alpha)_{t}=\rho_{0} + t \nu \, - \, \int_{0}^{t} D\alpha_{s}^{+}ds.
$$


Note that $\Delta$ maps $\Gamma$ into itself.  
Moreover, for $\alpha,\beta\in\Gamma$, 
$$
\| \Delta \alpha -\Delta \beta\|_{c}  \le  
     \sup_{0\le t \le T} e^{-ct} \int_{0}^{t} 
    \bigl\| D \alpha_{s}^{+} - D \beta_{s}^{+}\bigr\|_{\Was} \, ds.
$$

By Lemma~\ref{L:wasbound} below,  the norm inside the 
integral on the right has the following bound
$$
 \bigl\| D \alpha_{s}^{+} - D \beta_{s}^{+}\bigr\|_{\Was} \le 
  \sigma\bigl( 2+8\ls \alpha_{s}^{+}(\M) \wedge \beta_{s}^{+}(\M) \rs \bigr)
      \bigl\| \alpha_{s}-\beta_{s}\bigr\|_{\Was}.
$$
We conclude that 
\begin{align*}
& \| \Delta \alpha -\Delta \beta\|_{c}  
\le  \sup_{0\le t \le T} e^{-ct} \int_{0}^{t}  
    \sigma\bigl( 2+8\ls \rho_{0}(\M)+s\nu(\M) \rs \bigr) e^{cs}
    \bigl\| \alpha-\beta\bigr\|_{c} \, ds    \\
& \quad \le \sup_{0\le t \le T} e^{-ct}
     \Bigl[ \sigma(2+8\rho_{0}(\M))\frac{e^{ct}-1}{c}
     + 8\sigma\nu(\M) \frac{(ct-1)e^{ct}+1}{c^{2}}\Bigr] \|\alpha-\beta\|_{c}.
\end{align*}
Thus, $\Delta :\Gamma\to\Gamma$ is a contraction, provided $c$ 
is chosen sufficiently large.  

It follows from the Contraction Mapping Theorem that the equation
\index{contraction mapping theorem} 
\begin{equation}  \label{E:contsol}
\rho_{t}=\Delta \rho_{t} = \rho_{0} + t \nu \, - \, 
           \int_{0}^{t}D\rho_{s}^{+} \, ds 
\end{equation}
has a unique solution in $\Gamma$.  Furthermore, any function 
in $\H$ that is a solution to \eqref{E:contsol} must 
automatically be in $\Gamma$.  Therefore, the solution is unique.

It remains to show that it actually 
takes values in the subset $\H^{+}$ of nonnegative
measures.  For any Borel set $A\subseteq\M$,
$$
\rho_{t}(A) = \rho_{0}(A) + t \nu(A) \, - \int_{0}^{t}\int_{A} 
           F_{\rho_{s}^{+}}(x ) \, 
       \rho_{s}^{+}(dm) \, ds.
$$
In particular, $t\mapsto \rho_{t}(A)$ is continuous.  
For any Borel set $A\subset\M$ we have
$$
\rho_{t}^{+}(A) \le \rho_{0}(A) +t\nu(A).
$$

The Lipschitz condition on $S$ in terms of the Wasserstein metric
implies, via Lemma~\ref{L:wasbound} below,  that   
$F_{\pi}(m)\le \sigma$ for all $m \in \M $ and $\pi \in \H^{+}$. 
Then, 
\begin{align*}
\rho_{t}(A)&\ge \rho_{0}(A) -\sigma\int_{0}^{t} 
       \bigl( \rho_{0}(A) + s\nu(A)\bigr) \, ds + t\nu (A)   \\
&= (1-\sigma t) \rho_{0}(A) + t\left( 1-\frac{\sigma t}{2}\right) \nu(A).
\end{align*}
Hence, $\rho_{t}(A)\ge 0$ for $0\le t\le 1/\sigma $.  
Because this holds for all Borel sets $A \subseteq \M$, we 
have $\rho_{t}\in\H^{+}$ for $0\le t\le 1/\sigma $.  
Iterating this argument, with the time $0$  replaced successively 
by the times   $1/\sigma ,2/\sigma ,\dots$    gives the result.
\end{proof}

\section{Lemmas used in the proof of existence and uniqueness}
\label{SS:lemmas_ex_uniq}

We assume in this section that the hypotheses
of Theorem~\ref{T:existence} hold 
and prove the lemmas on which that theorem relies.

\begin{lemma}
\label{L:FLip}
The function $F_{\pi}$ is Lipschitz in the Wasserstein metric
for each finite measure $\pi \in \H^{+}$,
and, in terms of the Lipschitz constant $\sigma $ of the selective
cost function $S$,  
\[
\sup_{\pi \in \H^{+}} \|F_{\pi}(\cdot)\|_{\Lip} \le 2\sigma 
\]
and 
$$
\sup_{\pi \in \H^{+}, m \in \M} F_{\pi}(m) \le \sigma .
$$
Furthermore, for all $m \in \M$ and all $\pi \in \H^{+} $,  
\begin{equation}\label{E:Frecip}
\frac{1}{F_{\pi}(m)} \, \le  \, \frac{ \exp\{ \pi(\M) \} }{S(\delta_m)}.
\end{equation}
\end{lemma}

\begin{proof}
By definition,
\begin{align*}
\|F_\pi(\cdot)\|_{\Lip} 
  &= \sup_x \left |\bE[S(X^{\pi}+\delta_x) - S(X^{\pi})]\right| \\
  &+ \sup_{x \ne y} \left|\bE[S(X^{\pi} + \delta_x) - S(X^{\pi})] 
      - \bE[  S(X^{\pi} + \delta_y) -  S(X^{\pi})]\right|/d(x,y) \\
  &\le \sup_x \sigma \|\delta_x\|_{\Was} + \sup_{x \ne y} 
\sigma \|\delta_x-\delta_y\|_{\Was}/d(x,y) \\
&\le \sigma  + \sigma   =  2\sigma .
\end{align*}
Furthermore,  
\begin{align*}
F_\pi(m)
  &=    \bE \left[ \left( S(X^{\pi}+\delta_x) - S(X^{\pi})   \right)
           \left(  \indic_{ \{X^{\pi} = 0 \}}   +  
                        \indic_{ \{X^{\pi} > 0 \}} \right) \right]  \\ 
  &\ge   S(\delta_m) \, \bE [  \indic_{ \{X^{\pi} = 0 \}} ] 
         =  S(\delta_m)  \exp\{ -\pi(\M) \}, 
\end{align*}
implying the bound on $ 1/ F_{\pi} $.

\end{proof}


\begin{lemma}
\label{L:Fpipi}
For two finite measures $\pi',\pi'' \in \H^{+}$,
$$
\sup_{x \in \M} |F_{\pi'}(x) - F_{\pi''}(x)| \le 8\sigma \|\pi'-\pi''\|_{\Was}.
$$
\end{lemma}

\begin{proof}
Fix $x \in \M$.
Define $\Phi: \G \to \bR_+$
by $\Phi(g):= (S(g+\delta_{x})-S(g))/2\sigma  $.  
Then, $\|\Phi\|_{\on{Lip}}\le 1$ by the Lipschitz assumption on $S$.  
Note that $F_{\pi}(x) = 2\sigma  \Pi_{\pi} [\Phi]$.  
By definition of the Wasserstein metric,
$$
|F_{\pi'}(x) - F_{\pi''}(x)|
   =  2\sigma  \bigl|  \Pi_{\pi'}[\Phi]-\Pi_{\pi''}[\Phi]\bigr|
      \le  2\sigma    \was{\Pi_{\pi'}}{\Pi_{\pi''}}.
$$
The lemma now follows from Lemma~\ref{L:poisson}.
\end{proof}

\begin{lemma}
\label{L:wasbound}
For two finite signed measures $\alpha,\beta \in \H$,
$$
\|D\alpha^+ - D\beta^+\|_{\Was} \le \sigma (
              2 + 8\{\alpha^+(\M) \wedge \beta^+(\M)\})
              \, \|\alpha-\beta\|_{\Was}.
$$
\end{lemma}

\begin{proof}
Suppose without loss of generality that $\alpha^+(\M) \le \beta^+(\M)$.  
By Lemmas \ref{L:FLip} and \ref{L:Fpipi},
for any Lipschitz test function $f$,
\begin{align*}
\Bigl| \int f(x) &dD\alpha^+(x) - \int f(x) \, dD\beta^+(x)\Bigr|  \\
   & \le \left| \int F_{\alpha^+}(x) f(x) \, d\alpha^+(x) 
               - \int F_{\beta^+}(x) f(x) \, d\alpha^+(x)\right| \\
  & \quad + \left| \int F_{\beta^+}(x)f(x) \, d\alpha^+(x) 
                        - \int F_{\beta^+}(x)f(x) \, d\beta^+(x)\right| \\
  & \le   8\sigma \|\alpha^+-\beta^+\|_{\Was} \,  \|f\|_{\infty} \alpha^+(\M) 
      +  2\sigma \|f\|_{\Lip} \,  \|\alpha^+-\beta^+\|_{\Was} \\
  &\le (2\sigma  + 8\sigma \alpha^+(\M)) \quad \|f\|_{\Lip}  \|\alpha-\beta\|_{\Was}.
\end{align*}
We have used the fact 
that $\|f'f''\|_{\Lip} \le \|f'\|_{\Lip} \, \|f''\|_{\Lip}$ 
for $f',f'' \in \Lip$.
\end{proof}


\section{Density form of the dynamical system}
\label{SS:Linfinity}

The solutions provided by Theorem~\ref{T:existence} are 
absolutely continuous with respect to $ \rho_0 + \nu$ 
and so can be written out in terms of Radon-Nikodym
derivatives (that is, densities) with
respect to that reference measure.   We show below that, more generally, the
Radon-Nikodym derivatives with respect
to suitable reference measures $\zeta$
belong to the space $ L_{+}^{\infty}( \M, \zeta)  $
of nonnegative functions that are essentially-bounded for $\zeta$.   
This parallel approach of viewing our dynamical system as taking
values in a space of functions rather than
a space of measures was first developed in \cite{CE09}.  
It provides another route to defining the informal dynamical
system rigorously and establishing the existence and
uniqueness of solutions.

Of course our analysis as a whole cannot be
carried through in $ L^{\infty}( \M, \zeta)$.  The singleton
mutations $\delta_m$ in $\G$ with which selective
costs are measured can not be viewed as elements 
of $L^{\infty}( \M, \zeta)$ for some
measure $\zeta$ equivalent to $\rho_0 + \nu$ 
unless $\rho_0$ or $\nu$ has atoms, and the
simultaneous consideration of a continuum
of non-equivalent initial measures $\rho_0$
can not be accommodated within a single $L^{\infty}( \M, \zeta)$.
For the full theory, spaces of measures equipped
with Wasserstein metrics or 
tools of similar generality are required.
However, proofs of existence and uniqueness for fixed initial
conditions and mutation measures
can be carried out in the $L^{\infty}(\M, \zeta)$ setting.
Moreover, this approach facilitates an extension of the theory
to some cases where the mutation measure $\nu$ 
has infinite total mass and it also
justifies a step in the proof of 
Lemma~\ref{L:comparison} in our chapter on equilibria.
\index{comparisons lemma}
\begin{notation}
\label{N:L-infinity}
\par\noindent
\begin{itemize}
\item
The measure $\zeta$ on $\M$ is any measure with respect 
to which the initial state $\rho_0$ and the mutation 
measure $\nu$ are both absolutely continuous, with 
bounded Radon-Nikodym derivatives $r_0(m)$ and $q_{\nu}(m)$.   
(In practice, it is convenient to
take $\zeta$ to be the sum of $\nu$ and the part of $\rho_0$,
if any,  orthogonal to $\nu$).
\glossary{$\zeta$}
\glossary{$r_t(m)$}
\glossary{$q_{\nu}(m)$}
\item
Denote by
$ L^{\infty}(\M, \zeta)$
the Banach space of equivalence classes 
of $\zeta$-essentially bounded functions on $\M$ 
and by $ L_{+}^{\infty}(\M, \zeta)$ its subset 
consisting of equivalence classes of nonnegative functions.   
\item
Write $\K$ for the space of finite 
signed measures on $\M$ that are absolutely continuous 
with respect to $\zeta$ with $\zeta$-essentially bounded 
Radon-Nikodym derivatives,
equipped with the $L^{\infty}(\M, \zeta) $ norm via
the usual bijection between  $ L^{\infty}(\M, \zeta)$ and $\K$. 
\item
Write $\K_{+} $ for the subset of $\K$ consisting 
of nonnegative measures. 
\end{itemize}
\end{notation}
\glossary{$L^{\infty}(\M, \zeta)$}
\glossary{$L_{+}^{\infty}$}
\glossary{$\K$}
\glossary{$\K{+}$}

Instead of working forward from conditions on the selective 
cost $S$  via the lemmas of Section~\ref{SS:lemmas_ex_uniq},    
we impose a bound and Lipschitz condition directly on the
whole contribution from  $ D \rho $.  Anticipating the
remarks in Section~\ref{SS:infinitenu},  we offer proofs
that do not rely on the finiteness of $\nu(\M)$.

\begin{theorem} 
\label{T:L-inf-existence}
Fix a mutation measure $\nu$ and starting state $\rho_0$ in $\H^{+}$ 
and a selective cost $S:\G \to \bR_{+}$ 
such that $S(0)=0$, and $S(g)\le S(g+h)$ for all $g,h\in\G$.  
Let $\nu$ and $\rho_0$ be absolutely continuous with respect
to a measure $ \zeta $ in $\H^{+}$ with bounded Radon-Nikodym
derivatives $q_{\nu}(m)$ and $r_0(m)$.

Suppose  the 
function $(\pi,m) \mapsto F_{\pi}(m)$ defined from $S$ in Definition~\ref{D:F}
the following conditions.
\begin{itemize}  
\item
There is a constant $\sigma $ such that for 
all $\pi \in \H^{+}$ and $m \in \M$, $F_{\pi}(m) \le \sigma  $.
\item
For all pairs of measures $\alpha = a(m) \zeta(dm) $
and $\beta = b(m) \zeta(dm) $  in $\K_{+}$,
$$
 \|  F_{\alpha}(m) a(m)  - F_{\beta}(m) b(m)  \|_{\infty}  \le   
 \sigma   (  \| a  \|_{\infty} \,  \wedge \,  \| b \|_{\infty} ) \,  
      \|  \alpha  - \beta  \|_{\infty}.     
$$
\end{itemize}
Then, the dynamic equation \eqref{E:dynam} 
has a unique solution in $C( \bR_{+} , \K_{+}) $.
\end{theorem}

\begin{proof}

The proof mimics the proof of Theorem~\ref{T:existence}.  
The space $ \H $ is replaced by $\K$, $\H^{+}$ is replaced by $\K_{+}$,  
and the norm $ \| \cdot \|_{\Was} $ is replaced by $ \| \cdot \|_{\infty} $.
The analogue of the subset $\Gamma $  
consists of $\K$-valued functions $\alpha$
with $\alpha_0 = \rho_0$  and 
$$
\| \alpha^{+}_{t} \|_{\infty}  \le \| r_0(m) + t q_{\nu}(m) \|_{\infty}.  
$$
The assumptions imposed on $F$ 
take the place of the bounds from 
Lemmas \ref{L:wasbound}  and  \ref{L:FLip}.  
\end{proof} 

Theorem~\ref{T:L-inf-existence} gives us for each $t \ge 0$ 
a measure which, as an element of $\K_{+}$,   
has a Radon-Nikodym derivative with respect to $\zeta$.  
{\em A priori,} the Radon-Nikodym derivative is   
defined separately for each $t \ge 0$ as a function of $m \in \M$
and is only unique up to $\zeta$-null sets.
The behavior of a particular choice for the
 ensemble of Radon-Nikodym derivatives is,
for fixed $m \in \M$, not guaranteed to be even
a measurable function of $t \ge 0$.   
We now show that we can define versions of the
Radon-Nikodym derivatives that are in fact continuously differentiable 
functions of $t \ge 0$ for every $m \in \M$.   

\begin{theorem} 
\label{T:niceradon}
Suppose that $\rho_t$ is a solution to the dynamic
equation \eqref{E:dynam} in $ C( \bR_{+}, \K_{+})$
for selective costs satisfying the conditions 
of Theorem~\ref{T:L-inf-existence}.  
Then, there exists a Borel function $ (t,m) \mapsto r_t(m) $ 
on  $ \bR_{+} \times \M $ such that for all $ t \ge 0 $ 
the function $ m \mapsto  r_t(m) $ is a Radon-Nikodym
derivative of $\rho_t$ with respect to $\zeta$ 
and such that for all $m \in \M$  the function $ t \mapsto r_t(m) $ 
is continuously differentiable with a derivative which satisfies
$$
\dot{r}_t(m) = q_{\nu}(m) -  F_{\rho_t}(m) \, \, r_t(m).  
$$
\end{theorem}

\begin{proof}
We will construct a curve of {\em a priori} new measures $\xi_t$ with
Radon-Nikodym derivatives $ m \mapsto x_t(m) $ that have 
nice behavior in $t \ge 0$ for fixed $m \in \M$.   We then prove that,
as measures,  $\xi = \rho$.   Finally, we replace
the arbitrary choice of Radon-Nikodym derivative $r_t$ for $\rho_t$
whose existence is guaranteed by Theorem~\ref{T:L-inf-existence} with $x_t$. 

Because $t \mapsto \rho_t$ is continuous, it follows from
a monotone class argument that $t \mapsto \rho_t(A)$ is
Borel measurable for all $A \in \BM$ 
(recall that $\BM$ is the Borel $\sigma$-field on $\M$).  
We have $\BM = \bigvee_{k \in \bN} \mathcal{F}_k$,
where $\mathcal{F}_1 \subseteq \mathcal{F}_2 \subseteq \ldots$
is a suitable increasing sequence of finitely generated sub-$\sigma$-fields.
Suppose for $k \in \bN$
that $\mathcal{F}_k$ consists of the empty set and unions of collections
of sets drawn from the Borel partition 
$\{A_{k,1}, \ldots, A_{k,n(k)}\}$ of $\M$. Then, we may suppose that
\[
r_t(m) := 
\begin{cases}
\lim_{k \to \infty} 
\sum_{j=1}^{n(k)} \frac{\rho_t(A_{k,j})}{\zeta(A_{k,j})}
\mathbf{1}_{A_{k,j}}(m),& \quad \text{if the limit exists,} \\
0,& \quad \text{otherwise}
\end{cases}
\]
(see, for example,
Section~III-1 of \cite{MR0402915}).
In particular, we may suppose that the map 
$(t,m) \to r_t(m)$ is Borel measurable. 
 
When $\rho_t$ is already known,  for each fixed $m \in \M$ the
function $ F_{\rho_t}(m) $ is just some known Borel
function of $t$ and the Radon-Nikodym derivative $m \mapsto r_0(m)$
is just some known bounded function.   
Consider, then,  for each fixed $m \in \M$ the
following ordinary integral equation for an unknown 
function $x_t(m)$, analogous to our dynamic equation
\begin{equation}
\label{E:dynamdensity}
x_t(m) =  r_{0}(m)  +  q_{\nu}(m) \, t   
      -\int_{0}^{t} F_{\rho_{s}}(m)  \, x_{s}(m)\,   ds.  
\end{equation}

For every $m \in \M$, \eqref{E:dynamdensity} has the 
continuously differentiable solution 
%
\begin{equation}       
\label{E:RN}
\begin{split}
x_t(m)
& =
\exp\left\{ -\int_0^t F_{\rho_u}(m) \, du \right\}  \\
& \quad \times \left(x_0(m) + q_{\nu}(m) \int_0^t \exp\left\{\int_0^u F_{\rho_v}(m) \, dv \right\} \,
   du \right). \\
\end{split}
\end{equation}
For $ t \ge 0 $, define measures $\xi_t$ 
by $ \xi_t(dm) = x_t(m) \zeta(dm) $.  
For $\xi_t$, for any Borel set $A$,  
\begin{equation}
\label{E:xi_is_rho_1}
\begin{split}
\xi_t(A)  
    & = \int_A \, x_t(m) \, d\zeta(m)       \\ 
    & = \int_A
 \,  ( r_0(m) + q_{\nu}(m) \, t) \, d\zeta(m)
     - \int_A \left( \int_0^t  F_{\rho_s}(m) \, x_s(m) \, ds \right).
         \, d\zeta(m).
\end{split}
\end{equation}

For $\rho_t$, with Radon-Nikodym derivatives $r_t$,  we have
a similar equation but with the integral over $m \in \M$ 
inside the integral over $s \in \bR_+$, namely 
\begin{equation}
\label{E:rhovias}
\rho_t(A) = 
        \int_A \,  ( r_0(m) + q_{\nu}(m) \, t) \, d\zeta(m)
      - \int_0^t   \left(  \int_A \,  F_{\rho_s}(m) \, r_s(m) \, 
          d\zeta(m) \right) \, ds.  
\end{equation}

We want to reverse the order of integration in the first equation 
for $\xi$.  We can apply Fubini's Theorem 
so long as the function $ (s,m) \mapsto x_s(m) \in \bR $ is 
Borel measurable, which is the case (because of the equation
defining $x_t(m)$) when $ (s,m) \mapsto F_{\rho_s}(m) $ 
is  Borel measurable.   As we have already remarked,
the map $ s \mapsto \rho_s $
is continuous by construction,  and the 
map $ (\pi, m ) \mapsto  F_{\pi}(m) $ is always  Borel
measurable. (Under the conditions of Theorem~\ref{T:existence}
it is jointly continuous.)  Thus, we can apply Fubini's Theorem. 

After reversing the order of integration in \eqref{E:xi_is_rho_1},
we subtract \eqref{E:rhovias} from \eqref{E:xi_is_rho_1}: 
The terms in the initial state and the mutation measure cancel

$$
 \xi_t(A) - \rho_t(A)   =     \int_0^t  \, 
   \int_A  \,  F_{\rho_s}(m)\,  \left( x_s(m) - r_s(m) \right) \, 
             d\zeta(m) \, \,ds.   
$$

Since $ F_{\pi}(m) \le \sigma  $ for all $\pi$ in $\H^{+}$,  
$$
| \xi_t(A) - \rho_t(A) |  \le      \int_0^t  
   \sigma     \|  x_s - r_s \|_{\infty} \quad  
       \left|   \int_A   d\zeta(m) \right|   \,ds.         
$$
Thus,
\begin{equation}
\label{E:ximinusrho}
| \xi_t(A) - \rho_t(A) |  \le     \sigma  \, \left( \int_0^t  \,
        \|  x_s - r_s \, \|_{\infty} \, ds \,\right) \,   \zeta(A). 
\end{equation}
Define $ \beta_t  
=  \|  x_t - r_t \, \|_{\infty}$. 
For Borel sets $A$ with strictly positive measure,  and particularly
for sets $A_n$, $n \in \mathbb{N}$, defined by
$$
A_n :=  \{ m \in \M :  | x_t(m) - r_t(m) | \ge \beta_t - 1/n \}.
$$
we can divide both sides of \eqref{E:ximinusrho} by $\zeta(A)$.  
The resulting left-hand side is no less 
than  $  \beta_t - 1/n $ for all $n \in \mathbb{N}$.  
Taking the supremum over $n \in \mathbb{N}$,
we conclude for all $ t \ge 0 $ that 
$$
\beta_t \le \sigma  \int_0^t \, \beta_s \, ds. 
$$
Gronwall's Inequality (see Appendix~\ref{SS:Gronwall}) 
then forces $ \beta_t \equiv 0 $. 
\index{Gronwall's Inequality}
It follows that $ x_t(m) = r_t(m)$ outside of a set of $m \in \M$ with 
$\zeta$ measure zero  so that,  for every $t$,  $x_t(m)$ is itself
a Radon-Nikodym derivative of $\rho_t$ with respect to $\zeta$.
In other words,  $ \xi_t = \rho_t $ and $ \xi_t$ satisfies the
dynamic equation.  We replace our original 
function $(t,m) \mapsto r_t(m)$ with the function 
$(t,m) \mapsto x_t(m)$ with the property that
$t \mapsto x_t(m)$ is continuously differentiable 
for every $m \in \M$. 
To spare notation,  we henceforth write $r_t(m)$ for $x_t(m)$.  
When we differentiate this new family of Radon-Nikodym derivatives
with respect to $t$, we find, pointwise
for all $t \ge 0$ and $m \in \M$ that 
$$
\dot{r}_t(m) = q_{\nu}(m) -  F_{\rho_t}(m) \, r_t(m).  
$$
\end{proof} 

A continuously differentiable function 
$h: \bR_+ \to \bR$ may have a set of zeros that is
quite messy,  but its positive part can still be expressed
as an integral in a straightforward way. For applications of
Theorem~\ref{T:niceradon},  the following elementary 
fact from real analysis is helpful.

\begin{lemma}
\label{L:abcon}
Suppose that the function $h: \bR_+ \to \bR$ is an absolutely continuous function
with $h(0) = 0$  and derivative $\dot{h}$ defined Lebesgue-a.e.
Write $J$ for the indicator function of the set $\{t \in \bR_+ : h(t) > 0 \}$.
\glossary{$J(t)$}
Then, for all $t \in \mathbb{R}_+$,
$$
h^{+}(t) = h(t)J(t)  =  \int_0^t \, \dot{h}(s) \, J(s) \, ds.   
$$
\end{lemma} 

\begin{proof}
The fundamental theorem of calculus holds for 
absolutely continuous functions  and so
$h(t) = \int_0^t \, \dot{h}(s) \, ds$ for all $t \in \mathbb{R}_+$.
Note that the function $t \mapsto g(t) := h(t)^{+}$ is also absolutely continuous.
Therefore, the nonnegative function $g$ is differentiable
almost everywhere with respect to Lebesgue measure 
and $g(t) = \int_0^t \dot{g}(s) \, ds$ for all $t \in \mathbb{R}_+$.


Choose a point $t \in \mathbb{R}_+$ in the set
of full Lebesgue measure where $h$ is differentiable.
Consider the following four alternatives.
\begin{itemize}
\item[(1)]
If $h(t) > 0$, then $g(s) = h(s)$ for $s$ in a neighborhood of
$t$, and so $g$ is differentiable at $t$ and $\dot{g}(t) = \dot{h}(t)$.
\item[(2)]  
If $h(t) < 0$, then $g(s) = 0$
in a neighborhood of $t$, and so $g$ is differentiable at $t$
and $\dot{g}(t) = 0$.  
\item[(3)]
If $h(t) = 0$ and $t$ is not an isolated point
of the set $\{s \in \mathbb{R}_+: h(s) = 0\}$,  
then there exists a sequence $(t_n)_{n \in \bN}$ 
such that $t = \lim_{n \to \infty} t_n$ with $t_n \ne t$ and $h(t_n) = 0$. 
Then, 
$ \dot{h}(t) = \lim_{n \to \infty}  h(t_n)/(t_n - t) = \lim_{n \to \infty} g(t_n)/(t_n - t) = 0 $.
Since $h$ is differentiable at $t$,  $ h(t_j)/(t_j - t) $ converges
to zero for any sequence $t_j$ converging to $t$.
For any $j$, $ g(t_j)$ either equals $h(t_j)$ or equals zero,
so $ g(t_j)/(t_j - t) $ also converges to zero, 
meaning that $g$ is differentiable at $t$ and $\dot{g}(t)  = 0$.
\item[(4)]
If $h(t) = 0$ and $t$ is an isolated point
of the set $\{s \in \mathbb{R}_+: h(s) = 0\}$, then
$g$ need not differentiable at $t$, but the set 
of such $t$ is countable and hence Lebesgue null.
\end{itemize}
 
Thus, $\dot{g}(t) \, = \, \dot{h}(t) J(t)$ for Lebesgue-a.e. $t \in \bR_+$  
and we conclude that 
$$
h^{+}(t) = g(t) = \int_0^t \dot{g}(s) \, ds = \int_0^t \dot{h}(s) \, J(s) \, ds
$$
for all $t \in \mathbb{R}_+$.  
\end{proof}

\section{Mutation measures with infinite total mass}  
\label{SS:infinitenu}

The assumption that the mutation measure $\nu$ has a finite
total mass underlies our development of the model, our use
of Wasserstein metrics,  and our main  theorems
in Chapters \ref{Ch:hypotheses} through \ref{Ch:convergence}.  
However, the dynamic system itself, taken in isolation, can be defined 
for some cases when $\nu(M)$ is infinite and the selective cost
$S$ is chosen suitably.  For example,
the case where $\nu$ is a multiple of Lebesgue measure on the positive
reals is important for demographic applications.
This section sketches how the dynamical system for such measures can be
handled by viewing the state space of the dynamical system as 
a subset of $ L^{\infty}(\M, \zeta)$ for a suitable reference measure $\zeta$.

\begin{notation} \label{N:Hbar}
\par\noindent
\begin{itemize}
\item
Write $\bar{\H}$ for the space of signed measures on $\M$ 
such that the trace on any bounded set Borel set 
$A \subset \M$ belongs to $\H$.
The space $\bar{\H}$ is equipped with the $\sigma$-field
generated by the maps  $ \pi \mapsto \pi(A)$ as $A$
ranges over the bounded Borel subsets of $\M$.  
(The trace of a measure $\pi$ on a Borel set $A$ is the measure
$\pi( \cdot \cap A) $.)
\item
Denote the subset of nonnegative measures in $\bar{\H}$ by $\bar{\H}_{+}$.
\item
Denote the set of integer-valued elements of $\bar{\H}_{+}$ by $\bar{\G}$. 
\end{itemize}
\end{notation}
\glossary{$\bar{\H}$}
\glossary{$\bar{\G}$}

Fix $ \zeta \in \bar{\H}_{+} $,  and 
identify $ L^{\infty}_{+}(\M, \zeta) $ 
with $\bar{\K}_{+}$,  the set of elements of $\bar{H}_{+}$
that are absolutely continuous with respect to $\zeta$ with
essentially bounded Radon-Nikodym derivatives.  


If $ \pi \in \bar{\H}_{+}$, then the Poisson random 
measure $ X^{\pi} $ with intensity measure $\pi$
takes values in the set $ \bar{\G} $ almost-surely.
Suppose that the selective cost 
function $S: \bar{G} \to \bR \cup \{+\infty \} $
is Borel measurable with $S(0) = 0 $ and $S(g) \le S(g + h) $.
Then,  $ \pi \mapsto \bE [ S(X^{\pi}) ] $ is a Borel measurable
map from $ \bar{\K}_{+}$ to  $  \bR \cup \{+\infty \} $ 
and  the map $ (\pi, m) \mapsto \bE [ S( X^{\pi} + \delta_m) \ $
from $ \bar{\K}_{+} \times \M$ 
to  $  \bR \cup \{+\infty \} $ is also Borel measurable. 
\index{selective cost}

Suppose that the selective cost is such that  
$$ 
  \bE [ S( X^{\pi} )]  \le \bE [ S( X^{\pi} + \delta_m) ] < \infty  
$$
for all $ \pi  \in \bar{\K}_{+}$ and $m \in \M $. 
Then,  
$$
(\pi, m) \mapsto \bE [ S( X^{\pi} + \delta_m) - S(X^{\pi})] =: F_{\pi}(m) 
$$
is a well-defined and Borel measurable map 
from $ \bar{\K}_{+} \times \M $ to  $  \bR_{+} $.

\begin{remark}
An examination of the proofs in Section~\ref{SS:Linfinity} 
of Theorems \ref{T:L-inf-existence} and \ref{T:niceradon} 
shows that they apply in this broader setting and do
not depend on the finiteness of $\nu(\M)$.
\end{remark}


\chapter{Equilibria}
\label{Ch:equilibria}

One of the primary problems concerning our dynamical
systems is to understand their
asymptotic behavior over time. We begin the analysis of 
these asymptotics by identifying when fixed points
of the dynamical system exist, and then examining whether there
is convergence to these fixed points from suitable
initial conditions.  

{\bf We assume throughout this chapter that the assumptions
of Theorem~\ref{T:existence} always hold.}

\begin{definition}
A finite {\em fixed point} or {\em equilibrium} for the dynamical system 
(we use the terms  interchangeably) is
a measure $\rho_{*}\in \H^{+}$ at which the driving 
vector field vanishes. 
\index{equilibrium}
That is, $\rho_{*}$ is absolutely continuous with respect 
to $\nu$, with Radon-Nikodym derivative satisfying
\begin{equation} \label{E:defineequil}
F_{\rho_{*}} \frac{d\rho_{*}}{d\nu}=1.
\end{equation}
\glossary{$\rho_*$}

The equilibrium $\rho_{*}$ is called {\em stable} 
if for every neighborhood $V$ of $\rho_{*}$ there is
a neighborhood $U\subset V$, such that $\rho_{t}\in V$ 
for all times $t \ge 0$ if $\rho_{0}\in U$. 
It is called {\em attractive} if it is stable 
and there is a neighborhood $U_{0}$ of $\rho_{*}$ 
such that $\lim_{t\to\infty} \rho_{t}=\rho_{*}$ 
whenever $\rho_{0}\in U_{0}$.

We introduce the terms {\em box-stable} and {\em box-attractive} 
when the above definitions hold if ``neighborhoods'' 
in the above definitions are replaced by sets of the form 
\begin{equation} \label{E:definebox}
B(\tilde\rho,\tilde\rho'):=\bigl\{\rho\,:\, 
      \tilde\rho<\rho<\tilde \rho'\bigr\},
\end{equation}
where $\tilde\rho\le\rho_{*}\le\tilde\rho'$, and both
measures $\tilde\rho-\rho_{*}$ and $\tilde\rho'-\rho_{*}$
are mutually absolutely continuous with respect to $\nu$. 
\end{definition}
\index{stable equilibrium}
\index{attractive equilibrium}
\index{box stable}
\index{box attractive}
\index{Wasserstein metric}
\glossary{$B(\tilde\rho,\tilde\rho')$}

\begin{remark}
Note that box-stability (respectively, box-attractivity)
is a weaker condition than stability (respectively, attractivity), 
because boxes do not contain open neighborhoods in the topology
induced by the Wasserstein metric (that is, the topology
of weak convergence),
but open neighborhoods do contain boxes.
\end{remark}

Of course, it is not obvious that the dynamical system needs to 
have fixed points, since $\nu$ could
dominate all fitness costs. This possibility
is easy to see in the one-dimensional case (when $\M$ is a single
point), which we described in Section~\ref{SS:counting} and which
we revisit in Section~\ref{SS:onedimensional}. However, we show
in Section~\ref{SS:existequil} that at least one fixed point 
exists when the mutation
measure $\nu$ is small enough. In order to go further, 
we need to impose additional assumptions.
For the case of multiplicative selection costs, 
Section~\ref{SS:multiplicative_equilibrium} 
gives a complete description
of the fixed points, of which there can be $0$, $1$ or $2$. 
\index{equilibrium}

In the remainder of this chapter we then impose
a weaker assumption, namely that the selective cost is concave, 
in the sense that the marginal cost of
an additional mutation decreases as more mutations are added 
to the genotype. (The formal definition is given as
the last condition in Theorem~\ref{T:increasing_condition}.) 
Under this assumption, we show in
Section~\ref{SS:concave} that trajectories starting 
from $0$ increase monotonically, and we
give a sufficient condition for them to converge 
to a finite equilibrium. This equilibrium is dominated 
by any other equilibrium.
If the driving vector field at nearby points above this minimal
equilibrium are in the negative ``orthant'', 
then trajectories starting above the equilibrium converge down
to it, so the minimal equilibrium is box attractive.
We give an iterative procedure for
computing the minimal equilibrium that avoids following the
dynamical system to large times in Section~\ref{SS:short-cut}.  
Moreover, the iteration
is a useful theoretical tool in 
Section~\ref{SS:concaveE:ODE}, where we establish that the equilibria
shown to exist for suitably small $\nu$ in 
Section~\ref{SS:existequil} are in fact the same
as the minimal equilibria in the concave setting
and that these equilibria are then box stable. Finally, in 
Section~\ref{SS:equilibrium_demographic}
we apply the above results to the demographic selective cost 
introduced in Section~\ref{SS:mortality}

These results cover many cases of substantive interest,
although they do not settle all relevant questions.  
Cases with very small $\nu( \M ) $ do
not wholly put on display the rich array of
differences between the full non-linear model and
non-epistatic additive models.  However, 
the conditions of Corollary~\ref{C:convergence_finite_equilibrium} 
and Theorem~\ref{T:concave_attractive}
can often be verified in specific cases,
and we have found that they can hold when
$\nu( \M )$ is only moderately small.

\section{Introductory example: One-dimensional systems}  
\label{SS:onedimensional}
\index{one-dimensional systems}

\begin{figure}
	\centering
		\includegraphics[width=1.00\textwidth]{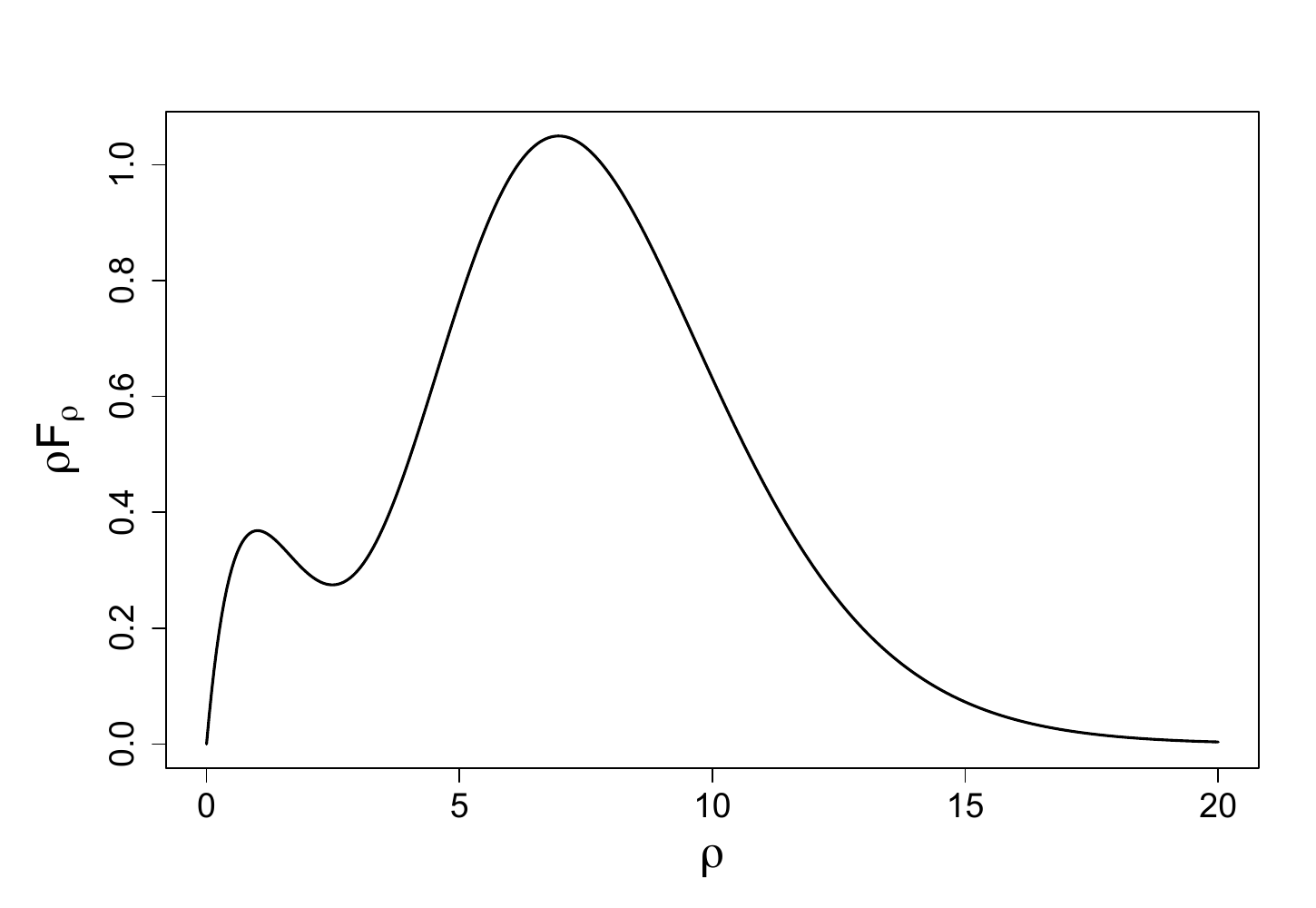}
	\caption{Plot of the function $- \rho F_{\rho}$ when $S(k)=\indic_{\{k\ge 1\}}+\indic_{\{k\ge 6\}}$.}
	\label{F:Frho}
\end{figure}

Suppose as in Section~\ref{SS:counting} 
that selective costs depend only on the number of mutations and
the original space of loci has been pushed forward into 
a space $\M$ consisting  of a single point. 
The space of genotypes $\G$ may be identified 
with nonnegative integers $\bN_0$,
the selective cost $S$ is simply an increasing function 
from $\bN_{0}$ to $\bR_{+}$, and the mutation measure $\nu$ 
is a positive constant. 
The space $\H^{+}$ of finite measures on $\G$ 
may also be identified with $\bR_{+}$. 
The dynamical system $(\rho_{t})_{t \ge 0}$ is $\bR_{+}$-valued and satisfies
\index{dynamical system}
\begin{equation} \label{E:1d}
\dot\rho_{t}=\nu - F_{\rho_{t}}\rho_{t},
\end{equation}
where
\begin{equation} \label{E:F1d}
F_{\rho} \rho = e^{-\rho}\sum_{k=1}^{\infty} \frac{\rho^k(k-\rho)}{k!} S(k).
\end{equation}
The system has equilibria at solutions to the equation
$\rho F_{\rho}=\nu$ and we
discussed some special cases in Section~\ref{SS:counting}.
In general, it is possible to construct selective costs
for which the number of equilibria is arbitrarily large 
for a given mutation rate. 
For example, suppose the selective cost has magnitude $1$ for
$1$ to $5$ mutations and magnitude $2$ for 
$6$ or more. The resulting function $-\rho F_{\rho}$ is shown in 
Figure~\ref{F:Frho}. 
The number of equilibria may be $0$, $1$, $2$, $3$, or $4$, 
depending on the value of $\nu$.

However, we can say quite generally the following things about one-dimensional systems.
\begin{itemize}
\item As long as $S$ is not identically $0$, the 
same is true for $F_{\rho}$,
so there is at least one equilibrium 
for $\nu$ sufficiently small.
\item The smallest equilibrium is attractive, 
unless it corresponds to a
local minimum of $-F_{\rho}\rho$, in which case it
attracts only trajectories
coming from below; trajectories starting above 
the equilibrium are repelled.
\item If $S$ is bounded above, then $-F_{\rho}\rho$ is bounded below, and
so there is no equilibrium for $\nu$ sufficiently large.
\end{itemize}
\index{attractive equilibrium}
\index{equilibrium}

\section{Introductory example: Multiplicative selective costs}  
\label{SS:multiplicative_equilibrium}
\index{multiplicative costs}

It is probably apparent that one
cannot hope to get anything like a closed-form solution
for the dynamical system \eqref{E:dynam} for general selective costs.
However, we show in this section that it is possible to
solve \eqref{E:dynam} explicitly when the selective cost has the 
multiplicative form
\begin{equation}  
\label{E:fc1}
S(g) = 1 - \exp\left\{-\int_\M \theta(m) \, dg(m)\right\}
\end{equation}
\glossary{$\theta(m)$}
for some $\theta:\M \to \bR_{+}$.  Moreover, in this case 
it is possible to analyze whether $\lim_{t \rightarrow \infty} \rho_t$
exists, and we couple this analysis with a comparison
argument in later sections to give sufficient conditions 
for the existence of a limit for more general selective costs.

Suppose that $\rho_0$ is absolutely continuous with respect to $\nu$.
It follows from Theorem~\ref{T:niceradon} that the measure $\rho_t$ has 
a Radon-Nikodym derivative $m \mapsto r_t(m)$ against $\nu$ for all $t \ge 0$ 
such that for each $m \in \M$ the function $t \mapsto r_t(m)$ is continuously
differentiable and solves the equation
$$
\frac{d r_t(m)} {dt} = 1 - \bE[S(X^{\rho_t} + \delta_m) - S(X^{\rho_t})] 
       \, r_t(m).
$$

Substituting from  \eqref{E:fc1}, we see that the
expected marginal cost is given by the product of  $  1 - e^{-\theta(m)} $
and    $ \bE  [ \exp\{ -\sum_{m' \in g} \, \theta(m') \}] $. 
The first factor $  1 - e^{-\theta(m)} $ is $S(\delta_m)$,
the selective cost of a singleton mutation,  which is the
same as $ F_0(m)$.  The second factor, the expectation, 
can be evaluated via the Poisson identity \eqref{E:Pidentity}
and it equals $ \exp\{ - \int ( 1 - e^{-\theta(m')}) \,  d\rho_t(m) \} $.
\index{selective cost}

We now define analogues of moments for $k \in \bN$ by the equations
$$
a_k  =  \int e^{-k \theta(m')}   \, d\nu(m')    
$$
\begin{equation}
\label{E:bkt}
b_k (t)  =  \int e^{-k \theta(m')}  \,  r_t(m') \, d\nu(m')   
\end{equation}
\glossary{$a_k$}
\glossary{$b_k$}

Our differential equation becomes
\begin{equation}
\label{E:ODE}
\frac{d r_t(m)} {dt} = 1 - (1 - e^{-\theta(m)}) \,
             e^{b_1(t) - b_0(t)} \,  r_t(m).
\end{equation}

This is an ordinary differential equation for each $m \in \M$ 
whose solution, a special case of  \eqref{E:RN},
is given by 
\[
\begin{split}
r_t(m) & =
\exp\left\{-\int_0^t (1 - \exp\{-\theta(m)\}) \exp\{b_1(s) - b_0(s)\} \, ds\right\} \\
       & \quad \times \left[ r_0(m) + \int_0^t 
\exp\left\{\int_0^s (1 - \exp\{-\theta(m)\}) \exp\{b_1(r) - b_0(r)\} \, dr\right\}
\, ds \right]. \\
\end{split}
\]
Thus, we have reduced what is in general an  infinite collection 
of coupled ordinary differential equations to the problem 
of finding two functions, $b_{0}$ and $b_{1}$.

Of course, this benefits us only if we have an 
autonomous system of equations in just the 
functions $b_{0}$ and $b_{1}$.  
Combining \eqref{E:ODE} with \eqref{E:bkt} we have
\begin{equation}
\label{integratedE:ODE}
\frac{d b_k(t)}{dt}
=
a_k + \exp\{b_1(t) - b_0(t)\} (b_{k+1}(t) - b_k(t)). 
\end{equation}
We introduce three  generating functions
\[
A(z)  := \sum_{k=0}^\infty a_k \, z^k \, / \, k!
\]
\[
B(z,t)  := \sum_{k=0}^\infty b_k(t) \, z^k \,  / \,  k!
\]
\[
C(z)  :=  B(z,0) = \sum_{k=0}^\infty b_k(0) \, z^k \,  / \,  k!.
\]
\glossary{$A(z)$}
\glossary{$B(z,t)$}
\glossary{$C(z)$}

The system of ordinary differential equations \eqref{integratedE:ODE} 
then becomes the first order, linear partial differential equation
\[
\frac{\partial B(z,t)}{\partial t}
=
A(z)
+
\exp\{b_1(t) - b_0(t)\}
\left[
\frac{\partial B(z,t)}{\partial z} - B(z,t)
\right].
\]

We may find the general solution of this PDE via the method
of characteristic curves.  Once we obtain the general solution 
(which involves the unknown functions $b_0$ and $b_1$), we have to
impose the boundary conditions 
\[
B(0,t) = b_0(t)
\]
and
\[
\frac{\partial B(0,t)}{\partial z} = b_1(t)
\]
to select the solution to our original differential equation.

Using {\em Mathematica}, the PDE is seen to have the solution
\[
\begin{split}
& B(z,t) \\
& \quad =
\int_0^t \exp\left\{-\int_u^t \exp\{b_1(s) - b_0(s)\}\, ds\right\}
    A\left(z +\int_u^t \exp\{b_1(s) -b_0(s)\} \, ds\right) \,
   du \\
& \qquad +
\exp\left\{- \int_0^t \exp\{b_1(s) - b_0(s)\} \, ds\right\} 
   C\left(z+\int_0^t \exp\{b_1(s) -b_0(s)\} \, ds\right) \\
\end{split}
\]

Therefore, the functions $b_0$ and $b_1$ solve the following 
system of equations expressed in terms of the generating
functions $A(z)$ and $C(z)$ and their derivatives $A'(z)$ 
and $C'(z)$ 
\[
\begin{split}
b_0(t) 
& =
\int_0^t \exp\left(-\int_u^t \exp\{b_1(s) - b_0(s)\}\, ds\right)
    A\left(\int_u^t \exp\{b_1(s) -b_0(s)\} \, ds\right) \,
   du \\
& \quad +
\exp\left(- \int_0^t \exp\{b_1(s) - b_0(s)\} \, ds\right) 
   C\left(\int_0^t \exp\{b_1(s) -b_0(s)\} \, ds\right) \\
\end{split}
\]
and
\[
\begin{split}
b_1(t) 
& =
\int_0^t \exp\left(-\int_u^t \exp\{b_1(s) - b_0(s)\}\, ds\right)
    A'\left(\int_u^t \exp\{b_1(s) -b_0(s)\} \, ds\right) \,
   du \\
& \quad +
\exp\left(- \int_0^t \exp\{b_1(s) - b_0(s)\} \, ds\right) 
   C'\left(\int_0^t \exp\{b_1(s) -b_0(s)\} \, ds\right). \\
\end{split}
\]
Note that this system of ordinary differential equations is autonomous.
It depends only on the unknown functions $b_{0}$ and $b_{1}$ themselves, 
together with the known functions $A$ (determined by the 
given mutation and selection) 
and $C$ (determined by the initial conditions). 
Hence, these equations could at least be solved numerically. 
(Of course, we started with a system with as many equations 
as there are points in $\M$. This reduction to a system 
of two coupled ordinary differential equations is only 
advantageous if the space $\M$ has more than two points.)

We can now obtain a necessary and sufficient condition
for the existence of an equilibrium for \eqref{E:dynam}
\index{equilibrium}
in $\H^{+}$ when the selection cost is the 
multiplicative one of \eqref{E:fc1}.
Note that if $\rho_*$ is an equilibrium, then 
we have
\begin{equation} \label{E:equilibrium}
\nu(dm) - \bE[S(X^{\rho_*} + \delta_m) - S(X^{\rho_*})] \, \rho_*(dm) = 0,
\end{equation}
and so $\rho_*$ has a Radon-Nikodym derivative $r_*$ with respect to $\nu$ that
satisfies
\[
r_*(m) = \frac{\exp\{b_0^* - b_1^*\}}{1 - \exp\{-\theta(m)\}}, 
\]
where $b_k^* := \int \exp\{-k \theta(m)\}  r_*(m) \, d\nu(m)$ for $k \in \{0,1\}$
(cf. \eqref{E:ODE}).  
\index{equilibrium}
Although the equation for $r_*$ sometimes makes sense
even when $\nu$ and $\rho$ have infinite total mass,
as discussed in Section~\ref{SS:infinitenu},
here we are only considering $\nu$ and $\rho$ in $\H^{+}$,
the space of finite positive measures.  
An equilibrium exists in this space if and only if, first, 
\[
\int \frac{1} {1 - \exp\{-\theta(m)\}} \, d\nu(m) < \infty
\]
and, second,  the conditions defining $b_0^{*}$ and $b_1^{*}$ 
are consistent with the formula for $r_{*}$, so that 
there is a constant $c > 0$ such that
\[
\begin{split}
c & = \int \frac{\exp\{c\}}{1 - \exp\{-\theta(m)\}}   \, d\nu(m)
- \int \exp\{- \theta(m)\} \frac{\exp\{c\}}{1 - \exp\{-\theta(m)\}}  \, d\nu(m) \\
& = \exp\{c\} \, \nu(\M), \\
\end{split}
\]
in which case $c = b_0^* - b_1^*$.
Such a constant exists if and only if 
\[
\nu(\M) \le \sup_{x \ge 0} x e^{-x} = e^{-1}.\] 
Note that there are three possible cases.
\begin{itemize}
\item 
If $\nu(\M)<e^{-1}$, then there are two equilibria, 
corresponding to the two distinct solutions of $c e^{-c}=\nu(\M)$. 
The equilibrium corresponding to the smaller
$c$ is attractive, while that corresponding to the larger $c$ is unstable.
\item 
If $\nu(\M)>e^{-1}$, then there is no equilibrium.
\item 
If $\nu(\M)=e^{-1}$, then there is a single equilibrium, 
which is unstable.
\end{itemize}
\index{attractive equilibrium}
\index{equilibrium}
\index{stable equilibrium}


We are now in a position to give an example which previews
the general approaches in the following sections.
Take $ \M = [1, \infty) \subset \bR_{+}$.
We generate multiplicative selective costs from
the function $ \theta(m) = \log(m/(m-1)) $,  so that
\index{multiplicative costs}
$$
S(\delta_m) = F_0(m) = 1 - e^{-\theta(m)}  =  1/m.     
$$
We take mutation measure to be the measure $\nu(dm) = u (2/m^3) \, dm $
with total mass $\nu(\M) = u $.
We have just seen that when $ u < e^{-1} $ there exists
an equilibrium whose Radon-Nikodym derivative with respect to $\nu$ is given by
$$
r_{*}(m)  =  e^{c(u)} \, m.  
$$
Here $c(u)$ is the smaller root of the equation $ c e^{-c} = u $. 

The equilibrium  Radon-Nikodym derivative is not bounded.  It increases
linearly with $m \in \M$, even though the equilibrium 
measure $\rho_{*}(dm) = r_{*}(m) \nu(dm) $ has finite total
mass and does belong to $\H^{+}$.   If we want 
a bounded continuous function of $m \in \M$,  we need 
to turn to the product   $ S(\delta_m) r_{*}(m) $.  
In this example the product equals  $ \exp\{c(u)\} \mathbf{1}$,
where $  \mathbf{1} $ is the function taking the value $1$
for every $m \in \M$.  

For every $m \in \M$, the product is a continuously differentiable function
of the total mass $u$ of the mutation measure and is
in fact for small $u$ the solution of the differential equation 
$$
\frac{d \phantom{u}}{du} S(\delta_m) r_*(m) 
= \mathbf{1} e^{c(u)} \frac{d c(u)}{du}.
$$
If we did not know $r_{*}$, we could arrive at this
equation by differentiating \eqref{E:equilibrium}.  
We exploit this approach to give a general construction
of equilibria for small $\nu(\M)$ in Section~\ref{SS:existequil}.

The scale factor in this example with multiplicative costs 
can be obtained by iteration of the map  $  c \mapsto  u e^c $
starting from $c = 0 $.   This scalar-valued iteration
corresponds to an iterative scheme for generating  $r_{*}$ itself,
namely  $  r  \mapsto  1/F_{r \nu} $.  For selective
costs which are concave, in a sense defined in Section~\ref{SS:concave},
an iterative scheme along these lines enables us to
prove stability results for eqilibria in Section~\ref{SS:concaveE:ODE}.

\section{Fr\'echet derivatives}  
\label{SS:frechet}
\index{Fr\'echet derivatives}

We need some machinery on derivatives of curves and
vector fields in order to analyze equilibria in more generality.
Consider a Banach space $(X, \| \cdot \|_X)$ and a closed convex cone
$X_+ \subseteq X$.  For $x \in X_+$, let $U_x$ be the closed convex
cone $\{t(x' - x) : t \ge 0, \, x' \in X_+\}$.
Consider another Banach space  $(Y, \| \cdot \|_Y)$.
Extending the usual definition slightly, we say that
a map $\Phi : X_+ \rightarrow Y$, is Fr\'echet differentiable 
at $x \in X_+$, if there is map $D_{x} \Phi : U_x \rightarrow Y$ 
with the properties
\[
\lim_{x' \rightarrow x, \, x' \in X_+} 
\|x'-x\|_X^{-1} \Bigl(\Phi(x')-\Phi(x)-D_{x} \Phi[x'-x] \Bigr)=0,
\]
$D_x \Phi[t z] = t D_x \Phi[z]$ for $t \ge 0$ and $z \in U_x$,
$D_x \Phi[z'+z''] = D_x \Phi[z'] + D_x \Phi[z'']$ for $z',z'' \in U_x$,
and, for some constant C, $\|D_x[z]\|_Y \le C \|z\|_X$ for $z \in U_x$.
\glossary{$D_x$}
It is not difficult to show that if $\Phi$ is differentiable at $x$,
then $D_x \Phi$ is uniquely defined.   
The definitions depend on the choice of cone.  We shall be 
applying them to the cone $\H^{+}$ and taking care to restrict 
directions to such positive measures.

As usual, we say that a curve $\psi: I \to X$, where $I \subseteq \bR$
is an interval, is differentiable at $t \in I$ if the limit
$$
\dot\psi_{t}=\lim_{t' \rightarrow t} (t'-t)^{-1}(\psi_{t'}-\psi_{t})
$$
exists.

For the sake of completeness, we record the following 
standard fundamental theorem of calculus and chain rule.

\begin{lemma}
\label{L:fundamental_theorem_calculus}
Consider an interval $I \subseteq \bR$
and a Banach space $(X, \| \cdot \|_X)$.
Suppose that a curve $\psi: I \rightarrow X$ is
differentiable at every $t \in I$ and the curve $t \mapsto \dot \psi_t$
is continuous.  Then,
\[
\psi_b - \psi_a = \int_a^b \dot \psi_t \, dt
\]
for all $a,b \in I$ with $a < b$.
\end{lemma}

\begin{lemma} 
\label{L:chainrule}
Consider an interval $I \subseteq \bR$,
two Banach spaces $(X, \| \cdot \|_X)$ and $(Y, \| \cdot \|_Y)$,
and a closed convex cone $X_+ \subseteq X$.
Suppose for some $t \in I$ that the 
curve $\psi: I \rightarrow X_+$ is differentiable at $t \in I$
and the map $\Phi:X_+ \rightarrow Y$ is differentiable at $\psi_t$.
Then, the curve $\Phi \circ \psi : I \rightarrow Y$ 
is differentiable at $t$ with
derivative $D_{\psi_t}\Phi[\dot \psi_t]$.
\end{lemma}

We also have a particular analogue of the product rule.
Recall that $\H$ is the Banach space of finite signed 
measures on $\M$ equipped with the Wasserstein metric,
and $\H^{+}$ is the cone of positive measures within $\H$.

\vskip 24pt
\begin{notation}
\label{N:tildeK}
\par\noindent
\begin{itemize}
\item
Write $C_{b}(\M,\bR)$ for the Banach space of bounded 
continuous functions from $\M$ to $\bR$
equipped with the supremum norm.
\item
For $\eta \in \H^{+}$ and $m',m'' \in \M$, define the kernel 
\[
\tilde{K}_{\eta}(m',m''):=
\mathbb{E}
\Bigl[ 
S(X^\eta +\delta_{m'}+\delta_{m''})
-S(X^\eta+\delta_{m''})
-S(X^\eta+\delta_{m'})
+S(X^\eta)
\Bigr].
\]
\end{itemize}
\end{notation}
\glossary{$C_{b}(\M,\bR)$}
\glossary{$\tilde{K}$}

\begin{lemma}
\label{L:prod}
Consider an interval $I \subseteq \bR$ and two curves
$\gamma: I \rightarrow \H^{+}$ and $f: I \rightarrow C_{b}(\M,\bR)$. 
Suppose that $\gamma$ and $f$ are differentiable at $t \in I$.
Define a curve $\beta: I \rightarrow \H^{+}$ by
$\beta_{u}:=f_{u}\cdot\gamma_{u}$, $u \in I$; 
that is, $\beta_u$ is the element of
$\H$ that has Radon-Nikodym derivative $f_u$ with respect to $\gamma_u$. 
Then, $\beta$ is differentiable at $t$ with 
\[
\dot \beta_{t}=\dot{f}_{t}\cdot \gamma_{t} + f_{t}\cdot \dot \gamma_{t}.
\]
\end{lemma}

\begin{proof}
This follows from Lemma~\ref{L:chainrule} 
with $X = C_{b}(\M,\bR) \times \H^{+}$,
$Y = \H$, $\Phi(e, \eta) = e \cdot \eta$, and $\psi = (f,\gamma)$ upon
showing that the map $\Phi$ is differentiable 
at any $(e, \eta) \in C_{b}(\M,\bR) \times \H$
with
\[
D_{e,\eta} \Phi[(e',\eta')] = e' \cdot \eta + e \cdot \eta'
\]
and the curve $\psi$ is differentiable at $t$ 
with $\dot \psi_t = (\dot f_t, \dot \gamma_t)$.
Both proofs are straightforward and we leave them to the reader.
\end{proof}

By our standing assumption,  the conditions of
Theorem~\ref{T:existence} are in place.
By the Lipschitz condition on $S$,  with constant $\sigma$, 
via Lemma~\ref{L:FLip},
the absolute value of the sum of the first two terms in $\tilde{K}$
is bounded by $ \sigma \| \delta_{m''} \|_{\Was} = \sigma $  and 
the absolute value of the sum of the second two terms 
is bounded by $ \sigma \| \delta_{m'} \|_{\Was} = \sigma  $.
No separate uniform bound on $S$ is required. 
Thus, the map $(\eta,m',m'') \mapsto \tilde{K}_{\eta}(m',m'')$
is bounded by $2\sigma $.  Ideas similar to those
behind  Lemma~\ref{L:poisson} and Lemma~\ref{L:Fpipi}
establish that this kernel gives the Fr\'echet derivative 
of the map $\eta \mapsto F_\eta(\cdot)$.
\index{Fr\'echet derivatives}

\begin{lemma} 
\label{L:Ffrechet}
The mapping $\eta \mapsto F_\eta(\cdot)$
from $\H^{+}$ to $C_b(\M,\bR)$
is Fr\'echet differentiable at every point $\eta' \in \H^{+}$
with derivative $D_{\eta'} F$ given by
\[
D_{\eta'}F [\eta''] (m')= \int_{\M} \tilde{K}_{\eta'}(m',m'') \, d\eta''(m'').
\]
\end{lemma}

Moreover, straightforward coupling arguments establish the
following bounds, where we recall that the constant $\sigma $
is such that $|S(g) - S(h)| \le \sigma  \|g - h\|_{\Was}$ for
all $g,h \in \G$.

\begin{lemma}
\label{L:Dbound}
For any $\rho,\rho',\eta\in\H^{+}$ with $\rho\le \rho'$,
\[
\bigr\| D_{\rho}F[\eta]\bigr\|_\infty
\le 
2\sigma  \eta(\M),
\]
and
\[
\bigr\| D_{\rho}F[\eta]-D_{\rho'}F[\eta]\bigr\|_\infty 
\le 
16 \sigma  \bigl(\rho'(\M)-\rho(\M)\bigr)\eta(\M).
\]
\end{lemma}


\section{Existence of equilibria via perturbation} 
\label{SS:existequil}


We now proceed to prove the existence of equilibria 
when the total mutation rate is sufficiently small.  
We fix a selective cost function $S$ and
define a family of mutation measures $\nu^{(u)}$ 
for $u\in \bR_{+}$ and some  $\zeta \in \H^{+}$ by
$$
\nu^{(u)}  =  u \, \zeta .  
$$
In other words, the mutation measures are scalar
multiples of each other differing only in total mass. 
A family of corresponding dynamical systems starting
from the null state is given by 
\index{dynamical system}
\begin{equation} \label{E:dynamu}
\rho^{(u)}_{t}=   u t \zeta  - \int_{0}^{t} D\rho^{(u)}_{s} ds.
\end{equation}
An equilibrium for any one of these systems has a 
Radon-Nikodym derivative with respect to $\zeta$.  

\begin{theorem} 
\label{T:existequil}
Consider a selective cost function $S$ with $S(0) = 0 $
that satisfies the monotonicity and Lipschitz conditions
of Theorem~\ref{T:existence}.  Suppose, moreover, that 
$$
\inf_{m \in \M} S(\delta_m) > 0.
$$ 
Then, there exists $U>0$ and a curve  
$(u,m) \mapsto p^{(u)}(m) $ for $(u,m) \in [0,U] \times \M $ 
such that the following hold.
\begin{itemize}
\item
For each $u \in [0,U] $ the function  $ m \mapsto p^{(u)}(m) $ 
is a bounded continuous function of $m \in \M$. 
\item
For each $m \in \M$,  the function $u \mapsto p^{(u)}(m)$ is a continuously differentiable
function of $u \in [0,U]$.
\item
The measure
$$
\rho^{(u)}(dm)  :=   p^{(u)}(m) \, \zeta (dm) 
$$ 
is a finite equilibrium for  \eqref{E:dynamu} in $\H^{+}$
for all $u \in [0,U]$.
That is, 
\begin{equation*}
\nu^{(u)} = u \zeta   =  F_{\rho^{(u)}} \cdot \rho^{(u)}.
\end{equation*}
\end{itemize}

\end{theorem}
\glossary{$p^{(u)}$}
\index{equilibrium}

\begin{proof}
If measures $\rho^{(u)}$ do exist satisfying the conditions
of the theorem,  then we expect their Radon-Nikodym derivatives to satisfy
an equation obtained by differentiating both sides of 
\[
\nu^{(u)} = u \zeta   =  F_{\rho^{(u)}} \cdot \rho^{(u)}. 
\]
Taking advantage of our
expression for the  Fr\'echet derivative of $F_{\rho}$ 
\index{Fr\'echet derivatives}
and applying Lemmas  \ref{L:prod} and \ref{L:Ffrechet}
with $\eta' = p^{(u)} \zeta  $
and $\eta'' = \frac{dp^{(u)}}{du} \zeta $, the
desired relationship comes out to be  
\begin{equation}
\label{E:afterfrechet}
\mathbf{1} =  F_{p^{(u)} \zeta}(m)  \frac{dp^{(u)}}{du}(m) 
       - p^{(u)}(m) \, \int_{\M} \, (-\tilde{K}_{p^{(u)} \zeta}(m, m')) 
             \frac{dp^{(u)}}{du}(m') \,  d\zeta(m'). 
\end{equation}
Here $\mathbf{1} \in C_{b}(\M,\bR_+)$ 
is the function with constant value $1$.
\glossary{$\mathbf{1}$}

Our strategy is to start with an operator equation of similar
form,  show that it does have solutions in $C_b(\M, \bR_{+})$,
derive a differential equation from them, solve it,
and identify $p^{(u)}$ with a suitable function of 
the solution.

For $p \in  C_b(\M,\bR)$, define the 
bounded linear operator $T_p: C_b(\M,\bR) \rightarrow C_b(\M,\bR)$ by
%
\begin{equation}
\label{E:Tpq} 
T_{p}(q):=  p_+(m') \, \left[\int_{\M} \tilde{K}_{p_+}(m',m'') 
              \, q(m'') \, d\zeta (m'')\right]
              + F_{p_+}(m') q(m').
\end{equation}
\glossary{$T_{p}$}
\glossary{$\D$}
\glossary{$L(p)$}

Let $\D$ be the set of functions $p \in C_b(\M, \mathbb{R})$
such that $T_{p}$ is invertible. It follows 
that  $0\in \D$, because $T_{0}(q)(m')=S(\delta_{m'}) q(m')$
and $\inf \{ S(\delta_{m}) : m \in \M \} > 0$ by assumption. 
A standard result in operator theory (see Lemma VII.6.1 of \cite{DS88}) 
tells us that the invertible operators form an open set 
in the operator norm topology,
so that $\D$ includes all $p$ such that $\|T_{p}-T_{0}\|$
is sufficiently small, where 
(here only) $\| \cdot \|$ denotes the operator norm.

By Lemmas \ref{L:Dbound} and \ref{L:Fpipi} 
we can bound $\|T_{p}-T_{0}\|$ in terms 
of the Lipschitz constant $\sigma $ on $S$ 
by 
\begin{align*}
|T_{p}-T_{0}\|  
    & \le 2 \sigma  \zeta(\M) \| p \|_{\infty} +  8 \sigma  \| p_+ \|_{\Was} \\  
    & \le 10 \sigma  \zeta(\M) \, \| p \|_\infty. 
\end{align*}
We conclude that $\D$ includes an open ball around $0$.

Define a map $L:\D \to C_{b}(\M,\bR_{+})$ by
$$
L(p):= T_{p}^{-1}(\mathbf{1}),
$$
\begin{align*}
0 &=  \mathbf{1} - \mathbf{1}  = 
             T_{p} [L(p)] - T_{0}[L(0)]       \\
  &= T_{0} [L(p)-L(0)]+ (T_{p}-T_{0}) [L(0)] + (T_{p}-T_{0})[L(p)-L(0)].
\end{align*}
Thus, 
$$
L(p) = L(0) - T_{0}^{-1} \, \Bigl(  (T_{p}-T_{0}) [L(0)]
          + (T_{p}-T_{0})[L(p)-L(0)] \,   \Bigr).
$$
Now  $ T_0^{-1} $ is the diagonal operator that multiplies a 
function of $m \in \M$ by the bounded function $m \mapsto 1/S(\delta_m) = L(0)$,
and we can combine our bound on the operator norm of $ T_p - T_0$
to show that for a suitable constant $c$ 
\begin{align*}
\| L(p) - L(0) \|_{\infty} 
     & \le  \| L(0) \|_{\infty} 
          \Bigl(  \| T_{p}-T_{0}\| \| L(0) \|_{\infty}
          + \| T_{p}-T_{0} \|  \| [L(p)-L(0)] \|_{\infty}\, \Bigr) \\ 
     & \le   
             c  \, \| p \|_{\infty}  \, \| L(0) \|^2_{\infty}  \,   
    +  c  \, \| p \|_{\infty} \,  \| L(0) \|_{\infty} 
           \| L(p) - L(0) \|_{\infty}.     
\end{align*}
By requiring $ \| p \|_{\infty} $ to be small enough, 
we can make the first term on the right arbitrarily small and
make  $  1 -  c \, \| p \|_{\infty} \,  \| L(0) \|_{\infty}  $
arbitrarily close to $1$, so we can 
make $ \| L(p) - L(0) \|_{\infty} $ arbitrarily small.     
Since $L(0)(m)$ is bounded away from 0, it follows 
that there is a neighborhood $\D' \subset \D $ of 0
such that $0 < \inf_{p\in \D'} \, \inf_{m\in \M} \, L(p)(m)$.

Furthermore, again by  Lemma VII.6.1 of \cite{DS88},  within
this neighborhood, the map from an operator to its inverse is
a homeomorphism  and  
$$
\| L(p') - L(p'') \|_{\infty} \le \frac{ \| L(p') \|_{\infty}^2 
              \| T_{p'} - T_{p''} \| }
              { 1 - \| T_{p'} - T_{p''} \| \| L(p') \|_{\infty} }. 
$$
It follows that $L(p)$ satisfies a Lipschitz condition.

Then, by standard results on existence and uniqueness of 
solutions to ordinary differential equations in a Banach space,  
the ordinary differential 
equation $\frac{dp^{(u)}}{du} = L(p^{(u)})$ 
with initial condition $p^{(0)}=0$ 
has a solution on an interval $[0,U]$ and this solution
takes values in $C_{b}(\M,\bR_+)$.
Thus, for $0 \le u \le U$,  $p^{(u)} = \int_0^u L(p^{(v)}) dv $ 
satisfies the requirements in the conclusion of the theorem.

\end{proof}

\section{Concave selective costs} 
\label{SS:concave}
\index{concave costs}

For an important class of examples, including the demographic example of 
Section~\ref{SS:mortality},   the selective costs are concave, 
in the sense that the marginal cost of adding a given 
mutation becomes smaller, the more other mutations are already present. 
\index{demographic costs}
Formally, this is stated in Definition~\ref{D:concave}. 
Under a few mild constraints, we show in
Theorem~\ref{T:increasing_condition} that concave selective costs yield
monotonic solutions $(\rho_{t})_{t \ge 0}$ when started from the
pure wild type population $\rho_0 = 0$, and hence such systems 
must either diverge to a measure with infinite total mass or converge
to an element of $\H^{+}$. 
Corollary~\ref{C:convergence_finite_equilibrium} gives a further 
condition that is sufficient to ensure that the limit is an 
element of $\H^{+}$. Our conditions for monotone increase in $\rho$ 
over time turn out to be satisfied quite generally for the applications
we have investigated.  The conditions for the existence of a
limit in $\H^{+}$, on the other hand,  are not always satisfied,  and there
are important cases (discussed in \cite{walls}) for which $\rho_t$ 
increases to a measure with infinite total mass.  

\begin{definition}
\label{D:concave}
A selective cost function $S$ is {\em concave} if
\begin{equation} 
\label{E:concave_definition}
S(g + h + k) - S(g+h) 
\le S(g+k) - S(g) \text{ for all }g,h, k\in \G.
\end{equation}
\end{definition}

\begin{lemma}
\label{L:concave}
A selective cost
$S$ is concave if and only if 
\[
S(g + \delta_{m} + \delta_{m'}) - S(g+\delta_{m}) \le S(g+\delta_{m'}) - S(g)
\] for all $g\in \G$
and $m,m'\in\M$.
%
Moreover, if $S$ is concave,  then 
\[
F_{\pi}(m)  \ge F_{\pi + \eta}(m)
\]
for $\pi, \eta \in \H^{+} $ and 
$m \in \M$.
\end{lemma}


\begin{proof}
Since elements of $\G$ have finite integer mass, 
we can prove \eqref{E:concave_definition} by induction on
$n:=h(\M) \vee k(\M)$. Our assumption is 
equivalent to \eqref{E:concave_definition} in the case
$n=1$. Assume now that \eqref{E:concave_definition} 
holds whenever $h(\M) \vee k(\M) \le n-1$. 
Suppose $h(\M)=n$ and $k(\M)\le n-1$. 
Let $m \in \M$ be in the support of $h$, and let $h=\tilde{h}+\delta_{m}$. Then
\begin{align*}
S(g + &h + k) - S(g+h) - S(g+k) + S(g)\\
&=S(g + k+\tilde{h}+\delta_{m}) - S(g+\tilde{h}+\delta_{m}) - S(g+k) + S(g)\\
&=\bigl[S(g + \delta_{m} + k+\tilde{h}) 
    - S(g+\delta_{m}+\tilde{h}) - S(g+\delta_{m}+k) + S(g+\delta_{m})\bigr]\\
 &\hspace*{2cm}+\bigl[ S(g+\delta_{m}+k) - S(g+\delta_{m})-S(g+k)+S(g) \bigr].
\end{align*}
Since $\tilde{h}$ and $k$ both have mass smaller than $n$, 
each of the terms in brackets is $\le 0$ by 
the induction hypothesis. To complete the induction, 
we need only address the case when $h(\M)=k(\M)=n$;
this case proceeds exactly as above.

Finally,  the Poisson random measures  $ X^{\pi + \eta} $ is
distributed like the sum of independent copies of the Poisson random
measures $X^{\pi}$ and $X^{\eta}$, 
and so     
\begin{align*}
F_{\pi + \eta}(m) 
  &=   \bE \left[ S( X^\pi + X^\eta + \delta_x) - S( X^\pi + X^\eta ) \right] \\
  &\le  \bE \left[ S(X^\pi  + \delta_x) - S( X^\pi  ) \right] \\
  &= \quad  F_{\pi}(m).    
\end{align*}
\end{proof}

\begin{theorem} 
\label{T:increasing_condition}
Fix a mutation measure $\nu\in\H^{+}$
and a selective cost $S:\G\to\bR_{+}$, 
that satisfies the conditions 
\begin{itemize}
\item
$S(0)=0$,
\item
$S(g)\le S(g+h)$ for all $g,h\in\G$,
\item
for some constant $\sigma $,
$\bigl| S(g)-S(h)\bigr| \le \sigma  \bigl\| g-h \bigr\|_{\Was}$,
for all $g,h\in \G$,
\item
$S(g + h + k) - S(g+h) \le S(g+k) - S(g)$
for all $g,h, k\in \G$.
\end{itemize}
If $\dot\rho_0 \ge 0 $
then the solution of \eqref{E:dynam}
guaranteed by Theorem~\ref{T:existence} satisfies $\rho_s \le \rho_t$
for all $0 \le s \le t < \infty$.
If $\dot\rho_0 \le 0 $, then  the solution satisfies
$\rho_{s} \ge \rho_{t}$ for all $0 \le s \le t < \infty$
\end{theorem}
\begin{proof}

By Definition~\ref{D:F}, 
$F_\eta(m) = \mathbb{E}[S(X^\eta + \delta_m) - S(X^\eta)]$
for $\eta \in \H^{+}$ and $m \in \M$.
By Lemmas \ref{L:FLip} and \ref{L:Fpipi}, 
thanks to the Lipschitz bound on $S$,  it is clear 
that  $\eta \mapsto F_\eta(\cdot)$  is a continuous map 
from $\H^{+}$ to $C_{b}(\M,\bR)$.

The curve $\rho$ is differentiable at each $t \ge 0$ and satisfies
\begin{equation} \label{E:keyequation}
\dot{\rho}_{t}=\nu - F_{\rho_{t}} \cdot \rho_{t}.
\end{equation}
The right-hand side is continuous in $t$. 
By Lemma~\ref{L:fundamental_theorem_calculus}, it then suffices
to show that $\dot\rho_{t} \ge 0$ for all $t \ge 0$.

By Lemma~\ref{L:chainrule} and Lemma~\ref{L:Ffrechet}, the curve
$t \mapsto F_{\rho_{t}}$, $t \in \bR_+$, is differentiable, with
\index{Fr\'echet derivatives}
$$
\frac{d\phantom{t}}{dt} F_{\rho_{t}} = D_{\rho_{t}}F[\dot{\rho}_{t}].
$$
The value of this derivative at $m'$ equals the integral
$  \int_{\M} \tilde{K}_{\rho_{t}}(m',m'') \, \dot{\rho}_{t}(dm'') $,
where the kernel $\tilde{K} $ is defined in Notation~\ref{N:tildeK}.
By the concavity condition on $S$, for all $m'$ and $m''$ in $\M$,
$$
0 \le  -\tilde{K}_{\rho_{t}}(m',m'') \le F_{\rho_{t}}(m').
$$
Furthermore, by Lemma~\ref{L:prod},
$$
\ddot\rho_{t}:=\frac{d\phantom{t}}{dt} \dot{\rho}_{t}
    = -\Bigl(D_{\rho_{t}}F[\dot{\rho}_{t}] \Bigr) \cdot \rho_{t} 
          - F_{\rho_{t}}\cdot \dot{\rho}_{t}.
$$

Suppose now that $\dot{\rho}_0 \ge 0$.
Define a negatively-rescaled version of $t \mapsto \dot{\rho}_t$ by 
$$
\gamma_{t}:=-\exp\left\{\int_{0}^{t}F_{\rho_{s}} \, ds\right\} \dot{\rho}_{t}.
$$
\glossary{$\gamma_{t}$}
Then,
\[
\begin{split}
\frac{d\gamma_{t}}{dt}
& =-\exp\left\{\int_{0}^{t}F_{\rho_{s}}ds\right\} \left(F_{\rho_{t}} \dot\rho_{t} + \ddot\rho_{t}\right) \\
& = -\exp\left\{\int_{0}^{t}F_{\rho_{s}}ds\right\} 
           \left(F_{\rho_{t}} \dot\rho_{t} 
          -\Bigl(D_{\rho_{t}} F[\dot{\rho}_{t}] \Bigr) \cdot \rho_{t} 
          - F_{\rho_{t}}\cdot \dot{\rho}_{t} \right) \\
  & = \exp\left\{\int_{0}^{t}F_{\rho_{s}}ds\right\} 
           \Bigl(D_{\rho_{t}}F\left[
           \dot{\rho}_{t}\right] \Bigr) \cdot \rho_{t}. \\
\end{split}
\]
%
%
By assumption, $\gamma_{0}=-\dot\rho_{0}\le 0$.
For any Borel set $B\subseteq\M$
$$
\gamma_{t}(B) \le \gamma_{t}(B)-\gamma_{0}(B) 
     =  \int_0^t  \,     
\frac{d\phantom{s}}{ds} \gamma_{s}(B) \,  ds.  
$$
Writing out the derivative in terms of the kernel $\tilde{K}$ and
expressing $\dot{\rho_t} $ in terms of $\gamma_t$, we
obtain 
\[
\begin{split}
\gamma_{t}(B)
    & \le
          \int_{0}^{t} 
          \Bigl(\int_B
          \Bigl[\int_{\M} -\tilde{K}_{\rho_{s}}(m',m'') 
           \exp\left\{-\int_{0}^{s}F_{\rho_{u}}(m'') \, du\right\} \, 
           d \gamma_{s}(m'') \Bigr]                       \\
   & \hspace*{2cm}\times
           \exp\left\{\int_{0}^{s} F_{\rho_{u}}(m') \, du\right\}
           \, d\rho_{s}(m') \Bigr) \, ds. 
\end{split}
\]
%
%
Let  $  \gamma_s  =  \gamma^+_s - \gamma^-_s $ be the
Hahn-Jordan decomposition of $\gamma_s $ into its positive and
negative parts.  
\index{Hahn-Jordan decomposition}
Since $-\tilde{K}$ and the exponential factors
are nonnegative, $\gamma_s$ can be replaced by its
positive part in the inequality for $\gamma_t(B)$
thanks to the upper bound on $-\tilde{K}$.  
The inner integral over $m''$ is bounded above
by $ \gamma^+_s(\mathcal{M}) F_{\rho_{s}}(m') $.
Hence, 
\begin{equation}
\gamma_{t}(B) \le  
          \int_{0}^{t} \gamma^+_s(\mathcal{M}) 
          \Bigl(\int_B F_{\rho_{s}}(m') 
          \exp\left\{\int_{0}^{s} F_{\rho_{u}}(m') \, du\right\}
           \, \rho_{s}(dm') \Bigr)  \, ds. 
\end{equation} 
For any positive $T$, set $C_T $ equal to the product of two bounds,
namely,  first,  the bound on the exponential multiplier  
$$
\exp\left\{\int_{0}^{s} F_{\rho_{u}}(m') \, du \right\} \, \le \, \sigma T 
$$
and, second, the bound of the integral 
$$
  \int_B  F_{\rho_{s}}(m') \, \rho_{s}(dm')  \le  \sigma  \rho_s(B) 
     \le \sigma  ( \rho_0(\M) + T \nu(\M) ).  
$$ 
Both bounds hold for $ s \le t \le T$ by Lemma~\ref{L:FLip},  
and $ \rho_s(\M) \le \rho_0(\M) + s \nu(\M) $ 
because $F$ and $\rho$ are nonnegative in \eqref{E:keyequation}.   
 
Since $C_T$ is finite, we have shown that
\begin{equation}
\gamma_{t}(B) \le
C_T \int_{0}^{t} 
      \gamma^+_s(\mathcal{M}) \, ds.
\end{equation}
Put $\beta_{t}:=\sup_{B \subseteq \M} \gamma_{t}(B)$,
where the supremum is taken over Borel sets including
the null set, so that $\beta_{t}$ is nonnegative.  
\glossary{$\beta_t$}
\glossary{$C_T$}
Now,   $ \gamma^+_s(\mathcal{M}) = \beta_s $.
Hence, we have shown that 
$$
\beta_{t}\le C_{T}\int_{0}^{t} \beta_{s}ds
$$
for $0\le t\le T$. By Gronwall's Inequality
(see Appendix~\ref{SS:Gronwall}), 
\index{Gronwall's Inequality}
this equation implies that $\beta_{t} \equiv  0$ for all $t$. 
It follows that the measure $\gamma_{t}$ is nonpositive.  
Thus, $\dot\rho_{t}$ (which differs from $\gamma_{t}$ 
by a strictly negative Radon-Nikodym factor) is nonnegative.
This finishes the proof of the claim for the case $\dot{\rho}_0 \ge 0$.

If $\dot\rho_{0}\le 0$, then we define $\gamma_{t}$ to 
be $+\exp\left\{\int_{0}^{t}F_{\rho_{s}}ds\right\}\dot{\rho}_{t}$, 
and the rest of the proof carries through as before. 
\end{proof}

\begin{corollary}
Suppose the conditions of Theorem~\ref{T:increasing_condition} 
hold and there exists $\rho_{**} \in \H^{+}$ satisfying the 
equilibrium condition 
\[
\nu(dm) = \bE\left[S(X^{\rho_{**}} + \delta_m) 
        - S(X^{\rho_{**}})\right] \, \rho_{**}(dm)
\]
for \eqref{E:dynam}. For the dynamic system 
$(\rho_t)_{t \ge 0}$ started at $\rho_{0}=0$,
$\rho_t \uparrow \rho_* \in \H^{+}$, where $\rho_* \le \rho_{**}$
and $\rho_*$ is also an equilibrium for  \eqref{E:dynam}.
\end{corollary}
\glossary{$\rho_{**}$}
\index{equilibrium}


The following Comparison Lemma, also proved via Gronwall's
Inequality, is a powerful tool for applications. 
\index{Gronwall's Inequality}
\index{comparisons lemma}
It treats pairs of solutions $ \rho'_t$ and $\rho''_t$ in 
which, informally speaking, $\rho'$ starts ahead of $\rho''$ 
and the marginal selective costs slowing the progress of $\rho'$ are 
always less than the marginal costs slowing $\rho''$. In such 
a race $\rho'$ always keeps the lead.  

\begin{lemma}
\label{L:comparison}
Consider two selective cost functions $S'$ and $S''$
that satisfy the conditions of Theorem~\ref{T:increasing_condition}.
Let $\rho'$ and $\rho''$ be the corresponding solutions of
\eqref{E:dynam}.
Suppose that $S'(g + \delta_m) - S'(g) \le S''(g + \delta_m) - S''(g)$
for all $g \in \G$ and $m \in \M$ 
and that $ \,  \rho_0' \, \ge \, \rho_0'' \, $.
Then, for all $ t \ge 0 $  we have $ \,  \rho_t' \, \ge \, \rho_t'' \, $.
\end{lemma}

\begin{proof}
%

Define a signed measure $\xi_{t}=\rho''_{t}-\rho'_{t}$ 
with Hahn-Jordan decomposition $ \xi_t = \xi^{+}_t - \xi^{-}_t$.
We seek to prove that the positive part $\xi^{+}_t$ is zero for all $t \ge 0$.

Set 
$$
\eta_{t} :=\rho'_{t}\wedge \rho''_{t} = \rho''_t - \xi^{+}_t 
           =  \rho'_t - \xi^{-}_t
$$
and
$$
\beta_{t} :=\sup_{A \subseteq \M}\{\xi_{t}(A)\}= \| \rho''_{t}- \eta_{t}\|_{TV},
$$
where the supremum is over the Borel subsets of $\M$.
Here $ \| \cdot \|_{TV} $ is the total variation norm, which, 
as remarked in Section~\ref{SS:wasserstein}, 
dominates the Wasserstein metric on $\H$.
\glossary{$ \| \cdot \|_{TV} $}
\glossary{$\beta_{t}$}
\index{total variation norm}
\index{Wasserstein metric}
The function $\beta$ is nonnegative,  and we proceed to show that
it is actually zero, so that there is no Borel set on which $\rho''$
is bigger than $\rho'$.

Let $(m,\rho) \mapsto F'_{\rho}(m)$ 
be the expected cost function corresponding to $S'$,
and $(m,\rho) \mapsto  F''_{\rho}(m)$ 
be the expected cost function corresponding to $S''$. 
As before, via Lemmas \ref{L:FLip} and \ref{L:Fpipi}, 
for fixed $\rho$,  $F'_{\rho}(\cdot) $ and $F''_{\rho}(\cdot) $ 
can be regarded as elements of the space 
$C_{b}(\M,\bR)$ of bounded continuous functions on $\M$
equipped with the supremum norm.  
Because $\eta \le \rho' $, our assumptions imply the inequalities 
$$
F'_{\rho'}(m) \le F''_{\rho'}(m) \le F''_{\eta}(m) 
$$
for all $m \in \M$.

By Lemma~\ref{L:Fpipi} we have the Lipschitz bound 
$$
\sup_{m \in \M}  \left(F''_{\rho''_{t}}(m) - F''_{\eta_{t}}(m)\right)   =   
      \left\|F''_{\rho''_{t}}-F''_{\eta_{t}}\right\|_{\infty}  
   \le 8 \sigma  \left\|\rho''_{t}-\eta_{t}\right\|_{\Was}   \le 8 \sigma \beta_t .
$$
By Lemma~\ref{L:FLip} we also have $ \| F'' \|  \le 2 \sigma  $.  
Bearing in mind that $ F''$ is nonnegative, we see that 
\begin{align}\label{E:xidot}
\dot\xi_{t} 
    &=    F'_{\rho'_{t}} \cdot \rho'_t   -  
         F''_{\rho''_{t}} \cdot \rho''_t               \\  \nonumber 
    &=  - (F''_{\rho^*_{t}}-F'_{\rho'_{t}})\cdot \rho'_{t} 
         +   (F''_{\rho^*_{t}}-F''_{\rho''_{t}})\cdot \rho''_{t} 
         +   (F''_{\rho^*_{t}})\cdot (-1) 
                     \cdot (\rho''_{t} - \rho'_{t} )    \\ \nonumber
    &\le     0   + 8 \sigma  \beta_t  \rho''_{t} 
                   + 2 \sigma  \, \xi^{-}_{t}.
\end{align}

Our assumption that $ \rho'_0 \ge \rho''_0 $
makes $\xi^{+}_{0}  =  0 $. 
Let $(s,m) \mapsto x_s(m) $ be the function supplied by Theorem~\ref{T:niceradon}
with the properties that the function $m \mapsto x_s(m)$
is a Radon-Nikodym derivative of $\xi_s$ with
respect to the measure $\zeta := \rho_0 + \nu$ for every $s \ge 0$ and
for every $m \in \M$ the function $s \mapsto x_s(m)$
is a continuously differentiable function. 
Write $J$ for the indicator function of the 
set of $\{(s,m) \in \bR_{+} \times \M : x_s(m) > 0\}$.
\glossary{$J(s,m)$}
For a Borel set $A$ we have, by Lemma~\ref{L:abcon},  
$$
\xi^{+}_t(A) =  \int_0^t  \left( \int_{A} \, J(s,m) \, 
            d \dot{x}_s(m) \,  d\zeta(m) \right) \, ds.
$$
Since $ J(s,m) \, x^{-}_s(m) $ vanishes, we also have 
$$
\int_A \, J(s,m) \dot{x}_s(m)  \, d\zeta(m)  \,
     \le  \, 8 K \beta_s  \int_A J(s,m) \, d \rho''_s(m) .
$$
Thus, the inequality \eqref{E:xidot}  on $\dot{\xi}_{s} $ implies 
$$
\xi^{+}_t(A)   \le   
      8 K  \, \int_0^t  \beta_s \, \rho''_{s}(\M) \,  ds.    
$$
Note that $\sup_{s \in [0,T]} \rho''_s(\M) \le \rho''_0(\M) + T \nu(\M) $.  Taking the supremum over Borel sets $A$
on the left-hand side, we have a new constant $ K' $
such that
$$
\beta_t  \,  \le \,   K' \,  \int_0^t  \beta_s  \, \, ds. 
$$ 
Gronwall's Inequality (Appendix~\ref{SS:Gronwall})
\index{Gronwall's Inequality}
then gives  $\beta_{t}  =  0$ for all $t$ in the interval $[0,T]$, 
and hence for all $t \ge 0$.  Thus,  $ \rho''_t \le \rho'_t $ for all $t \ge 0$.  
\end{proof}

An alternative proof along the same lines takes advantage of 
the $L^{\infty}$ norm.  It is crafted so that it does not
depend on the assumption of finite total mass for $\nu$.  

\begin{lemma}
\label{L:comparedensities}
Consider two selective cost functions $S'$ and $S''$ 
and corresponding solutions $\rho'$ and $\rho''$ under
the conditions of Theorem~\ref{T:L-inf-existence}.  
Suppose for all $g \in \G$ and $m \in \M$ that 
$$
S'(g + \delta_m) - S'(g) \le S''(g + \delta_m) - S''(g).
$$
Suppose also that 
$$
\rho_0' \, \ge \, \rho_0'' .  
$$
Then, for all $ t \ge 0 $  we have 
$$
 \rho_t' \, \ge \, \rho_t'' .
$$
\end{lemma}

\begin{proof}
Set $\eta_{t} = \rho'_{t}\wedge \rho''_{t}$. 
Theorem~\ref{T:niceradon} supplies functions
$(t,m) \mapsto r'_t(m) $ and $(t,m) \mapsto  r''_t(m)$ 
such that $m \mapsto r'_t(m) $ and $m \mapsto  r''_t(m)$
are the Radon-Nikodym 
derivatives  of $ \rho'_t $ and $ \rho''_t $ with
respect to $\zeta = \rho'_0 + \nu$ for each $t \ge 0$,
and $t \mapsto r'_t(m) $ and $t \mapsto  r''_t(m)$ 
are continuously differentiable for all $m \in \M$.
Write $q_{\nu}$ for the Radon-Nikodym 
derivative of $\nu$ with respect to $\zeta$.
For $\pi \in \K^{+}$, let $F'_{\pi}$ and $F''_{\pi} $ 
be the expected cost functions corresponding 
respectively to $S'$ and $S''$. 

Set 
$ x_t(m) =  r''_t(m) - r'_t(m)$.
Let $J$ be the indicator function of the subset 
of $\{(t,m) \in \bR_{+} \times \M : x_t(m) > 0\}$,
which, by Theorem~\ref{T:niceradon} is Borel measurable. 
We now appeal to Lemma~\ref{L:abcon}.
Separately for every $m \in \M$ for all $t \ge 0$ 
\begin{align*}
x_t(m) & J(t,m)  =    x_0(m)  \, 
        + \, \int_0^t \dot{x}_s(m) J(s,m)  ds         \\  
    &=  x_0(m) + \, \int_0^t  \, \left[ \,  F'_{\rho'_{s}}(m) \, r'_s(m)  
        \, - \,   F''_{\rho''_{s}}(m)  \, r''_s(m)  \right] 
          \, J(s,m) \, ds.   
\end{align*}

We write the integrand as the sum of three terms as follows
\begin{align*}
\label{E:xdotofs}
\dot{x}_{s}(m) J(s,m)   =  
     & + \, \left[ \, F''_{\eta_{s}}(m) 
              \, - \, F''_{\rho''_{s}}(m)\right] \,  r''_s(m) J(s,m) \\  
     & - \, \left[ \, F''_{\eta_{s}}(m)
              \,- \,  F'_{\rho'_{s}}(m) \right]  \, r'_s(m)  J(s,m) \\ 
     & + \, \left[ \, F''_{\eta_{s}}(m) \right] (-1) 
          \left( r''_{s}(m) - r'_{s}(m) \right) \,  J(s,m).   
\end{align*}

The third term is never positive,  since $J(s,m)$ vanishes 
whenever $  r''_{s}(m) - r'_{s}(m)  $ is negative. 
The second term is never positive,  since the
assumed inequality on the marginal costs 
makes $F'_{\rho'_s}(m) \le F''_{\rho'_s}(m)$ for all $m \in \M$
and the concavity condition arranges for $ \eta_s \le \rho'_s $
to imply  $F''_{\eta_s}(m) - F''_{\rho'_s }(m) \ge 0 $ for all $s \ge 0$ 
and $m \in \M$. 
In contrast, the first term is never negative and the factor $J(s,m)$
is redundant, by the same concavity argument applied to $\rho''_s$. 

The Lipschitz condition on $F''$ bounds the first term 
by the 
quantity 
\[
\sigma  \|  \rho''_s - \eta_s \|_{\infty} r''_s(m) 
=
\sigma \|  x_s(m) J(s,m ) \|_{\infty} r''_s(m).
\]
By assumption, the contribution of the starting state $x_0$ is negative,
so we conclude for all $m \in \M$ that 
$$
x_t(m) J(t,m)   \le  \int_0^t  \|  x_s(m) J(s,m ) \|_{\infty} r''_s(m) \, ds. 
$$ 

For $ t \in [0,T] $, the Radon-Nikodym derivative $r''_s(m)$ is bounded 
by $ r''_0(m) +  q_{\nu}(m) T $, so there is a new constant $K'$ such
that the essential supremum of the left hand side,  
namely  $ \beta(t) := \|  x_s(m) J(s,m ) \|_{\infty} $,
satisfies 
$$
\beta_t  \le  K' \, \int_0^t \, \beta_s  \, ds. 
$$
Gronwall's Inequality (Appendix~\ref{SS:Gronwall}) then 
forces $ \beta_t  \equiv 0 $ for $ t \in [0,T] $  and so for all $t \ge 0$.
\index{Gronwall's Inequality}
\end{proof}

\begin{remark}
The proof of  Lemma~\ref{L:comparedensities} 
does not depend on the finiteness of $\nu(\M)$ and the lemma 
remains valid in the broader setting developed
in Section~\ref{SS:infinitenu}. 
\end{remark}

\section{Concave selective costs: 
Existence and stability of equilibria} 
\label{SS:concave_stability}

If the conditions of Theorem~\ref{T:increasing_condition} hold, 
trajectories starting from $0$ either converge 
as time goes to infinity to an equilibrium state in $\H^{+}$ or diverge
to a measure with infinite total mass.
We therefore wish to consider conditions that ensure 
the existence of an equilibrium with finite total mass. 
One approach is to compare the concave
selective cost to a multiplicative selective cost. 
This produces the small benefit over the general existence result 
of Theorem~\ref{T:existequil} of providing an explicit value of $\nu(\M)$ 
that is small enough to guarantee the existence of finite equilibria.
\index{multiplicative costs}
\index{equilibrium}

\begin{corollary}
\label{C:convergence_finite_equilibrium}
\begin{itemize}
\item[(a)]
Suppose that the selective cost $S: \G \rightarrow \bR$
satisfies the conditions of Theorem~\ref{T:increasing_condition}
and also satisfies the bound
\begin{equation*} 
S(g+\delta_{m'}) - S(g) 
\ge 
\xi \left[1 - \exp\{-\tau(m')\}\right] 
      \exp\left(-\int_\M \tau(m'') \, dg(m'')\right)
\end{equation*}
for all $m' \in \M$ for some constant $\xi > 0$ 
and function $\tau:\M \rightarrow \bR_+$ such that
\[
\int_\M \frac{1}{1 - \exp\{-\tau(m)\}} \, d\nu(m) < \infty.
\]
Suppose also that $\rho_0$ is the null measure $0$ 
and $\nu(\M) \le e^{-1} \xi$.
Then, there exists
$\rho_* \in \H^{+}$ such that $\rho_t \uparrow \rho_*$ 
as $t \rightarrow \infty$.
\item[(b)]
Conversely, if there exist some constant $\xi$ and
function $\tau$ such that the reverse inequality 
\begin{equation*} 
S(g+\delta_{m'}) - S(g) 
\le 
\xi \left[1 - \exp\{-\tau(m')\}\right] \exp\left\{-\int_\M \tau(m'') \, dg(m'')\right\}
\end{equation*}
holds and $\nu(\M)> e^{-1}\xi$, then $\lim_{t\to\infty}\rho_t(\M) =\infty$.
\end{itemize}
\end{corollary}
\begin{proof}
Consider part (a). Let $\rho''$ be the solution of \eqref{E:dynam} with 
selective cost $S''(g) = \xi \int_\M(1 - \exp\{-\tau(m)\} \, dg(m)$ 
and initial condition
$\rho_0'' = \rho_{**}$, where
\[
\rho_{**}(dm) = \frac{\exp\{c\}}{1 - \exp\{-\tau(m)\}} \, \nu(dm)
\]
with $c \xi = \exp\{c\} \nu(\M)$ (such a $c$ exists by
the assumption that $\nu(\M) \le e^{-1} \xi$).  It follows
from the results of Section~\ref{SS:multiplicative_equilibrium} 
that $\rho_t'' = \rho_{**}$ for all $t \ge 0$.  

Apply Lemma~\ref{L:comparison} with $S' = S$ 
\index{comparisons lemma}
and $\rho_0' = 0$ to conclude that $\rho_t \le \rho_{**}$ for
all $t \ge 0$.  It follows from
Theorem~\ref{T:increasing_condition}
that $\rho_t \uparrow \rho_*$ as $t \rightarrow \infty$ 
for some $\rho_* \in \H^{+}$ with $\rho_* \le \rho_{**}$.

Now consider part (b). 
We define $\rho''$ as before, with selective cost $S''$, 
but with initial condition $\rho''_{0}=0$. 
Lemma~\ref{L:comparison} implies then that $\rho_{t}\ge \rho''_{t}$ 
for all $t\ge 0$. We know that $\rho''_{t}$ is increasing in $t$, 
and there is no finite equilibrium. 
Suppose $R:=\lim_{t\to\infty} \rho_{t}(\M) < \infty$. 
Then, for any Borel set $A$, the quantity $\rho''_{t}(A)$ 
is increasing in $t$ and bounded by $R$, 
so it converges to a limit $\rho_{*}''(A)$. 
It is easy to check that $A \mapsto \rho_{*}''(A)$
is a measure in $\H^{+}$ 
and that $\rho_t''$ converges to $\rho_*''$ in the Wasserstein metric 
(that is, in the topology of weak convergence). 
From \eqref{E:dynam} we know that 
for any Borel set $A$, and any $s<t$,
$$
0\le \rho_{t}''(A)-\rho_{s}''(A) 
=
\int_{s}^{t} \bigl( \nu(A)-D'' \rho_{u}'' (A) \bigr) \, du,
$$
where the operator $D''$ is the analogue of the operator $D$ 
when the selective cost $S$ is replaced by the selective cost $S''$.
We conclude that $\nu(A)-D'' \rho_{u}'' (A)\ge 0$ 
for all $u\ge 0$, and so
$$
\bigl|\rho_{*}''(A)-\rho_{t}''(A) \bigr| 
=
\left|\int_{t}^{\infty} \bigl( \nu(A)-D'' \rho_{u}'' (A) \bigr) \, du\right|
\downarrow 0 \text{ as } t \to \infty.
$$
Since $D'' \rho$ is continuous in $\rho$ by Lemma~\ref{L:FLip}, 
it follows that the integrand on the right-hand side
converges to $\nu(A)-D'' \rho_{*}''(A)$, which must then be $0$. 
Since this is true for all Borel sets $A$, 
it would follow that $\rho_{*}''$ would be an equilibrium for
the dynamical system with selective cost $S''$, 
contradicting the fact that no such equilibrium exists.
\end{proof}

The monotone growth of systems with concave fitness cost 
functions allows us to derive a simple sufficient condition for
stability of the ``minimum equilibrium'' $\rho_{*}$, that is,  
the fixed point to which the dynamical system converges when started from $0$. 
We know that all trajectories that start strictly 
below $\rho_{*}$ converge asymptotically to $\rho_{*}$. 
Since the vector field vanishes at $\rho_{*}$, 
this implies that the derivative of the vector 
field $-D_{\rho_{*}} F[\eta] \cdot \rho_{*}  - F_{\rho_{*}} \cdot \eta$ 
is nonpositive for all positive directions $\eta$. 
The equilibrium is box stable if 
this nonpositivity extends to a neighborhood 
of $\rho_{*}$, which can be guaranteed if the 
derivative is actually bounded away from 0, 
measured by its Radon-Nikodym derivative with respect to $\nu$.
\index{box stable}
\index{box attractive}
\index{equilibrium}

\begin{theorem}
\label{T:concave_attractive}
\begin{itemize}
\item[(a)] 
Suppose that the selective cost function satisfies the 
conditions of Theorem~\ref{T:increasing_condition},
and the curve $(\rho_{t})_{t \ge 0}$, started at $\rho_{0}=0$, 
converges to a finite fixed point $\rho_{*}$. 
Suppose further that
\begin{equation*}
C_{\rho{*}} := \inf_{m\in\M}D_{\rho_{*}} F[\nu] (m)\frac{d\rho_{*}}{d\nu}(m)  
     + F_{\rho_{*}}(m)>0.
\end{equation*}
Then, the fixed point is box stable.
\item[(b)] Moreover, if $\M$ is compact and the equation
\[
D_{\rho_{*}} F[\eta] \cdot \rho_{*}  + F_{\rho_{*}} \cdot \eta=0
\]
has no solution $\eta\in\H^{+}$ 
that is absolutely continuous with respect to $\nu$,
then $\rho_{*}$ is box attractive.
\end{itemize}
\end{theorem}
\index{attractive equilibrium}
\index{equilibrium}
\index{stable equilibrium}

%
\begin{proof} 
Consider part (a).  
Define an intensity measure $\psi_{s}$ which is offset above 
the equilibrium $\rho_{*}$ by a constant multiple of the 
mutation measure $\nu$.
That is, $\psi_{s}:=\rho_{*} + s\nu$. 
Let $\phi_{s} = \nu - F_{\psi_{s}}\cdot\psi_{s}$ 
be the driving vector field evaluated at the point $\psi_s$.
We express the derivative of $\phi_{s}$ with respect to $s$
in terms of the quantity whose infimum is bounded below
in the definition of $C_{\rho{*}}$.  This quantity would be
denoted by $ T_p [\mathbf{1}] \cdot \nu $ in the notation 
of  \eqref{E:Tpq} with $ \rho_* = p \cdot \nu$,
but $p$ does not necessarily belong to $C_b(\M, \bR_{+})$.

The derivative of $\phi_{s} $ with respect to $s$ is given by
\begin{align*}
- \, \frac{d \phi_{s}}{ds}  &= 
     (D_{\rho_{*} + s \nu} F)[\nu] \left(  \rho_{*} + s \nu \right)
             +   F_{\rho_{*} + s \nu} \cdot \nu        \\ 
       &= \left[ \, (D_{\rho_{*} + s \nu} F)[\nu]  
              \, - \, (D_{\rho_{*}} F)[\nu]  \right] 
              \,  \left(  p \, +  \,m s \right) \cdot \nu              \\   
       & \quad + \,  (D_{\rho_{*}} F)[\nu]  \, \rho_{*} 
         +  (D_{\rho_{*}} F)[\nu]  \, s \, \nu          \\ 
       & \quad + \,  ( F_{\rho_{*} + s \nu} - F_{\rho_{*}}) \cdot \nu  \\        
       & \quad + \,  (F_{\rho_{*} })  \cdot \nu.              
\end{align*}

Lemma~\ref{L:Dbound} guarantees that 
\[
\left|
\left[ \, (D_{\rho_{*} + s \nu} F)[\nu]  
              \, - \, (D_{\rho_{*}} F)[\nu]  \right]
\right|
\le
16 \sigma s \nu(\M)^2,
\]
where $\sigma$ is the
Lipschitz constant for the selective cost $S$. 
By the same lemma,  
\[
\left| D_{\rho_{*}} F[\nu]s \right| \le 2 \sigma s  \nu(\M).
\]
By Lemma~\ref{L:Fpipi},  
\[
\left| F_{\rho_{*} + s \nu} - F_{\rho_{*}} \right|
\le
8 \sigma s \nu(\M). 
\] 
The remaining two terms are the terms bounded below 
by  the assumption that
$C_{\rho{*}} >0$.     We have shown that
the derivative of $\phi_{s}$ with respect to $s$ has a Radon-Nikodym
derivative with respect to $\nu$ bounded below for all $m \in \M$ by 
$$
C_{\rho_{*}} - s( 10 \sigma \nu(\M) + 16 \sigma \nu(\M)^2 \,  ( \| p \|_{\infty} + s)).
$$ 
For small enough positive $s$ this lower bound is positive.
For such $s$, starting from $\phi_{0}=0$, 
$$
\phi_{s}  =\int_{0}^{s} \frac{ d\phi_{u}}{du}  \, du     \\
   =   -  \int_{0}^{s} \bigl(D_{\psi_{u}}F[\nu]+F_{\psi_{u}}\cdot \nu \bigr) \, du
       < 0. 
$$
Applying Theorem~\ref{T:increasing_condition}, 
we see that when $\rho_{0}=\psi_{s}$, 
the trajectory $\rho_{t}$ is monotonically decreasing. 
Thus, for all times $t\ge 0$ we have $\rho_{t}\in B(\rho_{*},\psi_{s})$,
where $ B(\rho_{*},\psi_{s}) $ is the box defined in
 \eqref{E:definebox}. 

Now consider part (b). 
Suppose that there is no $\eta \in \H^{+}$
that is absolutely continuous with respect to $\nu$ and satisfies
$D_{\rho_{*}} F[\eta] \cdot \rho_{*}  + F_{\rho_{*}} \cdot \eta=0$.
For all $s$ sufficiently small, the trajectory 
starting from $\psi_{s}$ is monotonically decreasing,
and converges to an equilibrium $\rho_{*}^{(s)}$. 
If $\rho_{*}$ were not box attractive, then these equilibria 
would be distinct from $\rho_{*}=\rho_{*}^{(0)}$. 
Since the $\rho_{*}^{(s)}$ are all between $\rho_{*}$ and $\rho_{*}+s\nu$,
we know that $\rho_{*}^{(s)}-\rho_{*}$ is absolutely continuous 
with respect to $\nu$ and belongs to $\H^{+}$.
Consider the measures $\pi^{(s)}$, defined by
$$
\pi^{(s)}= \frac{\rho_{*}^{(s)}-\rho_{*}}{\rho_{*}^{(s)}(\M)-\rho_{*}(\M)}.
$$
These are probability measures on $\M$. 
Since $\M$ is compact, the space of
probability measures on $\M$ is also compact (recall that
the Wasserstein metric induces the topology of weak convergence), 
so there is an accumulation point.
Let $s_{1},s_{2},\dots$ be a sequence converging to $u$, 
with $\pi^{(s_{i})}\to \eta$.
Since the vector field vanishes at all of these points, 
the derivative in direction $\eta$
vanishes as well, contradicting our assumption, 
and hence proving that $\rho_{*}$
is in fact box attractive.
\end{proof}

\section{Iterative computation of the minimal equilibrium}
\label{SS:short-cut}

The measure $\pi \in \H^{+}$ is an equilibrium 
for the dynamic equation \eqref{E:dynam} if
it is true that $F_{\pi} \cdot \pi = \nu$.  
Recall that $F_{\pi} \cdot \pi$ is the
measure that has Radon-Nikodym derivative
$F_\pi$ with respect to $\pi$.  
The equation implies that $\pi$ is absolutely continuous
with respect to $\nu$ with a Radon-Nikodym 
derivative $p$ that satisfies
\begin{equation}
\label{RN_equilibrium}
F_{p\nu}(m) p(m) = 1, \quad m \in \M.
\end{equation}
\index{equilibrium}

Under the conditions of Theorem~\ref{T:increasing_condition},
if \eqref{RN_equilibrium} has a solution, then it has a minimal
solution that arises as the Radon-Nikodym derivative 
with respect to $\nu$ of $\lim_{t \rightarrow \infty} \rho_t$ 
when $\rho$ is the solution of \eqref{E:dynam} with $\rho_0 = 0$.

If one is only interested in finding the minimal equilibrium 
numerically, then it would be desirable to be able 
to do so without having to solve \eqref{E:dynam}.  
An obvious approach to that problem is to define a sequence
of functions $p_n: \M \rightarrow \bR_+$ inductively by $p_0 = 0$ and
\[
p_{n+1} = \frac{1}{F_{p_n \nu}}, \quad n \ge 0.
\]
\glossary{$p_n$}

By Lemma~\ref{L:concave},  it is clear that if $S$ is concave, 
%
%
then $F_{p'\nu} \ge F_{p''\nu}$ for $p' \le p''$. 
Because $p_0 = 0 \le p_1$, it follows 
that $p_0 \le p_1 \le p_s \le \ldots$.
Moreover, if $p_{**}$ is any solution of  \eqref{RN_equilibrium} 
then $p_{**}$ has to be a fixed point of the 
map $ p \mapsto 1/F_{p \nu} $,   so  $ p_n  \le p_{**} $ for all $n$.
If there is a solution $p_{**}$ 
such that $\int_{\M}  p_{**}(m) \, \nu(dm) < \infty$,
then $p_n \uparrow p_* \le p_{**}$ as $n \rightarrow \infty$
for some function $p_*: \M \rightarrow \bR_+$ such that $p_{*}\nu$
assigns finite mass to $\M$. 
Thus, if a limit exists in $\H^{+}$ for $\rho_t$ 
started from $\rho_0 = 0 $ as time goes to infinity,  it is
greater than or equal to the iterative limit $p_{*} \nu $.
But such a limit over time is also less than or equal to any
other equilibrium by Lemma~\ref{L:comparison}, the Comparison Lemma.
So, starting from zero,  when the iterative limit and the limit over time
exist,  they are the same limit,  the so-called ``minimal equilibrium''.
\index{minimal equilibrium}

\begin{remark}
If $\pi \in \H^{+}$ is an equilibrium for \eqref{E:dynam} with
Radon-Nikodym derivative $p$ against $\nu$, then, from
\eqref{RN_equilibrium},
\[
\begin{split}
\pi(\M)
& = \int_\M p(m) \, d \nu(m)
  = \int_\M \frac{1}{F_p(m)} \, d\nu(m) \\
& \ge \int_\M \frac{1}{\sup_{g\in\G} S(g+\delta_{m})-S(g)} \, d\nu(m)
   = \int_\M \frac{1}{S(\delta_m)} \, d\nu(m) \\
\end{split}
\]
under the concavity assumption.
Thus, a necessary condition for the existence of
an equilibrium in $\H^{+}$ is that the last integral is finite.
\end{remark}

\section{Stable equilibria in the concave setting via perturbation} 
\label{SS:concaveE:ODE}

Suppose that the selective cost is concave.
In Section~\ref{SS:existequil} we constructed equilibria 
for sufficiently small mutation measures.
We know from Section~\ref{SS:concave} that all trajectories starting
below an equilibrium $\rho_{**}$ converge asymptotically 
to an equilibrium that is also dominated by  $\rho_{**}$. 
This leaves open the question of whether the equilibrium 
constructed by perturbing the dynamical
system away from $\nu\equiv 0$ is the same as the minimal 
equilibrium $\rho_{*}$ to which the
system converges when started from $0$.

\begin{theorem}
\label{T:concave_stability}
Suppose the selective cost function satisfies the Lipschitz condition,
the concavity condition, and the other conditions 
of Theorem~\ref{T:increasing_condition}.
Suppose also that $ \inf_{m \in \M} S(\delta_m) > 0$.  Then the following hold.
\begin{itemize}
\item[(a)]
For $U>0$ sufficiently small, there is a 
unique $p:[0,U]\to C_{b}(\M, \bR_+)$ solving the equation
\[
\left[\int_{\M} \tilde{K}_{p^{(u)}}(m',m'') \, \dot{p}^{(u)}(m'') 
           \, \nu(dm'')\right] p^{(u)}(m')
      + F_{p^{(u)}}(m')  \dot{p}^{(u)}(m') = 1,
\]
with $p^{(0)}\equiv 0$. 
The  measure $ p^{(u)}\nu \in \H^{+}$ is the minimal equilibrium for the
system with mutation measure $u \nu$ for all $u\in [0,U]$. Furthermore,
the minimal equilibria so realized for $u<U$ are box stable. 
\item[(b)]
Moreover, if $\M$ is compact and the equation
\[
 D_{\rho_{*}} F[\eta] \cdot \rho_{*}  + F_{\rho_{*}} \cdot \eta=0
\]
has no solution $\eta\in\H^{+}$ with $\eta$ absolutely 
continuous with respect to $\nu$,
then $\rho_{*}$ is box attractive.
\end{itemize}
\end{theorem}
\index{attractive equilibrium}
\index{box stable}
\index{box attractive}
\index{equilibrium}
\index{stable equilibrium}

\begin{proof}
Following the method of Section~\ref{SS:short-cut}, 
we may compute the minimal equilibrium as the
limit of the iteration
\[
p_{n+1} = \frac{u}{F_{p_{n}}},
\]
with $p_0 \equiv 0$. Let $q_n$ be the corresponding 
iterates for $v \nu$, where $0 \le u < v$.

The concavity of $S$ implies via Lemma~\ref{L:concave}
that $p_n \le q_n$ for all $n$ and
\[
\begin{split}
q_{n+1} - p_{n+1}
   & = \frac{v}{F_{q_{n}}} - \frac{u}{F_{p_{n}}} \\
%
%
   & = \frac{v - u }{F_{q_{n}}} 
    + \frac{u (F_{p_{n}} - F_{q_{n}})}{F_{p_{n}} F_{q_{n}}}. \\
\end{split}
\]
%
By Lemma~\ref{L:Fpipi} the factor $ F_{p_n} - F_{q_n}$
in the second term on the right is no greater 
than $ 8 \sigma \| q_n - p_n \|_{\infty} \, \nu(\M)$. 

We now need upper bounds on the reciprocals of $F_p$ and $F_q$.
Thanks to the lower bound on $S(\delta_m)$, the
conditions of  Theorem ~\ref{T:existequil} are satisfied.
That theorem provides a curve of equilibria for $u$ in $[0,U]$
for some $U > 0 $.  To avoid confusion with the minimal
equilibria that we are constructing here,
denote these equilibria from Theorem ~\ref{T:existequil}
as $ x^{(u)}(m) \nu(dm) $. 
By Lemma~\ref{L:comparison},  $ x^{(u)} \le x^{(U)} $,
and because $ x^{(u)} $ is a fixed point of the
map  $ p \to u/F_{p \nu} $ and $ 0 \le x^{(u)} $,
we have $ p_n  \le x^{(u)} \le x^{(U)}$ for all $n$.  
Similarly $q_n \le x^{(v)} \le x^{U)}$ for $ 0 < u \le v \le U$.
It follows from  \eqref{E:Frecip} that
\begin{equation}\label{E:recipbound}
\frac{1}{F_{p_n}(m)} \le \frac{1}{F_{q_n}(m)} 
        \le \frac{\exp\left\{ \int_{\M} x^{(U)}(m') \, d\nu(m') \right\} }
                 {\inf_{m' \in \M} S(\delta_{m'})}.   
\end{equation}
%

For fixed $U$, write $a$ for the right-hand side of the 
inequality \eqref{E:recipbound},  which is constant over values of $m \in \M$,
and write $ b = 8 \sigma \nu(\M) a^2  u $.   We now have 
\[
   \|q_{n+1} - p_{n+1}\|_\infty \le a |v - u| + b \|q_n - p_n\|_\infty.
\]
By choosing $ U' $ in $[0, U]$ small enough and 
requiring $ 0 < u \le v < U' $, we can make $ b $ less than $1$.
Iterating while bearing in mind that $ q_0 - p_0  = 0 $,
we find  $ \|q_{n+1} - p_{n+1}\|_\infty \le (a/(1-b)) \,  |v - u| $
for all $n$ Hence, with $c = a/(1-b)$,  the corresponding minimal 
equilibria, say $p_{*}$ and $q_{*}$, satisfy
$$
\|q_{*} - p_{*}\|_\infty \le c |v - u|. 
$$

We now drop the stars subscripting equilibria and
write $u \mapsto p^{(u)}$ for the curve of Radon-Nikodym derivatives of
minimal equilibria whose dependence on $u$ we have 
been investigating.  We have just verified the 
Lipschitz condition 
$$
0 \le p^{(v)}(m) - p^{(u)}(m) \le  c(v-u)
$$
for all $m \in \M$.  The condition guarantees 
that $u \mapsto p^{(u)}(m)$ is Lebesgue almost everywhere
differentiable, and is the integral of its derivative. 
Since $p^{(u)}$ satisfies the relation $p^{(u)} F_{p^{(u)}}=u$ 
for every $u < U'$, we see
by differentiating with respect to $u$ that $p^{(u)}$ satisfies the
differential equation in the statement of the theorem.
By standard uniqueness results for ordinary differential equations,
it is the unique solution with $p^{(0)} = 0$.  
It therefore agrees with the solution provided by 
Theorem~\ref{T:existequil}, here called $x^{(u)}$ but there 
(and henceforth) called $p^{(u)}$, defined for all $u$ in $[0,U]$.

It remains to verify box stability and box attractivity.
Fix $u\in [0,U)$. The dynamical system
with mutation measure $U\nu$ converges to a finite equilibrium 
$p^{(U)}\nu\in \H^{+}$, where $p^{(U)}\ge p^{(u)}$. Let $(\rho_{t})_{t \ge 0}$
be the dynamical system with mutation measure 
$U\nu$ started from $\rho_{0}=p^{(u)}\cdot \nu$
and let $(\rho'_{t})_{t \ge 0}$
be  the dynamical system with mutation measure 
$u\nu$ started from $\rho'_{0}=p^{(U)}\cdot \nu$. 
Thus, $(\rho_t)_{t \ge 0}$ starts below its equilibrium, 
and $(\rho_t')_{t \ge 0}$ starts above its equilibrium. 
We have
\begin{align*}
\dot{\rho'}_{0} &=u \nu-F_{p^{(U)}\nu} p^{(U)}\cdot \nu = (u-U)\nu\le 0,\\
\dot{\rho}_{0}  &=u \nu-F_{p^{(u)}\nu} p^{(u)}\cdot \nu = (U-u)\nu \ge 0.
\end{align*}
Therefore, $p^{(U)}>p^{(u)}$, and we can conclude from
Theorem~\ref{T:increasing_condition}
that the measure $\rho_t$ stays bounded 
between the measures $p^{(u)}\cdot \nu$ 
and $p^{(U)}\cdot \nu$ for all times $t \ge 0$, and hence 
the minimal equilibrium $p^{(u)}\cdot \nu$ is box stable.

Now consider part (b). 
We know that the system with mutation measure $u\nu$ converges monotonically
downward to an equilibrium when started from $p^{(v)}\nu$. 
The final part of the proof, showing that this 
equilibrium is in fact $p^{(u)}\cdot\nu$,
proceeds exactly as in the previous proof of Theorem~\ref{T:concave_attractive}.
\end{proof}

\section{Equilibria for demographic selective costs} 
\label{SS:equilibrium_demographic}
\index{demographic costs}


As an example of Corollary~\ref{C:convergence_finite_equilibrium},
suppose that $S$ is the demographic selective cost 
of Section~\ref{SS:mortality}, so that
\begin{equation}\label{E:demmargcost}
\begin{split}
& S(g + \delta_{m} ) - S(g) \\ 
     & \quad =  \int_0^{\infty} \left ( 1 - e^{-\theta(m,x)} \right) f_x 
\exp \left( -\lambda x - \int_{\M} \theta(m' ,x) \, dg(m') \right) \, dx. \\
\end{split}
\end{equation}
Here, as before,  $\theta$ specifies an increment to the 
cumulative hazard function.  
This selective cost is concave, allowing us to 
capitalize on the results of Sections \ref{SS:concave}
and \ref{SS:concaveE:ODE} so long as the $\theta$ 
are such that the required Lipschitz condition 
on the selective cost is satisifed,  a requirement that is
messy to formulate in general but easy to verify for
most specific applications.    
\index{hazard function}
\index{cumulative hazard}
%
\begin{lemma}\label{L:deminf}
Suppose that the selective cost function $S$ is the demographic
cost of  \eqref{E:demmargcost}.   
Suppose also the age-specific profiles $\theta$ satisfies the following two 
conditions.
\begin{itemize}
\item
The supremum $ \tau :=  \sup_{m \in \M, x \in \bR_+} \theta(m,x)$ is finite. 
\item
The infimum $\inf_{m \in \M, x \in B} \theta(m,x)0$ is strictly positive for
some Borel set $B \subseteq \bR_+$ such that the integral $\int_B f_x \, dx$ is strictly positive.
\end{itemize}
Then, 
$$
\inf_{m \in \M} S(\delta_m)  > 0 
$$  
and
$$
S(g + \delta_m) - S(g) \ge   
e^{- \tau g(\M)} \inf_{m \in \M} S(\delta_m)
$$
for all $g \in \G$.
\end{lemma}

\begin{proof}
For all $m \in \M$,
\begin{align*}
S(\delta_m) &= F_{0}(m) 
    = \int (1 - e^{-\theta(m,x)}) f_x e^{-\lambda x} \, dx \\ 
    &\ge \int_B \left( 1 - \exp\{-\inf_{m \in \M, y \in B} \theta(m,y)\} \right)  
                  f_x e^{-\lambda x } \, dx   \, > \, 0. 
\end{align*}
We also have
\[
\begin{split}
& S(g + \delta_m) - S(g) \\ 
& \quad \ge 
  \int_0^{\infty} \left ( 1 - e^{- \theta(m,x)} \right) 
       f_x \, e^{-\lambda x}  
       \exp \left\{  - \sup_{m'' \in \M, x'' \in \bR_+} \, \theta(m'' ,x'') g(\M) \right\} \, dx  \\
& \quad =  e^{- \tau  g(\M) } \,  
       \int_0^{\infty}  1 - e^{- \theta(m,x)} 
       \, f_x \, e^{-\lambda x}  \, dx       \\
& \quad \ge  e^{-\tau g(\M) }  \,   
            \inf_{m \in \M} S(\delta_m). \\
\end{split}
\]
\end{proof}

It follows from the first implication
in Corollary~\ref{C:convergence_finite_equilibrium} 
that the dynamical system started from the null state 
with these costs converges to a finite equilibrium $ \rho_{*}$
in $\H^{+}$. 
In other words, if there is a range of fertile ages over
which all deleterious mutations reduce survival by at
least some minimal amount,  then selection keeps the total intensity
from going to infinity provided $\nu(\M)\le e^{-1}\xi$.

%

The converse implication 
in Corollary~\ref{C:convergence_finite_equilibrium} 
cannot be readily applied to this setting, at least not with
a constant function $\tau$.  However, in some cases we can
give a direct proof that no equilibrium exists with finite
total mass. 

For example, suppose that $S$ is the demographic selective cost
of Section~\ref{SS:mortality}, with $\M = [\alpha,\beta]$
for $0 < \alpha < \beta < \infty$, 
$\nu$ is a constant multiple of Lebesgue measure on $\M$,
$f_x$ is constant, and $\theta(m,x) = \eta \indic_{[m,\beta]}(x)$
for some constant $\eta$.
Such a simplified model has featureless fertility between two
ages that represent the onset and end of reproduction, 
mutations associated with effects at specific ages, constant
mutation rate during the reproductive span, and 
equal increments to the hazard from all mutations.
Similar models were introduced and studied by 
Brian Charlesworth in \cite{bC01}.
\index{Charlesworth, Brian}
\index{hazard function}
For a suitable constant $K'$,
%
$$
S(g+\delta_m) - S(g)
  \le       K' \left ( 1 - e^{-\eta} \right)
  \left[ \exp \left( -\lambda m \right) 
       - \exp \left( -\lambda \beta \right) \right]
   \le K' (\beta - m ). 
$$

The bound on the right-hand side is independent of $g$,
so the expected value of the marginal cost has the same bound.
Any equilibrium $\rho_* $  must satisfy  $ \rho_* = \nu/F_{\rho_*}$.
Since
$$
\int_{\alpha}^{\beta}  \frac{1}{F_{\rho_*}(m)} dm   \ge 
     \int_\alpha^\beta \frac{1}{\beta - m } \, dm = \infty,
$$
an equilibrium with finite total mass does not exist.
Of course, \eqref{RN_equilibrium} may have a solution $p$
such that $p \cdot \nu$ has infinite total mass and so
does not belong to $\H^{+}$.  In the setting of 
Theorem~\ref{T:increasing_condition}, the measure $\pi = p \cdot \nu$
could still arise as the increasing limit of the solution $\rho_t$ to
the dynamic equation \eqref{E:dynam} and be such that the population 
average selective cost $\mathbb{E}[S(X^\pi)]$ would be finite.  
These more delicate possibilities are treated in \cite{walls}.


We now investigate the box stability and box attractivity 
of the minimal equilibrium.
\index{demographic costs}
\index{equilibrium}
\index{stable equilibrium}

\begin{theorem}\label{T:boxdemog}

Suppose the selective cost function $S$ has the demographic
form specified in  \eqref{E:demmargcost}, 
(or in \eqref{E:ellxofg} and \eqref{E:demogcosts})   
with $\rho_0 = 0$ and $\int f_x \, dx < \infty$.   
Suppose the age-specific effect profiles $\theta$ 
are such that the selective cost satisfies the Lipschitz
condition of Theorem~\ref{T:existence}  and also 
satisfies the following two conditions.
\begin{itemize}
\item
The supremum $ \tau := \sup_{m \in \M, x \in \bR_+} \theta(m,x)$ is finite. 
\item
The infimum $\inf_{m \in \M, x \in B} \theta(m,x)$ is strictly positive for
some Borel set $B \subseteq \bR_+$ such that the integral $\int_B f_x \, dx$
is strictly positive.
\end{itemize}
Set  $ \xi :=  \inf_{m \in \M} S(\delta_m))/( 1 - e^{-\tau})$.
Then, when $\nu(\M) <  (\xi / e )  \wedge (1/ e )  $
the measures $\rho_t$ converge monotonically as $t \to \infty$ to an 
equilibrium $\rho_{*}$ which is box stable.  
If $\M$ is also compact, this equilibrium is box attractive.
\end{theorem}
\index{attractive equilibrium}
\index{box stable}
\index{box attractive}

\begin{proof}
For a measure $\rho \in \H^{+}$, we 
write $A_{\rho}(x)$ for the aggregate population net maternity
function, that is, 
\[
\begin{split}
A_{\rho}(x) 
& =  
\mathbb{E}[f_x \ell_x(X^\rho)] \\
& =
f_x \, e^{-\lambda x}  \exp\left(- \int_{\mathcal M} 
             (1 - e^{-\theta(m,x)}) \, d\rho(m)  \right).
\end{split}
\]
\glossary{$A_{\rho}(x)$}
\index{net maternity function}

The expected marginal cost $ F_{\rho}$ is given by
$$
F_{\rho}(m')  = \int ( 1 - e^{-\theta(m',x)} ) \, A_{\rho}(x) \, dx,
$$
and the kernel $\tilde{K}$ for computing its Fr\'echet derivative 
defined in Notation~\ref{N:tildeK} turns out to be 
\index{Fr\'echet derivatives}
$$
\tilde{K}(m', m'') = - \, \int ( 1 - e^{-\theta(m', \, x)} )
                 ( 1 - e^{-\theta(m'', \, x)} ) \, A_{\rho}(x) \, dx.
$$

The existence of a finite minimal equilibrium $\rho_{*} \in \H^{+}$  
when $\nu(\M) < \xi / e $ is guaranteed,  
as we have seen, by  Corollary~\ref{C:convergence_finite_equilibrium}.
For sufficiently small values of $\nu(\M)$,  $\rho_{*}$ is the
same as the equilibrium guaranteed by Theorem~\ref{T:existequil}
because we can bound $S(\delta_m)$ from below thanks to 
Lemma~\ref{L:deminf}.   

We derive an upper bound on $\rho{*}$.  
By hypothesis,  $\nu(\M)$ is strictly less than the smaller
of $\xi/e$ and $1/e$. 
From the Comparison Lemma~\ref{L:comparison} and 
the discussion in Section~\ref{SS:multiplicative_equilibrium}
(substituting $\nu/\xi $ for $\nu$), 	
we know that $\rho_{*}$ is bounded above by an 
equilibrium solution with multiplicative costs.
Specifically,  
$$
( 1 - e^{-\tau} ) \rho_{*}  \le  e^{c} \cdot \nu 
$$
where $c$ is the unique solution in $[0,1]$ of the
equation  $ c e^{-c}  = z $ for $ z = \nu(\M)/\xi $. 
As we have seen in Section~\ref{SS:counting}, 
this solution exists when $ \nu(\M) \le \xi/e $ and
can be expressed in terms of Lambert's W-function,
dating back to 1758.  Moreover, the map $ z \mapsto c $
is convex,   lying below the line $ z \mapsto ez $,
so our hypothesis $\nu(\M) < 1/e $  
yields $ c \xi < 1 $, implying $ e^c \nu(\M) \le 1 - \epsilon $ 
for some $\epsilon > 0 $.   Since $\rho_{*}$ has
a bounded Radon-Nikodym derivative, call it $r_{*}$,
with respect to $\nu$, the function $ F_{\rho_{*}}(m)$ is bounded 
below, away from zero. 

To establish the claims about stability and attractivity,
we appeal to Theorem~\ref{T:concave_attractive}.
Consider any measure $\eta$ in $\H^{+}$ which is
absolutely continuous with respect to $\nu$ with
Radon-Nikodym derivative  $ q_{\eta}(m)$,
supplying a direction for a perturbation of $\rho_{*}$. 
As before, we put $\psi_s = \rho_{*} + s \eta $ 
and $\phi_s = \nu - F_{\psi_s} \cdot \psi_s $.
The derivative at $ s = 0 $ of $ -\phi_s $ with respect
to $s$ is the measure  $ T_{r_*}( q_{\eta}) \cdot \nu $
in the notation of \eqref{E:Tpq}. 
Expressing the Fr\'echet derviative of $ F_{\rho_{*}}$ in 
terms of the kernel $\tilde{K}$, this measure is given by 
%
\[
\begin{split}
& T_{r_{*}}( q_{\eta}) \cdot \nu (dm') \\
      & \quad = - \rho_{*}(dm') \, \int_{\M}  \, 
        \left[ \int ( 1 - e^{-\theta(m',x)} ) ( 1 - e^{-\theta(m'',x)} ) \, 
          A_{\rho_{*}}(x) \, dx \right] \, \eta(dm'')    \\  
      &\quad + \, \eta(dm') \, \int ( 1 - e^{-\theta(m', \, x)} ) 
            \, A_{\rho_{*}}(x) \, dx. \\  
\end{split}
\]
Examine the nonnegative function of $m'$ and $x$ 
multiplying $- \rho_{*}(dm')$ in the first term on the right-hand side.
Because $ 1 - \exp\{ - \theta(m'', x)\} \le 1 - \exp\{-\tau\} $ 
for all $m''$ and $x$,  this multiplier is bounded above 
by $ (1 - \exp\{-\tau\}) \eta(\M) F_{\rho_{*}}(m') $.  
The nonnegative function multiplying $\eta(dm')$
is $F_{\rho_{*}}(m')$.  
Hence,
\begin{equation}\label{E:etaminusrho}
T_{r_{*}}( q_{\eta}) \cdot \nu   \, 
     \ge \,  \left( F_{\rho_{*}} \right)  \left(  \, \eta  
         \, - \,  \eta(\M) (1 - e^{-\tau}) \rho_{*} \right).
\end{equation}
Taking advantage of the bound on $\rho_{*}$, in terms of
Radon-Nikodym derivatives, 
\begin{equation}\label{E:Trq} 
T_{r_{*}}( q_{\eta})             
     \ge   \, \left( F_{\rho_{*}} \right)  \left(  \, q_{\eta} \, 
         - \, (1 - \epsilon)    \eta(\M)/\nu(\M) \right).         
\end{equation}

We now use Theorem~\ref{T:concave_attractive}. 
For box stability we put $\eta = \nu $, so that $ q_{\eta} \equiv 1 $,
and, as required,  $\inf_{m \in \M}  T_{r_{*}}( \indic )(m) $ 
is bounded below by $ \inf_{m \in \M} F_{\rho_{*}}(m) \, \epsilon > 0$. 

For box attractivity, when $\M$ is compact, we let $\eta$ be
any probability measure in $\H^{*}$. As required, the left-hand side
of  \eqref{E:Trq} cannot be almost-surely zero,
since the integral over $\M$ of the right-hand side with respect
to $\nu$ is bounded below by $\inf_{m \in \M} F_{\rho_{*}}(m) \epsilon$.
\end{proof}  

With demographic costs,  assuming $\M$ to be compact
is not an onerous restriction.  Taking advantage of
the boundedness of demographic costs,   $\M$ can often
be embedded in a compact space without disrupting 
the Lipschitz condition on $S$,  and 
Theorem~\ref{T:boxdemog} then comes into force.

\chapter{Mutation-selection-recombination in discrete time}  
\label{Ch:discrete_time}

We devote the remainder of this work
to establishing rigorously the claim that
we argued heuristically in the Introduction: 
The continuous-time dynamical system \eqref{E:model_informal}, 
which we defined formally in \eqref{E:dynam}, 
is the limit of a sequence of discrete-time infinite
population mutation-selection-recombination
models in the standard asymptotic regime where 
selection and mutation are weak relative to recombination and
both scale at the same infinitesimal rate in the limit.

More specifically, we show that our continuous time model
is a limit of the sort of discrete generation, 
infinite population
mutation-selection-recombination models considered by
Barton and Turelli \cite{BT91} and Kirkpatrick, Johnson, 
and Barton \cite{KJB02}, once such models have been extended
to incorporate our more general definition of genotypes.
\index{Barton, Nicholas} 
\index{Turelli, Michael}
Such a result not only justifies our model as a tractable approximation 
to more familiar models in the literature, 
but, as we remarked in the Introduction,
it also marks out the range of relative strengths for mutation,
selection, and recombination where this approximation
can be expected to be satisfactory,
and thus where the resulting conclusions that underly applications to the
demographic study of longevity can be trusted.

\section{Mutation and selection in discrete-time}  
\label{SS:frame}


This section describes the operations of mutation and selection in
the discrete-time setting,  leaving the description of 
recombination to Section~\ref{SS:recom}.  

We consider a sequence of models indexed by the positive integers.
In the $n^{\mathrm{th}}$ model, time is sped up by a factor
of $n$, so that $n$ generations pass in $1$ unit of time.  
Unlike familiar diffusion approximations for genetic
processes,  all models in the sequence are infinite-population
models; $n$ does not have any reference to population size.  
The distribution of genotypes in the population is
described, as before, by a probability measure 
on $\G$, the space of finite integer-valued measures on the 
set $\M$ of loci.  

At each generation each new birth gets a random set
of extra mutations away from the ancestral wild type added to 
those mutant alleles it inherited from its parents.  
In the $n^{\mathrm{th}}$ model in our sequence, 
the loci at which these mutations occur are distributed as 
the Poisson random measure $X^{\nu/n}$ with intensity $\nu/n$.    
The fitness of a genotype $g \in \G$ is $e^{-S(g)/n}$.  
The $n$'s  in the denominators are what, in effect, speed up time.
Mutation and selective cost per unit time are maintained.  
In a finite, non-zero amount of time the population is asymptotically
subject to finite, non-zero ``amounts'' of mutation and selection.


We now define the four principal operators that act on
the space of probabilities on $\G$.  The 
operators $\mM_{n}$,  $\mS_{n}$, $\mR$, and $\mP$   
describe the transformation in genotype distribution by, 
respectively, one round of mutation,  one round of selection, 
one round of recombination,  and a process of complete
Poissonization in which a genotype with possibly
linked loci is replaced by one in which
there is no linkage.  The $\mM_n$,  $\mS_n$, and $\mP$  operators
are defined in this section. The $\mR$ operator, requiring
further structure, is defined in the following section. 
The intuition behind these definitions is described in the Introduction.
In this chapter we use the notation $F$ 
without a subscript to stand for a bounded Borel
test function.  There is no connection with the subscripted 
function $F_{\pi} $ defined in Definition~\ref{D:F}.  
We also use the letter $Q$ to stand for a generic probability
measure, so that $P$ remains available for Poisson probability
measures like the measures that describe the continuous-time
dynamical system.

\begin{notation}  \label{N:operators}
Given a probability measure $Q$ on $\G$, define new probability
measures $\mM_{n} Q$, $\mS_{n} Q$, and $\mP Q$  by their action on
a bounded Borel test function $F:\G \to \bR_+$ as follows 
\begin{equation}
\label{E:Mop}
\mM_{n} Q [F]:= \int_\G \bE[F(g + X^{\nu/n})] \, dQ(g)
\end{equation}
\begin{equation}
\label{E:Sop} 
\mS_{n} Q [F]:=\frac{\int_\G \exp\{-S(g)/n\} F(g) \, dQ(g)}
                    {\int_\G \exp\{-S(g)/n\} \, dQ(g)}
\end{equation}
$$
\mR Q [F] \hbox{ is defined in Notation~\ref{N:recom}  }
$$
\begin{equation}
\label{E:Pop}
\mP Q [F] :=\Pi_{\mu Q} [F].
\end{equation}      
\end{notation}
\glossary{$\mM_{n}$}
\glossary{$\mS_{n}$}
\glossary{$\mP$}
\index{complete Poissonization}
 
We now show that as $n$ becomes large,  our discrete-time operators,
considered separately,  come to agree with their continuous-time
counterparts. 
When $n$ is large,
\[
\bP\{X^{\nu/n} = 0\} = \exp\{-\nu(\M)/n\} \approx 1 - \nu(\M)/n,
\]
and conditional on the event $\{X^{\nu/n} \ne 0\}$ the random measure
$X^{\nu/n}$ is approximately a unit point mass with location in $\M$
chosen according to the probability measure $\nu(\cdot)/\nu(\M)$.
In other words,  $ X^{\nu/n}$ takes the value $\delta_m$ with
probability approximately $\nu(dm)$ and takes the value zero otherwise. 
For any probability measure $Q$ on $\G$ 
\begin{equation}
\label{E:mutation_becomes_linear}
\lnf n \bigl( \mM_{n} Q [F] -Q[F]\bigr)
=
\int_\G \left(\int_\M \bigl(F(g + \delta_m) - F(g) \bigr)\, d\nu(m)\right) \, dQ(g).
\end{equation}
In particular, if $F$ is of the form $F(g):=\exp\{-g[f]\}$
for some Borel function $f:\M\to \bR_{+}$, then,
by the Poisson identity \eqref{E:Pidentity},
\begin{equation}
\mM_{n} Q [F] = Q [F] \cdot \exp\{\nu[e^{-f}-1]/n\}
\end{equation}
and so
\begin{equation}
\lnf n \bigl( \mM_{n} Q [F] -Q[F]\bigr)
=
\nu\bigl[ e^{-f}-1 \bigr] Q[F].
\end{equation}
Note also that
$\exp\{-S(g)/n\} \approx 1 - S(g)/n$ when $n$ is large
and so, when $Q[S]$ is finite, 
%
\begin{equation}
\label{E:selection_becomes_linear}
\lnf n \bigl( \mS_{n} Q [F] -Q[F]\bigr)= - Q[ S \cdot F] + Q[S]Q[F].
\end{equation}

When we start with a population genotype distribution $Q$ and 
assume that selection precedes mutation, the 
population genotype distribution after one generation 
of mutation and selection  is $\mM_{n} \mS_{n} Q$.    
A trajectory of the
resulting discrete-time model is defined by iteration.  That is,
given an initial population genotype distribution $Q_{0}$, 
the population genotype distribution after $k$ generations 
is $\bigl(\mM_{n} \mS_{n}\bigr)^{k} Q_{0}$.

If, after speeding up time in the $n^{\mathrm{th}}$ model, 
the resulting sequence of trajectories has a 
continuously differentiable limit $P_{t}$, 
this limit should satisfy the equation
\begin{align*}
\lim_{\epsilon\downarrow 0} \epsilon^{-1}
        \bigl( P_{t+\epsilon}&[F] - P_{t}[F] \bigr) =
        \lnf n\bigl( \mathfrak{M}_{n} \mS_{n} P_t[F] - P_t[F] \bigr) \\
&=\lnf n\bigl( \mS_{n} P_t[F] - P_t[F] \bigr) + 
     \lnf n\bigl( \mathfrak{M}_{n} \mS_{n} P_t[F] - \mS_{n} P_t[F] \bigr)\\
&=P_t[ SF] - P_t[S]P_t[F] + \nu\bigl[ e^{-f}-1 \bigr] P_t[F]
\end{align*}
for test functions $F : \G \rightarrow \bR_+$ of the 
form $F(g) = \exp\{-g[f]\}$ for some Borel 
function $f: \M \rightarrow \bR_+$.
For this special choice of test function, this is 
precisely the equation defining dynamical system 
without recombination that we introduced 
in \cite{SEW05} and derived heuristically in
\eqref{E:mutation_and_selection}.  
In fact, these test functions are enough to consider, 
since they determine probability measures
on $\G$.  Formally, a proof that the discrete-time dynamical system 
converges to this continuous-time one 
in the absence of recombination would require 
that we prove the existence of the continuously differentiable 
limit, a fact that we assumed above.
We forgo such a proof, since we wish
to incorporate recombination.

\section{Recombination in discrete-time}  
\label{SS:recom}

We now introduce recombination.  Imitating \cite{BT91}, 
\index{Barton, Nicholas}
\index{Turelli, Michael}
we think of a recombination event as taking two 
genotypes $g^{(1)}, g^{(2)} \in \G$ from the population and
replacing the genotype $g^{(1)}$ in the population by the genotype $g$
defined by $g(A) := g^{(1)}(A \cap R) + g^{(2)}(A \cap R^\complement)$, 
where $R \subseteq \M$ is the
particular segregating set for the recombination event. That is,
the new individual with genotype $g$ has the same accumulated mutations
as the individual with genotype $g^{(1)}$ (respectively, $g^{(2)}$) 
for ``loci'' in the set $R$ (respectively, $R^\complement$). 
\glossary{$g^{(j)}$}

As described in the Introduction, we are working
in an abstract framework in which 
loci are just places at which
mutations from wild type can occur rather than 
concrete physical loci
{\em per se}.  From a biological point of view, 
the recombination process might reasonably be defined
on a linear sequence of loci, as in \cite{BT91},
with ends of chromosomes at pre-determined positions,
and with recombination caused by crossovers during meiosis.  
Our abstract space of loci accommodates such a concrete
representation but also leaves us freedom to specify 
loci using structures related to their demographic effects
rather than their chromosomal positions.
\index{Barton, Nicolas}
\index{Turelli, Michael}

Of course, we think of $g^{(1)}$ and $g^{(2)}$ 
as being chosen independently
at random according to the particular probability measure describing
the distribution of genotypes in the population.  We also imagine
that the segregating set is chosen at random via some suitable
mechanism.  In order to discuss random sets rigorously, we follow
formalism described in \cite{dK74} and define 
a $\sigma$-algebra on sets of Borel subsets of $\M$ 
by the requirement that all incidence functions with 
Borel subsets are measurable.  A consequence of this 
definition is that if $A$ is a random Borel set 
and $\kappa$ is a finite measure, then $\kappa(A)$ is 
a real-valued random variable.  
\index{segregating set}
We suppose that there is a 
probability measure $\mathcal{R}$ that
describes the distribution of the random set of loci
that segregate together.   We always assume, without 
loss of generality, that $\mathcal{R}$ is {\em symmetric} 
in the sense that
\begin{equation}
\label{E:symmetry_condition}
\mathcal{R}(A)=\mathcal{R}\bigl( \{ R^{\complement}\, : \, R\in A\}\bigr),
\end{equation}
where $R^{\complement}$ denotes the complement of the set $R$.

\begin{notation} \label{N:recom}
For any Borel measure $g$ on $\M$ and
Borel subset $R$ of $\M$, define
the Borel measure $g_{R}$ on $\M$ by
$$
g_{R}(A) :=g(A\cap R)
$$
for Borel subsets $A\subseteq\M$.  Given the (symmetric) probability measure
$\mathcal{R}$ of a random subset of $\M$, define the corresponding
recombination operator that maps the space of Borel probability measures
on $\G$ into itself by
\begin{equation}  \label{E:recomop}
\mR Q [F] :=\int_{\mathcal{B}(\M)} 
\int_\G \int_\G F\left(g^{(1)}_R + g^{(2)}_{R^{\complement}}\right) 
          \, dQ(g^{(1)}) \, dQ(g^{(2)})\, d\mathcal{R}(R),
\end{equation}
where $Q$ is a Borel probability measure on $\G$, the function $F:\G\to\bR$ 
is bounded Borel, and ${\mathcal{B}(\M)}$ is the 
collection of Borel subsets of $\M$.
\end{notation}
\glossary{$g_A$}
\glossary{$\mathcal{R}$}
\glossary{$\mR$}
\glossary{$\mathcal{B}$}
\glossary{$R^{\complement}$}
\index{recombination measure}
\index{recombination operator}

Thus, if $Q$ describes the distribution of genotypes in the population, then
$\mR Q$ describes the distribution of a genotype that is obtained by
picking two genotypes $g^{(1)}$ and $g^{(2)}$ independently according to $Q$,
picking a segregating set $R$ according to the probability measure $\mathcal{R}$,
and forming a composite genotype that agrees with $g^{(1)}$ on the
set $R$ and $g^{(2)}$ on the set $R^{\complement}$.  
Where confusion with subscripts might result, 
the customary notation $ g|_{R}$ is also used for the
restriction $g_R$ of $g$ to the set $R$.
\index{segregating set}
\index{intensity measure}

Recall from Notation~\ref{N:spaces} that $\mu Q$ 
denotes the intensity measure of $Q$; that is,
\begin{equation}  \label{E:definemu}
\mu Q(A) := \int_{\G}  g(A) \, dQ(g).
\end{equation}

%
\begin{remark}\label{R:muRP}
Note that $\mu \mR Q = \mu Q$; that is, the intensity measure of the
random measure describing the genotype of a randomly chosen individual
is left unchanged by recombination.  
Specifically, for Borel sets $B$,  
\begin{eqnarray*}
\mu \mR Q(B)  & = \int\int_{\G}\int_{\G}
           \bigl( g^{(1)}(R \cap B) + g^{(2)}(R^{\complement} \cap B) \bigr) \, 
                dQ(g^{(1)})  dQ(g^{(2)})  d\mathcal{R}(R)     \\  
       &=  \int  \bigl(  \mu Q (R \cap B) + \mu Q( R^{\complement} \cap B) \bigr) 
            \,   d\mathcal{R}(R)   =  \mu Q(B). 
\end{eqnarray*}
Moreover, if $P$ is the
distribution of a Poisson random measure, then $\mR P = P$.
\end{remark}


We are now in a position to put the operators for recombination,
mutation, and selection together to describe the whole
discrete-time dynamical system.

\begin{notation}
\label{N:Qk}
\qquad \\[-2mm]
\begin{itemize}
\item
$Q_0$ is the distribution of any simple, $\G$-valued random measure 
and describes the distribution of genotypes in the initial population;
\item
$P_0  = \mP Q_0 = \Pi_{\mu Q_0} = \Pi_{\rho_0}$ is the Poisson probability measure with the same intensity as $Q_0$;
\item
$Q_{k}:=\left( \mR\mM_{n}\mS_{n} \right)^k Q_{0}$ and
describes the distribution of
genotypes in the population after $k$ generations of selection,
mutation, and recombination; 
\item
$P_{k}:= \left(\mM_{n}\right)^{k} P_{0} $ is the Poisson
probability measure on $\G$ that results from $k$ generations of mutation
in the absence of selection starting from $P_0$;
\item
$O_{k} : = (\mP\mM_{n}\mS_{n})^{k} P_{0} $ is the Poisson probability measure
resulting from $k$ iterations of mutation and selection 
with intensity rescaled by $1/n$ and complete Poissonization (rather than
recombination) after every round.
\end{itemize}
\end{notation}
\glossary{$Q_{k}$}
\glossary{$P_{k}$}
\glossary{$O_{k}$}

%
\section{Recombination trees and annealed recombination}
\label{SS:trees} 

In order to establish the convergence of our
discrete-generation dynamical system
to our continuous-time dynamical system,  
we need to understand the effects of mutation, selection
and recombination
not only over a single generation, as we have done
in Sections \ref{SS:frame} and \ref{SS:recom}, but also over a
number of successive generations.  Over $k$ 
generations,  each of the operators $ \mM_n$, $\mS_n$,
and $\mR$ acts $k$ times.  However, these operators
are interleaved: there is first an $\mS_n$, then an $\mM_n$, then an $\mR$,  
then another $\mS_n$, then another $\mM_n$, and so on,
and the operators in this product do not all commute with each other.
For our convergence proof,  we want to be able to decompose the
impact of each kind of operator over $k$
generations into its impact on the mutant alleles that first appear
in a particular generation.
Separating out the $k$-fold effects of each operator 
requires further structure to keep track of the
generation in which a given mutant allele first appears.   In this section
we define structure for $k$-fold iterates of the  
recombination operator in terms of a version of
a genealogical tree.  In the next sections
we define structure for separating out effects
of $\mM_n$ and of $\mS_n$ over $k$ generations by
introducing a concept of ``vintage'',
tracking the vintages of mutant alleles,  and defining
associated ``starred'' versions of our basic
operators that are defined on genotypes that
have been decomposed according to the allele vintages.
\index{vintage}

The iterates of our recombination operator implicitly create
a genealogical tree.   The genotype of an individual sampled from
the population at generation $k$ is formed from the
genotypes of two parents sampled from
the population,  which are formed from the genotypes
of four grandparents,  and so on back to $ 2^{k} $
individuals sampled from the initial population 
at generation zero.  Taken together, these ${2^{k+1} - 1}$  
individuals comprise the nodes of a complete rooted
binary tree. The root stands for the individual
sampled at generation $k$ and the leaves stand for 
the earliest ancestors back in the initial population
at generation $0$.  
Each node except the leaves has two edges (branches)
directed in from parent nodes and each node except the root 
has one branch directed out to the offspring node.

For each choice of the time-scaling factor $n$, we have
a finite number of generations 
$k_{\mathrm{max}} = \lfloor nT \rfloor$ up to time $T$.
The tree of ancestors for an individual in generation $k$
just introduced may be envisioned as a subtree
of the tree built from a root node at generation $k_{\mathrm{max}}$.
Call this latter tree $\mathcal{L}$.
\glossary{$\mathcal{L}$}

The tree structure defines a partial ordering among nodes
in the usual fashion.  We label nodes in any convenient way,
and write  $ \upsilon  \prec \ell $  if node $\upsilon$ is
an ancestor of node $\ell$,  that is,  if there exists a path
from $\upsilon$ to the root without duplicate edges 
containing $\ell$.   Nodes prior in the ordering are prior in
time.   We count generations starting with the leaves
at generation zero. 
Since the letter ``$g$'' is already in use for genotypes,
we write  $V(\ell)$ for the generation (``vintage'') 
of node $\ell$.    
\index{recombination tree}
\index{segregating set}
\glossary{$ \upsilon  \prec \ell $}
\glossary{$V(\ell)$}

The recombination structure is implemented by 
assigning complementary segregating sets $B$ and
$B^\complement$ to the two branches connecting a node
to its parents: the genotype of the corresponding individual
in the set $B$ (respectively, $B^\complement$)
coincides with that of the first (respectively, second) parent
in the set $B$ (respectively, $B^\complement$).
More specifically, we first assign each node $\ell$ an 
independent set $ R_{\ell}$ with distribution $\mathcal{R}$.  
(Sets for the leaves go unused.)   Suppose that the parents
of node $\ell$ are $\ell^\prime$ and $\ell^{\prime\prime}$.
We label the branch connecting $\ell^\prime$ (respectively, 
$\ell^{\prime\prime}$) to $\ell$
with the label $\ell^\prime$ (respectively, 
$\ell^{\prime\prime}$).
The complementary segregating sets assigned to
the branches labeled $\ell^{\prime}$ and
$\ell^{\prime\prime}$ are, respectively, $B_{\ell'} = R_{\ell}$  
and $B_{\ell''} = R^{\complement}_{\ell}$.  There is some ambiguity in this
construction due to the choice of which branch receives
$R_\ell$ or its complement, and this can be resolved by
embedding the complete rooted binary tree in the plane
so that we can distinguish between the ``left'' and ``right''
ancestors of a node.  Probabilistically, however,
this ambiguity is immaterial thanks to the symmetry of the 
recombination measure -- any way of resolving it will lead
to an assignment of segregating sets to branches that
has the same distribution. 

\begin{figure}
	\centering
\includegraphics[width=.635\textwidth]{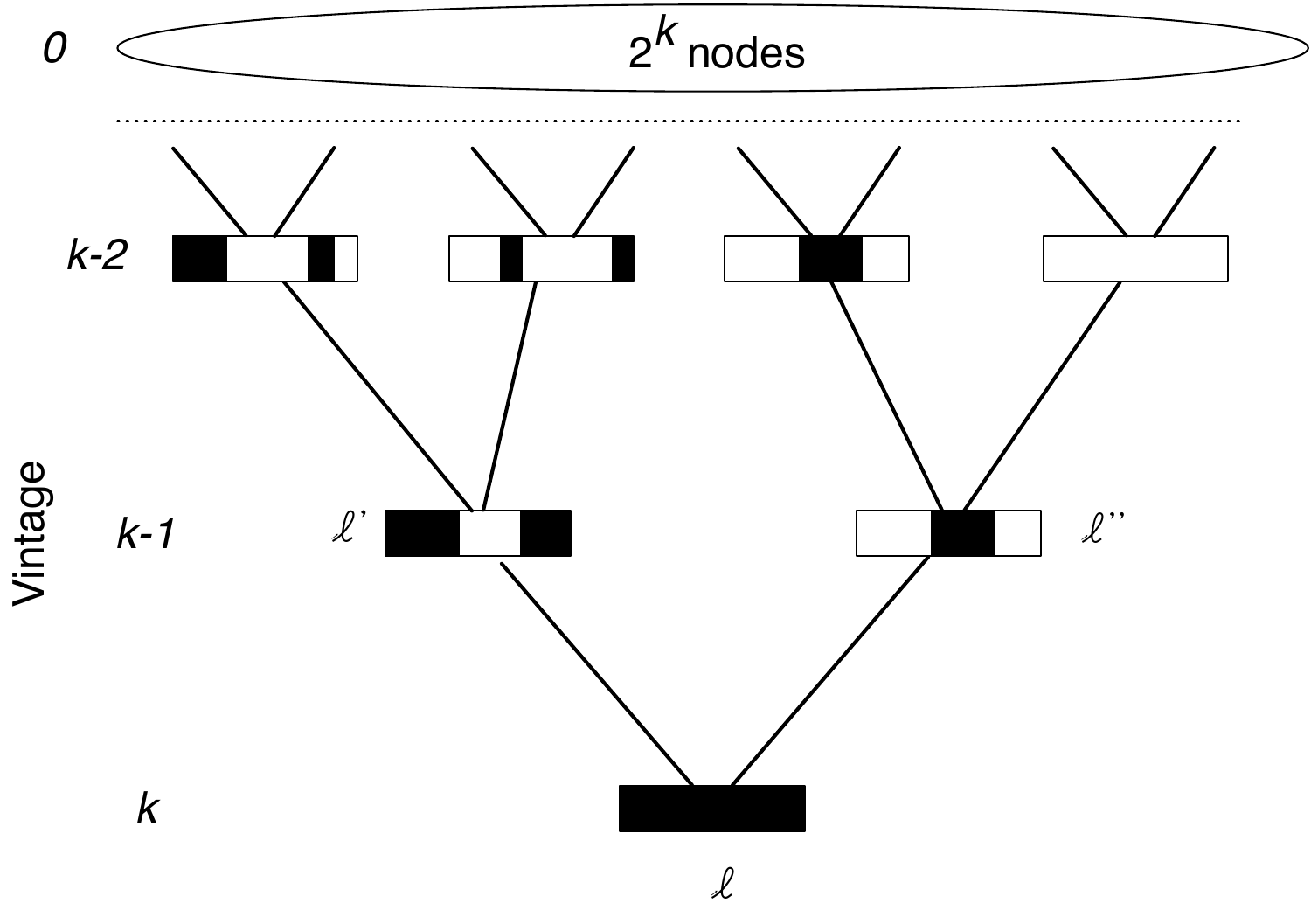}
\includegraphics[width=.365\textwidth]{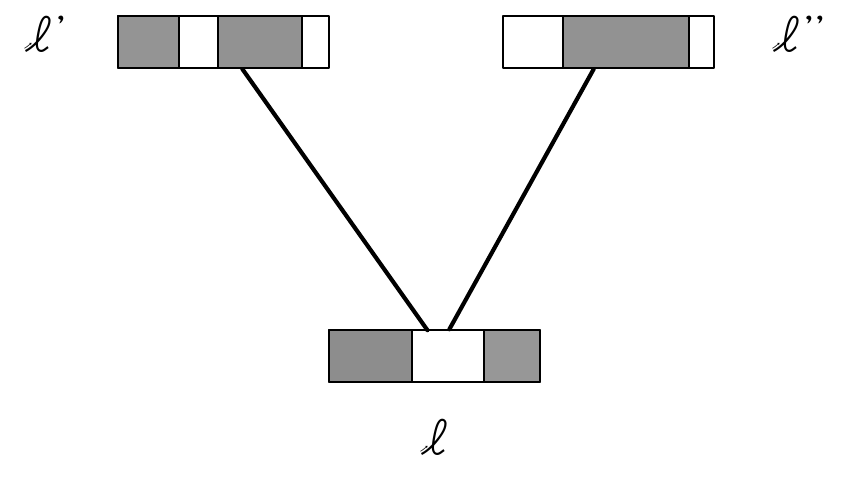}
	\caption{The ancestor tree for an individual of vintage $k$. For purposes of illustration,
	the mutation space $\M$ is  the unit interval. The first picture shows
	in black the sets $W(v,\ell)$ at the node $v$, for $v$ in the first two ancestor generations 
	preceding $\ell$, so vintages $k$, $k-1$, and $k-2$. The second picture shows in gray the
	recombination set $R_{\ell}$ (vintage $k$) and $R_{\ell'}$ and $R_{\ell''}$ (vintage $k-1$)
	that generated these partitions. Note that $W(\ell,\ell)=\M$. Note too that some sets in
	the partition will be empty, meaning that all genetic information from that ancestor
	has been lost through recombination.}
	\label{F:treepics}
\end{figure}

\begin{notation}
The genetic legacy $W$ for any pair of 
nodes satisfying   $  \upsilon  \preceq \ell $  is
the subset of $\M$ given by 
$$
W( \upsilon,  \ell  )  
    :=  \bigcap  \{ B_{\alpha} :  
               \upsilon \preceq \alpha  \prec \ell \} .
$$  
The intersection of the empty set is taken to be $\M$.
For other pairs $  \upsilon, \ell $ such that 
$  \upsilon  \not \preceq \ell $,  $W( \upsilon,  \ell  ) := \emptyset$.
See Figure~\ref{F:treepics}.
\end{notation}
\index{genetic legacy}
\glossary{$W( \upsilon,  \ell  )$}

When $\ell^{\prime} $ and $\ell^{\prime\prime}$ are ancestors
(not necessarily parents) of a node $\ell$ and belong to the
same generation,   the sets $W( \ell^{\prime},  \ell  )$  
and $W( \ell^{\prime\prime},  \ell  )$ are disjoint subsets of $\M$,  
since the branches leading to the first common descendant pass
through sets $R$ and $R^{\complement}$ that are complements of each other.
The sets $W( \upsilon,  \ell  )$ for fixed $\ell$  
over all $\upsilon $ 
satisfying  $\upsilon \prec \ell$ and $V(\upsilon) = j$ 
for some generation $j$ cover $\M$  and so form a partition of $\M$
(We adopt the convention that a partition may contain empty sets.)

We are now in a position to specify the combined action of $k$
applications of the recombination operator.

\begin{figure}
	\centering
		\includegraphics[width=1.00\textwidth]{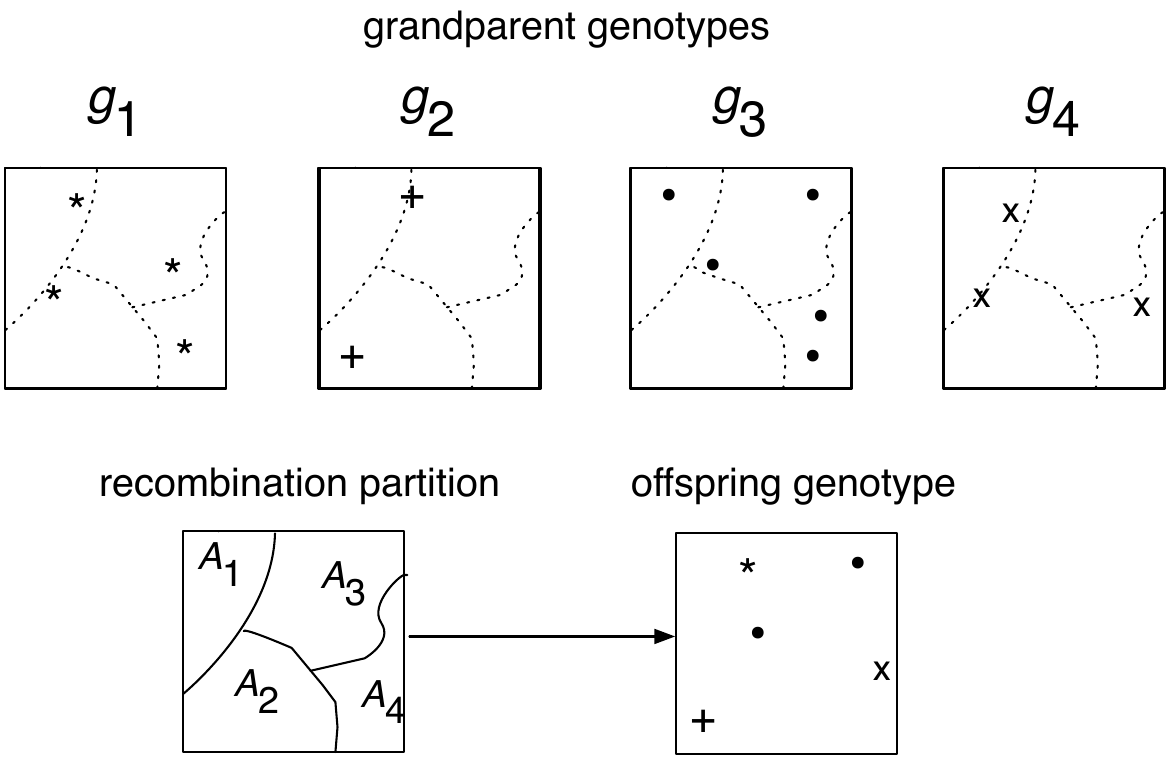}
	\caption{Illustration of the operation on genotypes induced by
	a partition $\mathcal{A}$ that underlies the action of the annealed recombination $\mathfrak{R}_{\mathcal{A}}$ operator on a collection of probability measures on $\G$. Here the mutation space $\M$ is the unit square, and the partition $\mathcal{A}$ is the collection of four sets $\{A_{1},A_{2},A_{3},A_{4}\}$ shown in the lower left. The mutations of four grandparent genotypes are shown at the top, and these genotypes (shown with a different symbol for each grandparent to clarify the origins of each one) are filtered through the partition to produce the offspring genotype at lower right.}
	\label{F:annealed}
\end{figure}

\begin{notation}  
For any finite partition $\mathcal{A} = \{A_1, \ldots, A_L\}$
of $\M$ into Borel sets, some possibly empty,  the
{\em annealed recombination operator} $\mR_{\mathcal{A}}$ is the
operator acting on probability measures $Q$ in $\G$ defined by
the condition
\glossary{$\mR_{\mathcal{A}}$}
\index{recombination}
\begin{equation}
\label{E:annealed_recombination_def}
\mathfrak{R}_{\mathcal{A}} Q [F] : =
     \int \cdots \int F\left( g^{(1)}|_{A_{1}} +\cdots+
      g^{(L)}|_{A_{L}} \right) \, dQ(g^{(1)})\cdots dQ(g^{(L)})
\end{equation}
for bounded Borel functions $F:\G\to\bR$.
Here, $g |_A $ is the restriction of the integer-valued
measure $g$ to the subset $A$, also written $g_A$ as in
Notation~\ref{N:recom} in contexts where no confusion
with indexing subscripts would result.
See Figure \ref{F:annealed}.  
\end{notation}

In other words,
if $Y_1, \ldots, Y_L$ are i.i.d. random measures with
common distribution $Q$, then $\mathfrak{R}_{\mathcal{A}} Q$
is the distribution of the random measure 
$Y_1 |_{A_1} + \cdots + Y_L |_{A_L}$. 

Via annealing,  the genetic legacies from ancestors at generation
zero generate the $k$-th iterate of the recombination operator as follows:


\begin{lemma}\label{L:anneal}
Let $\ell \in \mathcal{L} $ be a node with $V(\ell) = k $.  
Let   $ \mathcal{A} = ( A_1 \ldots A_{2^k} ) $ 
be the random partition of $\M$ consisting of the $2^k$ sets  
$ \{ W( \alpha, \ell) : \alpha \prec \ell \hbox { and } V(\alpha) = 0 \}$,
constructed from the random segregating
sets $ \{ R_{\upsilon} : \upsilon \preceq \ell \}$.
Then, for any probability 
measure $Q$ on $ \G$ and bounded Borel function $F : \G \to \bR $,
\begin{equation*}
\mathfrak{R}^k Q [F]  =      
       \bE\left[
      \mathfrak{R}_{\mathcal{A}} Q[F]
      \right].      
\end{equation*}
\end{lemma}

\begin{proof}
The claim holds trivially for $ k = 0 $.  We proceed by induction.
Suppose the claim holds up to generation $ k - 1 $ and thus
for the parent nodes $\ell^{\prime}$ and $\ell^{\prime\prime}$ 
of node $\ell$.  For all $\omega \prec \ell$,
either $\omega \preceq \ell^{\prime}$ 
or $\omega \preceq \ell^{\prime\prime}$, 
and  $ W(\omega, \ell) = W(\omega, \ell^{\prime}) \cap R_{\ell}$
or $ W(\omega, \ell) = W(\omega, \ell^{\prime\prime}) \cap R^{\complement}_{\ell}$,
respectively.
For the genotypes of the parent nodes, write
$$
\Gamma^{\prime}((g_{\upsilon})_{\upsilon \preceq \ell^{\prime}, V(\upsilon) = 0}) 
=  \sum_{\begin{smallmatrix}\omega \preceq \ell^{\prime}, \\ V(\omega) = 0\end{smallmatrix}}  (g_{\omega}) 
|_{W( \omega, \ell^{\prime}) }
$$
and
$$
\Gamma^{\prime\prime}((g_{\upsilon})_{\upsilon \preceq \ell^{\prime\prime}, V(\upsilon) = 0}) 
= \sum_{\begin{smallmatrix}\omega \preceq \ell^{\prime\prime}, \\ V(\omega) = 0\end{smallmatrix}}  (g_{\omega}) 
|_{W( \omega, \ell^{\prime\prime})}. 
$$
Then, by the induction hypothesis,
$$
\mathfrak{R}^{k-1} Q [F]  =    
\int \bE \left[  
         \,  F(\Gamma^{\prime}(\bg))   
\right] \, d \hspace*{-2mm}\bigotimes_{\begin{smallmatrix} \upsilon \preceq \ell^{\prime}, \\ V(\upsilon) = 0\end{smallmatrix}}\hspace*{-2mm} Q_\upsilon(\bg), 
$$
where each $Q_\upsilon$ is a copy of $Q$, 
and the analogous equation holds for $\Gamma^{\prime\prime}$ and $\ell^{\prime\prime}$.
Applying the operator $\mathfrak{R}$ to the expressions 
for $\mathfrak{R}^{k-1} $ gives
$$
\mathfrak{R}^{k} Q [F]  = 
\iint
\bE\left[     
F(\Gamma^{\prime}(\bg'|R_{\ell}) + \Gamma^{\prime\prime}(\bg''|R^{\complement}_{\ell}))
\right]
\, d\hspace*{-2mm}\bigotimes_{\begin{smallmatrix} \upsilon \preceq \ell^{\prime},\\ V(\upsilon) = 0\end{smallmatrix}} \hspace*{-2mm}Q_\upsilon(\bg')
\: d\hspace*{-2mm}\bigotimes_{\begin{smallmatrix}\upsilon \preceq \ell^{\prime\prime},\\ V(\upsilon) = 0\end{smallmatrix}}\hspace*{-2mm} Q_\upsilon(\bg''). 
$$ 
Substituting for $\Gamma^{\prime}$ and $\Gamma^{\prime\prime}$ shows this last
expression is equal to the claimed formula.
\end{proof}

We use the tree structure of $\mathcal{L}$
to construct a nested sequence of random partitions
$\mcA_0(\ell) \supseteq \mcA_1(\ell) \supseteq \ldots \supseteq \mcA_{V(\ell)} (\ell)$
of $\M$ for an arbitrary node 
$\ell \in \mathcal{L}$.  
There is one set in the partition $ \mathcal{A}_j(\ell) $  for each
ancestor of the individual $\ell$ in generation $j$.
This set represents the portion of that ancestor's genotype
inherited by the individual $\ell$ through the shuffling
together of genotypes due to recombination.


\begin{notation}\label{N:Rtwiddle}
For $\ell \in \mathcal{L} $ and $k \le V(\ell)$, define a random partition
$ \mathcal{A}_{k}(\ell) $  of $\M$ by
\[
\begin{split}
\mathcal{A}_{k}(\ell) 
& := 
\{ W(\upsilon, \ell) : V(\upsilon) = k \text{ with } \upsilon \prec \ell \} \\
& =
\left\{
\bigcap  \{ B_{\alpha} :  
               \upsilon \preceq \alpha  \prec \ell \}
: V(\upsilon) = k \text{ with } \upsilon \prec \ell
\right\}. \\
\end{split}
\]
\end{notation}
\glossary{$\mathcal{A}_{k}(\ell)$}

\begin{remark} \label{R:refinement}
Note that $\mcA_{k}(\ell)$ is
a refinement of $\mcA_{k+1}(\ell)$, 
and $\mcA_{V(\ell)}(\ell)$ is the trivial partition $\{\M\}$.
Note also that for $j \ge k$
the distribution of the sequence of partitions
$(\mcA_{0}(\ell), \mcA_{1}(\ell), \ldots, \mcA_{k}(\ell))$
is the same for all nodes $\ell$ with $V(\ell) = j$.
Recall that $R_\ell$ is the recombination set at node $\ell$.
Suppose that $\ell$ has parents 
$\ell'$ and $\ell''$ and that the sets $R_\ell$ and
$R_\ell^\complement$ are associated with the branches labeled
$\ell'$ and $\ell''$.  Then, for $k\le V(\ell)-1$,
$$
\mcA_{k}(\ell)=
 \bigl\{ A'\cap R_\ell \, :\, A'\in \mcA_{k}(\ell') \bigr\} \cup  \bigl\{ A''\cap R_\ell^\complement \, :\, A''\in \mcA_{k}(\ell'') \bigr\}.
$$
\end{remark}

\begin{notation}
Fix a node $\ell \in \mathcal{L}$ with $V(\ell) = k$.
Write $(\mathcal{A}_{k,0}, \mathcal{A}_{k,1}, \ldots, \mathcal{A}_{k,k})$
for the nested sequence of random partitions $(\mathcal{A}_{0}(\ell), \mathcal{A}_{1}(\ell), \ldots, \mathcal{A}_k(\ell))$.
\end{notation}

\begin{remark}
The distribution of $(\mathcal{A}_{k,0}, \mathcal{A}_{k,1}, \ldots, \mathcal{A}_{k,k})$ does not depend on the choice of the node $\ell$
with vintage $k$.  Note also that for $0 \le j \le k$ the sequence
$(\mathcal{A}_{k-j,0}, \mathcal{A}_{k-j,1}, \ldots, \mathcal{A}_{k-j,k-j})$
has the same distribution as the sequence
$(\mathcal{A}_{k,j}, \mathcal{A}_{k,j+1}, \ldots, \mathcal{A}_{k,k})$.
We can re-state the conclusion of Lemma~\ref{L:anneal} as
\begin{equation*}
\mathfrak{R}^k Q [F]  =      
       \bE\left[
      \mathfrak{R}_{\mathcal{A}_{k,0}} Q[F]
      \right].      
\end{equation*}
\end{remark}

\section{Vintages}
\label{SS:vintages} 

The annealed recombination operator allows us to condition on
choices of segregating sets while we isolate $k$-generational
effects of mutation and selection.  We next consider how 
the process of mutation relates to the recombination tree. 
Any element of $\M$ assigned mass by the genotype
for a node at generation $k$ has to have entered at some earlier
generation,  either in one of the nodes for ancestors in
the initial population
(distributed according to $Q_0$) or as a new mutation
at one of the intermediate nodes in the tree of ancestors
(distributed as a Poisson random measure with intensity measure $\nu/n$).
In our infinite-population setting,  selection reweights the
probabilities of outcomes. It does not remove elements of $\M$ or
alter their pattern of descent from nodes of the tree.
Although $Q_0$ may not be a Poisson probability measure,  one of 
our assumptions will require it to be absolutely continuous with respect
to the Poisson probability measure $P_0$ with the same intensity measure.
It turns out that the difference between $Q_0$ and $P_0$ can be
treated as a process of reweighting,  analogous to the reweighting 
resulting from selection.   With this strategy in mind,
we use Poisson random measures to build a joint distribution
of random genotypes across the tree,  a template to which
reweighting is applied. 


Specifically, we assign to each node $\upsilon$ in $\mathcal{L}$ an 
independent Poisson random measure $\Gamma_{\upsilon}$  
with distribution $P_0 = \Pi_{\rho_0}$ when $V(\upsilon) = 0 $ 
and with distribution $\Pi_{\nu/n}$ otherwise.
Given a choice of the segregating sets $R_\ell$, which define the
genetic legacies $W$ at each node, consider restrictions of
the Poisson measures to the legacy sets and put 
\begin{equation}
\label{E:Xi-ell}
\Xi_{\ell}   \,  =  \,  \sum_{\upsilon \preceq \ell} \, 
                    \Gamma_{\upsilon} \, |_{W(\upsilon, \ell)}.
\end{equation}
\glossary{$\Xi_{\ell}$}
\glossary{$\Gamma_{\upsilon}$}
%
These random elements of $\G$ are independent across nodes within any
one generation but are dependent across nodes down any line of
descent.   They correspond to the genotypes of the ancestors
of a randomly chosen individual  in generation $k_{\mathrm{max}}$
in the absence of selection starting from a population with
genotype distribution $P_0$. 
With this representation,  which underlies the crucial 
Lemma~\ref{L:exactRN} at the start of Chapter~\ref{Ch:techlem},
we may separate out the effects of mutation over $k$ generations from
the effects of the other operators.

Any mutant allele found in the genotype of  
a reference  individual sampled from the population at 
generation $k$  originates in the $\Gamma_{\upsilon} $ term
for some ancestral node.  However,  knowing the 
genotype at generation $k$ and knowing 
the segregating sets for the tree of ancestral
nodes is not enough to identify the origin of each mutation.
A mutant allele found in the subset of $\M$ inherited from a
particular great-grandparent might have originated from that
great-grandparent or it might come from a new mutation in
the same portion of the parental genome.  What is missing
is information on the generation, or vintage, from which the mutation
originates.  Because we need to be able to condition on the  
identities of the mutations that are still extant in the
genome at generation $k$, while we average over all the
other contributions to the $\Gamma_{\upsilon}$, 
we introduce additional structure to keep track of vintages.
\index{vintage}

We can rewrite \eqref{E:Xi-ell} as
\begin{equation}
\label{E:Xi-ell2}
\Xi_{\ell}   \,  =  \,  
\sum_{j \le V(\ell)}
\sum_{\begin{smallmatrix}\upsilon \preceq \ell, \\V(\upsilon) = j\end{smallmatrix}} \, 
                    \Gamma_{\upsilon} \, |_{W(\upsilon, \ell)}.
\end{equation}
The term
\[
\sum_{\begin{smallmatrix}\upsilon \preceq \ell, \\V(\upsilon) = j\end{smallmatrix}} \, 
                    \Gamma_{\upsilon} \, |_{W(\upsilon, \ell)}
\]
separates out the contribution to $\Xi_{\ell}$ due to mutant alleles
that first appeared in generation $j$ in the lineage of the 
individual associated with node $\ell$.  This suggests
that we imagine that the genotype $g$ of an individual in
generation $k$ is decomposed as a sum $g=g_{0}+\cdots+g_{k}$,
where $g_j$, $j < k-1$, is the contribution to $g$ that first appeared
in an ancestor in the past generation $j$ and $g_k$ is the
contribution of new mutations that first appeared 
in the individual herself.
We thus have a richer structure where a genotype $g \in \G$
is replaced by the $(k+1)$-tuple  
$ \mathbf{g} = (g_{0},g_{1},\dots,g_{k-1},g_{k})$ 
in the product space  $\G^{k+1}$ that decomposes $g$
according to vintage.  In most contexts there
is no danger of confusing the subscripts for vintage
components of $\bg$ with subscripts for Borel sets 
to which genotypes $g$ may be restricted.

The $k+1$-tuple  giving the decomposition according
to vintage of an individual drawn at random from the
population has a joint distribution which is
a probability measure on $\G^{k+1}$.  
This object describes the vintages of the
current stock of mutations retained
in the genomes of individuals in generation $k$.
The force of selection does not depend on the decomposition
by vintages -- it only matters what mutations are present
in an individual's genotype, not when they entered her lineage.    
But because selection alters the weights of whole
genotypes in the population it necessarily alters the
overall distribution of the decomposition by vintages.  
Recombination shuffles contributions within vintages.  
Mutation adds a new component to the decomposition
by vintages at each generation.  


In order to have a single space on which all our
vintage-specific distributions are defined,  
we form the disjoint union $\G^*$
of the spaces of tuples for successive generations
starting with the initial population for $ k = 0 $ on up.
That is,
\begin{equation}  
\label{E:gstar}
\G^{*}:=\G^{1} \sqcup \G^{2} \sqcup \G^{3} \sqcup\cdots,
\end{equation}
where $\sqcup$ denotes disjoint union.  
\glossary{$\G^{*}$}
\glossary{$\sqcup$}  
In principle,  a probability measure on this space
can assign positive probability to tuples
in several different generations and can thus be thought of
as a sum of sub-probability measures on 
the different disjoint pieces. 
In practice, the probability measures that 
we consider are always concentrated on the
tuples for some particular generation.  
We can always
return from our big tuple space $\G^*$ to $\G$ itself
by adding up the components associated with each vintage. 

The operators for mutation, selection, and 
recombination $\mM_n$, $\mS_n$, and $\mR$,
which we have defined to act on probability measures
on $\G$, all have counterparts acting on the 
probability measures on our big tuple space $\G^*$.
Just as $\mM_n$, $\mS_n$, and $\mR$ combine
to take the distribution of genotypes in one
generation to the distribution in the next, these
new operators combine to take the distribution of genotypes
decomposed by vintage
in each generation into the distribution in the next. 
We denote these new operators by adding a star to
the notation for the corresponding operators
acting on probability measures on $\G$.
  
%
%
%

In order to define these new operators, we first need to introduce
a little more notation.
We usually represent a generic element of $\G^{*}$ by the boldface $\bg$. 
There are natural ``projection'' operations $\Sigma: \G^* \rightarrow \G$
defined for $\bg = (g_0, g_1, \ldots g_k) \in \G^{k+1}$ by  
\glossary{$\bg$} 
\[
\Sigma(\bg) := g_0 + g_1 + \cdots + g_k
\]
and
\[
\Psi_j(\bg) := 
\begin{cases}
g_j, & \quad 0 \le j \le k,\\
0, & \quad j > k.
\end{cases}
\]
\glossary{$\Sigma(\bg)$}
\glossary{$\Psi_j(\bg)$}
In essence, $\Sigma$ removes the labels that record 
the vintages of the various mutations from the ancestral wild type, 
while $\Psi_j$ isolates mutations with vintage $j$.  
Note that, as operators, $\Sigma = \sum_{j=0}^\infty \Psi_j$. 


Given a probability measure $Q$ on $\G^{*}$,
we define, with a slight abuse of notation, 
probability measures $\Sigma Q$ and $\Psi_j Q$ on $\G$ by
\[
(\Sigma Q)[F] := Q[F \circ \Sigma] 
\quad \text{and} \quad
(\Psi_j Q)[F] := Q[F \circ \Psi_j] 
\]
for a Borel function $F : \G \rightarrow \bR_+$. 
\glossary{$\Sigma Q$}
\glossary{$\Psi_j Q$}

Although we shall only be employing probability measures 
on $\G^*$ that are in fact concentrated on some $\G^{k}$,  
we give a brief account of the structure of
a general probability measure on the disjoint union $\G^{*}$. 
As we remarked above,
a probability measure $Q$ on $\G^{*}$ may be thought of as a
sequence of sub-probability 
measures $(Q^{(k)})_{k=0}^\infty$, 
where $Q^{(k)}$ is the portion of $Q$ 
concentrated on $(k+1)$-tuples from $\G$.
Thus, $\sum_{i=0}^{\infty} Q^{(k)}(\G^{k+1})=1.$  
We interpret such a $Q$ as the distribution 
of a finite, random length sequence of random measures
$(X_0,X_{1}, \ldots, X_I)$, 
with  $Q^{(k)}(\cdot;\cdot) / Q^{(k)}(\G^{k+1})$ 
being the joint distribution 
of $(X_0,X_{1}, \ldots, X_k)$ 
conditional on the event $\{I=k\}$, and 
$Q^{(k)}(\G^{k+1})$  being the probability 
of the event $\{I=k\}$.



\begin{notation}  \label{N:starops}
Consider a probability measure $Q$ on $\G^*$
that is concentrated on $\G^{k+1}$ for some $k \in \bN$.   
Write $ ( X_{0} \ldots X_{k})$ 
for a $k+1$-tuple of random measures
with conditional distribution $Q^{(k)}$.

\begin{itemize}
\item
Define a new probability measure $\mM_n^{*} Q$ on $\G^*$ 
to be the distribution of $ ( X_{0} \ldots X_{k}, Y)$,
where $Y$ is an independent Poisson
random measure with intensity measure $\nu/n$.
\glossary{$\mM_n^{*} Q$}
\item
Define a new probability measure $\mS_n^* Q$ on $\G^*$
that is concentrated on $\G^{k+1}$ by
\[
\begin{split}
(\mS_n^{*} Q)^{(k)} [F] & : = 
   \frac{\int_{\G^{k}}\exp\{-S(\Sigma(g))/n\}F(g) \, dQ^{(k)}(g)}
        {\int_{\G^{*}}\exp\{-S(\Sigma(g))/n\} \, dQ^{(k)}(g)} \\
& =
\frac{ \bE[\exp\{-S(X_0 + \cdots + X_k)/n\} F(X_0, \ldots, X_k)]} 
     { \bE[\exp\{-S(X_0 + \cdots + X_k)/n\}] }\\
\end{split}
\]
for any bounded Borel function $F:\G^{*}\to \bR_+$.
\glossary{$\mS_n^* Q$}
\item 
Define a new probability measure $\mR^{*} Q$ on $\G^{*}$ 
that is concentrated
on $\G^{k+1}$ by
\[
\int F(\bg) \, d\mR^{*}Q(\bg)   :=
  \int \int F(\bg^{(1)}|A +  \bg^{(2)}|A^{\complement} ) \, 
               dQ(\bg^{(1)}) \,  dQ(\bg^{(2)}) \, d\mathcal{R}(A)
\]
\glossary{$\mR^{*} Q$} 
for any bounded Borel function $F:\G^{*}\to \bR_+$.
\item
Define a new probability measure $\mP^{*} Q$ on $\G^{*}$
that is concentrated
on $\G^{k+1}$ by
letting $\mP^{*} Q^{(k)}$ be
the distribution of $(Y_0, \ldots, Y_k)$, 
where the random measure $ Y_j$ is Poisson 
with the same intensity measure as $X_j$
and $Y_0, \ldots, Y_k$ are independent. 
\glossary{$\mP ^{*} Q$}
\end{itemize}
\end{notation}

%

Observe that we have the four intertwining relations
\[
\Sigma \mM_n^{*}=\mM_n \Sigma, \; 
\Sigma \mS_n^{*}= \mS_n \Sigma, \;
\Sigma \mR^{*}=\mR \Sigma \; 
\Sigma \mP^{*}=\mP \Sigma.
\]
These equalities confirm that starred operators agree 
with the unstarred ones once the labeling  by vintage
is removed. 


Our initial conditions $P_{0}$ and $Q_{0}$ can also be regarded as 
probability measures on $\G^*$ that are
concentrated on sequences of length $1$.

\begin{notation}   
\label{N:Qkstar_Pkstar}
Set
\begin{equation}  \label{E:defineqkstar}
\begin{split}
Q_{k}^{*} &:= \left(\mR^{*} \mM^{*}_{n}\mS^{*}_{n}\right)^k Q_{0}\\
P_{k}^{*} &: =\left(\mM^{*}_{n}\right)^{k} P_{0}  \\ 
O_{k}^{*} &:= (\mP^{*} \mS_n^{*}\mM_n^{*})^{k} P_{0}.
\end{split}
\end{equation}
\end{notation}

The operator $O_{k}^{*}$ is the analogue of $Q^{*}_{k}$ 
with the complete Poissonization operator $\mP^*$ replacing the
recombination operator $\mR^*$.
\glossary{$Q_{k}^{*}$}
\glossary{$P_{k}^{*}$}
\glossary{$O_k^{*}$}

Note that all three 
probability measures $O_{k}^{*}$, $Q_{k}^{*}$ and $P_{k}^{*}$ 
are concentrated on genotype sequences of length $(k+1)$.
Note also that if $(X_0, \ldots, X_k)$ is distributed 
according to $P_{k}^{*}$, then $X_1, \ldots, X_k$ are
independent Poisson random measures, each with intensity 
measure $\nu/n$, and $X_0$ is independent with
distribution $P_{0}$.  
As we expect, $Q_{k} = \Sigma Q_{k}^*$, $O_{k} = \Sigma O_{k}^{*}$, 
and $P_{k} = \Sigma P_{k}^*$,
where we recall that $Q_{k} =(\mR\mM_{n}\mS_{n})^k Q_{0}$,
$O_{k} =(\mP\mM_{n}\mS_{n})^k Q_{0}$, 
and $P_{k}:=(\mM_{n})^{k} P_{0}$.



\chapter{Shattering and the formulation of the convergence result}
\label{Ch:hypotheses}

\section{Shattering of random measures}
\label{SS:shattering} 

We expect recombination to break up
dependencies between different parts of the genome, so that
when $k$ is large  $\mR^k P$ should be approximately $\Pi_{\mu P}$,
the distribution of the Poisson random measure with intensity $\mu P$
in the notation established in Notation~\ref{N:spaces}.
\index{shattering}
\index{recombination}

In order that this approximation should hold 
for a given probability measure $P$, it must generically
be the case that there is a positive probability that
the segregating set and its complement both intersect
any set with positive $\mu P$ mass in two sets that each have
positive $\mu P$ mass.  The following condition (with
$\lambda = \mu P$) is key to establishing 
establishing quantitative bounds on the rate with which
$\mR^k P$ converges to $\Pi_{\mu P}$.

\begin{definition}  \label{D:shat}
Given a (symmetric) recombination measure $\mathcal{R}$ 
and a finite measure $\lambda$ on $\M$, we say that 
the pair $(\mathcal{R},\lambda)$ 
is  {\em shattering} if there 
is a positive constant $\alpha$ such that for any Borel set $A$,
\begin{equation}  \label{E:rsep}
\begin{split}
\lambda(A)^{3}
& \le 
\alpha\left[ \lambda(A)^{2} - 2\int \lambda(A\cap R)^{2} \, d\mathcal{R}(R)\right] \\
& = 
\alpha\left[ \lambda(A)^{2} - \int \lambda(A\cap R)^{2} \, d\mathcal{R}(R)
- \int \lambda(A\cap R^\complement)^{2} \, d\mathcal{R}(R) \right] \\
& =
2 \alpha \int \lambda(A\cap R) \lambda(A \cap R^\complement)  \, d\mathcal{R}(R).
\end{split}
\end{equation}
\end{definition}

We show in Section~\ref{SS:iter_recomb} 
that if the pair $(\mathcal{R}, \mu Q)$ is shattering 
and $P$ satisfies a further simple condition,  
then $\mR^{k}Q$ converges to $\Pi_{\mu Q}$ as $k \rightarrow \infty$.


The concept of shattering is illustrated by the following
elementary example.

\begin{remark}
Suppose that $N$ is a finite simple random measure on $\M = (0,1]$.
We think of $\M$ in this case as a physical chromosome and $N$
as the set of crossover points formed during meiosis.
Write $0 < T_1 < \ldots < T_L < 1$ for the successive points of $\M$.
Set $T_0 = 0$ and $T_{L+1} = 1$.  Let $Z$ be a $\{0,1\}$-valued random 
variable that is independent of $N$ with $\mathbb{P}\{Z=0\} = \frac{1}{2}$.
Define $\mathcal{R}$ to be the distribution of the random set given by
\[
(T_0, T_1] \cup (T_2, T_3] \cup \ldots, \quad \text{if $Z=0$},
\]
\[
(T_1, T_2] \cup (T_3, T_4] \cup \ldots, \quad \text{if $Z=1$}.
\]
Take $\lambda$ to be any diffuse probability measure (that is,
$\lambda$ has no atoms).
Suppose that there is a constant $c$ such that
$\mathbb{P}\{N(u,w] = 1\} \ge c \lambda((u,w])$ for $0 < u < w \le 1$.
This is the case, for example, if $N$ is Poisson random measure
with intensity measure bounded below by a positive
 multiple of $\lambda$ or $N$ consists
of a single point with distribution bounded below by
a positive multiple of $\lambda$.  Also, for most
reasonable simple random measures with intensity 
$C \lambda$ for some constant $C$, it is the case that
$\mathbb{P}\{N(u,w] = 1\} \approx \mathbb{E}[N(u,w]]
= C \lambda((u,w])$ when
$|u-w|$ is small, where the notation $\approx$ indicates that
the ratio of the two sides is close to $1$.

For the recombination measure $\mathcal{R}$ in this example 
we have for the right-hand side of  \eqref{E:rsep}  
\[
\begin{split}
& \lambda(A)^{2}-2\int \lambda(A\cap R)^{2} \, \mathcal{R}(dR) \\
&  \quad =
   2 \biggl[
    \int_0^1 \int_0^w \indic_A(u) \indic_A(w) 
       \, \lambda(du) \, \lambda(dw) \\
& \qquad   - 2 \int_0^1 \int_0^w \indic_A(u) \indic_A(w) 
   \mathbb{P}\{N((u,w]) = 0 \mod 2\} \, \lambda(du) \, \lambda(dw) \\
&  \quad =
   2 \int_0^1 \int_0^w \indic_A(u) \indic_A(w) 
   \mathbb{P}\{N((u,w]) = 1 \mod 2\} \, \lambda(du) \, \lambda(dw) \\
&  \quad \ge
   2 \int_0^1 \int_0^w \indic_A(u) \indic_A(w) 
   \mathbb{P}\{N((u,w]) = 1\} \, \lambda(du) \, \lambda(dw)\\
&  \quad \ge
      2 c  \int_0^1 \int_0^w \indic_A(u) \indic_A(w)  \lambda((u,w]) \, 
      \lambda(du) \, \lambda(dw). \\
\end{split}
\]
As for the left-hand side of  \eqref{E:rsep}
observe that
\[
\begin{split}
\lambda(A)^3 
& = 
3! \iiint_{\{0 <u<v<w \le 1\}} \indic_A(u) \indic_A(v) \indic_A(w) \, 
       \lambda(du) \, \lambda(dv) \, \lambda(dw)         \\
& \le
3! \iiint_{\{0 <u<v<w \le 1\}} \indic_A(u)  \indic_A(w) \, 
       \lambda(du) \, \lambda(dv) \, \lambda(dw)         \\
& =
3! \iint_{\{0 <u< w \le 1\}} \indic_A(u)  \indic_A(w) \lambda((u,w]) \, 
       \lambda(du) \,  \lambda(dw)         \\
& = 3!  \int_0^1 \int_0^w \indic_A(u) \indic_A(w)  \lambda((u,w]) \, 
       \lambda(du) \, \lambda(dw).         \\
\end{split}
\]
Thus, the pair $(\mathcal{R}, \lambda)$ is shattering 
with constant $\alpha = 3/c$.
\end{remark}



\begin{remark}
The meaning of the shattering condition may also be clarified 
by the following observation.  Suppose that
$\M$ is equipped with a metric $\delta$, that $\lambda$ is a probability
measure,  and for some constant $c > 0$
\[
\begin{split}
p(r) & := \inf\{\mathcal{R}\{R : m' \in R, \, m'' \in R^\complement\} : 
m',m'' \in \M, \, \delta(m',m'') \ge r\} \\
& \ge
c \sup\{\lambda\{m'' \in \M : \delta(m',m'') \le r\} : m' \in \M\} 
=: c \, \varphi(r) \\
\end{split}
\]
for all $r \in \bR_+$.  Loosely speaking, this condition says
that the probability that two loci inherit their contents 
from different parents dominates a multiple of the $\lambda$ mass 
of a ball with radius the distance between the two loci.
By a change of variables,
\[
\begin{split}
& \int \lambda(A\cap R) \lambda(A \cap R^\complement)  \, \mathcal{R}(dR)
 =
\int_A \int_A 
\mathcal{R}\{R : m' \in R, \, m'' \in R^\complement\} 
\, \lambda(dm'') \, \lambda(dm') \\
& \quad \ge
\int_A \int_A 
p(\delta(m',m''))
\, \lambda(dm'') \, \lambda(dm') \\
& \quad \ge
c \int_A \int_\M \indic\{\delta(m',m'') \le \varphi^{-1}(\nu(A))\}
\varphi(\delta(m',m''))
\, \lambda(dm'') \, \lambda(dm') \\
& \quad = c \lambda(A) \frac{1}{2} \lambda(A)^2, \\
\end{split}
\]
Thus, $(\mathcal{R},\lambda)$
is shattering with constant $\alpha = c^{-1}$.
\end{remark}


We collect some elementary observations about the shattering condition
in the following lemma.

\begin{lemma}
\label{L:must_be_diffuse}
Suppose that the pair $(\mathcal{R}, \lambda)$ is 
shattering with some constant $\alpha$.
\begin{itemize}
\item[(a)]
%
%
If $\lambda$ is non-zero, then $\alpha \ge 2 \lambda(\M)$.
\item[(b)]
The measure $\lambda$ is necessarily diffuse.
\item[(c)]
If the pair
$(\mathcal{R},\eta)$ is also shattering for the
same recombination measure and same shattering constant $\alpha$, 
then the pair $(\mathcal{R},\lambda + \eta)$  
is shattering with constant $ 4 \alpha$.
\item[(d)] 
For any $t\ge 0$ the pair $(\mathcal{R},t\lambda)$ 
is shattering with constant $t\alpha$.
\item[(e)] 
If $\eta$ is another measure on $\M$ 
with $\bar H:=\sup|\log d\eta/d\lambda|<\infty$,
then $(\mathcal{R},\eta)$ is shattering with constant $\alpha e^{5 \bar H}$.
\end{itemize}  
\end{lemma}

\begin{proof}
(a) Suppose that $\lambda(\M) > 0$. Then, from \eqref{E:rsep},
\[
\lambda(\M)
\le 
2 \alpha \int \frac{\lambda(R)}{\lambda(\M)} 
        \frac{\lambda(R^\complement)}{\lambda(\M)}  \, d\mathcal{R}(R)
\le
2 \alpha \frac{1}{4}.
\]

\par\noindent
(b) It is clear from \eqref{E:rsep} that $\lambda(\{m\}) = 0$ 
for all $m \in \M$.

\noindent
(c) By Jensen's Inequality, 
$ ( \lambda(A)/2 + \eta(A)/2)^3 \le \lambda(A)^3 /2 + \eta(A)^3 /2$, 
and the sum of two 
terms $ \lambda(A\cap R) \lambda(A \cap R^\complement) + \eta(A\cap R) \eta(A \cap R^\complement)$
is less than or equal to the sum of four 
terms resulting from the product
$ (\lambda + \eta)(A\cap R) \, (\lambda + \eta)(A\cap R^\complement) $.

\noindent
(d) For any Borel set $A$,
$$
(t\lambda)(A)^{3}=t^{3}\lambda(A)^{3}
    \le 2t\alpha \int (t\lambda)(A\cap R)(t\lambda)
     (A\cap R^{\complement}) d\mathcal{R}(R).
$$

\noindent
(e) For any Borel set $A$,
\begin{align*}
\eta(A)^{3}&\le e^{3 \bar H}\lambda(A)^{3}\\
&\le
e^{3 \bar H}\cdot 2\alpha \int \lambda(A\cap R)
         \lambda(A\cap R^{\complement}) \, d\mathcal{R}(R)\\
&\le
2\alpha e^{5 \bar H}\cdot \int \eta(A\cap R)
          \eta(A\cap R^{\complement}) \, d\mathcal{R}(R).
\end{align*}
\end{proof}

\begin{remark}
Most pairs  $(\mathcal{R},\lambda)$ that arise naturally in 
applications are shattering,  but there do exist pairs for
which the  recombination measure  $\mathcal{R}$ splits
every set with positive $\lambda$-measure into sets with
positive $\lambda$-measure and the pair nonetheless fails
to be shattering.  For example,  let $\M$ be the positive
real axis. Let $\lambda$ have a density with respect
to Lebesgue measure equal to $ 2^{-k} $ on $ [k, k+1)$.
Consider random partitions of the axis into two parts $[0,x)$
and $[x, \infty)$ (taken for the sake of symmetry in random order),   
where the recombination measure chooses $ x $  according to a 
probability density equal to  $ 2^{-k} (1/k) (x- k)^{-1 + 1/k} $
on $[k, k+1)$.   This density assigns the same measure to 
each whole interval $[k, k+1) $ as $\lambda$, namely $2^{-k}$. 

For each $k$, consider the set $A_k = [k, k+1)$.  Put $ y = x - k $.
If the pair was shattering, there would be some constant $\alpha$
such that for all $ k $   
\begin{equation}\label{E:noshatter}
\frac{1}{2 \alpha} 
 \le \frac{1}{\lambda(A_k)} 
  \int \frac{ \lambda(A\cap R)}{\lambda(A_k)}
        \frac{ \lambda(A_k \cap R^\complement)}{\lambda(A_k)} \, \mathcal{R}(dR). \\
\end{equation}
With probability $1 - \lambda(A_k) $, the partition does not
cut the interval $A_k$, that is, $x$ falls outside the interval,
and the integrand on the right-hand side vanishes.
The rescaled measure $\mathcal{R}(dR)/\lambda(A_k)$ restricted
to $A_k$ is a probability measure,  and  \eqref{E:noshatter}
can be rewritten 
$$
\frac{1}{2 \alpha}
 \le 
  \int_0^1  \, y \, ( 1 - y) \, (1/k) \,  y^{-1 + 1/k} \,  dy 
   =   \frac{1}{k + 1 } \, - \frac{1}{2k + 1}  
   \le  \frac{1}{k + 1 }.
$$
It follows that there is no finite shattering constant $\alpha$ 
which satisfies the required condition for all the sets $A_k$,
establishing that this pair  $ (\mathcal{R}, \lambda) $ is
not shattering.
\end{remark}


\section{Consequences of shattering}
\label{SS:shat_cons}
\index{shattering}

We are now in a position to state and prove a pair of results 
that quantify the extent to which $\mR^{k} Q$ 
for some probability measure $Q$ on $\G$ is a mosaic
of many small pieces, each taken from genomes
sampled independently from the population
described by $Q$, where we measure the size of a subset of $\M$
using its $\mu Q$ mass.

The following is a convenient way of measuring the 
extent to which a partition of $\M$ is made up
of sets that each have small mass with respect to some
reference measure. We apply it to the random nested partitions
defined in Notation~\ref{N:Rtwiddle}.

\begin{notation}
\label{N:shat}
Suppose that $\mathcal{A}=\{A_{1},\dots,A_{L}\}$ is a
partition of $\M$, $\lambda$ is a measure on $\M$, and $r>0$.
Set
$$
|\mathcal{A}|_{r}^{(\lambda)}:=\sum_{i=1}^{L}\lambda(A_{i})^{r}.
$$
\end{notation}
\glossary{$|\mathcal{A}|_{r}^{(\lambda)}$}
%


\begin{lemma}  \label{L:rsep}
Suppose that the pair $(\mathcal{R},\lambda)$ is shattering
with constant $\alpha$. Then, for all $k \in \bN_0$,
$$
\bE\left[|\mathcal{A}_{k,0}|^{(\lambda)}_{2}\right]
     \,   <  \, \frac{\alpha \lambda(\M)}{k+1}.
$$
\end{lemma}

\begin{proof}
Recall that $\mathcal{A}_{k+1,0}$ is the random partition
\[
\mathcal{A}_{0}(\ell) 
:= \{ W(\upsilon, \ell) : V(\upsilon) = 0 \text{ with } \upsilon \prec \ell \}
\]
for some
$\ell \in \mathcal{L}$ with $V(\ell) = k+1$ 
(the choice of $\ell$ is arbitrary insofar as the distribution 
of $\mathcal{A}_{k+1,0}$ is concerned).  By construction,
a set of the form $W(\upsilon, \ell)$ with $V(\upsilon) = 0$
and $\upsilon \prec \ell$ can be written as either
$W(\gamma,\ell) \cap R_\gamma$ or $W(\gamma,\ell) \cap R_\gamma^\complement$,
where $V(\gamma) = 1$ and $\gamma \prec \ell$, and so
\[
|\mathcal{A}_{k+1,0}|^{(\lambda)}_{2}
=
\sum_{V(\gamma)=1, \gamma \prec \ell}
\left[
\lambda(W(\gamma,\ell) \cap R_\gamma)^2
+
\lambda(W(\gamma,\ell) \cap R_\gamma^\complement)^2
\right].
\]

Set $X_{j} :=|\mathcal{A}_{j,0}|^{(\lambda)}_{2}$ for $j \in \bN_0$
and recall the notation
\[
\mathcal{A}_{1}(\ell) 
:= \{ W(\gamma, \ell) : V(\gamma) = 1 \text{ with } \gamma \prec \ell \}.
\]
From the above and the symmetry of $\mathcal{R}$,
\begin{align*}
\bE \left[ X_{k+1} \cond \mathcal{A}_{1}(\ell) \right]
   &=\sum_{A\in\mathcal{A}_{1}(\ell)} 2\int \lambda(A\cap R)^{2} \, d\mathcal{R}(R)\\
   &\le \sum_{A\in\mathcal{A}_{1}(\ell)} \lambda(A)^{2}
             (1-(1/\alpha) \lambda(A))    \\
   &= |\mathcal{A}_{1}(\ell)|^{(\lambda)}_{2}
               - (1/\alpha) |\mathcal{A}_{1}(\ell)|^{(\lambda)}_{3}.
\end{align*}

%
For any partition $\mathcal{A} = \{A_{1},\dots,A_{L}\}$ of $\M$, it
follows from the Cauchy-Schwarz inequality that 
\[
\begin{split}
|\mathcal{A}|^{(\lambda)}_{2}
& =
\sum_{i=1}^{L} \lambda(A_{i})^{2} \\
& =
\sum_{i=1}^{L} \lambda(A_{i})^{\frac{1}{2}} \lambda(A_{i})^{\frac{3}{2}} \\
& \le
\left(\sum_{i=1}^{L} \lambda(A_{i})^{\frac{2}{2}}\right)^{\frac{1}{2}}
\left(\sum_{i=1}^{L} \lambda(A_{i})^{\frac{6}{2}}\right)^{\frac{1}{2}} \\
& =
\left(|\mathcal{A}|^{(\lambda)}_{1}\right)^{\frac{1}{2}} \left(|\mathcal{A}|^{(\lambda)}_{3}\right)^{\frac{1}{2}} \\
& =
\left(\lambda(\M)\right)^{\frac{1}{2}}
\left(|\mathcal{A}|^{(\lambda)}_{3}\right)^{\frac{1}{2}}. \\
\end{split}
\]
Thus,
\[
|\mathcal{A}|^{(\lambda)}_{3}\ge
\lambda(\M)^{-1} \bigl(|\mathcal{A}|^{(\lambda)}_{2}\bigr)^{2}
\]
and so
%
%
$$
\bE \left[X_{k+1}\cond \mathcal{A}_{1}(\ell)\right]
        \le \tilde X_{k} (1- \tilde X_{k}/c ),
$$
where $c=\alpha\lambda(\M)$ and
$\tilde X_k = |\mathcal{A}_{1}(\ell)|^{(\lambda)}_{2}$.
By construction, the random partition
$\mathcal{A}_{1}(\ell)$
has the same distribution as $\mathcal{A}_{k,0}$.
Applying Jensen's inequality to the concave function $x(1-x/c)$,
we see that
\[
\begin{split}
\bE [X_{k+1}]
& \le \bE \bigl[ \tilde X_{k}(1-\tilde X_{k}/c)\bigr] \\
& \le \bE [\tilde X_{k}] \bigl( 1 - \bE [\tilde X_{k}]/c \bigr) \\
& = \bE [X_{k}] \bigl( 1 - \bE [X_{k}]/c \bigr), \\
\end{split}
\]
and so
\[
\frac{c}{\bE [X_{k+1}]} \ge
       \frac{c}{\bE [X_{k}]}   +   \frac{ c}{  c - \bE [X_{k}]}
       >   \frac{c}{\bE [X_{k}]}   +   1
\]
provided $\bE [X_{k}] < 1$.
%
%
We have  $X_{0}=\lambda(\M)^{2}$, which is strictly smaller
than $ c = \alpha \lambda(\M) $ since the shattering
constant $\alpha$ is necessarily greater than $\lambda(\M)$.
It follows that  $ \bE [X_{k}] <   c/(1+k) $ for all $ k \ge 0 $.
\end{proof}
%
%


%
%
%

\section{Convergence to Poisson of iterated recombination}
\label{SS:iter_recomb}

The purpose of the shattering condition is to guarantee that
recombination bring the population distribution of genotypes
back to that of a Poisson random measure, even in the face
of the countervailing force of selection.   However, before we proceed
to our main result in which recombination, mutation,
and selection are intermingled, it is instructive to see how
recombination does its job in the absence of selection.
This section shows that, under the shattering condition,
the recombination process acting alone rapidly shuffles
the distribution of a non-Poisson random measure on $\M$
to produce a probability measure on $\G$ that is close to the
distribution of a Poisson random measure on $\M$.
\index{recombination}
\index{shattering}

Along with the concept of annealed recombination, 
our result on convergence of iterated recombination in the
absence of mutation and selection also requires the following
condition on the initial distribution of genotypes in the population.

We cannot expect iterated recombination to to send an arbitrary
probability measure $Q$ on $\G$ to Poisson.  In the extreme situation
where the intensity measure $\mu Q$ is diffuse and yet
a realization of $Q$ is not simple, recombination is unable
to break up atoms of size greater than one, and so no amount
of recombination will produce a result that is close to
a Poisson random measure with intensity $\mu Q$.  We have, of course,
adopted the standing assumption that all of our probability
measures on $\G$ are the distributions of simple random measures, but
it is clear that if $Q$ is, in some sense, close to the
distribution of a random measure that is not simple, then
the convergence of $\mR^{k} Q$ to the Poisson 
measure $ \Pi_{\mu Q}$ can at best be arbitrarily slow.  The following
condition quantifies how far a probability measure $Q$ is from one
with a realization that has atoms and it will enable us to provide
a bound on the rate that $\mR^{k} Q$ converges to $ \Pi_{\mu Q}$.

\begin{definition}
A probability measure $Q$ on $\G$ is {\em dispersive} if there is a
constant $\beta$ such that for any Borel set $A\subseteq\M$,
$$
\int_\G g(A)\indic_{\{g(A)\ge 2\}} \, dQ(g) \le \beta \mu Q(A)^{2}.
$$
\end{definition}
\index{dispersive probability measure}
Of course, the distribution of a Poisson random measure is always dispersive.


\begin{theorem}  \label{T:tg}

Suppose that the pair $(\mathcal{R},\mu Q)$ is shattering 
and $Q$ is dispersive with constant $\beta$.
Then, $\mR^{k} Q$ converges to the Poisson 
measure $ \Pi_{\mu Q}$  as $k\to\infty$, with
%
%
\begin{equation}  \label{E:recombound}
\|\mR^{k}Q   -   \Pi_{\mu Q} \|_{\Was}
   \le (6\beta + 2)(\alpha  \mu Q(\M)) (k+1)^{-1}.
%
\end{equation}
\end{theorem}

\begin{proof}

Recall that for a measure $g \in \G$ and a Borel set $A \in \BM$ 
that $g_A \in \G$ is the measure $g(\cdot \cap A)$.  
Write $Q\bigl|_A$
for the push-forward of the probability measure $Q$ by the map
$g \mapsto g_A$.  (This notation is used only within the context
of this proof,  and is not to be confused with the use of $|$
elsewhere to denote the restriction of a measure to a set.) 

Let $\mathcal{A}=\{A_{1},\dots,A_{L}\}$ be a partition of $\M$.  
Then, by definition,
the push-forward of $\mR_{\mathcal{A}}Q$ by the map
$g \mapsto (g_{A_1}, \ldots, g_{A_L})$ is   
$Q\bigl|_{A_1} \otimes \cdots \otimes Q\bigl|_{A_L}$.  
Similarly, the push-forward of $\Pi_{\mu Q}$ by the map
$g \mapsto (g_{A_1}, \ldots, g_{A_L})$ is   
$\Pi_{\mu Q}\bigl|_{A_1} \otimes \cdots \otimes \Pi_{\mu Q}\bigl|_{A_L}$.
%
%
Suppose that $F: \G \to \bR$ is Lipschitz with Lipschitz norm
$\|F\|_{\Lip}$ at most $1$.  Define $\Phi:\G^L \to \bR$
by $\Phi(g_1, \ldots, g_L) = F(g_1 + \cdots + g_L)$.
Note for fixed $g_1, \ldots, g_{i-1}, g_{i+1}, \ldots, g_L$
that the function 
$g \mapsto \Phi(g_1, \ldots, g_{i-1}, g, g_{i+1}, \cdots, g_L)$
is Lipschitz with Lipschitz norm at most $1$ and so,
writing $\bigotimes$ for the product of measures,
\[
\begin{split}
& \left|
\left(\bigotimes_{i=1}^L Q\bigl|_{A_i} 
	- \bigotimes_{i=1}^L \Pi_{\mu Q} \bigl|_{A_i}
\right) [\Phi]
\right| \\
& \quad \le
\sum_{i=1}^L
\left|
\left(
\bigotimes_{j=1}^{i-1} Q\bigl|_{A_j} 
\otimes 		
\bigotimes_{k=i}^{L} \Pi_{\mu Q} \bigl|_{A_k}
-
\bigotimes_{j=1}^{i} Q\bigl|_{A_j} 
\otimes 		
\bigotimes_{k=i+1}^{L} \Pi_{\mu Q} \bigl|_{A_k}
\right)
[\Phi]
\right| \\
& \quad =
\sum_{i=1}^L
\left|
\left(
\bigotimes_{j=1}^{i-1} Q\bigl|_{A_j}
\otimes \left(Q\bigl|_{A_i} - \Pi_{\mu Q}\bigl|_{A_i}\right)
\otimes \bigotimes_{k=i+1}^{L} \Pi_{\mu Q} \bigl|_{A_k}
\right)
[\Phi]
\right| \\
& \quad \le 
\sum_{i=1}^{L} 
\bigl\|
\mR_{\mathcal{A}}Q\bigl|_{A_i}
-\Pi_{\mu Q}\bigl|_{A_i}\bigr
\|_{\Was}.\\
\end{split}
\]
Therefore, 
\begin{equation}  \label{E:thirdbound}
\begin{split}
\bigl\|\mR_{\mathcal{A}}Q-\Pi_{\mu Q}\bigr\|_{\Was}  
\le 
\sum_{i=1}^{L}\was{Q\bigl|_{A_{i}}}{\Pi_{\mu Q}\bigl|_{A_{i}}}.
\end{split}
\end{equation}

Given a Borel subset $A \subseteq \M$,
define $\hat{\pi}$ to be the measure on $\M$ given by
$$
\hat{\pi} (B) = \int_{\G} g(B \cap A)\indic_{\{g(A)=1\}} \, dQ(g).
$$
Clearly $\hat{\pi}\le \mu Q$.  Also, define the sub-probability 
measure $\hat{Q}$ on $\G$ by 
$$
\hat{Q}[F]=\int_{\G}F(g\bigl|_{A}) \indic_{\{g(A)\le 1\}} \, dQ(g).
$$
Observe that $\hat{\pi}$ is the intensity measure 
of $\hat{Q}$.  Note also that
\begin{equation}  \label{E:muA}
\mu Q(A)=\hat{\pi}(A)+\int_{\G}g(A)\indic_{\{g(A)\ge 2\}} \, dQ(g).
\end{equation}

We have
\begin{equation}  \label{E:allbound}
\begin{split}
\bigl\|Q\bigl|_{A}& - \Pi_{\mu Q}\bigl|_{A}\bigr\|_{\Was}\\
&\le  \bigl\| Q\bigl|_{A} - \hat{Q}\bigr\|_{\Was}+\bigl\|\hat{Q} 
   - \Pi_{\hat{\pi}}\bigr\|_{\Was}+\bigl\|\Pi_{\hat{\pi}} 
   - \Pi_{\mu Q}\bigl|_{A}\bigr\|_{\Was}.
\end{split}
\end{equation}
%
We bound the first term on the right by 
\begin{equation}  \label{E:bound1}
\was{Q\bigl|_{A}}{\hat{Q}}\le Q\ls g \in \G :g(A)\ge 2\rs \le \beta \mu Q(A)^{2},
\end{equation}
since $Q$ is dispersive with constant $\beta$.
\index{dispersive probability measure}

For a bound on the second term on the right of  \eqref{E:allbound},
for any  $F:\G\to\bR$ with $\|F\|_{\Lip}\le 1$,
we can write $ \hat{\pi}[F (\delta_{\cdot})] $ 
for $\int F(\delta_{m}) \, d\hat{\pi}(m)$, and we have
\[
\begin{split}
\bigl| \hat{Q}[F] - \Pi_{\hat{\pi}}[F]\bigr| 
&\le \bigl| Q\ls g \in \G: g(A)=0\rs 
      - \Pi_{\hat{\pi}}\ls g \in \G:g(A)=0\rs\bigr| \\
& \quad +  \bigl| \hat{Q} [F\indic_{\{g \in \G:g(A)=1\}}] 
      - \Pi_{\hat{\pi}} [F\indic_{\{g \in \G :g(A)=1\}}]\bigr| \\
& \quad +   \Pi_{\hat{\pi}}\ls g \in \G:g(A)\ge 2\rs\\
&\le \left| 1-\hat{\pi}(A)+Q\ls g \in \G:g(A)\ge 2\rs - e^{-\hat{\pi}(A)} \right|\\
& \quad + \left|\hat{\pi}[F (\delta_{\cdot})]\left( 1 - e^{-\hat{\pi}(A)}\right)\right| 
         +  \frac{\hat{\pi}(A)^{2}}{2}\\
&\le Q\ls g \in \G:g(A)\ge 2\rs + \frac{\mu Q(A)^{2}}{2} 
          +\mu Q(A)^{2}  +\frac{\mu Q(A)^{2}}{2}. \\
\end{split}
\]
Thus,
\begin{equation}  \label{E:bound2}
\was{\hat{Q}}{\Pi_{\hat{\pi}}}\le \bigl( \beta +2\bigr) \mu Q(A)^{2}.
\end{equation}

Finally, for our bound on the third term of  \eqref{E:allbound},
we have, by Lemma~\ref{L:poisson},
$$
\bigl\|\Pi_{\hat{\pi}} - \Pi_{\mu Q}\bigl|_{A}\bigr\|_{\Was}
\le 4 \bigl\|\hat{\pi} - \mu Q(\cdot \cap A)\bigr\|_{\Was},
$$
because the intensity measure of  $\Pi_{\mu Q}\bigl|_{A}$ is 
$\mu Q(\cdot \cap A)$. For any $f:A\to\bR$ with $\|f\|_{\Lip}\le 1$,
\begin{align*}
\bigl| \hat{\pi}[f] - \mu Q [f]\bigr| 
   &= \left| \int_{\G} g[f] \, d\hat{Q}(g) 
   	- \int_{\G} g[f \indic_A] \, dQ(g) \right|\\
   &= \left| \int_{\G} g[f \indic_A]\indic_{\{g(A)=1\}} \, dQ(g)
   	- \int_{\G} g[f \indic_A] \, dQ(g) \right|\\
   &= \left|\int_{\G} g[f \indic_A] \indic_{\{g(A)\ge 2\}} \, dQ(g) \right|\\
   &\le Q\ls g \in \G:g(A)\ge 2\rs \\
   & \le \beta \mu Q(A)^{2}.
\end{align*}

Thus,
\begin{equation}  \label{E:bound3}
\was{\Pi_{\hat{\pi}}}{\Pi_{\mu Q}\bigl|_{A}}\le 4\beta \mu Q(A)^{2}.
\end{equation}
Putting \eqref{E:bound1}, \eqref{E:bound2} and \eqref{E:bound3} 
into \eqref{E:allbound}, we get
\begin{equation}  \label{E:onesetbound}
\was{Q\bigl|_{A}}{\Pi_{\mu Q}\bigl|_{A}}\le (6\beta +2) \mu Q(A)^{2}.
\end{equation}
By \eqref{E:thirdbound}, then, 
$$
\bigl\|\mR_{\mathcal{A}}Q-\Pi_{\mu Q}\bigr\|_{\Was}  
     \le  (6 \beta + 2 ) \, \sum_{i=1}^{L}  \mu Q(A_{i})^{2}.
$$
It follows, via Jensen's Inequality, that 
\begin{equation}  \label{E:donebound}
\was{\mR^{k}Q}{\Pi_{\mu Q}}\
    = \was{ \bE \left[\mR_{\mathcal{A}_{k,0}}Q\right]}{\Pi_{\mu Q}}
    \le (6\beta+2)\, \bE \left[|\mathcal{A}_{k,0}|_2^{(\mu Q)}\right]. 
\end{equation}
Applying Lemma~\ref{L:rsep} completes the proof of \eqref{E:recombound}.
\end{proof}

Note that Theorem ~\ref{T:tg} is essentially a 
random measure version of Le Cam's  Poisson convergence 
result of \cite{LC60}.
\index{LeCam, Lucien} 


Theorem~\ref{T:tg} is about the situation where the initial 
distribution of genotypes in the population
satisfies both a shattering condition and a dispersive
condition and recombination acts alone.
Our main convergence theorem in Chapter~\ref{Ch:convergence}
posits an initial distribution of genotypes that satisfies a shattering
condition along with an hypothesis on its Radon-Nikodym derivative
with respect to the distribution of the Poisson measure
with the same intensity measure, but no dispersive condition.
When the selective cost $S$ and the mutation rate $\nu$ are set
to zero,  the main convergence theorem also covers the situation
in which recombination acts alone.  The relationship between
these two results on iterated recombination is clarified by
the following observation, where we recall the definition
of the quantity $\Theta(P,Q)$ from \eqref{E:ThetaPQ}.

\begin{proposition}
Suppose that $ P = \Pi_\pi$ and 
$Q$ is a probability measure on $\G$ such that $ \mu Q = \pi $ 
and $\Theta(P,Q) < \infty$.  Then, there exists a positive constant $\beta$
for which the dispersive condition
$$ 
\int_\G g(A) \mathbf{1}_{\{g(A) \ge 2\}} \, dQ(g) \le \beta \mu Q(A)^2
$$
holds.
\end{proposition}

\begin{proof}
Set $\hat H := \Theta(P,Q)$.
Lemma~\ref{L:RNcompare}, to be proved in the appendix,
shows that the Lipschitz condition implies that 
$$
\int_\G g(A) \mathbf{1}_{\{g(A) \ge 2\}} \, dQ(g)
\le
\int_\G g(A) \mathbf{1}_{\{g(A) \ge 2\}} \exp\{\hat H (g(\M) + \pi(\M)) \} \, dP(g).
$$

Splitting $\M$ up into two sets $A$ and $A^\complement$ and using the independence
properties of the Poisson probability measure $P$, we can bound this quantity
by a constant multiple of
\begin{align*}
& \int_\G g(A) \mathbf{1}_{\{g(A) \ge 2\}} \exp\{\hat H g(A)\} \, dP(g)  \\ 
& =
\sum_{k \ge 2} k e^{\hat H k} e^{-\pi(A)} \frac{\pi(A)^k}{k!}  \\ 
& =
e^{-\pi(A)} e^{\hat H} \pi(A) [\exp\{e^{\hat H} \pi(A)\} - 1]  \\ 
& \le
e^{-\pi(A)} e^{\hat H} \pi(A) e^{\hat H} \pi(A) \exp\{e^{\hat H} \pi(A)\}.
\end{align*}
The last expression 
is bounded by a constant multiple of $ \pi(A)^2 $, so the
dispersive condition
$$
\int_\G g(A) \mathbf{1}_{\{g(A) \ge 2\}} \, dQ(g) \le \beta \mu Q(A)^2
$$
holds.

\end{proof}

\section{Atoms in the initial intensity}
\label{SS:atomstart}

We have discussed the shattering condition in the context
of the mutation measure $\nu$.  We shall also be assuming
that the initial intensity measure $\rho_0$ for the starting
state $P_0$ is shattering.  This hypothesis excludes initial
intensity measures with atoms.   The question 
arises as to whether  atoms could be allowed.  The answer
is no.  

The reason for excluding atoms is illustrated by an
artificial case in which the mutation measure $\nu$
is zero and $\rho_0$ has a single atom at a point $m \in \M$.
As time passes $m$ remains the only mutation present 
in the population, since recombination does not introduce 
mutations not already present.  Each occurring element 
of $\G$ consists of an integer mass at $m$.
We identify $\G$ with the nonnegative integers $\bN_0$
and simply write $g$ for $g(m)$.  We can then
think of $S$ as a function
from $\bN_0$ to $\bR_+$.  
Take the initial probability measure $Q_0$ to be the distribution of
a Poisson random variable with mean $\lambda = \rho_0(m)$.
As usual, the notation $\lfloor t \rfloor $ denotes the greatest integer
less than or equal to the real number $t$.
\glossary{$\lfloor \cdot  \rfloor $}
 
After $\lfloor tn \rfloor $ generations of selection,
in the notation of Notation~\ref{N:Qk}, the population is
described by a measure $Q$ which, by   \eqref{E:Sop},
equals 
\begin{align*}
Q_{\lfloor tn \rfloor}(dg)
     & = \mS_n^{\lfloor tn \rfloor} \, dQ_0 (g)   \\ 
     & = \frac{ \exp\{- {\lfloor tn \rfloor} S(g)/n\} \, dQ_0(g) } 
              { \int_\G \exp\{- {\lfloor tn \rfloor} S(h)/n\} \, dQ_0(h)}.  
\end{align*}
This mass converges as $ n \to \infty $ to 
$$
\exp\{- t S(g) \} Q_0(dg)
  \Big / \int_\G \exp\{- t S(h) \} Q_0(dh), 
$$
which, in our case, is a 
probability measure on the nonnegative integers that assigns to 
each integer $g$ the mass 
$$
\left( \exp\{- t S(g)\} e^{-\lambda} (\lambda^g \, / \, g! )\right)  
\; \bigg / \; 
\left( \sum_{k=0}^\infty \exp\{- t S(k)\} e^{-\lambda} 
                 (\lambda^k \, / \, k! ) \right).
$$
This probability measure is not Poisson except
in the non-epistatic case case where $S(g) = c g$ for some constant
$c > 0$.   In short, in the presence of an atom in the initial
intensity,  the operation of selection drives the distribution
of genotypes in the population away from 
that of a Poisson random measure,  and 
recombination is powerless to restore its Poisson character.
For this reason, something akin to the shattering condition 
for the recombination measure and the initial
intensity is essential.

\section{Preview of the main convergence result}
\label{SS:conv_state}


We can now state the hypotheses and conclusions 
of our main result,  the result  that justifies
the continuous-time model \eqref{E:dynam} as a limit
of discrete-time models in which recombination acts
on a faster time scale than mutation and selection.
The formal statement of the theorem and its proof are 
given in Chapter~\ref{Ch:convergence}, Theorem~\ref{T:limit}.
The proof is based on supporting results established in
Chapters \ref{Ch:complete_Poisson}  and \ref{Ch:techlem}.  
The statement of the theorem is given here by way of
preview,  so as to make the purposes served by the  
supporting theorems and lemmas in those chapters easier to discern.

We continue to use the notation $\lfloor x \rfloor$ to denote
the greatest integer less than or equal to the
real number $x$.

\bigskip
\noindent
{\bf Hypotheses:}
\par\noindent
Let $(\rho_t)_{t \ge 0}$ be the measure-valued
dynamical system of \eqref{E:dynam} whose
existence is guaranteed by Theorem~\ref{T:existence}.
Suppose that the selective cost function $S$ satisfies the
hypotheses of Theorem~\ref{T:existence}, namely 
\begin{itemize}
\item
$S(0)=0$,
\item
$S(g)\le S(g+h)$ for all $g,h\in\G$,
\item
for some constant $\sigma$,
$\bigl| S(g)-S(h)\bigr| \le \sigma \bigl\| g-h \bigr\|_{\Was}$,
for all $g,h\in \G$.
\end{itemize}
\index{recombination}
\index{selective cost}
In addition, suppose that the following assumptions
are in force.
\begin{itemize} 
\item
The pair $(\mathcal{R},\nu)$
consisting of the recombination measure and the 
mutation measure is shattering.
\item
The pair  $(\mathcal{R}, \rho_0)$ 
consisting of the recombination measure and the 
initial intensity is shattering.
\item
The initial measure $Q_{0}$ is equivalent to
its Poissonization $P_{0} := \mP Q_{0}= \Pi_{\mu Q_0} = \Pi_{\rho_{0}}$,
and $\Theta(P_0,Q_0)<\infty$.
\end{itemize}

\bigskip
\noindent
{\bf Conclusions:}
\par\noindent
Then, for any $T >\epsilon> 0$,
\[  
\lnf \, \sup_{\epsilon \le t\le T} \, 
     \left\| \Pi_{\rho_{t}}- Q_{\lfloor tn \rfloor} \right\|_{\Was} \, = \, 0.
\]
If, in addition, the initial measure $Q_{0}=P_{0}$ is Poisson, then
this equation holds for $\epsilon=0$.

\begin{remark}
We have assumed throughout, for notational convenience, 
a particular order of operations -- 
in each generation there is first selection, 
then mutation, then recombination.  
This order has  no special significance, 
and the proofs hold
equally well for another order, or if the same total 
amounts of mutation and selection were split up into 
multiple bouts within a generation, whether before or after recombination.
\end{remark}
\index{recombination}

\begin{remark}
Because of Lemma~\ref{L:must_be_diffuse}, 
the hypotheses of Theorem~\ref{T:limit}
imply that the mutation intensity measure $\nu$ 
and the initial intensity $\rho_0$ are both diffuse.
It follows easily from this that each probability measure
$Q_{k}$ assigns all of its mass to the set of elements of $\G$ that 
have atoms of mass one; that is, every $Q_{k}$ is the 
distribution of a {\em simple} integer-valued
random measure. 
 
{\bf From now on, we assume without further comment
that all the probability measures
on $\G$ we consider are distributions 
of simple integer-valued random measures.}
\end{remark}

Before beginning the formal presentation in the following chapters,
we outline the strategy of the proof.  The distance in the 
Wasserstein metric between 
our continuous-time dynamical system and our discrete-generation
system at any point in time for any given value of the 
generation-scaling parameter $n$ can be broken into three pieces:
\index{Wasserstein metric}
\begin{enumerate}
\item
the distance between the dynamical system and (with a suitable
re-scaling of time) the discrete-generation
system $O_{k}$ with complete Poissonization after every generation;
\item
the distance between $O_{k}$ and the discrete-generation system
with Poissonization
only at the end of $k$ steps,  $ \mathfrak{P} Q_{k} $, which we
may call {\em end-state Poissonization};
\item
the distance between $ \mathfrak{P} Q_{k} $ and the discrete-generation
system $Q_{k}$ itself.
\end{enumerate}

Chapter~\ref{Ch:complete_Poisson} is devoted to showing that
the first distance,  
the one between $\Pi_{\rho_t}$  and $O_{\lfloor kt \rfloor}$,  is of
order $1/n$.   This part of the proof is the only one in which
the dynamical system comes into play,  and it is also the only
one in which recombination does not have a role because both systems
are built (implicitly for the first and explicitly for the second)
with complete Poissonization rather than a finite amount of recombination.  
Here we must focus primarily on understanding 
the operation of selection when the selective cost is scaled by $1/n$.

Chapter~\ref{Ch:techlem}  lays the groundwork for bounds on the
second distance and the third distance.
The second distance is a distance between the distributions
of two Poisson random 
measures, which, as we know from Lemma~\ref{L:poisson},  
can be bounded in terms of the distance
between their intensity measures.  The probability 
measure $ \mathfrak{P} Q_{k} $
has the same intensity measure as $Q_{k}$, so the quantity to be
bounded is the distance between $ \mu O_{k}$ and $\mu Q_{k}$.
We show in Chapter~\ref{Ch:convergence} that such a bound
can be derived from a bound on the third distance, that
between the probability measures $ \mathfrak{P} Q_{k} $ and $Q_{k}$.  

The required bound on the third distance has to hold for every
generation between zero and $\lfloor Tn  \rfloor$.   
The approach in the proof is to pick any intermediate generation
as a provisional starting point and look at the dynamical systems
after $k$ steps beyond such a starting point.  The dynamical systems
differ by some terms that involve the initial states
and some terms which primarily involve the selection costs
accumulated over $k$ steps.  Chapter~\ref{Ch:techlem} is 
largely devoted to showing that recombination drives down the
difference terms involving the starting states by a factor of $k+1$,
while terms involving selection costs remain bounded by 
a constant multiple of $k/n$. It follows that taking $ k = \sqrt{n}$
steps beyond any starting point makes the former kind of terms
drop by $1/\sqrt{n}$ before the latter kind of terms grow by
more than $1/\sqrt{n}$.   The dynamical system are getting closer to
each other over a time interval of length $ k/n = 1/\sqrt{n}$ 
which becomes infinitesimal as $n$ goes to infinity.      

In our infinite-population setting, as we have said, the
process of selection is a process which reweights the probabilities
of finding various combinations of mutant alleles in the genotype
of a randomly selected individual.  The tree structure of 
recombination events can be specified prior to the introduction
of selection.   Because the transformation of
probability measures by reweighting is at issue,
it makes sense to represent the probability measures $ Q_{k} $ and $O_{k}$
by their Radon-Nikodym derivatives with respect to a suitable underlying
probability measure.  The Poisson probability measure $P_{k}$ serves this
purpose.  These probability measures, and their starred counterparts
that keep track of the vintages of mutations from wild type,
are separate objects for each separate generation.  
Joint distributions of variables across generations and
sample paths of time-serial processes are not in the foreground.
However, in the background lie the tree-structure of 
recombination and the accumulating effects of selection which 
do reach across generations and do lead to such joint distributions.
The decomposition of genotypes in terms of vintages allows
the relevant features of these background joint distributions 
to be inferred from current states,  and so facilitates the proof.


\chapter{Convergence with complete Poissonization}  
\label{Ch:complete_Poisson}   

%

We embark on the proof of our main convergence result,
Theorem~\ref{T:limit},   by establishing in this chapter
a result, important in its own right, which provides
the first step in that proof.  For this result,
Theorem~\ref{T:pim},  the recombination operator $\mR$ is replaced
by the operator $\mP$ defined in \eqref{E:Pop} 
that immediately replaces a probability measure
on $\G$ by the distribution of a Poisson random measure
with the same intensity measure.
In place of  $Q_{k}:=(\mR \mM_{n}\mS_{n})^k Q_{0}$, 
we study the sequence of 
probability measures $O_{k}:=(\mP \mM_{n}\mS_{n})^k P_{0}$, 
defined in Notation~\ref{N:Qk}.
Both the measures $O_{k}$ and the measures $ \Pi_{\rho_t}$ 
that define our continuous-time dynamical system are
Poisson measures,  determined by their intensity measures.
It is simpler to work with the intensity measures, defined
on $\M$, than with probability measures defined on $\G$. The
relationship between discrete generations and continuous
time, mediated by the scaling parameter $n$, has to 
figure prominently in the proof.
Theorem~\ref{T:pim} is essentially a {\em shadowing} result, 
but none of the standard shadowing theorems, 
such as those in \cite{CKP95}, seem to cover this case.
%

\begin{remark}
In this and the following chapters we frequently make use of 
multiple ``constants'', numbered successively 
as $c_{1},c_{2},\dots$ or $C_{1},C_{2},\dots$. 
These constants may depend on on the time horizon $T$, 
the Lipschitz constant $\sigma$ for the selective cost $S$, 
the total mass $\nu(\M)$ of the mutation measure, 
and the total mass $\rho_{0}(\M)$ of the intensity measure
for the initial population.
\end{remark}

\begin{notation}
Write $\pi_{k} := \mu O_{k} $ for the intensity measure of the
Poisson probability measure $O_{k}$.
\end{notation}
\glossary{$\pi_k$}

\begin{theorem}
\label{T:pim}
There are constants $A$ and $B$, depending 
on $\sigma$, $\nu(\M)$ and $\rho_{0}(\M)$, 
such that for every $T > 0$,
\[
\sup_{0\le t\le T} \left\|\pi_{\lfloor tn\rfloor}-\rho_{t} \right\|_{\Was} 
\le 
\frac{A}{n} (1+T) T \exp\left\{B (1+T) T \right\},
\]
and hence
\[  
\sup_{0\le t\le T} \left\| O_{\lfloor tn \rfloor}    
          - \Pi_{\rho_{t}} \right\|_{\Was} 
\le
4 \frac{A}{n} (1+T) T \exp\left\{B (1+T) T \right\}.
\]
\end{theorem}

\begin{proof}
It suffices by Lemma~\ref{L:poisson} to establish the first inequality.
Note that
\begin{equation}
\label{E:discrete_plus_interpolate}
\begin{split}
\sup_{0\le t\le T} \left\| \pi_{\lfloor t n \rfloor}-\rho_{t} \right\|_{\Was}
& \le
\sup_{0 \le k \le \lfloor T n \rfloor} 
\Bigl[
\left\|\pi_{k}-\rho_{k/n}\right\|_{\Was} \\
& \quad +
\sup_{k/n \le t \le (k+1)/n \wedge T} 
\left\|\rho_{t}-\rho_{k/n}\right\|_{\Was}
\Bigr]. \\
\end{split}
\end{equation}

We obtain a bound on the first term 
on the right-hand side of \eqref{E:discrete_plus_interpolate},
namely, 
\begin{equation}
\label{E:discrete_plus_interpolate.1}
\sup_{0 \le k \le \lfloor T n \rfloor} 
\left\|\pi_{k}-\rho_{k/n}\right\|_{\Was} 
\end{equation}
by establishing a bound for $\|\pi_{k+1}-\rho_{(k+1)/n}\|_{\Was}$
in terms of    $ \|\pi_{k}-\rho_{k/n}\|_{\Was}$
and then iterating that bound.

Given $\pi \in \H^{+}$ and the associated Poisson random 
measure $ X^{\pi}$ with intensity $\pi$, 
define $H_\pi : \M \to \bR_+$ by 
\[
H_\pi(x) 
:= \frac{\bE \left[\exp\{-S(X^{\pi}+\delta_{x})/n\}\right]}
    {\bE \left[\exp\{-S(X^{\pi})/n\}\right]}.
\]
\glossary{$H_{\pi}$} 
By the assumptions on the selective cost $S$ and the inequality
\[
e^{-b}-e^{-c} = \int_b^c e^{-z} \, dz \le (c-b) e^{-a}
\]
for $a \le b \le c$, we have  
\begin{equation}
\label{E:H_bounded}
0 \le H_\pi(x) \le 1, \quad x \in \M,
\end{equation}
and
\begin{equation}
\label{E:H_Lipschitz}
\begin{split}
|H_\pi(x) - H_\pi(y)| 
& \le
\frac{\bE 
\left[
\exp\{-S(X^{\pi})/n\} 
|S(X^{\pi}+\delta_{x})/n - S(X^{\pi}+\delta_{y})/n|\right]}
{\bE \left[\exp\{-S(X^{\pi})/n\}\right]} \\
& \le
\frac{\sigma}{n} \|\delta_x - \delta_y\|_{\Was} \\
& =
\frac{\sigma}{n} d(x,y), \\
\end{split}
\end{equation}
where $d$ here is the metric on $\M$. 
We take advantage of the Lipschitz constant $\sigma$ for the
selective cost $S$ to write $0\le S(g)\le \sigma g(\M)$, which implies
for any $\pi\in \H^{+}$ that 
$$
\bE\left[ \exp\left\{-S(X^{\pi})/n\right\}\right]
   \ge \bE\left[ \exp\left\{-\frac{\sigma}{n}X^{\pi}(\M)\right\}\right]
   =\exp\left\{e^{-\sigma \pi(\M)/n}\right\}.
$$
Thus, for any $\pi', \pi'' \in \H^{+}$,
\begin{equation}
\label{E:H_two_measures}
\begin{split}
& |H_{\pi'}(x) - H_{\pi''}(x)| \\
& =
\biggl|
\bE \left[\exp\{-S(X^{\pi''})/n\}\right] 
     \bE \left[\exp\{-S(X^{\pi'}+\delta_x)/n\}\right] \\
& \quad -
\bE \left[\exp\{-S(X^{\pi'})/n\}\right] 
      \bE \left[\exp\{-S(X^{\pi''}+\delta_x)/n\}\right]
\biggr| \\
& \qquad \bigg/ \biggl(\bE \left[\exp\{-S(X^{\pi'})/n\}\right] 
        \bE \left[\exp\{-S(X^{\pi''})/n\}\right] \biggr) \\
& \le
e^{\sigma (\pi'(\M)+\pi''(\M)) / n}
\biggl (
\Bigl|
\bE \left[\exp\{-S(X^{\pi''})/n\}\right] 
      \bE \left[\exp\{-S(X^{\pi'}+\delta_x)/n\}\right] \\
& \qquad -
\bE \left[\exp\{-S(X^{\pi'})/n\}\right] 
      \bE \left[\exp\{-S(X^{\pi'}+\delta_x)/n\}\right]
\Bigr| \\
& \quad +
\Bigl| 
\bE \left[\exp\{-S(X^{\pi'})/n\}\right] 
      \bE \left[\exp\{-S(X^{\pi'}+\delta_x)/n\}\right] \\
& \quad \qquad -
\bE \left[\exp\{-S(X^{\pi'})/n\}\right] 
      \bE \left[\exp\{-S(X^{\pi''}+\delta_x)/n\}\right]
\Bigr| 
\biggr )\\
\end{split}
\end{equation}
and
\begin{equation}
\label{E:H_Lipschitz2}
\begin{split}
& |H_{\pi'}(x) - H_{\pi''}(x)| \\
& \le
e^{\sigma (\pi'(\M)+\pi''(\M)) / n}
\biggl (
\left|
\bE \left[\exp\{-S(X^{\pi''})/n\}\right] 
-
\bE \left[\exp\{-S(X^{\pi'})/n\}\right] 
\right| \\
& \quad +
\left| 
\bE \left[\exp\{-S(X^{\pi'}+\delta_x)/n\}\right]
-
\bE \left[\exp\{-S(X^{\pi''}+\delta_x)/n\}\right]
\right|
\biggr ) \\
& =
e^{\sigma (\pi'(\M)+\pi''(\M)) / n}\\
&\quad\times
\biggl (
\left|
\bE \left[1 - \exp\{-S(X^{\pi''})/n\}\right] 
-
\bE \left[1 - \exp\{-S(X^{\pi'})/n\}\right] 
\right| \\
& \qquad+
\left| 
\bE \left[1 - \exp\{-S(X^{\pi'}+\delta_x)/n\}\right]
-
\bE \left[1 - \exp\{-S(X^{\pi''}+\delta_x)/n\}\right]
\right|
\biggr ). 
\end{split}
\end{equation}
Now we may apply Lemma~\ref{L:poisson2} with $\beta=0$ and $C=\sigma/n$,
followed by Lemma~\ref{L:poisson}, to obtain
$$
|H_{\pi'}(x) - H_{\pi''}(x)| 
    \le e^{\sigma (\pi'(\M)+\pi''(\M)) / n} \cdot 
    \frac{8\sigma}{n} \was{\pi'}{\pi''}.
$$

Now, returning to the task of bounding the quantity in
\eqref{E:discrete_plus_interpolate.1},
\begin{equation}
\label{E:discrete_split}
\begin{split}
\left\| \pi_{k+1}-\rho_{(k+1)/n} \right\|_{\Was}
& \le
\left\| \pi_{k+1} - \frac{\nu}{n} 
\, - \, 
H_{\pi_{k}} \cdot \pi_{k} \right\|_{\Was} \\
& \quad +
\left\| H_{\pi_{k}} \cdot \pi_{k} 
\, - \, 
H_{\rho_{k/n}} \cdot \pi_{k}  \right\|_{\Was} \\
& \quad +
\left\| H_{\rho_{k/n}} \cdot \pi_{k}  
\, - \, 
H_{\rho_{k/n}} \cdot \rho_{k/n}  \right\|_{\Was} \\
& \quad +
\left\|
\rho_{(k+1)/n}  - \frac{\nu}{n} \, - \, H_{\rho_{k/n}} \cdot \rho_{k/n}\right \|_{\Was}. \\
\end{split}
\end{equation}

Consider the first term on the right-hand side of
\eqref{E:discrete_split}. By definition,
\[
\pi_{k+1} = \mu O_{k+1} = \mu \mP\mM_{n}\mS_{n} O_{k}.
\]
Thus,
\[
\pi_{k+1} - \frac{\nu}{n} 
= 
\mu \mP\mM_{n}\mS_{n} O_{k} - \frac{\nu}{n}
=
\mu \mM_{n}\mS_{n} O_{k} - \frac{\nu}{n}
=
\mu \mS_{n} O_{k}
=
\mu \mS_{n} \Pi_{\pi_{k}}.
\]
Hence, for a bounded Borel function $f: \M \to \bR$, 
\[
\begin{split}
\int_\M f(x) \, d\left(\pi_{k+1} - \frac{\nu}{n}\right)(x)
& =
\frac{
\int_\G g[f] \exp\{-S(g)/n\} \, d\Pi_{\pi_{k}}(g)
}
{
\int_\G \exp\{-S(g)/n\} \, d\Pi_{\pi_{k}}(g)
} \\
& =
\frac{
\int_\M f(x) \int_\G \exp\{-S(g + \delta_x)/n\} 
      \, d\Pi_{\pi_{k}}(g) \, d \pi_{k}(x)
}
{
   \int_\G \exp\{-S(g)/n\} \, d\Pi_{\pi_{k}}(g)
}, \\
\end{split}
\]
where we have again made use of Campbell's Theorem  
(see Proposition~\ref{P:Campbell}).
\index{Campbell's Theorem}

Equivalently,
\begin{equation}  \label{E:dpim}
\pi_{k+1} - \frac{\nu}{n} 
=
H_{\pi_{k}} \cdot \pi_{k},
\end{equation}
and so the first term on the right-hand side of
\eqref{E:discrete_split} is zero.

It follows from \eqref{E:dpim} and \eqref{E:H_bounded} that
\begin{equation}
\label{E:absolute_bd_pi}
\pi_{k}(\M) \le \rho_0(\M) + \frac{k}{n} \nu(\M), \quad k \ge 0.
\end{equation}
Combining \eqref{E:absolute_bd_pi} with \eqref{E:H_two_measures} 
establishes that the second term on the right-hand side of
\eqref{E:discrete_split} is bounded by
\begin{equation}  
\label{E:difdipim2}
\begin{split}
& \frac{8\sigma}{n} e^{\sigma(\pi'(\M)+\pi''(\M)) / n}
\left\|\pi_{k} - \rho_{k/n} \right\|_{\Was} \pi_{k}(\M) \\
& \quad \le
 \frac{8\sigma}{n} e^{\sigma(\pi'(\M)+\pi''(\M)) / n}
\left(\rho_0(\M) + \frac{k}{n} \nu(\M) \right)
\left\|\pi_{k} - \rho_{k/n} \right\|_{\Was}. \\
\end{split}
\end{equation}

From \eqref{E:H_bounded} and \eqref{E:H_Lipschitz}, the third term 
on the right-hand side of \eqref{E:discrete_split} is
bounded by
\begin{equation}  
\label{E:difdipim3}
\left\| H_{\rho_{k/n}} \right\|_{\Lip}  \left\|\pi_{k} - \rho_{k/n} \right\|_{\Was}
\le
\left(1 + \frac{\sigma}{n} \right) \left\|\pi_{k} - \rho_{k/n} \right\|_{\Was}.
\end{equation}

For the last term on the right-hand side of \eqref{E:discrete_split} we have
\begin{equation}  
\label{E:difdipim4}
\begin{split}
& \left\|
\rho_{(k+1)/n}  - \frac{\nu}{n} \, - \, H_{\rho_{k/n}} \cdot \rho_{k/n}\right \|_{\Was} \\
& \quad =
\left\|
\rho_{k/n} +\frac{1}{n} \nu + \int_0^{\frac{1}{n}} F_{\rho_{s+k/n}} \cdot \rho_{s+k/n} \, ds
- \frac{\nu}{n} \, - \, H_{\rho_{k/n}} \cdot \rho_{k/n}\right \|_{\Was} \\
& \quad \le
\left\|
\left( 1 - H_{\rho_{k/n}} + \frac{1}{n} F_{\rho_{k/n}} \right) \cdot \rho_{k/n} 
\right\|_{\Was} \\
& \qquad +
 \int_0^{\frac{1}{n}}\left\| F_{\rho_{s+k/n}} \cdot \rho_{s+k/n} - F_{\rho_{k/n}} \cdot \rho_{k/n} \right\|_{\Was}\, ds. \\
\end{split}
\end{equation}

By Lemma~\ref{L:expbound} we know for any $\rho \in \H^+$ that
\[
\begin{split}
-\frac{1}{n}\bE[S(X^{\rho})]
& \le
\log \bE\left[ \exp\left\{-\frac{1}{n} S(X^{\rho})\right\}\right] \\
& \le -\frac{1}{n} \bE[S(X^{\rho})]
	+ 
  \frac{1}{2n^2} \bE[S(X^{\rho})^{2}]
   \exp\left\{\frac{1}{n}\bE[S(X^{\rho})]\right\}, \\
\end{split}
\]
and a similar bound holds with $X^{\rho}$ replaced by $X^{\rho}+\delta_{x}$. 
Note also that
\[
\bE[S(X^{\rho})]
\le 
\bE[S(X^{\rho}+\delta_{x})] 
\le 
\sigma \bE[ (X^{\rho}(\M)+1)]
=
\sigma \left(  \rho(\M) +1 \right)
\]
and
$$
\bE[S(X^{\rho})^{2}]
\le 
\bE[S(X^{\rho}+\delta_{x})^{2}] 
\le 
\sigma^{2}\bE[ (X^{\rho}(\M)+1)^{2}]
=
\sigma^{2}\left(  \rho(\M)^{2}+3\rho(\M) +1 \right).
$$
Applying the obvious bound
\begin{equation}
\label{absolute_bd_rho}
\rho_t(\M) \le \rho_0(\M) +t \nu(\M), \quad t \ge 0
\end{equation}
and the inequality $1-y\le e^{-y}\le 1-y+\frac{y^{2}}{2}$ (for $y\ge 0$), it follows that 
\begin{equation}
\label{E:H_F_comparison}
\begin{split}
& \left| 1 - H_{\rho_{k/n}}(x) + \frac{1}{n} F_{\rho_{k/n}}(x) \right| \\
& \quad =
\biggl| 
1- \exp\left\{ \bE[S(X^{\rho_{k/n}})-S(X^{\rho_{k/n}}+\delta_{x}]\right\} \\
& \quad +
\frac{1}{n} 
\bE \left[S(X^{\rho_{k/n}}+\delta_x) -S(X^{\rho_{k/n}}) \right]
\biggr| + \frac{C_{1}}{n^{2}}\\
& \quad \le
\frac{C_2}{n^2}.
\end{split}
\end{equation}
Thus,
\begin{equation}
\label{E:difdipim4.1}
\begin{split}
\left\|
\left( 1 - H_{\rho_{k/n}} + \frac{1}{n} F_{\rho_{k/n}} \right) \cdot \rho_{k/n} 
\right\|_{\Was}
& \le
\frac{C_2}{n^2} \rho_{k/n}(\M) \\
& \le \frac{C_3}{n^2}
\end{split}
\end{equation}
for a suitable constant $C_3$.

From Lemma~\ref{L:FLip} and \eqref{absolute_bd_rho}, for any $0\le u \le v$, 
\begin{equation}
\label{E:discretediscrep}
\begin{split}
\left\| \rho_v-\rho_u \right\|_{\Was} 
& \le 
(v-u) \nu(\M) + 2 \sigma \int_u^v  \left( \rho_0(\M) + w \nu(\M) \right) \, dw \\
& \le
(v-u) \bigl( (1 + 2 \sigma v) \nu(\M) + 2 \sigma \rho_0(\M) \bigr). \\
\end{split}
\end{equation}

In particular, for $0 \le s \le \frac{1}{n}$,
\begin{equation}
\begin{split}
& \left\| F_{\rho_{s+k/n}} \cdot \rho_{s+k/n} - F_{\rho_{k/n}} \cdot \rho_{k/n} \right\|_{\Was} \\
& \quad \le
\left\| F_{\rho_{s+k/n}} \cdot \rho_{s+k/n} - F_{\rho_{s+k/n}} \cdot \rho_{k/n} \right\|_{\Was} \\
& \qquad +
\left\| F_{\rho_{s+k/n}} \cdot \rho_{k/n} - F_{\rho_{k/n}} \cdot \rho_{k/n} \right\|_{\Was} \\
& \quad \le
\left\| F_{\rho_{s+k/n}} \right\|_{\Lip} \left\| \rho_{s+k/n} - \rho_{k/n} \right\|_{\Was} \\
& \qquad + 
\sup_{x \in \M} \left |F_{\rho_{s+k/n}} - F_{\rho_{k/n}} \right| \rho_{k/n}(\M) \\
& \quad \le 
\frac{2\sigma}{n} \bigl( (1 + 2 \sigma (k+1)/n) \nu(\M) + 2 \sigma \rho_0(\M) \bigr) \\
& \qquad + 
 \frac{8 \sigma}{n} \bigl( (1 + 2 \sigma (k+1)/n) \nu(\M) + 2 \sigma \rho_0(\M) \bigr) \\
& \quad \qquad \times \left( \rho_{0}(\M) +\frac{k}{n} \nu(\M) \right), \\
\end{split}
\end{equation}
where we combined \eqref{E:discretediscrep} 
and \eqref{absolute_bd_rho} with Lemma~\ref{L:FLip} 
and Lemma~\ref{L:Fpipi} in the last inequality.

Hence, 
\begin{equation}
\label{E:difdipim4.2}
\int_0^{\frac{1}{n}}\left\| F_{\rho_{s+k/n}} \cdot \rho_{s+k/n} - F_{\rho_{k/n}} \cdot \rho_{k/n} \right\|_{\Was}\, ds
\le
\frac{C_{4}}{n^2},
\end{equation}
for a suitable constant $C_4$.

Combining \eqref{E:difdipim4.1} and \eqref{E:difdipim4.2}, 
we see from \eqref{E:difdipim4} that
\begin{equation}
\label{E:difdipim4.sum}
\left\|
\rho_{(k+1)/n}  - \frac{\nu}{n} \, - \, H_{\rho_{k/n}} \cdot \rho_{k/n}\right \|_{\Was} 
\le 
 \frac{C_{5}}{n^2}
\end{equation} 
for a suitable constant $C_5$.

Recall from \eqref{E:dpim} that the first term on the
right-hand side of \eqref{E:discrete_split} is zero and
bound the remaining terms using \eqref{E:difdipim2}, \eqref{E:difdipim3} 
and \eqref{E:difdipim4.sum} to obtain
\begin{equation}
\label{E:iterative_bd_diff}
\left\| \pi_{k+1} - \rho_{(k+1)/n} \right\|_{\Was}
\le
\left(1 +  \frac{C_6}{n} \right) \left\| \pi_{k} - \rho_{k/n} \right\|_{\Was} + \frac{C_7}{n^2}
\end{equation}
for $(k+1) \le n T$, where $C_6$ and $C_7$ are constants.
Iterating \eqref{E:iterative_bd_diff} leads to
\begin{equation}
\label{E:non-iterative_bd_diff}
\begin{split}
\left\| \pi_{k+1} - \rho_{(k+1)/n} \right\|_{\Was}
& \le
\sum_{j=0}^k \left(1 +  \frac{C_6}{n} \right)^j \frac{C_7}{n^2} \\
& \le
n T \exp\left\{\frac{C_6}{n}\right\}^{n T} \frac{C_7}{n^2} \\
& =
\exp\left\{C_8 \right\} \frac{C_9}{n} \\
\end{split}
\end{equation}
for $(k+1) \le n T$.

The result now follows by using
\eqref{E:non-iterative_bd_diff} and \eqref{E:discretediscrep}
 to bound the first and second terms,
respectively, on the right-hand side of \eqref{E:discrete_plus_interpolate}.
\end{proof}


\chapter{Supporting lemmas for the main convergence result}  
\label{Ch:techlem}     

\section{Estimates for Radon-Nikodym derivatives}
\label{SS:RN_est}


In this chapter we establish a set of lemmas which provide 
the central technical tools used in the proof of the
main convergence result in Chapter~\ref{Ch:convergence}. 
As previewed in Chapter~\ref{Ch:hypotheses},  
we work with weights in the form of Radon-Nikodym derivatives, 
because selection, in our infinite population, is a 
process of  reweighting of population frequencies
of genotypes.  Furthermore, we suppose that
our initial population  $Q_0$ can be treated as
a reweighted version of the Poisson probability measure $ P_0  = \mP Q_0 $.
Thus, it is natural to express our full discrete-generation 
system in terms of Radon-Nikodym derivatives 
with respect to a family of Poisson probability measures.

The substantive effects of selection have been treated
in Chapter~\ref{Ch:complete_Poisson}.  That chapter 
shows how, in the presence of selection,
the completely Poissonized version of the
discrete-generation system aligns with the continuous-time 
dynamical system.  From this point on in the story,
attention focuses on short stretches of time which become 
infinitesimal as the scaling parameter $n$ goes to infinity. 
Over these short stretches,  we show that the effect of
selection can be treated as negligible whereas the total effect
of recombination tracks closely that of complete Poissonization.
   
We thus require bounds on the overall effects of selection,
acting along with mutation and recombination,  over $k$
generations.  We are interested in the outcome at 
the $k$-th generation,  but the effects of selection 
accumulate across generations and depend on the genotypes
of the ancestors,  portions of whose genomes are recombined
into the genotype of any individual  sampled from the population
at generation $k$.  Reconstruction of the history of
accumulating selection is made possible by the 
decomposition of elements of $\G$ according to vintage 
and thus by the starred operators which update in each generation
the population distribution
of mutations from ancestral wild type labeled according to vintage.

The capstone of this chapter, Lemma~\ref{L:Tcomparednew}
is a bound on the difference between the starred operator
$ Q^{*}_{k}$ and its own Poissonization $\mP Q^{*}_{k} $.
This bound is obtained by combining an approximation
for $ d Q^{*}_{k}/d P^{*}_{k} $ derived in this section 
with an approximation for $ d \mP Q^{*}_{k}/d P^{*}_{k} $ 
derived in the following section.  In both approximations,
terms that are retained arise from contributions of
the initial probability measures.  Contributions from selection
are absorbed into an error term bounded by a multiple of $k/n$.


We begin with an elementary observation concerning how
the action of the starred recombination operator $\mR^{*}$ 
may be represented in terms of Radon-Nikodym derivatives.

\begin{lemma} \label{L:starqualities}

%
%
Suppose that $P^*$ is a Poisson random measure.
Then, for $\bg\in \G^{k+1}$,
\begin{equation*} 
\frac{d\mR^* Q^*}{dP^*}(\bg)
    =\hspace*{-1mm}
\int_{\BM} \int_{\G^{k+1}} 
\frac{dQ^*}{dP^*}(\bg|_{A}+\bg'|_{A^{\complement}})
\frac{dQ^*}{dP^*}(\bg'|_{A} + \bg|_{A^{\complement}}) 
 dP^{*}(\bg') d\mathcal{R}(A).
\end{equation*}
\end{lemma}

\begin{proof}

By Corollary~\ref{C:intindeppoisson},
for any bounded Borel function  $F:\G^{k+1}\to \bR$,
\[
\begin{split}
& \mR^* Q^*[F]  = 
\int_{\BM} \int_{\G^{k+1}} \int_{\G^{k+1}} 
     F(\bg|_{A}+\bg'|_{A^\complement}) \, dQ^*(\bg) dQ^*(\bg') \, d\mathcal{R}(A)\\
& = \int_{\BM}\int_{\G^{k+1}} \int_{\G^{k+1}} F(\bg|_{A}+\bg'|_{A^\complement})
 \frac{dQ^*}{dP^*}(\bg) \frac{dQ^*}{dP^*}(\bg') \, 
           dP^*(\bg) dP^*(\bg') d\mathcal{R}(A)\\
& = \int_{\BM}\int_{\G^{k+1}} \int_{\G^{k+1}} F(\bg)
        \frac{dQ^*}{dP^*}(\bg|_{A} + \bg'|_{A^\complement}) 
\frac{dQ^*}{dP^*}(\bg'|_{A} +\bg|_{A^\complement}) 
\, dP^*(\bg')dP^{*}(\bg) d\mathcal{R}(A). \\
\end{split}
\]
This is equivalent to the claim.
\end{proof}

We next develop apparatus for reconstructing the ancestral genotypes 
that are consistent with a particular genotype (decomposed by vintages) 
that is observed at the end of $k$ generations.  
At each ancestral generation, there is a partition of $\M$ 
(generally including empty sets), as presented in 
Section~\ref{SS:trees}.  Each set in this partition corresponds
to an ancestor in the given generation.
Mutant alleles within the set are retained 
from the genotype of the ancestor while mutant alleles outside the set 
are lost through recombination.   We work with functions 
that assign to each set a surrogate for the missing
piece of the ancestral genome.  This additional bookkeeping
facilitates our development of bounds on
the ``de-Poissonizing'' effects of selection.

\begin{definition} \label{D:Stild}
Let $\mcA$ be a partition of $\M$ into Borel sets 
and let $\xi:\mcA\to \G$ be a family of 
genotypes indexed by the sets of $\mcA$. 
Given $F:\G\to\bR$ and $g\in \G$, set
$$
\widetilde{F}_{\mcA}(g,\xi) = \sum_{A\in \mcA} 
 F(g|_{A}+\xi(A)|_{A^{\complement}}).
$$
\end{definition}

Note that if $\mcA$ is the trivial partition $\{\M\}$, 
then $\widetilde{F}_{\mcA}(g,\xi) = F(g)$ for any $\xi:\mcA\to \G$.
\glossary{$\widetilde{F}_{\mcA}$}
\index{total variation norm}

\begin{lemma}
\label{L:shatbound1}
Fix  a Borel partition $\mcA$ of $\M$ and
a family of genotypes $\xi:\mcA\to \G$.  Then,
for any $F:\G\to\bR$,
\[
\Bigl|\widetilde{F}_{\mcA}(g,\xi)-\widetilde{F}_{\mcA}(g',\xi) \Bigr| 
             \le \Lip_{\on{TV}} F\cdot \|g-g'\|_{\on{TV}}.
\]
\end{lemma}

\begin{proof}
We have
\begin{align*}
\bigl|\widetilde{F}_{\mcA}(g,\xi)-\widetilde{F}_{\mcA}(g',\xi) \bigr| 
&\le
        \sum_{A\in \mcA}\left| F(g|_{A}+\xi(A)|_{A^{\complement}})
         - F(g'|_{A}+\xi(A)|_{A^{\complement}})) \right|\\
&\le \sum_{A\in \mcA} \Lip_{\on{TV}} F \cdot \bigl\|g|_A-g'|_A\bigr\|_{\on{TV}}\\
&= \Lip_{\on{TV}} F \cdot \bigl\|g-g'\bigr\|_{\on{TV}}.
\end{align*}
\end{proof}


\begin{notation}
Recall the ancestral tree $\mathcal{L}$
and the random recombination sets $R_\beta$ 
and random genotypes $\Gamma_{\beta}$
indexed by the nodes $\beta \in \mathcal{L}$ 
introduced in Section~\ref{SS:vintages}.
Fix a node $\ell \in \mathcal{L}$ with vintage $V(\ell)=k$.
Any set $A$ in the random partition $\mcA_{i}(\ell) = \mcA_{k,i}$ 
corresponds to a node $\omega \preceq\ell$ with
vintage $V(\omega) =i$. Define $\xi_{k,i} : \mcA_{k,i} \to \G$ by
$\xi_{k,i}(A) :
=\Xi_{\omega}
=\sum_{\upsilon \preceq \omega} \Gamma_{\upsilon}|_{W(\upsilon,\omega)}$.
\end{notation}
\glossary{$\xi_{k,i}$}


%

We now state and prove a lemma which gives an exact representation
of the Radon-Nikodym derivative of the starred measure $ Q^{*}_{k}$
with respect to $ P^{*}_{k} $ at generation $k$ in terms of ancestral
episodes of selection.   The factors that make the total probability
integrate to $1$ are absorbed into a normalization constant $C$.



\begin{lemma}
\label{L:exactRN}
Set $H :=\log dQ_{0}/dP_{0}$. 
Suppose for $k \in \bN_0$ that the partitions $\mcA_{k,0}, \ldots, \mcA_{k,k}$
are defined using the node $\ell \in \mathcal{L}$ with vintage
$V(\ell) = k$.
Then,
\[
\begin{split}
\frac{dQ_{k}^{*}}{dP^{*}_{k}}(g_{0},\dots,g_{k}) 
& = C\bE\left[\exp\left\{\wt{H}_{\mcA_{k,0}}
         (g_{0},\xi_{k,0})
    -\rec{n} \sum_{i=0}^{k-1} 
         \wt{S}_{\mathcal{A}_{k,i}} (g_{0}+\cdots+g_{i},\xi_{k,i})\right\}
              \right] \\
&= C\bE\left[\exp\left\{
   \sum_{\omega\prec\ell} J_{V(\omega)}\Bigl((g_{0}+\cdots+g_{V(\omega)})|_{W(\omega,\ell)}
      +(\Xi_{\omega})|_{W(\omega,\ell)^{\complement}} \Bigr)  \right\}\right],
 \end{split}
\]
where $C$ does not depend on $(g_0, \ldots, g_k)$
and the functions $J_{i}:\G\to\bR$ are given
by $J_{0}(g):=H(g)-S(g)/n$ and $J_{i}(g):=-S(g)/n$ for $1\le i\le k-1$. 
 The expectation is taken with respect
to the random collection of genotypes $\xi_{k,i}$ (or $\Xi_{\ell}$)
and the random partitions $\mcA_{k,i}$ (or $W(\omega,\ell)$).
\end{lemma}

\begin{proof}
By definition,
\[
\begin{split}
\wt{H}_{\mcA_{k,0}}
         (g,\xi_{k,0})
& =
\sum_{A\in \mcA_{k,0}} 
 H(g|_{A}+\xi_{k,0}(A)|_{A^{\complement}}) \\
& =
\sum_{\omega \preceq \ell, V(\omega) = 0}
 H(g|_{W(\omega,\ell)}+(\Xi_\omega)|_{W(\omega,\ell)^{\complement}}) \\
\end{split}
\]
and
\[
\begin{split}
\wt{S}_{\mathcal{A}_{k,i}} (g,\xi_{k,i})
& =
\sum_{A\in \mcA_{k,i}} 
 S(g|_{A}+\xi_{k,i}(A)|_{A^{\complement}}) \\
& =
\sum_{\omega \preceq \ell, V(\omega) = i}
 S(g|_{W(\omega,\ell)}+(\Xi_\omega)|_{W(\omega,\ell)^{\complement}}), \\
\end{split}
\]
so
\begin{equation*}  
\begin{split}
& \wt{H}_{\mcA_{k,0}}
         (g_{0},\xi_{k,0})
    -\rec{n} \sum_{i=0}^{k-1} 
         \wt{S}_{\mathcal{A}_{k,i}} (g_{0}+\cdots+g_{i},\xi_{k,i})\\
& \quad =
\sum_{\omega\prec\ell} J_{V(\omega)}\Bigl((g_{0}+\cdots+g_{V(\omega)})|_{W(\omega,\ell)}+(\Xi_{\omega})|_{W(\omega,\ell)^{\complement}} \Bigr).
\end{split}
\end{equation*}
The equality of the two expectations follows.

The remainder of the proof is by induction on $k$. For $k=0$, the 
partition $\mcA_{0,0}$ is the trivial partition consisting 
of the single set $\M$, and $\wt{H}_{\mcA_{0,0}}(g,\xi)=H(g)$, 
and so the result is obvious. Now suppose the result is true for $k-1 \ge 0$.

By definition,
$Q_{k}^{*}=\mR^* \mM^*_n \mS^*_n Q_{k-1}^*$
and $P_{k}^*=\mM^*_n P_{k-1}^*$. Thus,
\begin{equation} \label{E:Qkstarstep}
\begin{split}
&\frac{dQ_{k}^{*}}{dP^{*}_{k}}(g_{0},\dots,g_{k}) \\
&=
\frac{d\mR^* \mM^*_n \mS^*_n Q_{k-1}^*}{d\mM^{*}_{n}P^{*}_{k-1}}(g_{0},\dots,g_{k})\\
&=
\int_{\BM}\int_{\G^{k+1}} 
\frac{d\mM^*_n \mS^*_n Q^*_{k-1}}{d\mM^*_n P_{k-1}^*}
      (\bg|_{A} + \bg|_{A^{\complement}}')\\
&\quad
\times\frac{d\mM^*_n \mS^*_n Q^*_{k-1}}{d\mM^*_n P_{k-1}^*}
      (\bg|_{A}' + \bg|_{A^{\complement}}) 
\, d\mM^*_n P_{k-1}^{*}(\bg')d\mathcal{R}(A)\\
&= \int_{\BM} 
\int_{\G^{k+1}} 
\frac{d\mS^*_n Q^*_{k-1}}{d P_{k-1}^*}\bigl( (g_{0},\dots,g_{k-1})|_{A} 
   + (g'_{0},\dots,g'_{k-1})|_{A^{\complement}}\bigr)\\
&\quad \times 
\frac{d\mS^*_n Q^*_{k-1}}{d P_{k-1}^*}\bigl( (g'_{0},\dots,g'_{k-1})|_{A} 
     + (g_{0},\dots,g_{k-1})|_{A^{\complement}}\bigr)
\, dP_{k-1}^{*}(\bg') d\mathcal{R}(A),
\end{split}
\end{equation}
by Lemma~\ref{L:starqualities} and the observation that for any
two probability measures $Q^*$ and $P^*$ on $\G^{k-1}$ we have
\[
\frac{d\mM^*_n Q^*}{d\mM^*_n P^*}(g_0, \ldots, g_k)
=
\frac{d (\Pi_{\nu/n} \otimes Q^*)}{d (\Pi_{\nu/n} \otimes P^*)}(g_0, \ldots, g_k)
=
\frac{dQ^*}{dP^*}(g_0, \ldots, g_{k-1}).
\]

Let the parents of the node $\ell$ be the two nodes
$\ell', \ell'' \in \mathcal{L}$ with common vintage $k-1$.
Suppose first of all that the random partitions 
$\mcA_{k-1,0}, \ldots, \mcA_{k-1,k-1}$
are defined using the node $\ell'$.
By the induction hypothesis,
\begin{equation} \label{E:ellprimeterm}
\begin{split}
&\frac{d\mS_{n}^{*}Q^{*}_{k-1}}{dP^{*}_{k-1}}(g_{0}, 
              \ldots, g_{k-1}) \\
&= C_{1}\bE\left[\exp\left\{
   \sum_{\omega\prec\ell'} J_{V(\omega)}\Bigl((g_{0}+\cdots+g_{V(\omega)})|_{W(\omega,\ell')}+(\Xi_{\omega})|_{W(\omega,\ell')^{\complement}} \Bigr)
\right\}\right] \\
&\quad \times\exp\left\{-\rec{n}S(g_{0}+\cdots+g_{k-1})\right\}\\
&= C_{1}\bE\left[\exp\left\{
   \sum_{\omega\preceq\ell'} J_{V(\omega)}\Bigl((g_{0}+\cdots+g_{V(\omega)})|_{W(\omega,\ell')}+(\Xi_{\omega})|_{W(\omega,\ell')^{\complement}} \Bigr)
\right\}\right] \\
\end{split}
\end{equation}
for some constant $C_{1}$. 
Of course, the same equation holds if $\ell'$ is replaced
by $\ell''$.

Recall that $R_\ell$ is a $\BM$-valued random variable with
distribution $\mathcal{R}$ and the collection of random sets
$\{R_\omega : \omega \preceq \ell\} 
= \{R_\ell\} 
\cup \{R_\omega : \omega \preceq \ell'\} 
\cup \{R_\omega : \omega \preceq \ell''\}$
is independent.  By convention,
the genotype of the individual $\ell$ coincides with that
of the parent $\ell'$ on the set $R_\ell$ and with that of the parent $\ell''$
on the complementary set $R_\ell^\complement$.
For $0\le i\le k-1$, define a random genotype by
$$
Y_{i}:=\sum_{\begin{smallmatrix}\omega\preceq \ell',\\
V(\omega)=i\end{smallmatrix}} (\Gamma_{\omega})|_{W(\omega,\ell')\cap R_\ell^{\complement}}
 +
\sum_{\begin{smallmatrix}\omega\preceq \ell'',\\
V(\omega)=i\end{smallmatrix}} (\Gamma_{\omega})|_{W(\omega,\ell'')\cap R_\ell}.
$$
Note that the conditional distribution of
$\mathbf{Y}:=(Y_{0},\dots,Y_{k-1})$ given $R_\ell$ is
$P_{k-1}^{*}$, so  $P_{k-1}^{*}$ is also the unconditional distribution of
$\mathbf{Y}$ and $\mathbf{Y}$ is independent of $R_\ell$.
Thus, the distribution of
\[
\left(R_\ell, (Y_{0},\dots,Y_{k-1})|_{R_\ell^\complement}, R_\ell^\complement, (Y_{0},\dots,Y_{k-1})|_{R_\ell}\right)
\]
is the push-forward of the probability measure 
$\mathcal{R} \otimes P_{k-1}^*$ by the map
\[
(A,(g_0, \ldots, g_{k-1})) 
\mapsto
\left(A, (g_0, \ldots, g_{k-1})|_{A^\complement}, A^\complement, (g_0, \ldots, g_{k-1})|_A\right),
\]
which is in turn that of the push-forward of the probability measure
$\mathcal{R} \otimes P_{k-1}^* \otimes P_{k-1}^*$ by the map
\[
(A, (g_0', \ldots, g_{k-1}'), (g_0'', \ldots, g_{k-1}''))
\mapsto
\left(A, (g_0', \ldots, g_{k-1}')|_{A^\complement}, A^\complement, (g_0'', \ldots, g_{k-1}'')|_A\right).
\]
Therefore, substituting \eqref{E:ellprimeterm} and
the analogous equation with $\ell'$ replaced by
$\ell''$ into \eqref{E:Qkstarstep} gives
\begin{equation*} 
\begin{split}
&\frac{dQ_{k}^{*}}{dP^{*}_{k}}(g_{0},\dots,g_{k}) \\
&=
\bE \biggl[
\frac{d\mS^*_n Q^*_{k-1}}{d P_{k-1}^*}\bigl( (g_{0},\dots,g_{k-1})|_{R_\ell} 
   + (Y_{0},\dots,Y_{k-1})|_{R_\ell^{\complement}}\bigr)\\
&\quad \times 
\frac{d\mS^*_n Q^*_{k-1}}{d P_{k-1}^*}\bigl( (Y_{0},\dots,Y_{k-1})|_{R_\ell} 
     + (g_{0},\dots,g_{k-1})|_{R_\ell^{\complement}}\bigr)
\biggr] \\
&=
C_1^2
\bE \Biggl[
\bE\biggl[\exp\biggl\{
   \sum_{\omega\preceq\ell'}
   J_{V(\omega)}\Bigl((g_{0}+\cdots+g_{V(\omega)})|_{W(\omega,\ell') \cap R_\ell} \\
& \quad \quad 
   +(Y_{0}+\cdots+Y_{V(\omega)})|_{W(\omega,\ell') \cap R_\ell^{\complement}}
   +(\Xi_{\omega})|_{W(\omega,\ell')^{\complement}} \Bigr)
\biggr\} \, \bigg |  \, R_\ell \biggl] \\
& \quad \times
\bE\biggl[\exp\biggl\{
	\sum_{\omega\preceq\ell''}
	J_{V(\omega)}\Bigl((g_{0}+\cdots+g_{V(\omega)})|_{W(\omega,\ell'') \cap R_\ell^\complement} \\
& \quad \quad \quad
	+(Y_{0}+\cdots+Y_{V(\omega)})|_{W(\omega,\ell'') \cap R_\ell}
	+(\Xi_{\omega})|_{W(\omega,\ell'')^{\complement}} \Bigr)
\biggr\} \, \bigg | R_\ell \biggr]	
\Biggr] \\
& =
C_{1}^{2}\bE\Biggl[\exp\biggl\{
     \sum_{\omega\preceq\ell'} J_{V(\omega)}\Bigl((g_{0}+\cdots+g_{V(\omega)})|_{W(\omega,\ell')\cap R_\ell}\\
     & \quad \quad +(Y_{0}+\cdots
        +Y_{V(\omega)})|_{W(\omega,\ell')\cap R_\ell^{\complement}}
         +  (\Xi_{\omega})|_{W(\omega,\ell')^{\complement}} \Bigr)\\
     & \quad 
     + \sum_{\omega\preceq\ell''} J_{V(\omega)}\Bigl((g_{0}
     +\cdots+g_{V(\omega)})|_{W(\omega,\ell'')\cap R_\ell^{\complement}}\\
     &\quad \quad \quad +(Y_{0}+\cdots+Y_{V(\omega)})|_{W(\omega,\ell'')\cap R_\ell}
        +  (\Xi_{\omega})|_{W(\omega,\ell'')^{\complement}}  \Bigr)
     \biggr\}\Biggr].\\
\end{split}
\end{equation*}

%
%
%

Now, $\omega \prec \ell$ if and only if either $\omega \preceq \ell'$
or $\omega \preceq \ell''$.  If $\omega \preceq \ell'$, then, by definition,
$W(\omega, \ell) = W(\omega,\ell') \cap R_\ell$. 
Also by definition,
\[
\begin{split}
&(Y_{0}+\cdots+Y_{V(\omega)})|_{W(\omega,\ell')\cap R_\ell^{\complement}} \\
& \quad =
\sum_{i=0}^{V(\omega)}
\left[
\sum_{\begin{smallmatrix}\upsilon\preceq \ell',\\
V(\upsilon)=i\end{smallmatrix}} (\Gamma_{\upsilon})|_{W(\upsilon,\ell')\cap R_\ell^{\complement}}
+
\sum_{\begin{smallmatrix}\upsilon\preceq \ell'',\\
V(\upsilon)=i\end{smallmatrix}} (\Gamma_{\upsilon})|_{W(\upsilon,\ell'')\cap R_\ell}
\right]
\Biggl|_{W(\omega,\ell')\cap R_\ell^{\complement}} \\
& \quad =
\sum_{\upsilon\preceq \omega} 
(\Gamma_{\upsilon})|_{W(\upsilon,\ell')\cap R_\ell^{\complement}}, \\
\end{split}
\]
since for $\omega \preceq \ell'$
and $\upsilon$ with $V(\upsilon) \le V(\omega)$, 
\[
(W(\upsilon,\ell'')\cap R_\ell) \cap (W(\omega,\ell')\cap R_\ell^{\complement})
= \emptyset
\]
and
\[
(W(\upsilon,\ell')\cap R_\ell^{\complement}) \cap (W(\omega,\ell')\cap R_\ell^{\complement})
=
\begin{cases}
W(\upsilon,\ell')\cap R_\ell^{\complement}, & \quad \text{if $\upsilon \preceq \omega$,} \\
\emptyset, & \quad \text{otherwise.} \\
\end{cases}
\]
Lastly,
\[
(\Xi_{\omega})|_{W(\omega,\ell')^{\complement}}
=
\left(\sum_{\upsilon \preceq \omega} (\Gamma_{\upsilon})|_{W(\upsilon,\omega)}\right)
\bigg|_{W(\omega,\ell')^{\complement}}
=
\sum_{\upsilon \preceq \omega} (\Gamma_{\upsilon})|_{W(\upsilon,\omega) \cap
W(\omega,\ell')^{\complement}}
\]
and
\[
(\Xi_{\omega})|_{W(\omega,\ell)^{\complement}}
=
\sum_{\upsilon \preceq \omega} (\Gamma_{\upsilon})|_{W(\upsilon,\omega) \cap
W(\omega,\ell)^{\complement}}.
\]
Thus,
\[
\begin{split}
(Y_{0}+\cdots+Y_{V(\omega)})|_{W(\omega,\ell')\cap R_\ell^{\complement}}
+
(\Xi_{\omega})|_{W(\omega,\ell')^{\complement}}
& =
\left(\sum_{\upsilon \preceq \omega} \Gamma_{\upsilon}\right)
\bigg|_{W(\omega,\ell)^\complement} \\
& =
(\Xi_\omega)|_{W(\omega,\ell)^\complement} \\
\end{split}
\]
because
\[
W(\omega,\ell)^\complement
=
(W(\omega,\ell') \cap R_\ell)^\complement
=
W(\omega,\ell')^\complement \cup R_\ell^\complement.
\]
Similarly, if $\omega \preceq \ell''$, then 
$W(\omega, \ell) = W(\omega,\ell'') \cap R_\ell^\complement$
and
\[
(Y_{0}+\cdots+Y_{V(\omega)})|_{W(\omega,\ell'')\cap R_\ell}
        +  (\Xi_{\omega})|_{W(\omega,\ell'')^{\complement}}
=
(\Xi_\omega)|_{W(\omega,\ell)^\complement},
\]
completing the induction.

\end{proof}

%
%


Recall from \eqref{E:ThetaPQ} that 
for two probability measures $P$ and $Q$ we denote
the total-variation
Lipschitz constant of $\log dQ/dP$ by
$\Theta(P,Q)$.

\begin{lemma}  \label{L:LipRN}
Suppose that $\hat H := \Theta(P_0,Q_0) < \infty$.
Then,  
$\Theta(P_{k}^{*}, Q_{k}^{*})$ and $\Theta(P_{k}, Q_{k})$ 
are both bounded by $\hat{H}+\frac{k}{n}\sigma$ for all $k \in \bN_0$.  
\end{lemma}
\glossary{$\hat{H}$}

\begin{proof}
Consider first the claim for $\log dQ_{k}^{*}/dP_{k}^{*}$.
It suffices to check for any $\bg\in\G^{k+1}$,  $x\in\M$,
and $0\le j\le k$
that 
\begin{equation}
\label{E:logRNstarLip}
\left|\log\frac{dQ_{k}^{*}}{dP_{k}^{*}}(\bg+\delta_{x}^{(j)})- 
    \log\frac{dQ_{k}^{*}}{dP_{k}^{*}}(\bg) \right| 
     \le \hat{H}+\frac{k}{n}\sigma,
\end{equation}
where $\delta_{x}^{(j)} := (0, \ldots, 0, \delta_x, 0, \ldots, 0) \in \G^{k+1}$
with $\delta_x$ in the $j^{\mathrm{th}}$ coordinate.

Suppose first that $j=0$. By Lemma~\ref{L:exactRN},
\[
\frac{dQ_{k}^{*}}{dP^{*}_{k}}(\bg)  \quad = C \bE\left[\exp\left\{\wt{H}_{\mcA_{k,0}}
         (g_{0}+\delta_{x},\xi_{k,0})
    -\rec{n} \sum_{i=0}^{k-1} 
         \wt{S}_{\mathcal{A}_{k,i}} (g_{0}+\cdots+g_{i}+\delta_{x},\xi_{k,i})\right\}
              \right].
\]
It follows from Lemma~\ref{L:shatbound1} that
\[
\begin{split}
&\wt{H}_{\mcA_{k,0}}
         (g_{0},\xi_{k,0}) -\hat{H}
    -\rec{n} \sum_{i=0}^{k-1} 
         \wt{S}_{\mathcal{A}_{k,i}} (g_{0}+\cdots+g_{i},\xi_{k,i}) - \frac{k}{n} \sigma \\
& \quad \le
\wt{H}_{\mcA_{k,0}}
         (g_{0}+\delta_{x},\xi_{k,0})
    -\rec{n} \sum_{i=0}^{k-1} 
        \wt{S}_{\mathcal{A}_{k,i}} (g_{0}+\cdots+g_{i}+\delta_{x},\xi_{k,i})\\
& \quad \le
\wt{H}_{\mcA_{k,0}}
         (g_{0},\xi_{k,0}) +\hat{H}
    -\rec{n} \sum_{i=0}^{k-1} 
         \wt{S}_{\mathcal{A}_{k,i}} (g_{0}+\cdots+g_{i},\xi_{k,i}) + \frac{k}{n} \sigma, \\
\end{split}
\]
and \eqref{E:logRNstarLip} follows immediately. 
The proof for $1 \le j \le k$ is similar.

In order to establish the claim for $\log dQ_{k}/dP_{k}$, 
it suffices to check for any $g\in\G$ and $x\in\M$ that
\begin{equation}
\label{E:logRNLip}
\left|\log\frac{dQ_{k}}{dP_{k}}(g+\delta_{x})- 
    \log\frac{dQ_{k}}{dP_{k}}(g) \right| \le \hat{H}+\frac{k}{n}\sigma.
\end{equation}

Define $\bX$ to be the canonical random variable 
on $\G^*$.  Observe that
\[
\frac{dQ_{k}}{dP_{k}}(g)
=
P_k^*\left[
\frac{dQ_{k}^{*}}{dP^{*}_{k}}(\bX)
\, \Big |
\sum \bX = g.
\right]
\]
Under $P_k^*$, the
conditional distribution of $\bX$ given the event
$\{\Sigma \bX=g=\sum_{j=1}^{g(\M)} \delta_{x_{j}}\}$, 
is the distribution of the $\G^{k+1}$-valued random variable
$$
\sum_{j=1}^{g(\M)} \delta_{x_{j}}^{(\mathbf{i}_{j})},
$$
where 
%
%
the $\{0,1, \ldots,k\}$-valued random 
variables $\mathbf{i}_{j}$, $1 \le i \le g(\M)$, 
are independent with distribution given by 
\glossary{$\mathbf{i}_{j}$}
%
%
$$
\mathbb{P}\{\mathbf{i}_{j}=0\}=1-k\pi(x_{j}),\qquad 
    \mathbb{P}\{\mathbf{i}_{j}=i\}=\pi(x_{j})\text{ for } 1 \le i \le k,
$$
with
$$
\pi(x)
:=\frac{d\nu/n}{d\mu P_{k}}(x) 
= \frac{d\nu/n}{d[\rho_0 + k \nu/n]}(x).
$$
Consequently, for a bounded measurable 
function $F:\G^{k+1} \to \mathbb{R}$,
we have 
\[
P_k^*[F(\bX)\cond \Sigma \bX=g+\delta_{x}\bigr]= \bE\left[P_k^*\bigl[F(\bX+\delta_{x}^{(\mathbf{i})})\cond \Sigma \bX=g\bigr]\right],
\]
where $\mathbf{i}$ is a $\{0,1, \ldots,k\}$-valued random variable with
distribution
$$
\mathbb{P}\{\mathbf{i}=0\}=1-k\pi(x),\qquad 
    \mathbb{P}\{\mathbf{i}=i\}=\pi(x)\text{ for } 1 \le i \le k.
$$
Thus,
$$
\frac{dQ_{k}}{dP_{k}}(g+\delta_{x})
=
\bE\left[P_k^*\biggl[ \frac{dQ^{*}_{k}}{dP^{*}_{k}}(\bX) \cdot 
   \frac{dQ_{k}^{*}/dP_{k}^{*}(\bX+\delta_{x}^{(\mathbf{i})})}{dQ_{k}^{*}/dP_{k}^{*}(\bX)}\,\biggl|\, \Sigma \bX=g \biggr] \right].
$$
It is immediate from \eqref{E:logRNstarLip} that
$$
\exp \left\{-\hat{H}-\frac{\sigma k}{n} \right\} 
\le 
\frac{dQ_{k}^{*}/dP_{k}^{*}(\bg+\delta_{x}^{(i)})}{dQ_{k}^{*}/dP_{k}^{*}(\bg)}
\le \exp \left\{\hat{H}+\frac{\sigma k}{n} \right\},
$$
for $0 \le i \le k$, and so \eqref{E:logRNLip} holds.
\end{proof}



\begin{corollary} \label{C:boundedRN}
Suppose that $\hat H := \Theta(P_0,Q_0) < \infty$.
Fix $k \in \bN_0$ and set $c=\hat{H} +\frac{k}{n}\sigma$.  Then,
\[
\exp\{-c(g(\M)+e^{c}\mu P_{k}(\M))\}
\le 
\frac{dQ_{k}}{dP_{k}}(g)
\le 
\exp\{c(g(\M)+\mu P_{k}(\M))\}
\]
for all $g\in\G$,
and 
\[
\exp\{-c(\Sigma\bg(\M)+e^{c}\mu P_{k}(\M))\}
\le 
\frac{dQ_{k}^{*}}{dP_{k}^{*}}(\bg)
\le 
\exp\{c(\Sigma\bg(\M)+\mu P_{k}(\M))\}
\]
for all $\bg\in \G^{k+1}$.
\end{corollary}

\begin{proof}
By Lemma~\ref{L:LipRN} we have that  $\Theta(P,Q) \le c$,   
and the inequalities for $dQ_{k}/dP_{k}$ follow from 
Corollary~\ref{C:RNcompare}.

The proof of the inequalities for $dQ_{k}^{*}/dP_{k}^{*}$
is similar.
\end{proof}


\begin{corollary} \label{C:shattrap}
Suppose that $\hat H := \Theta(P_0,Q_0) < \infty$. 
Then, there are constants $c$ and $\epsilon>0$ (depending on $\hat{H}$) 
such that for any $k \le \epsilon n$ and any $\bg\in\G^{k+1}$,
\begin{equation*} 
\left|
\log \frac{dQ^{*}_{k}}{dP^{*}_{k}}(\bg) 
 -\log\bE\left[ \prod_{A\in \mcA_{k,0}}
P_0\left[\exp\bigl\{H\bigl(g_{0}|_{A}+X|_{A^{\complement}}\bigr)\bigr\}\right]\right]\right|
 \le c\frac{k}{n} \left(1 + \Sigma \bg(\M)\right).
\end{equation*}
%
%
%
%
\end{corollary}

%
\begin{proof}
%
%
Set $H = dQ_0/dP_0$.
From Lemma~\ref{L:exactRN} and Lemma~\ref{L:shatbound1}, we know that 
there is a constant $C$ (independent of $\bg$) such that, 
for the random map $\xi_{k,0} : \mcA_{k,0} \to \G$
introduced in Lemma~\ref{L:exactRN},
\[
\begin{split}
& C 
\bE \left[
\exp
\left\{
\wt{H}_{\mcA_{k,0}}(g_{0},\xi_{k,0}) 
-\frac{\sigma k}{n} \Sigma \bg (\M) 
\right\} 
\right]\\
& \quad \le 
\frac{dQ_{k}^{*}}{dP_{k}^{*}}(\bg)\\
& \quad \le 
C \bE\left[
\exp\left\{
\wt{H}_{\mcA_{k,0}}(g_{0},\xi_{k,0}) 
+ \frac{\sigma k}{n} \Sigma \bg (\M) 
\right\} 
\right].\\
\end{split}
\]
(Note: We include the expectation with respect to the random partition
$\mcA_{k,0}$ in the $\bE[\cdot]$.)
Substituting in the definition of $\wt{H}$, we find that there is a
constant $C'$ such that
\[
\begin{split}
& C' \exp\Bigl\{-\frac{\sigma k}{n} \Sigma \bg(\M)\Bigr\}\bE\biggl[\exp\biggl\{
   \sum_{A\in \mcA_{k,0}} H\bigl(g_{0}|_{A}+\xi_{k,0}(A)|_{A^{\complement}}\bigr)
      \biggr\} \biggr]\\
      & \quad \le \frac{dQ_{k}^{*}}{dP_{k}^{*}}(\bg)\\
& \quad \le C' \exp\left\{\frac{\sigma k}{n} \Sigma \bg(\M)\right\}\bE\biggl[\exp\biggl\{
   \sum_{A\in \mcA_{k,0}} H\bigl(g_{0}|_{A}+\xi_{k,0}(A)|_{A^{\complement}}\bigr)
      \biggr\} \biggr]. \\
\end{split}
\]
Since the random measures $\{\xi_{k,0}(A) : A \in \mcA_{k,0}\}$ 
are i.i.d. with common distribution $P_0$, this becomes
\[
\begin{split}
&C' \bE \biggl[
\prod_{A\in \mcA_{k,0}}
P_0\left[\exp\biggl\{ -\frac{\sigma k}{n} \Sigma \bg(A)+
   H\bigl(g_{0}|_{A}+X|_{A^{\complement}}\bigr)
      \biggr\} \right]\biggr]\\
      & \quad \le \frac{dQ_{k}^{*}}{dP_{k}^{*}}(\bg)\\
& \quad \le C'\bE\biggl[\prod_{A\in \mcA_{k,0}} P_0\left[\exp\biggl\{ \frac{\sigma k}{n} \Sigma \bg(A)
    + H\bigl(g_{0}|_{A}+X|_{A^{\complement}}\bigr)
      \biggr\} \right] \biggr]. \\
\end{split}
\]

By Lemma~\ref{L:RNcompare},
\[
\begin{split}
&\left(
\int
\bE\biggl[\prod_{A\in \mcA_{k,0}} 
      P_0\left[\exp\biggl\{ \frac{\sigma k}{n} \Sigma \bg'(A)
    + H\bigl(g_{0}'|_{A}+X|_{A^{\complement}}\bigr)
      \biggr\} \right]
      \, dP_{k}^{*}(\bg') \biggr]
      \right)^{-1}                       \\
& \qquad \times \bE \biggl[
\prod_{A\in \mcA_{k,0}}
P_0\left[\exp\biggl\{ -\frac{\sigma k}{n} \Sigma \bg(A)+
   H\bigl(g_{0}|_{A}+X|_{A^{\complement}}\bigr)
      \biggr\} \right]\biggr]\\
& \quad \le 
\frac{dQ_{k}^{*}}{dP_{k}^{*}}(\bg)\\
& \quad \le
\left(
\int
\bE\biggl[ \prod_{A\in \mcA_{k,0}}
P_0\left[\exp\biggl\{ -\frac{\sigma k}{n} \Sigma \bg'(A)+
   H\bigl(g_{0}'|_{A}+X|_{A^{\complement}}\bigr)
      \biggr\} \right]
\, d(P_{k}^{*})(\bg')\biggr]
\right)^{-1} \\
& \qquad \times \bE \biggl[
\prod_{A\in \mcA_{k,0}} P_0\left[\exp\biggl\{ \frac{\sigma k}{n} \Sigma \bg(A)
    + H\bigl(g_{0}|_{A}+X|_{A^{\complement}}\bigr)
      \biggr\} \right] \biggr]. \\
\end{split}
\]

We have for any set $A \in \BM$ that
\[
\begin{split}
& \int
P_0\left[\exp\biggl\{ \frac{\sigma k}{n} \Sigma \bg(A) + H\bigl(g_{0}|_{A}+X|_{A^{\complement}}\bigr)
      \biggr\} \right] 
\, dP_{k}^*(\bg)\\
& \quad =
\int
P_0\left[\exp\left\{ 
\frac{\sigma k}{n} (g_1 + \cdots + g_k)(A) + 
\frac{\sigma k}{n} g_0(A)
+ H\bigl(g_{0}|_{A}+X|_{A^{\complement}}\bigr)
      \right\} \right] 
\, dP_{k}^*(\bg)\\
& \quad =
\int 
\exp\left\{ \frac{\sigma k}{n} (g_1 + \cdots + g_k)(A) \right\}
\, dP_{k}^*(\bg)
\cdot P_0\left[\exp\left\{  
H(X)
+
\frac{\sigma k}{n} X(A)
\right\}
\right] \\
& \quad = \exp\left\{\frac{\nu(A)k}{n} \bigl( e^{\sigma k/n}-1\bigr) \right\}
       \cdot P_0 \left[ \exp\left\{H(X)+\frac{\sigma k}{n} X(A)\right\} \right] \\
\end{split}
\]
because $\bg \mapsto (g_1 + \cdots + g_k)(A)$ is a Poisson random
variable with mean $\frac{k}{n} \nu(A)$ under $P_{k}^*$
and $\bg \mapsto g_0$ is a Poisson random measure with the
distribution $P_0$ under $P_{k}^*$.

Using the inequality $e^{x}-1\le xe^{x}$ for $x \ge 0$, we have
\[
\begin{split}
\exp\left\{\frac{\nu(A)k}{n} \bigl( e^{\sigma k/n}-1\bigr) \right\}
& \le
\exp\left\{\frac{\sigma \nu(A)k^{2}}{n^{2}} e^{\sigma k/n} \right\} \\
& \le
\exp\left\{\frac{c_1 \nu(A)k^{2}}{n^{2}} \right\}, \\
\end{split}
\]
where $c_{1}$ is a constant depending on $\sigma$ and $T$
and $k/n \le \epsilon$.
By the same inequality, the fact that 
$P_{0}[e^{H(X)}] = Q_0[1] =1$, and Corollary~\ref{C:RNcompare},
\[
\begin{split}
& P_0\left[\exp\left\{  
H(X)
+
\frac{\sigma k}{n} X(A)
\right\}
\right] \\
& \quad \le 
P_0\left[
\exp\{H(X)\} 
\left(1+\frac{\sigma k}{n} X(A) \exp\left\{\frac{\sigma k}{n} X(A)\right\}\right) 
\right] \\
& \quad \le
1
+
\frac{\sigma k}{n}
\mu P_0(A) \exp\left\{2\mu P_0(A^{\complement}) \hat{H} 
 +\hat{H}+\frac{\sigma k}{n}+\mu P_0(A)\left( e^{\hat{H}+\sigma k/n}-1\right)\right\} \\
& \quad \le
\exp\left\{
c_{2}\frac{\mu P_0(A)k}{n} \right\},
\end{split}
\]
where $c_{2}$ is a constant depending on $\sigma$, $\hat{H}$, $T$, $\nu(\M)$, and $\mu P_0(\M) = \rho_{0}(\M)$.

Thus, for any partition $\mcA$,
\[
\begin{split}
&  \int \prod_{A\in \mcA} 
P_0\left[ 
\exp\biggl\{   
\frac{\sigma k}{n} \Sigma \bg(A) + H\bigl(g_{0}|_{A}+X|_{A^{\complement}}\bigr) 
\biggr\} 
\right] 
\, dP_{k}^*(\bg)\\
& \quad \le \exp\left\{c_{1}\frac{\nu(\M)k^{2}}{n^{2}}  + c_{2}\frac{\rho_{0}(\M)k}{n} \right\} \\
& \quad \le \exp\left\{c_{3}\frac{k}{n} \right\}, \\
\end{split}
\]
for a constant $c_{3}$ depending on $\sigma$, $\hat{H}$, $T$, $\nu(\M)$, and $\rho_{0}(\M)$.

On the other hand, using similar arguments and
the inequality $e^{-x}\ge 1-x$, $x \ge 0$, we get for any
$A\in \BM$ that
\[
\begin{split}
& \quad \int P_0\left[ 
\exp\biggl\{
-\frac{\sigma k}{n} \Sigma \bg(A) + H\bigl(g_{0}|_{A}+X|_{A^{\complement}}\bigr)
\biggr\} 
\right] 
\, dP_{k}^*(\bg) \\
& \quad =
\exp\left\{\frac{\nu(A)k}{n} \bigl( e^{-\sigma k/n}-1\bigr) \right\}
       \cdot P_0 \left[ \exp\left\{H(X)-\frac{\sigma k}{n} X(A)\right\} \right] \\
& \quad \ge \exp\left\{-c_{4}\frac{\nu(A)k^{2}}{n^{2}}\right\} - c_{5}\frac{\rho_{0}(A)k}{n},
 \end{split}
\]
where $c_{4}$ and $c_{5}$ are constants depending on $\hat{H}$, $T$, $\nu(\M)$, and $\rho_{0}(\M)$.

We may then find constants $c_{6}$ and $\epsilon>0$ such that 
\[
\exp\left\{-c_{4}\frac{\nu(A)k^{2}}{n^{2}}\right\} - c_{5}\frac{\rho_{0}(A)k}{n}
\ge \exp\left\{-c_{6} \frac{k}{n} (\nu(A)+\rho_{0}(A))\right\}
\]
for all $k\le \epsilon n$, and so
\[
\begin{split}
& \int \prod_{A\in \mcA}
P_0\left[ 
\exp\biggl\{
-\frac{\sigma k}{n} \Sigma \bg(A) + H\bigl(g_{0}|_{A}+X|_{A^{\complement}}\bigr)
\biggr\} 
\right] 
\, dP_{k}^*(\bg) \\
& \quad \ge
\exp\left\{-c_{6} \frac{k}{n} (\nu(\M)+\rho_{0}(\M))\right\}. \\
\end{split}
\]

Combining the upper and lower bounds from above, we
see that there is a constant $c_{7}$ such that
\[
\begin{split}
& -c_{7}\frac{k}{n}\bigl(1+ \Sigma \bg(\M)\bigr)
+ \log
\bE\left[\prod_{A\in\mcA_{k,0}} P_0\left[ \exp\left\{H\bigl(g_{0}|_{A}+X|_{A^{\complement}}\bigr)\right\}\right] \right] \\
& \quad \le 
\log \frac{dQ_{k}^{*}}{dP_{k}^{*}}(\bg) \\
&\quad \le c_{7}\frac{k}{n}\bigl(1+ \Sigma \bg(\M) \bigr) 
+
\log \bE\left[\prod_{A\in\mcA_{k,0}} 
P_0\left[ \exp\left\{H\bigl(g_{0}|_{A}+X|_{A^{\complement}}\bigr)\right\}\right] \right]\\
\end{split}
\]
for $k\le \epsilon n$.
\end{proof}

\begin{remark}
The approximation to $\log dQ_{k}^{*}/dP_{k}^{*}$ 
in Corollary~\ref{C:shattrap}
involves a product over the (potentially very large) partition $\mcA_{k,0}$
and looks rather unwieldy.
We may rewrite the approximation as follows to make it apparent where the
nontrivial contributions to the product arise:
\begin{equation} \label{E:shattrap}
\begin{split}
\prod_{A\in \mcA_{k,0}}
& P_0\left[\exp\left\{H\bigl(g_{0}|_{A}+X|_{A^{\complement}}\bigr)\right\}\right]\\
&=\prod_{A\in \mcA_{k,0}}
\frac{P_0\left[\exp\left\{H\bigl(g_{0}|_{A}+X|_{A^{\complement}}\bigr)\right\}\right]}
{P_0\left[\exp\{H(X_{A^{\complement}})\}\right]}
 \times
\prod_{A\in \mcA_{k,0}} 
 \frac{P_0\left[\exp\bigl\{H\bigl(X|_{A^{\complement}}\bigr)\bigr\}\right]}
 {P_0\left[\exp\{H(X)\}\right]}\\
 &=\prod_{
\begin{smallmatrix}
 A\in \mcA_{k,0}\\
 g(A)\ge 1
\end{smallmatrix}
}
\frac{P_0\left[\exp\{H\bigl(g_{0}|_{A}+X|_{A^{\complement}}\bigr)\}\right]}
{P_0\left[\exp\{H(X|_{A^{\complement}})\}\right]}
 \times
\prod_{A\in \mcA_{k,0}} 
 \frac{P_0\left[\exp\{H(X|_{A^{\complement}})\}\right]}
 {P_0\left[\exp\{H(X)\}\right]}.
\end{split}
\end{equation}
This is a trivial consequence of the fact that $P_{0}[e^{H(X)}]=1$.
\end{remark}

\section{Comparisons with complete Poissonization}
\label{SS:comp_Poisson}

Recall from Notation~\ref{N:operators} that the 
Poissonization operator $\mP$ acts on a probability 
measure $P$ on $\G$ by $\mP P :=\Pi_{\mu P}$.
Also, recall from Notation~\ref{N:Qk} 
that $O_{k}:=(\mP \mS_{n}\mM_{n})^{k}P_{0}$
is the analogue of the sequence of probability measures
$Q_{k} := (\mR \mS_{n}\mM_{n})^{k}P_{0}$ from Notation~\ref{N:Qk}
that is of primary interest to us in Theorem~\ref{T:limit}, 
with the recombination operator $\mR$ 
replaced by the complete Poissonization operator $\mP$.
Finally, recall from Notation~\ref{N:Qkstar_Pkstar} that 
the counterpart of $Q_{k}$ for $\G^*$
(that is, for the setting in which we keep track of the generation in which
mutations from the ancestral wild type occurred) is
$Q^{*}_{k} =(\mR^{*} \mS_n^{*}\mM^{*})^{k}P_{0}$.

\begin{lemma} 
\label{L:pnqk1}
Suppose that $P = \Pi_\pi$ for some finite measure $\pi\in \H^{+}$
and $Q$ is an equivalent measure.  Put $H := \log dQ/dP$.
Set
\begin{equation*} 
\tau(m) := \log \int_{\G} \exp\{H(g+\delta_{m})\} \, dP(g).
\end{equation*}
\glossary{$\tau(m)$}
\begin{itemize}
\item[(a)]
The probability measure $\mP Q$ has intensity 
measure $\mu Q$ that is absolutely continuous with respect to
$\pi$ with Radon-Nikodym derivative 
\begin{equation*}  
\frac{d\mu\mP Q}{d \pi} (x)=\frac{d\mu Q}{d \pi} (x)=\exp\{\tau(x)\}.
\end{equation*}
\item[(b)]
The probability measure $\mP Q$ is absolutely continuous
with respect to $P$, with Radon-Nikodym derivative satisfying
\begin{equation*}  
\begin{split}
\log\frac{d\mP Q}{d P} (g)
& = g[\tau]- \int_{\M}\biggl[\int_{\G} \exp\{H(g+\delta_{x})\} \, dP(g) \biggr]
          \,  d\pi(x)  + \pi(\M) \\
& = g[\tau]- \int_{\G} g(\M) \exp\{H(g)\} \, dP(g) + \pi(\M) . \\
\end{split}
\end{equation*}
\end{itemize}
\end{lemma}

\begin{proof}
For a Borel function $f:\M\to\bR_+$,
\begin{align*}
  \int_{\M}f(x) d(\mu\mP Q)(x) &= \int_{\M} f(x) \, d(\mu Q)(x)\\
&= \int_{\G} g[f] \, dQ (g)\\
&=\int_{\G}g[f] \exp\{H(g)\} \, dP(g)\\
&=\int_{\M} f(x) \exp\{\tau(x)\} \, d\pi(x),
\end{align*}
where we used Campbell's Theorem (see Proposition~\ref{P:Campbell}) 
for the penultimate equality. Part (a) follows.
\index{Campbell's Theorem}
It then follows immediately from Lemma~\ref{L:PiPi} 
that $\mP Q$ is absolutely continuous with respect to $P$, with
$$
\log\frac{d\mP Q}{d P} (g)
     = g[\tau]- \int_{\M}\biggl[\int_{\G} \exp\{H(g+\delta_{x})\} \, dP(g) \biggr]
          \,  d\pi(x)  + \pi(\M) ,
$$
giving the first claim in part (b).
The second claim in part (b) follows by another 
application of Campbell's Theorem.
\end{proof}  
\index{Campbell's Theorem}

\begin{lemma}  \label{L:Qkprimebound}
Suppose that $Q_{0}=O_{0}=P_{0}$. 
\begin{itemize}
\item[(a)]
For all $x\in\M$ and $k \in \bN_0$,
\begin{equation*} 
\left| \log \frac{d\mu O_{k}}{d\mu P_{k}}(x)\right|\le \frac{\sigma k}{n}.
\end{equation*}
\item[(b)]
For all $g\in\G$ and $k \in \bN_0$,
\begin{equation*} 
\left|\log\frac{dO_{k}}{dP_{k}}(g)-\log\frac{dO_{k}}{dP_{k}}(0)\right| \le 
   \frac{k\sigma}{n} g(\M).
\end{equation*}
\item[(c)]
For all $g\in\G$ and $k \in \bN_0$,
\begin{equation*} 
\left|\log\frac{dO_{k}}{dP_{k}}(g)\right| \le 
   \frac{k\sigma}{n}\left(g(\M)+e^{k\sigma/n}\mu P_{k}(\M)\right).
\end{equation*}
\end{itemize}
\end{lemma}

\begin{proof}
The proof is by induction. 
We start with $O_{0}=P_{0}$, so $dO_{0}/dP_{0}=1$.
Let $\pi_{k}:=\mu P_{k}=\rho_{0}+\frac{k}{n}\nu$,  
and $\pi'_{k}:=\mu O_{k}$. Put
$\alpha_{k}:=\sup_{x\in\M} |\log d\pi'_{k}/d\pi_{k}(x)|.$ 

By Lemma~\ref{L:PiPi} and Lemma~\ref{L:pnqk1} we know 
that $d \mu \mP \mS_n O_{k}/d\mu P_{k}(x)=e^{\tau_{k}(x)}$, 
where $\tau_{k}(x):= \log P_{k}[\exp\{H(X+\delta_{x})\}]$ and
%
%
%
%
%
\begin{align*}
H(g)&:= \log \frac{d \mS_{n}  O_{k}}{dO_{k}}(g)+ \log \frac{dO_{k}}{dP_{k}}(g)\\
&=\log \frac{\exp\{-S(g)/n\}}{O_{k}[\exp\{-S/n\}]} 
   +g \left[ \log \, \frac{d\mu O_{k}}{d\mu P_{k}} \right]
    -\mu P_{k} \left[ \frac{d\mu O_{k}}{d\mu P_{k}} -1 \right]\\
&=-\frac{S(g)}{n}-\log O_{k}\left[ \exp\{-S(X)/n\}\right] 
      + g\left[\log \frac{d\pi'_{k}}{d\pi_{k}}\right] -
  \pi_{k} \left[\frac{d\pi'_{k}}{d\pi_{k}}-1\right].
\end{align*}
For any $g\in\G$ and $x\in\M$ we have
$$
\bigl| H(g+\delta_{x})-H(g)\bigr| \le \frac{\sigma}{n} + \alpha_{k}.
$$
Thus,
\[
\begin{split}
\tau_{k}(x)
&\le 
\log P_{k}\left[ \exp\{H(X)+\alpha_{k}+\sigma/n\}\right] \\
&\le 
\log P_{k} \left[ \exp\{H(X)\}\right]
    +\alpha_{k}+\frac{\sigma}{n}=\alpha_{k}+\frac{\sigma}{n}\\
\end{split}
\]
and
\[
\begin{split}
\tau_{k}(x)
&\ge 
\log P_{k}\left[ \exp\{H(X)-\alpha_{k}-\sigma/n\}\right] \\
&\ge 
\log P_{k} \left[ \exp\{H(X)\}\right]-\alpha_{k}
   -\frac{\sigma}{n}=-\alpha_{k}-\frac{\sigma}{n}, \\
\end{split}
\]
%
%
implying that $\sup |t_{k}(x)|\le \alpha_{k}+\sigma/n$. 
By Lemma~\ref{L:RNsums}, it follows 
that $\alpha_{k+1}\le \alpha_{k}+\sigma/n$, and so 
$\alpha_{k}\le k\sigma/n$ for all $k \in \bN_0$, establishing part (a).
%
%

Applying Lemma~\ref{L:PiPi}, we see that
$$
\left|\log\frac{dO_{k}}{dP_{k}}(g) 
   -\log\frac{dO_{k}}{dP_{k}}(0)\right|\le \alpha_{k} g(\M)
  \le \frac{k\sigma}{n}g(\M)
$$
and
$$
\left|\log\frac{dO_{k}}{dP_{k}}(g) \right|
    \le \alpha_{k} g(\M)+ \left( e^{\alpha_{k}}-1\right) \pi_{k}(\M)
  \le \frac{k\sigma}{n}\left(g(\M)+e^{k\sigma/n}\mu P_{k}(\M)\right),
$$
establishing parts (b) and (c).
\end{proof}
We now extend Lemma~\ref{L:pnqk1} to probability measures on $\G^{*}$.
Recall that a $(k+1)$-tuple 
of independent Poisson random measures on $\M$
with intensities $\pi_{0},\dots,\pi_{k}$, 
may, as discussed in Section~\ref{SS:vintages}, 
be thought of as a single Poisson random measure on the set
$\M\times\{0, \ldots, k\}$ 
with intensity $\sum \pi_{i}\otimes \delta_{\{i\}}$.
\begin{corollary} 
\label{C:tnqk}
Suppose that $P^* := \Pi_{\pi_0} \otimes \cdots \otimes \Pi_{\pi_k}$
for $\pi_0, \ldots, \pi_k \in \H^+$ and
and $Q^*$ is any probability measure on $\G^{k+1}$ such that
$\mu \Psi_{j}Q^{*}$ is absolutely continuous with
respect to $\mu \Psi_{j}P^{*} = \pi_j$ for $0 \le j \le k$.
Then, $\mP^{*} Q^{*}$ is absolutely continuous
with respect to $P^{*}$, with Radon-Nikodym derivative satisfying
\begin{equation*}  
\log\frac{d\mP^{*} Q^*}{d P^{*}} (\bg)
=\sum_{i=0}^{k}g_{i}[\tau^{(i)}]-
           \sum_{i=0}^{k} \pi_i\left[\exp\{ \tau^{(i)}\}-1\right] 
\end{equation*}
where
\begin{equation*}  
\tau^{(i)}(x) :=\log \int_{\G^{k+1}} \frac{dQ^*}{dP^*}(h_{0},\dots,h_{i-1},h_{i}+
       \delta_{x},\dots,h_{k}) \, dP^*(h_{0},\dots,h_{k}).
\end{equation*}
\end{corollary}

\begin{proof}
By definition $\mP^{*} Q^{*} 
=  \mP \Psi_{0} Q^{*} \otimes \cdots \otimes \mP \Psi_{k} Q^{*}$
with $\mP \Psi_{j} Q^{*} = \Pi_{\mu \Psi_j Q^*}$ for $0 \le j \le k$.
Thus,
$$
\log\frac{d\mP^{*} Q^{*}}{d P^{*}}(g_{0},\dots,g_{k})=
    \sum_{j=0}^{k}  \log\frac{d\mP \Psi_{j}Q^{*}}{d\Psi_{j}P^{*}}(g_{j})
$$
and the result follows then directly from Lemma~\ref{L:pnqk1}.
\end{proof}


\begin{lemma} \label{L:tshat}
Suppose that
$\hat H := \Theta(P_0, Q_0) < \infty$.  
Set $H := \log dQ_0/dP_0$.
Let $\tau^{(i)}$, $0 \le i \le k$, be
as in Corollary~\ref{C:tnqk}
with $Q^{*}=Q^{*}_{k}$ and $P^{*}=P^{*}_{k}$.
Then, there are positive constants $\epsilon$ and $c$, 
depending on $\hat{H}$ but not depending on $k$ or $n$, 
such that 
\[
\biggl|\tau^{(0)}(x)-\log P_0\left[\exp\{H(X+\delta_{x})\}\right]\biggr| 
\le c\frac{k}{n}
\]
and
\[
\left| \tau^{(i)}(x)\right| \le c\frac{k}{n} 
\]
for $1 \le i \le k \le n \epsilon$.
\end{lemma}
%

\begin{proof}
We apply Corollary~\ref{C:shattrap}. 
Since the Poisson random measures
$\xi_{k,0}(A)$ are independent as $A$ ranges 
over the random partition
$\mcA_{k,0}$, there are constants $c_{1}$ and $\epsilon$ 
such that for $k\le \epsilon n$,
\[
\begin{split}
& Ce^{-c_{1} k(1+\Sigma \bg(\M))/n} 
\bE\biggl[\prod_{A\in \mcA_{k,0}}P_0
     \left[\exp\{H\bigl(g_{0}|_{A}+X|_{A^{\complement}}\bigr)\}\right]\biggr]\\
&\quad \le
\frac{dQ^{*}_{k}}{dP^{*}_{k}}(\bg)\\
& \quad \le 
Ce^{c_{1} k(1+\Sigma \bg(\M))/n}\bE
   \biggl[\prod_{A\in \mcA_{k,0}}P_0\left[
    \exp\{H\bigl(g_{0}|_{A}+X|_{A^{\complement}}\bigr)\}\right]\biggr] , \\
\end{split}
\]
where $C$ does not depend on $\bg$, but may depend on
$k$ and $n$. We recombine terms to obtain
\begin{align*}
&C\exp\Bigl\{-c_{1} \frac{k}{n}(1+\sum_{i=1}^{k} g_{i}(\M))\Bigr\}
\bE\biggl[\prod_{A\in \mcA_{k,0}}P_0
     \left[\exp\left\{H\bigl(g_{0}|_{A}+X|_{A^{\complement}}\bigr) 
     -\frac{c_{1}k}{n}g_{0}(A)\right\}\right]\biggr]\\
&\le
\frac{dQ^{*}_{k}}{dP^{*}_{k}}(\bg)\\
&\le 
C\exp\Bigl\{c_{1} \frac{k}{n}(1+\sum_{i=1}^{k} g_{i}(\M))\Bigr\}
  \bE\biggl[\prod_{A\in \mcA_{k,0}}
  P_0\left[\exp\left\{H\bigl(g_{0}|_{A}+X|_{A^{\complement}}\bigr) 
     +\frac{c_{1}k}{n}g_{0}(A)\right\}\right]\biggr].
\end{align*}

Note that  $\bg \mapsto \sum_{i=1}^{k}g_{i}(\M)$
and $\bg \mapsto g_{0}$ are independent under $P_{k}^{*}$,
with the former having a Poisson distribution with parameter $k\nu(\M)/n$. 
%
A bound on the constant $C$ can be imported from Lemma~\ref{L:RNcompare},
leading to the following inequality
%
\begin{align*}
&\tau^{(0)}(x)= \log\int \frac{dQ^*_{k}}{dP^*_{k}}
(g_{0}+\delta_{x},\dots,g_{k}) \, dP^{*}_{k}(\bg)\\
&\le \log \bE \biggl[\int \exp\Bigl\{c_{1} 
          \frac{k}{n}(1+\sum_{i=1}^{k} g_{i}(\M))\Bigr\}\\
&\qquad\times  \prod_{A\in \mcA_{k,0}}
P_0\left[\exp\left\{H\bigl((g_{0}+\delta_{x})|_{A}+X|_{A^{\complement}}\bigr) 
       +\frac{c_{1}k}{n}(g_{0}+\delta_{x})(A)\right\}\right] \, 
             dP^{*}_{k}(\bg)\biggr]\\
&\quad - \log \bE \biggl[\int \exp\Bigl\{-c_{1} 
              \frac{k}{n}(1+\sum_{i=1}^{k} g_{i}(\M))\Bigr\}\\
&\qquad\times\prod_{A\in \mcA_{k,0}} \hspace*{-2mm}
P_0\left[\exp\left\{H\bigl(g_{0}|_{A}+X|_{A^{\complement}}\bigr) 
      -\frac{c_{1}k}{n}g_{0}(A)\right\}\right] \, dP^{*}_{k}(\bg) \biggr]
\end{align*}
\begin{align*}
&=\frac{3c_{1}k}{n}+2\frac{k\nu(\M)}{n} (e^{c_{1}k/n}-1)\\
&\qquad + \log \bE \biggl[\int \prod_{ A\in \mcA_{k,0}}
  P_0\left[\exp\left\{H((g'+\delta_{x})|_{A}+X|_{A^{\complement}}) 
        +\frac{c_{1}k}{n}g'(A)\right\}\right] \, dP_{0}(g') \biggr]\\
&\qquad\qquad - \log \bE \biggl[\int \prod_{ A\in \mcA_{k,0}}\hspace*{-2mm}
  P_0\left[\exp\left\{H(g'|_{A}+X|_{A^{\complement}}) 
       -\frac{c_{1}k}{n}g'(A)\right\}\right] \, dP_{0}(g') \biggr].
\end{align*}
%
%
Applying Corollary~\ref{C:intindeppoisson}, 
we see that there is a constant $c_{2}$ such that
\begin{align*}
\tau^{(0)}(x)&\le \frac{c_{2}k}{n}
 + \log \bE \biggl[\prod_{ A\in \mcA_{k,0}} 
     P_0\left[\exp\left\{H\bigl(X+(\delta_{x})_{A}) 
     +\frac{c_{1}k}{n}X(A)\right\}\right] \\
&\hspace*{3cm} \times P_0\left[\exp\left\{H(X) 
      -\frac{c_{1}k}{n}X(A)\right\}\right]^{-1} \biggr] \\
&\le \frac{c_{2}k}{n}
 + \log \bE \biggl[P_0\left[\exp\left\{H\bigl(X+\delta_{x}) 
           +\frac{c_{1}k}{n}X(A(x))\right\}\right] \\
&\hspace*{3cm}\times\prod_{ A\in \mcA_{k,0}}
             \frac{P_0\left[\exp\{H(X)+c_{1} k X(A)/n\}\right]} 
   {P_0\left[\exp\{H(X)-c_{1} k X(A)/n\}\right]} \biggr], 
\end{align*}
where $A(x)$ denotes the element of $\mcA_{k,0}$ containing $x \in \M$.

By Lemma~\ref{L:poisindicbound} there are 
constants $\epsilon_{2}$ and $c_{3}$ such that for $k/n\le \epsilon_{2}$ 
the product is bounded by
$$
\exp\biggl\{\sum_{A\in \mcA_{k,0}} c_{3}\frac{2c_{1} k}{n} \rho_{0}(A)\biggr\}\le \exp\biggl\{2c_{1}c_{3} \frac{k}{n} \rho_{0}(\M)\biggr\}.
$$
Thus, there is a constant $c_{4}$ such that 
for $k\le \epsilon' n$, where $\epsilon'= \epsilon \wedge \epsilon_{2}$,
\begin{equation}
\tau^{(0)}(x)\le \frac{c_{4}k}{n}+\log \bE
       \left[P_0\left[\exp\left\{H(X+\delta_{x})
          +\frac{c_{1} k}{n} X(A(x))\right\}\right]\right].
\label{E:t0xsum}
\end{equation}

We proceed by writing
\begin{equation} \label{E:P0H}
\begin{split}
 P_0&\left[\exp\left\{H(X+\delta_{x})+\frac{c_{1} k}{n} X(A(x))\right\}\right]\\
& =
   P_0\left[ \exp\{H(X+\delta_{x})\}\right] \\
& \quad \times \left(1+\frac{P_0\left[\exp\{H(X+\delta_{x})\}
            \left(\exp\{c_{1} k X(A(x))/n\}-1\right) \right]}
{P_0\left[\exp\{H(X+\delta_{x})\} \right]} \right). \\
\end{split}
\end{equation}
Using again the relation $e^{x}-1\le |x|e^{|x|}$, 
and the fact that $P_{0}$ is Poisson, hence the restrictions 
of $X$ to disjoint sets are independent under $P_0$, the expectation in the 
numerator above may be rewritten as
\[
\begin{split}
& P_0\left[\exp\left\{H\bigl(X|_{A(x)^{\complement}}\bigr)\right\} 
   \cdot \exp\left\{H(X+\delta_{x})-H(X|_{A(x)^{\complement}})\right\}
   \left(\exp\{c_{1} k X(A(x))/n\}-1\right) \right]
\\
& \le P_0\left[ \exp\left\{H\bigl(X|_{A(x)^{\complement}}\bigr)\right\} 
   \right] \cdot P_0\left[ \frac{c_{1} k X(A(x))}{n} 
    \exp\left\{\left(\hat{H}+c_{1} \frac{k}{n}\right)\cdot X(A(x))+\hat{H}
   \right\} \right]. \\
&  \le \frac{c_{1} k }{n} P_0\left[ \exp\left\{
        H\bigl(X|_{A(x)^{\complement}}\bigr)\right\} \right] 
        \cdot \rho_{0}(A(x)) c_{5}e^{(c_{5}-1)\rho_{0}(A(x))}.
\end{split}
\]
where $c_{5}:= \exp\left\{\hat{H}+c_{1} T\right\}$.

For the denominator, we have
$$
P_0\left[\exp\{H(X+\delta_{x})\} \right]
   \ge P_0\left[ \exp\left\{H\bigl(X|_{A(x)^{\complement}}\bigr)\right\} \right]
    \cdot \exp\left\{ -\hat{H}\bigl(\rho_{0}(A(x))+1\bigr)   \right\}.
$$
Combining these bounds and using the inequality $\log(1+x)\le x$ 
for $x\ge 0$, we may find a constant $c_{6}$ 
(depending on $\rho_{0}(\M)$ and $\hat{H}$) such that
$$
\log\left(1+\frac{P_0\left[\exp\{H(X+\delta_{x})\}
        \left(\exp\{c_{1} k X(A(x))/n\}-1\right) \right]}
{P_0\left[\exp\{H(X+\delta_{x})\} \right]} \right)
\le c_{6}\rho_{0}\bigl(A(x)\bigr)\frac{k}{n}\, .
$$
Combining this with \eqref{E:t0xsum} and \eqref{E:P0H} 
gives us a constant $c_{7}$ such that
\begin{equation} \label{E:t0xupperfinal}
\tau^{(0)}(x)\le \frac{c_{7}k}{n}+\log P_0\left[
             \exp\left\{H(X+\delta_{x})\right\}\right]
\end{equation}
for $k/n\le \epsilon'$. Nearly identical calculation give us 
a lower bound, completing the proof of the first claim.

The second claim
follows from similar, but simpler, arguments.
\end{proof}

%
%

\begin{corollary}\label{C:tautogether}
Under the hypotheses of Lemma~\ref{L:tshat}, 
\begin{equation*} 
\biggl| \log \frac{d\mP^{*}Q^{*}_{k}}{dP^{*}_{k}}(\bg)
     - \, \int \log P_{0} \left[ \exp\{H(X+\delta_{x})\}\right] \, dg_{0}(x) \biggr| \\
\le  \,  \frac{ck}{n}  \, ( 1 +  \Sigma \, \bg ). 
\end{equation*}
\end{corollary}

\begin{proof}
By Corollary~\ref{C:tnqk}, the quantity on the left-hand side of
the claimed inequality is at most 
\begin{equation} \label{E:PQfivepart}
\begin{split}
&  \left| \int \, \tau^{(0)}(m) \, dg_{0}(m)   
     -  \int \log P_{0} \left[ \exp\{H(X+\delta_{m})\}\right] \, dg_{0}(m) 
                  \right|     \\
  &+ \quad \left|\sum_{j=1}^{k} \int  \tau^{(j)}(m) \, dg_{j}(m) \right| \\
  &+ \quad \left| (-1)\int  ( \exp\{ \tau^{(0)}(y)\} - 1  ) \,  
                      d \pi_{0}(y) \right|   \\    
  &+ \quad \left|(-1)\sum_{j=1}^{k} \int  \exp\{ \tau^{(j)}(y)\}  
  \,  d \pi_{j}(y)  \,  \right|   \\    
  &+ \quad   k \nu(\M)/n. \\
\end{split}
\end{equation}

We bound the quantities on each of the five lines of \eqref{E:PQfivepart}
separately.
For the first line, the first part of Lemma~\ref{L:tshat} 
provides a bound of $ (ck/n) g_{0}(\M)$.
For the second line, the second part of Lemma~\ref{L:tshat}
provides a bound of $(ck/n) \sum_{j=1}^k g_{j}(\M)$
For the third line,  the same lemma lets us write 
$$
P_{0} \left[ \exp\{H(X+\delta_{y})\}\right] e^{-ck/n} 
    \le  \,  \exp\{ \tau^{(0)}(y)\}  
    \le  P_{0} \left[ \exp\{H(X+\delta_{y})\}\right] e^{ck/n}.
$$ 
\index{Campbell's Theorem}
By Campbell's Theorem~\ref{P:Campbell},  the integral 
with respect to $\pi_{0}(dy)$ of the left-hand side of this
last expression is $ \pi_{0}(\M) \exp\{-ck/n\} $ and the
corresponding integral on the right-hand side
is $ \pi_{0}(\M) \exp\{+ck/n\} $.  
Subtracting $\pi_{0}(\M)$ and making use once more 
of the inequality $ | e^{x} - 1 | \le |x| e^{|x|}$,
we conclude that the quantity on the third line is bounded 
by $ \pi_{0}(\M) (ck/n) e^{c \epsilon} $ when $ k/n \le \epsilon $.

For the fourth line, again by Campbell's Theorem, the integral is the
sum of the total intensities of $g_{1} $ to $g_{k}$ 
with respect to $Q_{k}^{*}$, that is to say,
$$
\int \,  ( g_{1}(\M) + \ldots g_{k}(\M) )  \,  
           \frac{d Q_{k}^{*}} {d P_{k}^{*}} \, d P_{k}^{*}(\bg).
$$ 
By Corollary~\ref{C:boundedRN}, this nonnegative quantity
is bounded above by
\[
\int ( g_{1}(\M) + \cdots + g_{k}(\M) ) 
\exp ( c g_{0}(\M) + \cdots + c g_{k}(\M) + c\mu P_{k}(\M)) )  \, d P_{k}^{*}(\bg). 
\]
Under $P_{k}^{*}$, the random measures
$\bg \mapsto g_0$ 
and $\bg \mapsto g_{1} + \ldots g_{k}$
are independent Poisson random measures
with respective intensity measures $\rho_{0}$ and    
$ k \nu/n $, and so  
\[
 \exp\{\pi_{0}(\M) (e^{c} -1 +c)\} 
\times
 \exp\{ c + (k/n)\nu(\M) (e^{c} -1 +c )\} 
\times \nu(\M) k/n,
\]  
or, in effect, a constant multiple of $k/n$.
The fifth line is already so bounded.  
Gathering terms and setting a new constant equal to the 
maximum of the constants from all the lines,  the bound 
in the corollary is established.  
\end{proof}

\begin{lemma} \label{L:Tcomparednew}
Suppose that $\hat H := \Theta(P_0, Q_0) < \infty$ and
and the pair $(\mathcal{R},\mu P_{0})$ is shattering 
with constant $\alpha$. Then, there exists a constant $c$ 
(depending on $T$, $\alpha$, $\rho_0(\M) = \mu Q_0(\M) = \mu P_{0}(\M)$, 
and $\hat{H}$), such that
\begin{equation*} 
\int\int\biggl|\frac{d\mP^{*}Q^{*}_{k}}{dQ^{*}_{k}}(\bg) 
              -1 \biggr|\, dQ^{*}_{k}(\bg)
\le c\left(\rec{k+1} \vee \frac{k}{n}\right).
\end{equation*}
\end{lemma}

\begin{proof}
Combining Corollary~\ref{C:shattrap} 
(in the form of \eqref{E:shattrap}), 
with Corollary~\ref{C:tautogether},
we see that there are positive constants $\epsilon$ and $c_{1}$, 
depending on $\hat{H}$, such that for $k/n\le \epsilon$,
\begin{equation} \label{E:pstarqstark}
\begin{split}
&-\log \frac{d\mP^{*}Q^{*}_{k}}{dQ^{*}_{k}}(\bg)
   =\log\frac{dQ^{*}_{k}}{dP^{*}_{k}}(\bg)
     - \log\frac{d\mP^{*}Q^{*}_{k}}{dP^{*}_{k}}(\bg) \\
&= \Biggl[\int \log P_{0} \left[ \exp\{H(X+\delta_{x})\}\right] \, dg_{0}(x) \\
&\qquad- \log\bE\biggl[ \prod_{A\in \mcA_{k,0}}
\frac{P_0\left[\exp\left\{H\bigl(g_{0}|_{A}+X|_{A^{\complement}}\bigr)\right\}\right]}
{P_0\left[\exp\left\{H(X|_{A^{\complement}})\right\}\right] }\Biggr]
+C + \gamma(\bg)\frac{k}{n} \Sigma \bg(\M),
\end{split}
\end{equation}
where $|\gamma(\bg)|\le c_{1}$ 
and 
$$
C  =    - \log \bE \left[ \prod_{A\in \mcA_{k,0}}
       P_0\left[\exp ( H(X|_{A^{\complement}}) ) \right] \right]  
$$ 
is a quantity not depending on $\bg$. 

Observe that the product in the second term
of \eqref{E:pstarqstark} is universally bounded above and below 
as follows
\begin{equation} \label{E:produniversal}
\exp\{-\hat{H}g_{0}(\M)\} \le \prod_{A\in \mcA_{k,0}}
\frac{P_0\left[\exp\{H((g_{0})_{A}+X_{A^{\complement}})\}\right]} 
  {P_0\left[\exp\{H(X_{A^{\complement}})\}\right] }
\le \exp\{\hat{H}g_{0}(\M)\} .
\end{equation}

We decompose the product as
\begin{equation} \label{E:Rsplit}
\begin{split}
&\bE\biggl[ \prod_{A\in \mcA_{k,0}}
  \frac{P_0\left[\exp\left\{H\bigl(g_{0}|_{A}+X|_{A^{\complement}} \bigr) \right\}\right]}
         {P_0\left[\exp\left\{H\bigl(X|_{A^{\complement}}\bigr)\right\}\right] } \biggr]\\
&\quad = \bE\left[ \prod_{A\in \mcA_{k,0}}
\frac{P_0\left[\exp\left\{H\bigl(g_{0}|_{A}+X|_{A^{\complement}} \bigr) \right\}\right]}
{P_0\left[\exp\left\{H\bigl(X|_{A^{\complement}}\bigr)\right\}\right] }
  \indic\left\{\bigvee_{A\in\mcA_{k,0}}g_{0}(A)\le 1\right\}\right]\\
&\qquad + \bE\left[ \prod_{A\in \mcA_{k,0}}
\frac{P_0\left[\exp\left\{H\bigl(g_{0}|_{A}+X|_{A^{\complement}} \bigr) \right\}\right]}
       {P_0\left[\exp\left\{H\bigl(X|_{A^{\complement}}\bigr)\right\}\right] }
  \indic\left\{\bigvee_{A\in\mcA_{k,0}}g_{0}(A) \ge 2\right\}\right].
\end{split}
\end{equation}
%
%
The logarithm of the product in the first expectation on the right is
\begin{equation}
\label{E:log_prod_less_1}
\begin{split}
&\left(
 \log \prod_{A\in \mcA_{k,0}}
\frac{P_0\left[\exp\left\{H\bigl(g_{0}|_{A}+X|_{A^{\complement}} \bigr) \right\}\right]}
{P_0\left[\exp\left\{H\bigl(X|_{A^{\complement}}\bigr)\right\}\right] }
\right)
  \indic\left\{\bigvee_{A\in\mcA_{k,0}}g_{0}(A)\le 1\right\} \\
& \quad =\int \log \frac{P_{0}\left[\exp\bigl\{H(\delta_{x}
          +X|_{A(x)^{\complement}})\bigr\}\right]}
    {P_{0}\left[\exp\left\{H\bigl(X|_{A^{\complement}}\bigr)\right\}\right]} 
             \, dg_{0}(x)\\
& \quad =\int \log P_{0} \left[ \exp\{H(X+\delta_{x})\}\right] \, dg_{0}(x)\\
&  \qquad +
\int \log \frac{P_{0}\left[\exp\bigl\{H(\delta_{x}
        +X|_{A(x)^{\complement}})\bigr\}\right]}
           {P_{0}\left[\exp\bigl\{H(\delta_{x}+X)\bigr\}\right]} \, dg_{0}(x) \\
& \qquad -
    \int \log \frac{P_{0}\left[\exp\bigl\{H(X|_{A(x)^{\complement}})\bigr\}
          \right]}{P_{0}\left[\exp\bigl\{H(X)\bigr\}\right]} \, dg_{0}(x). \\
\end{split}
\end{equation}

We next find bounds for the second and
third terms on the right of 
\eqref{E:log_prod_less_1}.
For any set $A\in\BM$,
\begin{align*}
 P_0\left[ \exp\{H(\delta_{x}+X)\}\right]
     &=  P_0\left[ \exp\{H(X|_{A^{\complement}}+\delta_{x})\}
    \exp\{H(X+\delta_{x})-H(X|_{A^{\complement}}+\delta_{x})\}\right]\\
 &\le  P_0\left[ \exp\{H(X|_{A^{\complement}}+\delta_{x})\}e^{\hat{H} X(A)}\right]\\
 &=P_0\left[\exp\{H(X|_{A^{\complement}}+\delta_{x})\}\right] 
                 \cdot P_0\left[ \exp\{\hat{H} X(A)\} \right]\\
 &= P_0\left[ \exp\{H(X|_{A^{\complement}}+\delta_{x})\}\right] 
          \cdot \exp\left\{  \rho_{0}(A)\left( e^{\hat{H}}-1\right)\right\}\\
 &\le  P_0\left[\exp\{H(X|_{A^{\complement}}+\delta_{x})\}\right] 
           \cdot \exp\left\{  \rho_{0}(A)\hhH\right\},
\end{align*}
where $\hhH:=\hat{H} e^{\hat{H}}$.
A similar calculation gives an analogous lower bound
for the second term on the right of \eqref{E:log_prod_less_1}, yielding
$$
\biggl|
\int \log \frac{P_{0}\left[\exp\bigl\{H(\delta_{x}
      +X|_{A(x)^{\complement}})\bigr\}\right]}
      {P_{0}\left[\exp\bigl\{H(\delta_{x}+X)\bigr\}\right]} dg_{0}(x)
\biggr|
\le
  \hhH\int \rho_{0}(A(x)) dg_{0}(x).
$$
The third term on the right of \eqref{E:log_prod_less_1}
may be bounded in the same way.  Thus,
\begin{align*}
&\Biggl| \log\prod_{A\in \mcA_{k,0}}
\frac{P_0\left[\exp\{H((g_{0})_{A}+X_{A^{\complement}})\}\right]}
     {P_0\left[\exp\{H(X_{A^{\complement}})\}\right] }
     \indic\left\{\bigvee_{A\in\mcA_{k,0}}g_{0}(A)\le 1\right\} \\
& \qquad  -
\int \log P_{0} \left[ \exp\{H(X+\delta_{x})\}\right] \, dg_{0}(x) \Biggr|\\
&\quad
\le
2\hhH \int \rho_{0}(A(x)) \, dg_{0}(x).
\end{align*}
Equivalently,
\begin{align*}
 &\bE\left[ \prod_{A\in \mcA_{k,0}}
  \frac{P_0\left[\exp\{H((g_{0})_{A}+X_{A^{\complement}})\}\right]}
       {P_0\left[\exp\{H(X_{A^{\complement}})\}\right] }
  \indic\left\{\bigvee_{A\in \mcA_{k,0}} g_{0}(A)\le 1\right\} \right]\\
 &= \exp \left\{\int \log P_{0} \left[ \exp\{H(X+\delta_{x})\}\right]
                 \, dg_{0}(x)\right\} \\
 & \quad \times \bE\left[ \exp\{R(g_{0},\mcA_{k,0})\}\indic\left\{\bigvee_{A\in \mcA_{k,0}} 
      g_{0}(A) \le 1\right\} \right],
\end{align*}
where $R(g,\mcA_{k,0})$ is a random variable satisfying
$$
|R(\bg,\mcA_{k,0})|\le 2\hhH \sum_{
\begin{smallmatrix}
 A\in \mcA_{k,0}\\
 g_{0}(A)= 1
\end{smallmatrix}
} \rho_{0}(A).
$$

We see from Lemma~\ref{L:expbound2} that
\begin{align*}
-\bE\biggl[2\hhH \sum_{
\begin{smallmatrix}
 A\in \mcA_{k,0}:\\
 g_{0}(A)= 1
\end{smallmatrix}
} \rho_{0}(A)  \biggr]&\le \log\bE\biggl[ \exp\biggl\{ 2\hhH \sum_{
\begin{smallmatrix}
 A\in \mcA_{k,0}:\\
 g_{0}(A)= 1
\end{smallmatrix}
} \rho_{0}(A) \biggr\} \biggr]\\
&\le
2\hhH e^{2\hhH\rho_{0}(\M)}\bE\biggl[ \sum_{
\begin{smallmatrix}
 A\in \mcA_{k,0}:\\
 g_{0}(A)= 1
\end{smallmatrix}
} \rho_{0}(A) \biggr].
\end{align*}

Combining this with \eqref{E:produniversal} 
and \eqref{E:Rsplit} we may find
constants $c_{2},c_{3}$ (depending on $\hat{H}$ 
and $\rho_{0}(\M)$) such that
\begin{align*}
\biggl|\int \log P_{0} \Bigl[ \exp\{H(X+\delta_{x})\}&\Bigr]\, dg_{0}(x)\, 
     - \,\log\bE\biggl[ \prod_{A\in \mcA_{k,0}}
  \frac{P_0\left[\exp\left\{H\bigl(g_{0}|_{A}+X|_{A^{\complement}}\bigr)\right\}\right]}
       {P_0\left[\exp\left\{H(X|_{A^{\complement}})\right\}\right] } \biggr]\biggr|\\
  &\le c_{2}\bE \biggl[\sum_{
\begin{smallmatrix}
 A\in \mcA_{k,0}:\\
 g_{0}(A)= 1
\end{smallmatrix}
} \rho_{0}(A) \biggr]
+
c_{3} \bP\ls \bigvee_{A\in\mcA_{k,0}} g_{0}(A)\ge 2\rs.
\end{align*}

We may now apply part (b) of 
Lemma~\ref{L:RNcompare} with
$$
f_{2}(\bg):= \gamma(\bg) \frac{k}{n} \Sigma \bg(\M)
+ c_{2}\bE \biggl[\sum_{
\begin{smallmatrix}
 A\in \mcA_{k,0}:\\
 g_{0}(A)= 1
\end{smallmatrix}
} \rho_{0}(A) \biggr]
+
c_{3} \bP\left\{ \bigvee_{A\in\mcA_{k,0}} g_{0}(A)\ge 2\right\},
$$
and $f_{1}(\bg)=0$, using the fact that $f_{2}(\bg)$ 
is uniformly bounded by a constant $c_{4}$, to obtain
%
$$
\int\biggl|\frac{d\mP^{*}Q^{*}_{k}}{dQ^{*}_{k}}(\bg) -1 \biggr|\, 
              dQ^{*}_{k}(\bg)
  \le 2 e^{2c_{4}}Q^{*}_{k}\left[ f_{2}(X)\right].
  $$

By Corollary~\ref{C:boundedRN} there are positive 
constants $c_{5}$ and $c_{6}$ such that
$$
\frac{dQ_{k}^{*}}{dP_{k}^{*}}(\bg)\le c_{6}e^{c_{5}\Sigma \bg(\M)}.
$$
Thus,
\begin{equation} \label{E:fgAbound}
\begin{split}
Q_{k}^{*}[f_{2}(X)]
    \le c_{7}\Biggl( \frac{k}{n} P_{k}[X(\M)] 
  &+\int  e^{c_{5}g'(\M)}\bE \biggl[\sum_{
\begin{smallmatrix}
 A\in \mcA_{k,0}:\\
 g'(A)= 1
\end{smallmatrix}
} 
\rho_{0}(A) \biggr] \, dP_{0}(g')\\
  &+ \int e^{c_{5}g'(\M)}\bP\left\{ \bigvee_{A\in \mcA_{k,0}} g'(A)
             \ge 2 \right\} \, dP_{0}(g') \Biggr)
\end{split}
\end{equation}
for a constant $c_{7}$. 
Note that $\Sigma \bg = g_0 + (g_1+ \cdots +g_k)$ and
$\bg \mapsto g_1+ \cdots +g_k$ is independent of 
of $\bg \mapsto g_{0}$ under $P_k^*$, 
and so the resulting integral is simply included in $c_{7}$.

We may bound the indicator of the event $\{\bigvee_A g'(A)\ge 2\}$ from above 
by the sum $\sum_{A\in\mcA_{k,0}} \indic\{g'(A)\ge 2\}$.
Exchanging the order of the integrals in the second and third terms, 
we then obtain
\begin{align*}
& \int  e^{c_{5}g'(\M)}\bE \biggl[\sum_{
\begin{smallmatrix}
 A\in \mcA_{k,0}:\\
 g'(A)= 1
\end{smallmatrix}
} \rho_{0}(A) \biggr] \, dP_{0}(g')
    + \int e^{c_{5}g'(\M)}\bP\left\{ \bigvee_{A\in \mcA_{k,0}} 
       g'(A)\ge 2 \right\} \, dP_{0}(g')\\
&  \quad \le
\bE \biggl[ \sum_{A\in \mcA_{k,0}} \int  e^{c_{5}g'(\M)} 
           \biggl(\indic_{\{g'(A)\ge 1\}}\rho_{0}(A) +
           \indic_{\{g'(A)\ge 2\}} \biggr) \, dP_{0}(g') \biggr]\\
& \quad =
\bE \biggl[ \sum_{A\in \mcA_{k,0}} e^{\rho_{0}(A^{\complement})
      (e^{c_{5}}-1)} \int e^{c_{5}g'(A)}\biggl(\indic_{\{g'(A)\ge 1\}}\rho_{0}(A) 
      + \indic_{\{g'(A)\ge 2\}} \biggr) \, dP_{0}(g') \biggr]\\
& \quad \le 
\bE \biggl[ 
      \sum_{A\in \mcA_{k,0}} 
      e^{\rho_{0}(A^{\complement})(e^{c_{5}}-1)} 
      (\rho_{0}(A) e^{c_{5}})\exp\{(e^{c_{5}}-1)\rho_{0}(A)\}\rho_{0}(A) \\
& \qquad 
+\rec{2} (\rho_{0}(A) e^{c_{5}})^{2}\exp\{(e^{c_{5}}-1)\rho_{0}(A)\}\biggr]\\
& \quad =\bE \biggl[ 
      e^{\rho_{0}(\M)(e^{c_{5}}-1)} \sum_{A\in \mcA_{k,0}} \rho_{0}(A)^{2}(e^{c_{5}}+e^{2c_{5}}/2) \biggr].
\end{align*}
%
This allows us to conclude for some constant $c_{8}$ that
$$
Q_{k}^{*}[f_{2}(X)]  \le  c_{7}   \frac{k}{n}   P_{k}[X(\M)]
      +  c_{8} 
     \bE \biggl[ \sum_{A\in \mcA_{k,0}} \rho_{0}(A)^{2}\biggr].
$$
%

We apply Lemma~\ref{L:rsep} to the second term on the right-hand
side and find a constant $c_{9}$, which now depends on the
shattering constant $\alpha$,  such that 
$$
Q_{k}^{*}[f_{2}(X)]  \le  c_{7} \frac{k}{n} P_{k}[X(\M)]
 +  \frac{c_{9}}{k+1}. 
$$
We also have $P_{k}[X(\M)] $ equal to $(\rho_{0}(\M)+k\nu(\M)/n)$, 
which is bounded by a constant for $k/n \le T$.
It follows that 
$$
\int\biggl|\frac{d\mP^{*}Q^{*}_{k}}{dQ^{*}_{k}}(\bg) -1 \biggr|\,
              dQ^{*}_{k}(\bg)
   \le \frac{c_{10}  k}{n}  + \frac{c_{11}}{k+1},
$$
completing the proof.
\end{proof}

\chapter{Convergence of the discrete-generation system}  
\label{Ch:convergence}  

\section{Outline of the proof}
\label{S:outline_convergence}

We have now assembled all the ingredients for our proof of
our main convergence result previewed in Chapter~\ref{Ch:hypotheses}. 
This result establishes the convergence of our discrete-generation
dynamical system to our continuous-time dynamical system as the scaling
parameter $n$ goes to infinity. This parameter $n$ 
governs the rate of recombination relative
to rates of selection and mutation.
The overall strategy behind the proof has been described at the end
of Chapter~\ref{Ch:hypotheses}.   Here we outline some more detailed
considerations that enter into the proof.

Thanks to the result on complete Poissonization proved in
Chapter~\ref{Ch:complete_Poisson},  the task that remains
involves only two Wasserstein distances which must be
bounded uniformly with bounds that go
to zero as $n$ goes to infinity.  One is the distance
within the Poisson family between the probability measures $O_{k}$ 
and $ \mP Q_{k}  $.  The other is the distance between
the Poisson family and the discrete-generation system $Q_{k}$
itself, gauged by the distance between $ Q_{k}$ and
its own Poissonizaton $\mP Q_{k}$.

The distance within the Poisson family can be bounded
by the Wasserstein distance between the corresponding intensity
measures. A critical observation, encapsulated in
the inequality \eqref{E:ab}, is that the intensity 
measures deviate from each other appreciably 
only when $Q_{k}$ deviates from Poisson.
This reduces the problem to bounding the distance
from the Poisson family,
the distance between $Q_{k}$ and $\mP Q_{k}$.

This problem is addressed in two parts, by parts requiring
results established in Chapter~\ref{Ch:techlem}. 
First we adduce the formula for the Radon-Nikodym derivative
of $Q_{k}$ with respect to $ P_{k}$ 
in Lemma~\ref{L:exactRN}. 
This formula is exact, in principle,
but is probably impossible to compute exactly in most cases. 
Next we rely on two approximations derived
in subsequent lemmas. 
\begin{itemize}
\item 
Uniformly on $k\le Tn$, there is a bound on $\Theta(P_0,Q_0)$.
\item 
For $k \ll n$, there is a bound on $\was{\mP Q_{k}}{Q_{k}}$
of the form $c(1/k)\vee(k/n)$, where the constant $c$ depends on
$\Theta(P_0,Q_0)$.
\end{itemize}

We bring these consequences to bear by noting that 
there is no reason why we need to start 
at the original $P_{0}$ and $Q_{0}$.
Since $\Theta(P_L,Q_L)$ 
is bounded for any $L\le Tn$, we
have a bound $\was{\mP Q_{L+\sqrt{n}}}{Q_{L+\sqrt{n}}}\le c /\sqrt{n}$ 
for any $L$.

This approach covers all $k$ in the desired range, except in a 
small neighborhood of $0$.  If the initial probability measure $Q_0$ is Poisson
(that is, $Q_{0}=P_{0}$),
then we have a bound $\was{\mP Q_{k}}{Q_{k}} \le ck/n$ for small $k$,
and this covers the gap at $0$.
If $Q_0$ is not Poisson, then there is a failure of uniform
convergence in a neighborhood of $0$: as $n \to \infty$, 
the non-Poisson probability measure at time $0$ jumps
immediately to its Poissonization at time $0+$.

\section{The convergence theorem}
\label{SS:proofconverge}

\begin{theorem}
\label{T:limit}
Let $(\rho_t)_{t \ge 0}$ be the measure-valued
dynamical system of \eqref{E:dynam} whose
existence is guaranteed by Theorem~\ref{T:existence}.
Suppose that the selective cost function $S$ satisfies the
hypotheses of Theorem~\ref{T:existence}, namely 
\begin{itemize}
\item
$S(0)=0$,
\item
$S(g)\le S(g+h)$ for all $g,h\in\G$,
\item
for some constant $\sigma$,
$\bigl| S(g)-S(h)\bigr| \le \sigma \bigl\| g-h \bigr\|_{\Was}$,
for all $g,h\in \G$.
\end{itemize}
\index{recombination}
\index{selective cost}
In addition, suppose that the following assumptions
are in force.
\begin{itemize} 
\item
The pair $(\mathcal{R},\nu)$
consisting of the recombination measure and the 
mutation measure is shattering. 
\item
The pair  $(\mathcal{R}, \rho_0)$ 
consisting of the recombination measure and the 
initial intensity is shattering.
\item
The initial measure $Q_{0}$ is equivalent to
its Poissonization $P_{0}:=\mP Q_{0}=\Pi_{\rho_{0}}$,
and $\Theta(P_0,Q_0) < \infty$.
\end{itemize}
Then, for any $T >\epsilon> 0$,
\[  
\lnf \, \sup_{\epsilon \le t\le T} \, 
     \left\| \Pi_{\rho_{t}}- Q_{\lfloor tn \rfloor} \right\|_{\Was} \, = \, 0.
\]
If, in addition, the initial measure $Q_{0}=P_{0}$ is Poisson, then
this equation holds for $\epsilon=0$.
\end{theorem}

\begin{proof}

The proof proceeds by establishing a bound for the  
total distance $ \left\| \Pi_{\rho_{t}}- Q_{k} \right\|_{\Was} $,
a bound uniform across generations $k$ in the 
set $ \{ k \in \bN_0 : \lfloor \epsilon n \rfloor \le k \le \lfloor Tn \rfloor \} $
which goes to zero as the scaling parameter $n$ goes to infinity.
By the triangle inequality we have
\begin{equation}\label{E:threenorms}
\left\| \Pi_{\rho_{t}}- Q_{k} \right\|_{\Was} 
     \le \left\| \Pi_{\rho_{t}}- O_{k}   \right\|_{\Was} 
      +  \left\|    O_{k}     - \mP  Q_{k} \right\|_{\Was} 
      +  \left\|   \mP Q_{k}  -   Q_{k}    \right\|_{\Was} 
\end{equation}

The first of the three distances on the right-hand side
has already been shown in Theorem~\ref{T:pim} 
to be bounded by $1/n$ times
a constant depending on $T$. 
In other words,
the measure describing the dynamical system is close 
to the  measure $O_{k} = (\mP\mM_{n}\mS_{n})^{k} Q_{0}$,
the Poisson probability measure that describes a population
after $k$ rounds of selection, mutation,
and complete Poissonization.  

What remains is to bound the second and third of the distance terms.  
We need to show that the Poisson measure $O_{k}$ 
is close to the Poisson measure $\mP  Q_{k}$,
and that $\mP  Q_{k} $ is close to $Q_{k}$ itself,  
where $Q_{k} = (\mR \mM_{n}\mS_{n})^{k} Q_{0}$ is
the probability measure that describes a population
after $k$ rounds of selection, mutation, and recombination.  

By Lemma~\ref{L:poisson}
the second term on the right in \eqref{E:threenorms} is
bounded by a multiple of the Wasserstein distance 
between the intensities $ \mu O_{k} $  and $ \mu Q_{k} $
(the latter of which is the same as $ \mu \mP Q_{k} $):
\begin{equation}  \label{E:intensenorm}
\left\|    O_{k}    - \mP  Q_{k} \right\|_{\Was}
   \le 4 \, \left\|    \mu O_{k} \, - \,   \mu Q_{k}  \right\|_{\Was}
\end{equation}

We introduce symbols for the two terms for which we are
now seeking bounds. Set
\[
a_{k}:=   \left\|   \mP Q_{k}  -   Q_{k}    \right\|_{\Was}
\]
and 
\[
b_{k} ;=  \left\|    \mu Q_{k}  -  \mu O_{k} \right\|_{\Was}.
\]  

We treat $ b_{k} $ first,  eventually obtaining a recursion
equation for it that turns out to involve $a_k$.  

Given a Borel function $f:\mM\to [-1,1]$ 
with $\|f\|_{\Lip}\le 1$, 
define the function $F:\G \rightarrow \mathbb{R}$ 
by $F(g):= g[f] = \int f(m) \, dg(m)$.
The Lipschitz condition on $f$ implies 
$$
\bigl|F(g')-F(g'')\bigr|= \left| \int_\M f(m) \, dg'(m)- \int_\M f(m) \, dg''(m) \right|
         \le \was{g'}{g''}.
$$
By definition, the integral $ \mu Q [f] $ is the 
expected value $ Q[F] $ for any probability measure $Q$.

By definition of the family $(Q_k)_{k \ge 0}$,
$$
\mu Q_{k+1}[f] = \mu (\mR \mM_{n}\mS_{n} Q_{k})[f].
$$
The recombination operator $\mR$ leaves intensities
unchanged,  and the mutation operator $\mM_{n}$ adds the measure $\nu/n$
to the intensity of any probability measure on $\G$.  
Thus,
$$
\mu Q_{k+1}[f] = \frac{\nu[f]}{n} + \mS_{n}Q_{k}[F].
$$
The analogous expression holds for $ \mu O_{k+1} $ and $O_{k}$.
By definition of the Wasserstein metric,
\begin{equation}  \label{E:ab}
b_{k+1}\le \sup\left\{\bigl| \mS_{n}Q_{k}[F] - 
       \mS_{n}O_{k} [F]\bigr| :\, \|f\|_{\Lip}\le 1 \right\}.
\end{equation}


Use of this equation requires an expression for the
difference $\bigl| \mS_{n}Q_{k}[F] - \mS_{n}O_{k} [F]\bigr| $.  
For any Borel function $F:\G\to \mR $,
%
%
\begin{equation}  \label{E:bkbound1}
\begin{split}
&\bigl|\mS_{n}Q_{k}[F] - \mS_{n}O_{k} [F]\bigr| \\
& =\left| \frac{Q_{k}\left[F\exp\{-S/n\}\right]}{Q_{k}\left[\exp\{-S/n\}\right]}-
\frac{O_{k}\left[F\exp\{-S/n\}\right]}{O_{k}\left[\exp\{-S/n\}\right]}\right|\\
& =Q_{k}\left[\exp\{-S/n\}\right]^{-1}O_{k}\left[\exp\{-S/n\}\right]^{-1}\\
& \quad \times \biggl| Q_{k}\left[\exp\{-S/n\}\right]
O_{k}\left[F\exp\{-S/n\}\right] \\
& \qquad -
    Q_{k}\left[F\exp\{-S/n\}\right] O_{k}\left[\exp\{-S/n\}\right]\biggr|.    \\
\end{split}
\end{equation}

The product of the two reciprocal factors on the right-hand side 
of \eqref{E:bkbound1} is bounded, via Jensen's Inequality, 
by $\exp ( (Q_{k}[S] + O_{k}[S])/n ) $.   
The remaining factor,  with all its terms,  is easier to handle 
if we rewrite it in terms of
expected values of  $ 1 - \exp\{-S(g)/n\} $  
and $ F(g)(1-\exp\{-S(g)/n\})$, quantities that are 
small when $n$ is large.  

Set $ H(g) := F(g)(1-\exp\{-S(g)/n\})$.
Then, the final factor in \eqref{E:bkbound1} equals
\begin{equation*}
\begin{split}
( 1 &- Q_{k}[1-\exp\{-S/n\}]) O_{k}[H-F]  
          -  ( 1 - O_{k}[1-\exp\{-S/n\}]) Q_{k}[H-F]         \\ 
  &= Q_{k}[F] - O_{k}[F] - Q_{k}[H] + O_{k}[H]          \\  
      &\quad
      + \, Q_{k}[H-F] O_{k}[1-\exp\{-S/n\}] \, -\, O_{k}[H-F] Q_{k}[1-\exp\{-S/n\}] \\ 
  &= O_{k}[H-F]\Bigl(\mP Q_{k}[1-\exp\{-S/n\}] -  Q_{k}[1-\exp\{-S/n\}]\Bigr)\\
  &\quad  + Q_{k}[H-F]\Bigl(O_{k}[1-\exp\{-S/n\}] - \mP Q_{k}[1-\exp\{-S/n\}]\Bigr)\\
  &\quad +\Bigl( Q_{k}[F]-O_{k}[F] \Bigr) \mP Q_{k}[\exp\{-S/n\}]\\
    &\quad +\Bigl( \mP Q_{k}[H]- Q_{k}[H] \Bigr) \mP Q_{k}[\exp\{-S/n\}]\\
  &\quad +\Bigl( O_{k}[H]- \mP Q_{k}[H] \Bigr) \mP Q_{k}[\exp\{-S/n\}].
\end{split}
\end{equation*}
We have  $ |Q_{k}[H-F]| \le Q_{k}[|F|] $  and $ |O_{k}[H-F]| \le O_{k}[|F|] $,
and $\mP Q_{k}[\exp\{-S/n\}]\le 1$. We may then bound \eqref{E:bkbound1} by

%
\begin{equation}  \label{E:bkbound2}
\begin{split}
& \bigl| \mS_{n}Q_{k}[F] - \mS_{n}O_{k} [F]\bigr|\\
&\le \exp\{(Q_{k}[S]+O_{k}[S])/n\} \Biggl(
 \Bigl| Q_{k}[F]-O_{k}[F] \Bigr|  \\
&\qquad + 
  \Bigl| \mP Q_{k}[H] - O_{k}[H]\Bigr| \\
&\qquad + 
Q_{k}[|F|] \cdot \Bigl| \mP Q_{k}[1-\exp\{-S/n\}] 
  - O_{k}[1-\exp\{-S/n\}]\Bigr| \\
&\qquad + \Bigl|\mP Q_{k}[H]-Q_{k}[H]\Bigr| \\
&\qquad +  O_{k}[|F|] \cdot \Bigl| \mP Q_{k}[1-\exp\{-S/n\}]  - Q_{k}[1-\exp\{-S/n\}]\Bigr|
\Biggr).\\
\end{split}
\end{equation}
%
%
Using Lemma~\ref{L:LipRN} and Lemma~\ref{L:Qkprimebound}, 
we may find positive constants $c_{1},c_{2}$ such that 
\begin{equation} \label{E:dQkbound}
c_{1}e^{-c_{2}g(\M)}\le \frac{dQ_{k}}{dP_{k}}(g)\le c_{1}e^{c_{2}g(\M)},\quad
c_{1}e^{-c_{2}g(\M)}\le \frac{dO_{k}}{dP_{k}}(g)\le c_{1}e^{c_{2}g(\M)}.
\end{equation}
If we write $c_{3}:= c_{1}(\sigma \vee 1) \mu P_{\lceil Tn\rceil}(\M) e^{(e^{c_{2}}-1)\mu P_{\lceil Tn\rceil}(\M)}$, then
\begin{equation} \label{E:QkOkbounds}
\begin{split}
Q_{k}[S]\le \int c_{1}e^{c_{2}g(\M)} &\sigma g(\M)dP_{k}(g)\le c_{3},\\
\mP Q_{k}[S]&\le c_{3},\qquad Q_{k}[|F|]\le c_{3},\qquad\\
O_{k}[S]\le c_{3},\qquad \mP O_{k}[S]&\le c_{3},\qquad O_{k}[|F|]\le c_{3}.
\end{split}
\end{equation}

As $F(g) = g[f]$,
\begin{equation} \label{E:firstQkbound}
\Bigl| Q_{k}[F]-O_{k}[F]\Bigr|=
  \Bigl| \mu Q_{k}[f]- \mu O_{k}[f]\Bigr|\le
  \was{\mu Q_{k}}{\mu O_{k}}
  \le b_{k}.
\end{equation}

Observe that $|H(g)|\le \sigma g(\M)^{2}/n$ and for $g,g'\in\G$,
\begin{align*}
\bigl| H(g')&-H(g)\bigr|\le \bigl| g[f] \bigr| 
         \left| \exp\{-S(g)/n\}-\exp\{-S(g')/n\} \right| \\
& \quad
         + \bigl| g[f]-g'[f] \bigr| \left( 1-\exp\{-S(g')/n\} \right) \\
&\le (g(\M)\vee g'(\M)) \cdot \frac{\sigma}{n} \was{g}{g'} 
         + \was{g}{g'} \cdot \frac{(g(\M)\vee g'(\M))\sigma}{n}\\
&\le \frac{2\sigma (g(\M)\vee g'(\M))}{n} \was{g}{g'}.
\end{align*}
%
%
We may then apply Lemma~\ref{L:poisson2} 
with $\beta=1$ and $C=4\sigma/n$ to obtain%
\begin{equation*}  \label{E:secondQkbound}
\Bigl| \mP Q_{k}[H] - O_{k}[H]\Bigr|\le 64\sigma n^{-1} (\mu Q_{k}(\M) \vee \mu O_{k}(\M)) b_{k}.
\end{equation*}
A simpler version of this computation completes the bound of the third line in
\eqref{E:bkbound2} by 
\begin{equation} \label{E:thirdQkbound}
\begin{split}
& \Bigl| \mP Q_{k}[H] - O_{k}[H]\Bigr| \\
& \qquad +
   Q_{k}\bigl[|F|\bigr]\Bigl| \mP Q_{k}[1-\exp\{-S/n\}] - O_{k}[1-\exp\{-S/n\}]\Bigr| \\
& \quad    \le
   c_{4}\frac{b_{k}}{n}.
\end{split}
\end{equation}
for an appropriate constant $c_{4}$.

We would like to bound the remaining terms with respect to the
Wasserstein distances among the measures involved, but this
is not possible directly, because the function $H$ is neither
bounded nor Lipschitz. To get around this problem we truncate
$F$ and $S$, and then make use of the Poisson bounds that we have
on the probability of large values of these functions.
Fix some positive $M$, and define 
\begin{align*}
F^{(M)}(g)&:=\on{sgn}(F(g)) (|F(g)| \wedge M),\\
S^{(M)}(g)&:=S(g) \wedge (\sigma M),\\
H^{(M)}(g)&:=F^{(M)}(g)\cdot(1-\exp\{-S^{(M)}(g)/n\}).
\end{align*}
The function $n(3M^{2}\sigma)^{-1}H^{(M)}$ is bounded by $1/3$ 
and has Lipschitz constant bounded by $2/3$, 
so $\|n(3M^{2}\sigma)^{-1}H^{(M)}\|_{\Lip}\le 1$.

Then
\begin{align*}
\Bigl|&Q_{k}[H]-\mP Q_{k}[H]\Bigr|=\, \Bigl| Q_{k}[F(1-\exp\{-S/n\})] - \mP Q_{k}[F(1-\exp\{-S/n\})]\Bigr| \\
&\le
   \Bigl| Q_{k}[F^{(M)}(1-\exp\{-S^{(M)}/n\})] 
     - \mP Q_{k}[F^{(M)}(1-\exp\{-S^{(M)}/n\})]\Bigr|\\
   &\hspace*{1cm} +\mP Q_{k}\bigl[\bigl(|F|-M\bigr)\indic_{\{|F|>M\}}\bigr] 
      + Q_{k}\bigl[\bigl(|F|-M\bigr)\indic_{\{|F|>M\}}\bigr]\\
   &\hspace*{1.5cm} +\mP Q_{k}\bigl[|F|\frac{|S-\sigma M|}{n} 
                e^{-\sigma M}\indic_{\{|S|>\sigma M\}}\bigr] \\
   &\hspace*{2.5cm}+ Q_{k}\bigl[|F|\frac{|S-\sigma M|}{n} 
                e^{-\sigma M}\indic_{\{|S|>\sigma M\}}\bigr]\\
   &\le \frac{3M^{2}\sigma}{n} \was{Q_{k}}{\mP Q_{k}}
      +4 P_{k}\bigl[\bigl(g(\M)-M\bigr)^{2}\indic_{\{g(\M)\ge M\}}\cdot 
           c_{1}e^{c_{2}g(\M)}\bigr] \\
      &\le \frac{3M^{2}\sigma}{n} \was{Q_{k}}{\mP Q_{k}}
         + c_{1}e^{c_{2}e^{c_{2}}\mu P_{k}(\M)}
         \bigl( \mu P_{k}(\M)e^{c_{2}} \bigr)^{M}\cdot \rec{(M-2)!}
\end{align*}
by Lemma~\ref{L:poisbound2}, where we used the fact that
\[
|F(g)|e^{-\sigma M}\le g(\M)e^{-\sigma M}
        \le (e\sigma)^{-1}+ (g(\M)-M)\text{ for }g(\M)\ge M
\]
in the penultimate inequality.
A similar (but simpler) calculation together
with Stirling's formula shows that we may find
constants $c_{5}$ and $c_{6}$ such that the fourth line of \eqref{E:bkbound2} is bounded by
\begin{equation} \label{E:fourthQkbound}
\begin{split}
& \Bigl| Q_{k} [H] - \mP Q_{k}[H]\Bigr| \\
& \qquad         +O_{k}\bigl[|F|\bigr]\Bigl| Q_{k}[1-\exp\{-S/n\}] 
         - \mP Q_{k}[1-\exp\{-S/n\}]\Bigr|\\
& \quad \le \frac{3M^{2}\sigma}{n}a_{k}
              + c_{5}\left(\frac{c_{6}}{M}\right)^{M-2}. \\
\end{split}
\end{equation}

Taking $M=\lceil \log n\rceil$ and combining this 
with \eqref{E:ab}, \eqref{E:firstQkbound}, \eqref{E:thirdQkbound} and \eqref{E:fourthQkbound}, we may find 
constants $c_{7}, c_{8},c_{9}$ such that
\begin{equation} \label{E:bk1}
b_{k+1}\le e^{c_{9}/n}b_{k}+\frac{c_{7}\log^{2}n}{n} a_{k}
        +\frac{c_{8}}{n^{\log\log n}}.
\end{equation}

We now take the important step mentioned in 
Section~\ref{S:outline_convergence} of considering
intermediate states of the systems $(Q_{k})_{k \ge 0}$ and 
$(O_{k})_{k \ge 0}$ as starting states for the purposes
of applying the bounds from Chapter~\ref{Ch:techlem} over
time intervals that are short in the scaling limit.
The starting state $Q_{0}$ from which $(Q_{k})_{k \ge 0}$ and 
$(O_{k})_{k \ge 0}$ 
are derived can be quite general,  although there are 
a few conditions that must be satisfied for the bounds in
Chapter~\ref{Ch:techlem} to apply.     
If an intermediate state $Q_{L} $ can be shown to satisfy these conditions,  
then we can treat $Q_{L}$ as a new starting state,
count generations forward from $L$,  and apply the results
of Chapter~\ref{Ch:techlem}.   Recall that some of the bounds in these
lemmas go down like $ 1/(k+1)$ and some of them go up like $k/n$.
The best bounds are achieved by going 
forward on the order of $ k=\lfloor \sqrt{n}\rfloor$ 
steps from an intermediate
state at generation $L$ with $0\le L\le Tn-k$.  The result will be
a bound on $a_{L+k}$.

Define 
\begin{align*}
\wt{Q}_{k}     &:=Q_{L+k},  \\ 
\wt{P}_{k}     &:=P_{L+k},  \\
\wt{Q}_{k}^{*} &:=
       \left(\mR^{*} \mM^{*}_{n}\mS^{*}_{n}\right)^k Q_{L},\\
\wt{P}_{k}^{*}  &: =\left(\mM^{*}_{n}\right)^{k} P_{L} .
\end{align*}

In order to apply the lemmas of Chapter $\ref{Ch:techlem}$,
it is necessary to check a Lipschitz condition and a shattering
condition.
With regard to the Lipschitz condition, 
Lemma~\ref{L:LipRN} guarantees the constant upper bound
$$
\Theta(\wt{P}_{0}, \wt{Q}_{0}) \le \Theta(P_0, Q_0) + T\sigma.
$$
With regard to the shattering condition, 
it is necessary for the recombination measure $\mR$ to be
shattering with respect to the initial intensity,
which for $\wt{Q}$ is the measure $\mu Q_{L}$. 

We verify the shattering condition by an argument
analogous to the proof of Lemma~\ref{L:must_be_diffuse}. 
Set $\pi_{L} = \mu P_{L}$ and choose any 
Borel set $ A$ of $\M$.
According to the upper bound on Radon-Nikodym derivatives 
in Corollary~\ref{C:boundedRN},
there is a positive constant $c$ such that 
$$
\mu Q_{L}(A)   =  \int  g(A) \,  \frac{dQ_{L}}{dP_{L}}(g) \, dP_{L}(g) 
  \le  \int g(A) \exp\{ c g(\M) + c \pi_{L}(\M)\} \, dP_{L}(g). 
$$
The right-hand side can be evaluated via Campbell's 
Theorem, Proposition~\ref{P:Campbell},  and equals   
\index{Campbell's Theorem}
$$
\exp \left( c + \pi_{L}(\M)( e^c - 1 + c )\right) \,  \pi_{L}(A). 
$$
Thus, $\mu Q_{L}(A) $ is bounded above by $ \pi_{L}(A) $ times
a constant depending on $\pi_{L}(\M)$ but not on $A$.
Similarly the lower bound on Radon-Nikodym derivatives
in Corollary~\ref{C:boundedRN} provides lower bounds 
on $ \mu Q_{L}( A \cap R ) $ and $ \mu Q_L (A \cap R^{\complement} ) $
for any segregating set $R$, bounds which are multiples 
of $ \pi_{L}( A \cap R ) $ and $ \pi_{L}(A \cap R^{\complement} ) $. 
Because $\pi_{0}$ is assumed to be shattering with
respect to $\mathcal{R}$, so is $\pi_{L} $ by 
Lemma~\ref{L:must_be_diffuse}.  Hence, we can find a
new shattering constant $\alpha^{\prime}$ such
that 
$$
( \mu Q_{L}(A))^3  \le 
2 \alpha^{\prime}
   \int \mu Q_{L}(A\cap R) \mu Q_{L}(A \cap R^\complement)  \, d\mathcal{R}(R),
$$
and $ \mu Q_L $ is shattering. 


A bound can now be established on $\alpha_{L+k}$. 
Recall that $P_{k}^{*} $ is the Poisson measure on $\G^{*} $
obtained by starting with the Poisson measure $P_{0}$ and
solely applying the mutation operator $\M_{n}^{*}$ (without
selection or recombination) to the
distribution of genotypes broken down by vintages.

It follows from the relations 
$\Sigma \wt{Q}_k^* = \wt{Q}_{k}$, $\Sigma \mP^* = \mP \Sigma$,
and $\Sigma \wt{P}_k^* = \wt{P}_{k}$ 
that 
\[ 
\begin{split}
a_{L+k}=\was{\wt{Q}_{k}}{\mP \wt{Q}_{k}}
& = \sup\{ |\wt{Q}_k[\phi] - \mP \wt{Q}_k[\phi]| : \|\phi\|_{\Lip} \le 1\} \\
& \le \sup\{ |\wt{Q}_k[\phi] - \mP \wt{Q}_k[\phi]| : \|\phi\|_{\infty} \le 1\} \\
& = \sup\{ |\wt{Q}_k^*[\phi \circ \Sigma] - 
    \mP^* \wt{Q}_k^*[\phi \circ \Sigma]| : \|\phi\|_{\infty} \le 1\} \\
& \le \sup\{ |\wt{Q}_k^*[\phi^*] - \mP^* \wt{Q}_k^*[\phi^*]| : 
                \|\phi^*\|_{\infty} \le 1\} \\
& =
\int \left | 1 - \frac{d \mP^* \wt{Q}_k^*}{d\wt{Q}_k^*}(\bg)
\right|
\, d\wt{Q}_k^*(\bg) \\
&\le c_{9}\left(\frac{1}{k+1} \vee \frac{k}{n}\right)
\end{split}
\]
for some constant $c_{9}$, by Lemma~\ref{L:Tcomparednew}. 

Setting $k=\lfloor \sqrt{n}\rfloor$ and combining this 
with the bound \eqref{E:bk1}, we obtain for all $nT\ge j\ge \sqrt{n}$,
\begin{equation} \label{E:bk2}
\begin{split}
b_{j+1}&\le e^{c_{6}/n} b_{j} +c_{7} \frac{\log^{2} n}{n}a_{j}
       +c_{8}n^{-\log\log n}\\
a_{j}&\le c_{9} n^{-1/2},
\end{split}
\end{equation}
Thus, there are constants $c_{6}$ and $c_{10}$ such that
\begin{equation} \label{E:bk3}
b_{j+1}\le e^{c_{6}/n}b_{j}+c_{10}\frac{\log^{2}n}{n^{3/2}}.
\end{equation}
For $0\le j\le \sqrt{n}$,
\begin{equation} \label{E:bk4}
\begin{split}
b_{0}&=0,\\
b_{j+1}&\le e^{c_{6}/n} b_{j} +c_{7} \frac{\log^{2} n}{n}a_{j}+c_{8}n^{-\log\log n},\\
a_{j}&\le c_{9} (j+1)^{-1},
\end{split}
\end{equation}
implying that $b_{j}\le c_{11} \log^{3}n/n$ for some constant $c_{11}$.
Taking this bound on $b_{\lfloor \sqrt{n}\rfloor}$ as a starting point for
iterating \eqref{E:bk3} we conclude that there is a constant $c_{12}$ such that
for $nT\ge j\ge \sqrt{n}$,
$$
b_{j}\le c_{12} \frac{\log^{2}n}{n^{1/2}}.
$$
Returning to \eqref{E:intensenorm}, this implies that
$$
\was{Q_{k}}{O_{k}}\le c \frac{\log^{2}n}{n^{1/2}}.
$$
Combined with Theorem~\ref{T:pim}, this immediately implies
\[  
\lnf \, \sup_{\epsilon \le t\le T} \, 
     \left\| \Pi_{\rho_{t}}- Q_{\lfloor tn \rfloor} \right\|_{\Was} \, = \, 0
\]
for $\epsilon >0$.

When $Q_{0}$ is already Poisson, we begin with $a_{0}=0$,
allowing us to extend \eqref{E:bk2} to $k$ in the range $0\le k<\sqrt{n}$,
completing the proof.
\end{proof}


\appendix
\chapter{Results cited in the text}
%

\section{Gronwall's Inequality}
\label{SS:Gronwall}

We make frequent use of the following version of
the inequality discovered by Thomas Hakon Gronwall in 1919
and extended by Richard Bellman in 1943, so we include
a statement for easy reference.  A proof
may be found in Appendix~5.1 of \cite{EK86}. 
\index{Gronwall's Inequality}
\index{Gronwall, Thomas H.}
\begin{proposition}
Let $f$ be a nonnegative Borel measurable 
function on $\bR_{+}$ that is bounded on bounded intervals.
Suppose there exists a constant $L > 0 $ such that
for all $ t \ge 0 $ 
$$
f(t) \le L \, \int_0^t \, f(s) \, ds.  
$$
Then,  $  f(t) = 0 $ for all $t$. 
\end{proposition}

\section{Two expectation approximations}
\label{SS:exp_approx}


The following two lemmas give 
give error bounds for approximations used several times in this work.

\begin{lemma}  
\label{L:expbound}
Let $Y$ be a nonnegative random variable with finite second moment.  Then,
\[
-\bE [Y]  \le \log\bE \left[\exp\{-Y\}\right]   
          \le -\bE [Y] + (1/2) \on{Var}(Y)\, \exp\{\bE[Y]\}
\]
and so
\[
0 \le \log \bE \left[\exp\{-Y\}\right] + \bE [Y] \le (1/2) \on{Var}(Y)\exp\{\bE[Y]\}.
\]
In particular, if $Y$ is bounded by a constant $\tau$, then
\[
0 \le \log \bE \left[\exp\{-Y\}\right] + \bE [Y] \le (1/2) \tau^2 e^{\tau}.
\]
\end{lemma}

\begin{proof}
Jensen's Inequality applied to the convex function $y \mapsto \exp\{-y\}$
implies $ -\bE [Y]  \le \log\bE \left[\exp\{-Y\}\right]$.

For nonnegative $y$, the function $y \mapsto (1/2)y^2 -\exp\{-y\} $ 
is also convex.  Jensen's Inequality implies
$$
(1/2) ( \bE [Y])^2 - \exp\{-\bE [Y] \}  \le \bE \left[ (1/2)Y^2 - \exp\{-Y\} \right] 
$$
Consequently,
\[
\begin{split}
\bE \left[ \exp\{-Y\}\right] & \le \exp\{-\bE[Y]\}+ (1/2) \, \on{Var}(Y) \\
& \quad =\exp\{-\bE[Y]\}\left(1+ (1/2) \, \on{Var}[Y]\exp\{\bE[Y]\}\right). \\
\end{split}
\]
Taking logarithms of both sides and using the bound 
$\log(1+x)\le x$, $x \ge -1$, 
complete the proof.
\end{proof}

\begin{lemma}
\label{L:expbound2}
Let $Y$ be a random variable with finite first moment.  Then,
\[
\bE[Y] \le \log \bE[\exp\{Y\}] \le \bE[|Y| \exp\{|Y|\}].
\]
\end{lemma}

\begin{proof}
The first inequality is immediate from Jensen's inequality and
the convexity of $y \mapsto \exp\{y\}$.  For the second inequality,
use the inequalities $\log(1+y) \le y$, $y \ge -1$, and
$\exp\{x\} - 1 \le x \exp\{x\}$, $x \ge 0$, to get
\[
\begin{split}
\log \bE[\exp\{Y\}] & \le \bE[\exp\{Y\}] - 1  \\
& \le \bE[\exp\{|Y|\}] - 1 \\
& \le \bE[|Y| \exp\{|Y|\}]. \\
\end{split}
\]
\end{proof}

\section{Identities for Poisson random measures}
\label{SS:Campbell}


The following elementary lemma is well-known and follows readily from the
fact that the conditional distribution of the Poisson random measure $X^\pi$
given the event $\{X^\pi(\M)=n\}$ is the distribution of the random measure
$\sum_{i=1}^n \delta_{Z_i}$, where $Z_1, \ldots, Z_n$ 
are i.i.d. $\M$-valued random variables with common distribution $\pi/\pi(\M)$.

\begin{lemma}
\label{L:PiPi}
Suppose that $\pi', \pi'' \in \H^+$ and $\pi'$
is absolutely continuous with respect to $\pi''$.
Then the Poisson probability measure
$\Pi_{\pi'}$ is absolutely continuous with respect to $\Pi_{\pi''}$ 
with Radon-Nikodym
derivative

\begin{equation*} 
\begin{split}
\frac{d \Pi_{\pi'}}{d \Pi_{\pi''}}(g)
&= \exp\left\{g\left[\log\left(\frac{d\pi'}{d\pi''}\right)\right]
 - \pi'(\M) + \pi''(\M) \right\}\\
 &=\exp\left\{g\left[\log\left(\frac{d\pi'}{d\pi''}\right)\right]
 - \int_{\M} \left(\frac{d\pi'}{d\pi''}(x)-1\right) d\pi''(x)\right\}.
 \end{split}
\end{equation*}
\end{lemma}


\begin{lemma}  \label{L:intindeppoisson2}
Let $X',X''$ and $Y', Y''$ be 
be two i.i.d. pairs of Poisson random measures on $\M$
with common intensity measure $\pi \in \H^+$.
Fix a partition $\{A, A^\complement\}$ of $\M$.
Then, the vector of random measures
$(X_A' + X_{A^\complement}'', X', X'')$
has the same distribution as 
$(Y', Y_A' + Y_{A^\complement}'', Y_A'' + Y_{A^\complement}')$.
Consequently, for bounded Borel functions
$F,H' ,H'':\G \to\bR$ 
\[
\begin{split}
& \int_{\G} \int_{\G} F(g_{A}'+g''_{A^{\complement}}) H'(g') H''(g'') \, d\Pi_\pi(g') d\Pi_\pi(g'')\\
& \quad = \int_{\G} \int_{\G} F(h)  
	H'(h_{A}'+h_{A^\complement}'')H''(h_{A}'' + h_{A^\complement}') \, d\Pi_\pi(h') d\Pi_\pi(h'').
   \end{split}
\]
\end{lemma}

\begin{proof}
Note that the vector of random measures
$(X_A', X_{A^\complement}', X_A'', X_{A^\complement}'')$
has the same distribution as 
$(Y_A', Y_{A^\complement}'', Y_A'', Y_{A^\complement}')$.
Then observe that
\begin{itemize}
\item
$Y' = Y_A' + Y_{A^\complement}'$,
\item
$X' = X_A' + X_{A^\complement}'$,
and
$X'' = X_A'' + X_{A^\complement}''$
\item
$(X_A' + X_{A^\complement}'')_A = X_A'$ 
and 
$(X_A' + X_{A^\complement}'')_{A^\complement} = X_{A^\complement}''$.
\end{itemize}
\end{proof}

An important property of Poisson random measures, 
which we use in our discussion of the starred probability measures, 
is that we can construct
a Poisson random measure on a product space by starting 
with a Poisson random measure
that lays down point on one component, 
and then conditional on that scatter of points 
choosing the corresponding second coordinates 
independently according to some Markov kernel. 
One clear statement of this property is Theorem 6.3.2 of \cite{eC08}.  
A version tailored to our particular application is the following.

\begin{proposition}
\label{P:lifting}
Suppose that  $\pi_{0},\ldots,\pi_{k} \in \H^+$.
Then, the push-forward of the probability measure
$\Pi_{\pi_0} \otimes \Pi_{\pi_1} \cdots \otimes \Pi_{\pi_k}$ on $\G^{k+1}$
by the map
\[
(g_{0},g_{1},\dots,g_{k}) \mapsto g_{0}\otimes \delta_{0}+ g_{1}\otimes \delta_{1}+\cdots+g_{k}\otimes \delta_{k}
\]
is the distribution of a Poisson random measure on $\G \times \{0,1, \ldots, k\}$ with intensity measure
$\sum_{j=0}^{k} \pi_{j} \otimes \delta_{j}$.
\end{proposition}

For the sake of completeness, we record the following
result that follows from combining Lemma~\ref{L:intindeppoisson2} 
and Proposition~\ref{P:lifting}.
Recall that for $\bg=(g_{0},\dots,g_{k})\in \G^{k+1}$ and $A\in\BM$ we define 
$\bg_{A}:=\bigl((g_{0})_{A},\dots,(g_{k})_{A}\bigr)$. 
We can identify $\bg$ with the measure on $\M\times \{0,\dots,k\}$
that assigns mass $g_i(B)$ go the set $B \times \{i\}$, and 
$\bg_{A}$ can be identified in the same way with the
restriction of that measure to the set $A\times\{0,\dots,k\}$.

\begin{corollary}  \label{C:intindeppoisson}
Let $P$ be the distribution of a Poisson random measure on $\M\times \{0,1,\dots,k\}$.
Fix a Borel set $A \subseteq \M$. 
Identifying $P$ with a probability measure on $\G^{k+1}$, we have for bounded Borel functions 
$F,H',H'':\G^{k+1} \to\bR$ that
\[
\begin{split}
& \int_{\G^{k+1}} \int_{\G^{k+1}} F(\bg_{A}'+\bg_{A^{\complement}}'') H'(\bg')H''(\bg'') 
\, dP(\bg') dP(\bg'')\\
& \quad  = \int_{\G^{k+1}} \int_{\G^{k+1}} F(\bh)  
	H'(\bh_{A}'+\bh_{A^\complement}'')H''(\bh_{A}'' + \bh_{A^\complement}') 
	\, dP(\bh') dP(\bh'').
   \end{split}
\]
\end{corollary}

The following version of Campbell's Theorem is 
one of a cluster of results relating to Palm
probabilities inspired by 
1909 work of N. R. Campbell.   
It follows immediately from two results proved  
in \cite{DVJ07},  
Proposition 13.1.IV (an equality involving general Palm
kernels)  and Example 13.1c (an expression
for the Palm kernel of a Poisson Process).
\index{Campbell's Theorem}
\index{Campbell, N.R. }

\begin{proposition}
\label{P:Campbell}
Suppose that $\pi \in \H^+$.
For bounded Borel functions $f:\M \to \bR$
and $F: \G \to \bR$  
\[
\begin{split}
\int_\G F(g) g[f] \, d\Pi_\pi(g)
& =
\int_\M \int_\G F(g + \delta_m) \, d\Pi_\pi(g) f(m) \, \pi(dm) \\
& =
\int_\M \Pi_\pi[F(\cdot + \delta_m)] f(m) \, \pi(dm). \\
\end{split}
\]
\end{proposition}

\begin{lemma}
For a bounded Borel function $h:\M\to\bR$ and $c \in \bR$,
\[
\int_\G g[h] \exp\{c g(\M)\} \, d\Pi_\pi(g)
=
\exp\left\{(e^{c}-1) \pi(\M)+c\right\} 
\int_\M h(x) \, d \pi(x).
\]
\end{lemma}

\begin{proof}
If $c \le 0$, then the function $F(g)=e^{c g(\M)}$ is bounded,
so this is a fairly direct consequence of Campbell's Theorem,
Proposition~\ref{P:Campbell}. 
The formula cannot be immediately applied to positive $c$ 
because the function $e^{cg(\M)}$ would not be bounded. 
But if we take some positive integer $B$ and 
define $F_{B}(g):=e^{c(g(\M)\wedge B)},$ then
\begin{align*}
 \int_\G F_{B}(g) g[h] \, d\Pi_\pi(g)
&=
\int_\M \Pi_\pi[F_{B}(\cdot + \delta_x)] h(x) \, d\pi(x)\\
&=\pi[h]\cdot \sum_{k=0}^{\infty} e^{-\pi(\M)}\frac{\pi(\M)^{k}}{k!} 
e^{c((k+1)\wedge B)}\\
&=\pi[h]\cdot e^{c}e^{\pi(\M)(e^{c}-1)}+
\sum_{k=B}^{\infty} e^{-\pi(\M)}\frac{\pi(\M)^{k}}{k!} 
\left(e^{c B}-e^{ck}\right)\\
&\to \pi[h]\cdot e^{c}e^{\pi(\M)(e^{c}-1)} \text{ as } B\to\infty.
\end{align*}
The result then follows by Dominated Convergence.
\end{proof}

\section{Bounds for Poisson random measures}

\begin{lemma}  \label{L:poisbound2}
For any nonnegative constant $c$, and positive integers $s>r$,
\begin{align*}
\int_\G \bigl(  g(\M)-s \bigr)^{r} &\exp\{c g(\M)\} \indic_{\{g(\M)\ge s\}} \, d\Pi_\pi(g)\\
&\le
\frac{1}{(s-r)!}\left( \pi(\M) e^{c} \right)^{s} \exp\left\{(e^{c}-1) \pi(\M)\right\}.
\end{align*}
\end{lemma}

\begin{proof}
We write
\[
\begin{split}
& \int_\G \bigl(  g(\M)-s \bigr)^{r} \exp\{c g(\M)\} \indic_{\{g(\M)\ge s\}} \, d\Pi_\pi(g)\\
& \quad = 
e^{-\pi(\M)}\sum_{j=0}^{\infty} \frac{(\pi(\M)e^{c})^{j+s}}{(j+s)!} j^{r}\\
& \quad = 
e^{-\pi(\M)} (\pi(\M)e^{c})^{s}
\cdot
\sum_{j=0}^{\infty} \frac{(\pi(\M)e^{c})^{j}}{j!} \frac{j!}{(j+s)!} j^r. \\
\end{split}
\]
Now,
\[
\begin{split}
\frac{j!}{(j+s)!} j^r
& =
\frac{j^r}{(j+1)(j+2) \cdots (j+s)} \\
& =
\frac{1}{(j+1)(j+2) \cdots (j+s-r)} \cdot \frac{j^r}{(j+s-r+1) \cdots (j+s)}
\le \frac{1}{(s-r)!}, \\
\end{split}
\]
and the result follows.
\end{proof}

The next result says that if the intensities
of two Poisson random measures are close in
the Wasserstein sense, then the same is true
of their distributions.
In the statement of the result, the Wasserstein
distance on the left-hand side of the inequality is
between probability measures on the space $\G$, while
the Wasserstein distance on the right-hand side is
between finite measures on $\M$.
\index{Wasserstein metric}

\begin{lemma}
\label{L:poisson2}
Consider a function $H:\G\to\bR$ for which
there are constants $C \in \bR_+$ 
and $\beta \in \bN_0$ 
such that for all $k \in \bN$
\begin{equation*} 
\sup_{\begin{smallmatrix}
g\ne g'\in\G\\
g(\M)\vee g'(\M)\le k
\end{smallmatrix}}\frac{\left| H(g)-H(g')\right|}{\was{g}{g'}} +
\sup_{\begin{smallmatrix}
g''\in\G:\,
g''(\M)\le k\\
z\in\M
\end{smallmatrix}} |H(g''+\delta_{z})-H(g'')|
   \le C k^{\beta}.
\end{equation*}
Then, for $\pi',\pi'' \in \H^{+}$,
\begin{equation*} 
\bigl| \Pi_{\pi'}[H]-\Pi_{\pi''}[H] \bigr|\le 4C(2\beta (\pi'(\M) \vee \pi''(\M) \vee 1))^{\beta}\ \|\pi' - \pi''\|_{\Was}.
\end{equation*}
\end{lemma}

\begin{proof}
If $\pi' = \pi'' = 0$ there is nothing 
to prove.  Therefore,
suppose without loss of generality that $\pi' \ne 0$ and $\pi'(\M) 
\ge \pi''(\M)$.  Set
$\pi^* = (\pi''(\M)/\pi'(\M))\pi' \in \H^+$.  We have
$$
\bigl| \Pi_{\pi'}[H]-\Pi_{\pi''}[H] \bigr| \le \bigl| \Pi_{\pi'}[H]-\Pi_{\pi^{*}}[H] \bigr| +\bigl| \Pi_{\pi^{*}}[H]-\Pi_{\pi''}[H] \bigr|. 
$$

Note that if $X^{\pi'}$ and $X^{\pi^{*}}$ are any two
Poisson random measures on the same probability
space with distributions
$\Pi_{\pi'}$ and $\Pi_{\pi^{*}}$, respectively, then
\[
\begin{split}
\bigl| \Pi_{\pi'}[H]-\Pi_{\pi^{*}}[H] \bigr|
& =
\bigl| \bE \left[H(X^{\pi'}) -  H(X^{\pi^{*}})\right] \bigr| \\
& \le 2C \bE\bigl[ (X^{\pi'}(\M) \vee X^{\pi^{*}}(\M))^{\beta} \indic\{X^{\pi'} \ne X^{\pi^{*}}\} \bigr] .
\end{split}
\]
In particular, if we first build  $X^{\pi'}$
and then construct $X^{\pi^{*}}$ by the usual
``thinning'' procedure of independently keeping
each point of $X^{\pi'}$ with probability 
$\pi''(\M)/\pi'(\M)$ and discarding it with
the complementary probability, we have
%
\[
\begin{split}
\bigl| \Pi_{\pi'}[H]-\Pi_{\pi^{*}}[H] \bigr| 
 &\le    2C\sum_{k=0}^{\infty}
           e^{-\pi'(\M)}\frac{\pi'(\M)^{k}}{k!} k^{\beta}
         \left[
               1 -\left(\frac{\pi''(\M)}{\pi'(\M)}\right)^{k}
         \right]\\
 &\le 2 C\sum_{k=1}^{\infty}
         e^{-\pi'(\M)}\frac{\pi'(\M)^{k-1}}{(k-1)!}
         k^{\beta}\left[\pi'(\M) - \pi''(\M)\right] \\
 &= 2 C \bE\left[(X^{\pi'}(\M)+1)^\beta\right] 
 		\left[\pi'(\M) - \pi''(\M)\right], \\
\end{split}
\]
\index{thinning}
where we used the inequality $a^k - b^k \le k a^{k-1} (a - b)$
for $a \ge b \ge 0$.
Note that if  $Y$ is an integer-valued Poisson 
random variable with mean $\lambda$, then writing
$S(\beta,k)$ for the number of partitions of a set 
of size $\beta$  into $k$ blocks (that is,
the Stirling number of the second kind) 
and $B_\beta = \sum_{k=0}^\beta S(\beta,k)$
for the number of partitions of a set with $\beta$ members
(that is, the Bell number)
\index{Stirling number}
\index{Bell number}
\glossary{$S(\beta,k)$}
\glossary{$B_{\beta,k}$}
\[ 
\bE[Y^{\beta}] 
= \sum_{k=0}^\beta S(\beta,k) \lambda^k
\le B_\beta (\lambda^\beta \vee 1) 
\le \beta^\beta (\lambda^\beta \vee 1),
\]
because $B_\beta \le \beta^\beta$ (any partition of
a set with $\beta$ elements that has $k$ blocks defines
$\beta(\beta - 1) \cdots (\beta - k + 1)$ different functions
from the set into itself by assigning a distinct element
of the set to be the image of all the elements in each block,
and there are $\beta^\beta$ functions from the set into itself).
Therefore,
\[
\begin{split}
\bE\left[(Y+1)^{\beta}\right] 
& \le 
2^{\beta}\left(\bE[Y^{\beta}] \vee 1^\beta\right) \\
& \le
2^{\beta}(\beta^\beta(\lambda^\beta \vee 1) \vee 1)\\
& = (2 \beta)^{\beta} (\lambda \vee 1)^\beta.\\
\end{split}
\]
Thus, 
\[
\begin{split}
\bigl| \Pi_{\pi'}[H]-\Pi_{\pi^{*}}[H] \bigr| 
& \le 2C  (2 \beta)^{\beta} (\pi'(\M) \vee 1)^\beta \bigl|\pi'(\M)-\pi''(\M)\bigr|\\
& \le 2C ((2 \beta)(\pi'(\M) \vee 1))^\beta 
   \| \pi'-\pi'' \bigr\|_{\on{Was}}. \\
\end{split}
\]

On the other hand, setting $r = \pi^*(\M) = \pi''(\M)$, 
\[
\begin{split}
   & \bigl| \Pi_{\pi^{*}}[H]-\Pi_{\pi''}[H] \bigr| \\
   & \quad \le \sum_{k=0}^{\infty} \frac {e^{-r}}{k!} 
     \left| \int \dots \int
     H\left(\sum_{\ell=1}^k \delta_{y_\ell}\right) \, \pi^*(dy_1)\cdots 
     \pi^*(dy_k)\right.\\
   &\qquad \left. 
     -\int \dots \int H\left( \sum_{\ell=1}^k
     \delta_{y_\ell}\right)\, \pi''(dy_1)\cdots
     \pi''(dy_k)\right| \\
   & \quad \le \sum_{k=0}^{\infty} \frac {e^{-r}}{k!} 
     \sum_{m=0}^{k-1}\left| \int \dots \int
     H\left( \sum_{\ell=1}^k \delta_{y_\ell}\right)
     \pi^*(dy_1)\cdots
     \pi^*(dy_{m})\pi''(dy_{m+1})\cdots \pi''(dy_k)\right. \\
   & \qquad \left. -\int \dots \int H\left(\sum_{\ell=1}^k 
     \delta_{y_\ell}\right) \, \pi^*(dy_1) \cdots
     \pi^*(dy_{m+1})\pi''(dy_{m+2})\cdots\pi''(dy_k)\right|. \\
\end{split}
\]
Observe that the function 
$z\mapsto H(g+\delta_{z})/C(1+g(\M))^{\beta}$ is Lipschitz
with Lipschitz constant $1$. Hence, by arguments similar
to those above,
\[
\begin{split}
\bigl| \Pi_{\pi^{*}}[H] &-\Pi_{\pi''}[H] \bigr| \\   
 &\quad \le 
\sum_{k=0}^{\infty} \frac {e^{-r}}{k!} r^{k-1} k \, \sup_{g : g(\M)=k} 
     \left| \int H(g +
     \delta_z)\pi^*(dz) - \int H(g + \delta_z)\pi''(dz)\right| \\
     &\quad \le \sum_{k=0}^{\infty} \frac {e^{-r}}{k!} 
           r^{k-1} k C(k+1)^{\beta}\was{\pi^{*}}{\pi''} \\
   & \quad \le C((2 \beta)(r \vee 1))^\beta \|\pi^* - \pi''\|_{\Was} \\
   & \quad \le C((2 \beta)(r \vee 1))^\beta\Bigl(\|\pi' - \pi''\|_{\Was} 
   	+ \bigl|\pi'(\M)-\pi''(\M)\bigr|\Bigr)\\
   & \quad \le 2 C ((2 \beta)(\pi'(\M) \vee 1))^\beta \|\pi' - \pi''\|_{\Was}. \\
\end{split}
\]
Putting these bounds together and recalling we had
assumed provisionally that $\pi'(\M) \ge \pi''(\M)$ yields
$$
\bigl| \Pi_{\pi'}[H]-\Pi_{\pi''}[H] \bigr|\le 4C(2\beta (\pi'(\M) \vee \pi''(\M) \vee 1\})^{\beta}\ \|\pi' - \pi''\|_{\Was},
$$
as required.
\end{proof}

The next result,
is a special case of Lemma~\ref{L:poisson2}.
It probably already exists
in some form in the literature, but we have
been unable to find a reference.  

\begin{lemma}
\label{L:poisson}
For two finite measures $\pi',\pi'' \in \H^{+}$ 
\[ 
\bigl\|\Pi_{\pi'}-\Pi_{\pi''}\bigr\|_{\Was} 
      \le 4\bigl\|\pi' - \pi''\bigr\|_{\Was}.
\]
\end{lemma}

\section{Bounds for Radon-Nikodym derivatives}

\begin{lemma}  \label{L:RNcompare}
Let $Q$ and $Q'$ be equivalent probability measures 
on some measurable space $\X$.
Suppose there are functions and $f_{1},f_{2}:\X\to\bR$ such that
$$
C\exp\{f_{1}(x)-f_{2}(x)\}\le \frac{dQ'}{dQ}(x) 
\le C\exp\{f_{1}(x)+f_{2}(x)\}.
$$
\begin{itemize}
\item[(a)]
Then,
\begin{equation*} 
 \frac{\exp\{f_{1}(x)-f_{2}(x)\}}{\int \exp\{f_{1}(y)+f_{2}(y)\} \, dQ(y)}\le \frac{dQ'}{dQ}(x) \le \frac{\exp\{f_{1}(x)+f_{2}(x)\}}{\int \exp\{f_{1}(y)-f_{2}(y)\} \, dQ(y)}.
\end{equation*}
\item[(b)]
Moreover, if $f_{1},f_{2}:\X\to\bR_+$, then
\begin{equation*} 
\int_{\X} \left|\frac{dQ'}{dQ}(x) -1\right| \, dQ(x)
\le 
2\exp\{Q[f_{2}]\} Q\left[ \bigl(f_{1}+f_{2} \bigr) \exp\{f_{1}+f_{2}\}\right].
\end{equation*}
\end{itemize}
\end{lemma}

\begin{proof}
We have
$$
C\int_{\X}\exp\{f_{1}(x)-f_{2}(x)\} \, dQ(x) \le 1 \le C\int_{\X }\exp\{f_{1}(x)+f_{2}(x)\} \, dQ(x).
$$
So,
$$
Q\left[\exp\{f_{1}+f_{2}\}\right]^{-1}
\le 
C\le Q\left[\exp\{f_{1}-f_{2}\}\right]^{-1}
$$
and (a) follows.

Now,
\[
\begin{split}
\int_{\X} \left|\frac{dQ'}{dQ}(x) -1\right| dQ(x) 
& \le
\int_{\X} \left|\frac{dQ'}{dQ}(x) -C\right| dQ(x)+|C-1| \\
& = \int_{\X} \left|\frac{dQ'}{dQ}(x) -C\right| dQ(x)
	+ \left|C - \int_\X \frac{dQ'}{dQ}(x) \, dQ(x)\right| \\
& \le 2 \int_{\X} \left|\frac{dQ'}{dQ}(x) -C\right| dQ(x).
\end{split}
\]
By Jensen's Inequality,
\[
C
\le 
Q\left[\exp\{f_{1}-f_{2}\}\right]^{-1}
\le
Q\left[\exp\{-f_{2}\}\right]^{-1}
\le
\exp\{Q\left[f_{2}\right]\}.
\]
Note also that if $a,b \ge 1$, then $|\frac{a}{b} - 1| \le ab - 1$;
this is obvious when $a \ge b$, and when $a < b$ the inequality
is equivalent to $ab^2 - 2b + a = a(b-1)^2 + 2(a-1)b > 0$.
Thus,
\[
\begin{split}
\left|\frac{dQ'}{dQ}(x) -C \right|
& \le 
C \left(\exp\{f_{1}(x) +f_{2}(x)\} - 1 \right) \\
& \le  
\exp\{Q\left[f_{2}\right]\}
	\bigl(f_{1}(x)+f_{2}(x) \bigr) \exp\{f_{1}(x)+f_{2}(x)\}, \\
\end{split}
\]
where we have used the inequality $e^y - 1 \le y e^y$ for $y \ge 0$,
and the result follows.
\end{proof}

\begin{corollary} \label{C:RNcompare}
Set $Q = \Pi_\pi$ for some $\pi \in \H^+$ and
suppose that the probability measure $Q'$ is equivalent to $Q$
with $\hat{H} := \Theta(Q,Q') < \infty$. 
\begin{itemize}
\item[(a)]
Then, writing $H := dQ'/dQ$,
\begin{equation*} 
-\hat{H} g(\M) - \pi(\M) (e^{\hat{H}}-1)
\le 
H(g)
\le 
\hat{H} g(\M) - \pi(\M) (e^{-\hat{H}}-1).
\end{equation*}
\item[(b)]
Hence,
\begin{equation*}
-\hat{H}(g(\M)+\pi(\M)e^{\hat{H}})
\le H(g)
\le
\hat{H}(g(\M)+\pi(\M)).
\end{equation*}
\item[(c)]
For any $\beta\in\bR$,
\begin{equation*} 
\left|\log\int \exp\{H(g)+\beta g(\M)\} \, dQ(g)\right|
\le \pi(\M)\left(\hat{H}+\exp\{\hat{H}+\beta\}-1\right).
\end{equation*}
\item[(d)]
For any Borel set $A\subseteq \M$,
\begin{equation*} 
\begin{split}
\int \exp\{H(g)+\beta g(A)\}&g(A) \, dQ(g)\\
&\le \pi(A) \exp\left\{2\pi(A^{\complement}) \hat{H} 
 +\hat{H}+\beta+\pi(A)\left( \exp\{\hat{H}+\beta\}-1\right)\right\}.
\end{split}
\end{equation*}
\end{itemize}
\end{corollary}

\begin{proof}

Apply Lemma~\ref{L:RNcompare} with $f_{1}(g)=0$ and $f_{2}(g)=\hat{H} g(\M)$
to get
\[
\frac{\exp\{-\hat H g(\M)\}}{\int \exp\{\hat H g(\M)\}  \, Q(dg)}
\le
H(g)
\le
\frac{\exp\{\hat H g(\M)\}}{\int \exp\{-\hat H g(\M)\}  \, Q(dg)}.
\]
Since
\[
\int \exp\{\hat H g(\M)\}  \, Q(dg) = \exp\{\pi(\M)(e^{\hat H} - 1)\}
\]
and
\[
\int \exp\{-\hat H g(\M)\}  \, Q(dg) = \exp\{\pi(\M)(e^{-\hat H} - 1)\},
\]
(a) follows immediately.
Part (b) comes from combining part (a)
with the inequality $|e^y - 1| \le |y| (e^{y} \vee 1)$ for all real $y$.

Hence,
\begin{align*}
\int_\G \exp\{H(g)+\beta g(\M)\} \, d Q(g)
& \le 
\int_\G e^{\hat H \pi(\M)} e^{(\hat{H}+\beta) g(\M)} \, d Q(g)\\
&=\exp\left\{ \pi(\M)\left(\hat{H}+e^{\hat{H}+\beta}-1\right)  \right\}.
\end{align*}
The lower bound in part (c) is obtained similarly.

Finally, since $g \mapsto g(A^{\complement})$ and $g \mapsto g(A)$ are independent under $Q$,
\[
\begin{split}
& \int_\G \exp\{H(g)+\beta g(A)\}g(A) \, d Q(g) \\
& \quad =
\int_\G \exp\{H(g_{A^{\complement}})\} \,
dQ(g) \cdot \int_\G \exp\{H(g)-H(g_{A^{\complement}})+\beta g(A)\}g(A) \, d Q(g)\\
& \quad \le \exp\left\{\pi(A^{\complement})\left(\hat{H}+e^{\hat{H}}-1\right)\right\} \cdot \int_\G \exp\{(\hat{H}+\beta)g(A)\} g(A) \, d Q(g).\\
\end{split}
\]
By Campbell's theorem, Proposition~\ref{P:Campbell},
\[
\begin{split}
& \int_\G \exp\{(\hat{H}+\beta)g(A)\} g(A) \, d Q(g) \\
& \quad =
\int_\M \int_\G \exp\{(\hat{H}+\beta)(g + \delta_m)(A)\} \indic_A(m) \, d Q(g) \, d \pi(dm) \\
&  \quad =
\int_\M \exp\{(\hat{H}+\beta) \delta_m (A)\} \indic_A(m) \, d \pi(dm)
\cdot
\int_\G \exp\{(\hat{H}+\beta) g(A)\} \, d Q(g) \\
& \quad \le
\pi(A) e^{(\hat{H}+\beta)}
\exp\left\{\pi(A)\left( e^{\hat{H}+\beta}-1\right)\right\}, \\
\end{split}
\]
and part (d) follows.
\end{proof}

\begin{lemma} \label{L:poisindicbound}
Suppose that $Q = \Pi_\pi$ for some $\pi \in \H^+$
and the probability measure $Q'$ is equivalent to $Q$
with $\hat{H} := \Theta(Q,Q') < \infty$. 
\begin{itemize}
\item[(a)]
For any real $\beta$,
\begin{equation*} 
\begin{split}
-\pi(A)&\le\log \int \exp\{\indic_{\{g(A)\ge 1\}\}(H(g)+\beta g(A))} \, dQ(g) \\
&\le\pi(A)\left(\hat{H}+e^{\beta+\hat{H}}-1\right).
\end{split}
\end{equation*}
\item[(b)]
There are positive constant $c$ and $\epsilon$, depending only on $\hat{H}$ and $\pi(\M)$, such that for any $\beta,\beta'$ with $-\epsilon\le \beta'\le 0\le \beta\le \epsilon$,
\begin{equation*} 
\log \frac{\int \exp\{H(g)+\beta g(A)\} \, dQ(g)}{\int \exp\{H(g)-\beta' g(A)\} \, dQ(g)}
  \le c(\beta+\beta')\pi(A).
\end{equation*}
\end{itemize}
\end{lemma}

\begin{proof}
For $i\ge 0$ let $\gamma_{i}:=Q[\exp\{H(X)\}\cond X(A){=}i].$ Then
$$
Q\left[ \exp\{H(X_{\Acom})\}\right] e^{-i\hat{H}}\le \gamma_{i}\le Q\left[ \exp\{H(X_{\Acom})\}\right] e^{i\hat{H}}.
$$
We also have
\[
\begin{split}
1
& = Q\left[ \exp\{H(X)\}\right] \\
& \ge  Q\left[ \exp\{H(X_{\Acom})\}\right] Q\left[ \exp\{-\hat{H}X(A)\}\right] \\
& =Q\left[ \exp\{H(X_{\Acom})\}\right] e^{-\hat{H}\pi(A)}\\
\end{split}
\]
and
\[
\begin{split}
1
& =
Q\left[ \exp\{H(X)\}\right]\le  Q\left[ \exp\{H(X_{\Acom})\}\right] Q\left[ \exp\{\hat{H}X(A)\}\right] \\
& \le 
Q\left[ \exp\{H(X_{\Acom})\}\right] e^{\check{H}\pi(A)}, \\
\end{split}
\]
where $\check{H}:=\hat{H}e^{\hat{H}}$. Moreover,
\begin{align*}
\int \exp\{\indic_{\{g(A)\ge 1\}\}(H(g)+\beta g(A))} \, dQ(g)&= e^{-\pi(A)}\sum_{i=0}^{\infty}
  \gamma_{i}\frac{(\pi(A)e^{\beta})^{i}}{i!}\\
&\le e^{\pi(A)(\hat{H}-1)}\sum_{i=0}^{\infty}
  \frac{(\pi(A)e^{\hat{H}+\beta})^{i}}{i!}\\
&\le\exp\left\{ \pi(A)\left( \hat{H}+e^{\beta+\hat{H}}-1\right)\right\}.
\end{align*}

We now proceed to estimate
\begin{align*}
\sum_{i=0}^{\infty}\gamma_{i}\pi(A)^{i}/i!&= \sum_{i=0}^{\infty} Q[\exp\{H(X)\}\cond X(A){=}i] Q\{X(A){=}i\}=1\\
\sum_{i=1}^{\infty}\gamma_{i}\pi(A)^{i}(e^{i\beta}-1)/i!&\le
  e^{\pi(A)\check{H}}\beta\sum_{i=1}^{\infty} \left( \pi(A)e^{\beta+\hat{H}} \right)^{i}/(i-1)!\\
  &= \beta \pi(A)  \exp\left\{\beta+\hat{H}+\pi(A)\left(\check{H}+e^{\beta+\hat{H}}\right)\right\},\\
\sum_{i=1}^{\infty}\gamma_{i}\pi(A)^{i}(e^{-i\beta'}-1)/i!&\ge
  -e^{\pi(A)\check{H}}\beta'\sum_{i=1}^{\infty} \left( \pi(A)e^{\hat{H}} \right)^{i}/(i-1)!\\
&=  -\beta' \pi(A) \exp\left\{\hat{H}+\pi(A)\left(\check{H}+e^{\hat{H}}\right)\right\}.
 \end{align*}
We can then estimate the ratio
\begin{align*}
\frac{\int \exp\{H(g)+\beta g(A)\} \, dQ(g)}{\int \exp\{H(g)-\beta' g(A)\} \, dQ(g)}
  &= \frac{\sum_{i=0}^{\infty}\gamma_{i}(\pi(A)e^{\beta})^{i}/i! }{\sum_{i=0}^{\infty}\gamma_{i}(\pi(A)e^{-\beta'})^{i}/i! }\\
  & =\frac{1+\sum_{i=1}^{\infty}\gamma_{i}\pi(A)^{i}(e^{i\beta}-1)/i!}
  {1+\sum_{i=1}^{\infty}\gamma_{i}\pi(A)^{i}(e^{-i\beta'}-1)/i!}\\
&\le \frac{1+C\beta\pi(A)}{1-C\beta'\pi(A)}
\end{align*}
for a positive constant $C$ (depending on $\hat{H}$ and $\pi(\M)$),
part (a) follows.

For $i\ge 1$ let $\gamma_{i}:=Q[\exp\{H(X)\}\cond X(A){=}i],$ and let $\gamma_{0}:=1$. Then, for $i\ge 1$,
$$
Q\left[ \exp\{H(X_{\Acom})\}\right] e^{-i\hat{H}}\le \gamma_{i}\le Q\left[ \exp\{H(X_{\Acom})\}\right] e^{i\hat{H}}.
$$
We also have
\[
\begin{split}
1 
& =Q\left[ \exp\{H(X)\}\right] \\
&\ge  Q\left[ \exp\{H(X_{\Acom})\}\right] Q\left[ \exp\{-\hat{H}X(A)\}\right] \\
& =Q\left[ \exp\{H(X_{\Acom})\}\right] e^{-\hat{H}\pi(A)} \\
\end{split}
\]
and
\[
\begin{split}
1
& =Q\left[ \exp\{H(X)\}\right] \\
& \le  Q\left[ \exp\{H(X_{\Acom})\}\right] Q\left[ \exp\{\hat{H}X(A)\}\right] \\
& =Q\left[\exp\{H(X_{\Acom})\}\right] e^{\hat{H}e^{\hat{H}}\pi(A)}. \\
\end{split}
\]
Thus,
\begin{equation} \label{E:QAcombound}
\left| \log Q\left[ \exp\{H(X_{\Acom})\}\right] \right| \le \check{H}\pi(A),
\end{equation}
where $\check{H}:=\hat{H}e^{\hat{H}}$. We have
\begin{align*}
\int \exp\{\indic_{\{g(A)\ge 1\}\}(H(g)+\beta g(A))} \, dQ(g)&= e^{-\pi(A)}\sum_{i=0}^{\infty}
  \gamma_{i}\frac{(\pi(A)e^{\beta})^{i}}{i!}\\
&\le e^{\pi(A)(\hat{H}-1)}\sum_{i=0}^{\infty}
  \frac{(\pi(A)e^{\hat{H}+\beta})^{i}}{i!}\\
&\le\exp\left\{ \pi(A)\left( \hat{H}+e^{\beta+\hat{H}}-1\right)\right\}.
\end{align*}

We now proceed to estimate
\begin{align*}
\sum_{i=1}^{\infty}\gamma_{i}\pi(A)^{i}(e^{i\beta}-1)/i!&\le
  e^{\pi(A)\check{H}}\beta\sum_{i=1}^{\infty} \left( \pi(A)e^{\beta+\hat{H}} \right)^{i}/(i-1)!\\
  &= e^{\pi(A)\check{H}}\beta \pi(A)  \exp\left\{\beta+\hat{H}+\pi(A)\left(\check{H}+e^{\beta+\hat{H}}\right)\right\},\\
\sum_{i=1}^{\infty}\gamma_{i}\pi(A)^{i}(e^{-i\beta'}-1)/i!&\ge
  -e^{\pi(A)\check{H}}\beta'\sum_{i=1}^{\infty} \left( \pi(A)e^{\hat{H}} \right)^{i}/(i-1)!\\
&=  -\beta' \pi(A) \exp\left\{\hat{H}+\pi(A)\left(\check{H}+e^{\hat{H}}\right)\right\}.
 \end{align*}
We can then estimate the ratio
\begin{align*}
\frac{\int \exp\{\indic_{\{g(A)\ge 1\}\}(H(g)+\beta g(A))} \, dQ(g)}{\int \exp\{\indic_{\{g(A)\ge 1\}\}(H(g)-\beta' g(A))} \, dQ(g)}
  &= \frac{\sum_{i=0}^{\infty}\gamma_{i}(\pi(A)e^{\beta})^{i}/i! }{\sum_{i=0}^{\infty}\gamma_{i}(\pi(A)e^{-\beta'})^{i}/i! }\\
  & \hspace*{-2cm}=\frac{\sum_{i=0}^{\infty}\gamma_{i}\pi(A)^{i}/i!+\sum_{i=1}^{\infty}\gamma_{i}\pi(A)^{i}(e^{i\beta}-1)/i! }
  {\sum_{i=0}^{\infty}\gamma_{i}\pi(A)^{i}/i!+\sum_{i=1}^{\infty}\gamma_{i}\pi(A)^{i}(e^{-i\beta'}-1)/i!}.
\end{align*}

Thus, there is a positive constant $C$ (depending on $\hat{H}$ and $\pi(\M)$) such that
$$
R \le \frac{1+C\beta\pi(A)}{1-C\beta'\pi(A)}.
$$
Part (b) follows immediately.
\end{proof}

\begin{lemma} \label{L:RNsums}
Let $\mu,\mu'\in\H^{+}$ be equivalent measures on $\M$ with 
\[
\alpha:=\sup_{m \in \M} \left|\log \frac{d\mu'}{d\mu}(m)\right|<\infty. 
\]
Then,
for any measure $\xi \in \H^+$,
\[
\sup_{m \in \M} \left|\log \frac{d[\mu'+\xi]}{d[\mu+\xi]}(m)\right| \le \alpha.
\]
\end{lemma}

\begin{proof}
Since $e^{\alpha}\ge 1$,
for any nonnegative function $F:\M\to\bR_{+}$,
\begin{align*}
\int_\M F(x) \, d[\mu'+\xi](x)
&= \int_\M F(x) \frac{d\mu'}{d\mu}(x)d\mu(x)+\int F(x) \, d\xi(x)\\
&\le \int_\M F(x) e^{\alpha} \, d\mu(x)+ e^{\alpha} \int F(x) \, d\xi(x) \\
&\le e^{\alpha} \int F(x) \, d[\mu+\xi](x).
\end{align*}
Thus, 
\[
\frac{d[\mu'+\xi]}{d[\mu+\xi]}(x)\le e^{\alpha} \quad \text{for 
$[\mu+\xi]$-a.e. $x \in \M$.}
\]
An essentially identical calculation gives a corresponding lower bound $e^{-\alpha}$ and completes the proof.
\end{proof}


\backmatter

\bibliographystyle{amsalpha}
\bibliography{biblio}

\printnotation


\addcontentsline{toc}{chapter}{Index}
%
%
%
%
\begin{theindex}

  \item antagonistic pleiotropy, 1
  \item attractive equilibrium, 11, 27, 29, 32, 43, 46, 49

  \indexspace

  \item B\"{u}rger, Reinhard, 2, 5, 7, 12
  \item Barton, Nicholas, 51, 53
  \item Barton, Nicolas, 3, 53
  \item Bell number, 115
  \item biodemography, 2, 3
  \item box attractive, 27, 43, 46, 49
  \item box stable, 27, 43, 46, 49

  \indexspace

  \item Campbell's Theorem, 12, 77, 92, 97, 109, 113
  \item Campbell, N.R. , 113
  \item Charlesworth, Brian, 2--4, 48
  \item comparisons lemma, 21, 39, 43
  \item complete Poissonization, 52
  \item concave costs, 36
  \item contraction mapping theorem, 19
  \item cumulative hazard, 11, 47
  \item Curtsinger, James, 2

  \indexspace

  \item Dawson, Kevin, 2
  \item demographic costs, 11, 36, 47, 49
  \item dispersive probability measure, 68, 69
  \item dynamical system, vii, 3, 8, 10, 17, 28, 35

  \indexspace

  \item epistatic effects, 7, 8
  \item equilibrium, 1, 10--12, 27, 29, 31, 32, 35, 39, 42, 43, 45, 46, 
		49

  \indexspace

  \item Fr\'echet derivatives, 33--35, 38, 49

  \indexspace

  \item genetic legacy, 56
  \item genotype, 1, 3, 4, 15
  \item Gompertz hazards, 4
  \item Gronwall's Inequality, 24, 39, 41, 42, 111
  \item Gronwall, Thomas H., 111

  \indexspace

  \item Hahn-Jordan decomposition, 15, 18, 38
  \item Haldane's Principle, 3
  \item Haldane, John B. S., 9
  \item hazard function, 2, 11, 47, 48
  \item Horiuchi, Shiro, 3

  \indexspace

  \item intensity measure, vii, 3, 6, 8, 54

  \indexspace

  \item Kantorovich-Rubinstein distance, 16
  \item Kimura, Motoo, 2

  \indexspace

  \item L\'{e}vy process, 5
  \item Lagrange inversion formula, 10
  \item Lambert's W function, 10
  \item LeCam, Lucien, 70
  \item longevity, 3

  \indexspace

  \item Maruyama, Takeo, 2
  \item Medawar, Peter, 1
  \item minimal equilibrium, 45
  \item Monge-Kantorovich distance, 16
  \item Mueller's ratchet, 7
  \item multiplicative costs, 10, 29, 32, 42
  \item mutation accumulation, 1
  \item mutation counting, 9

  \indexspace

  \item net maternity function, 49
  \item net reproduction ratio, 11

  \indexspace

  \item one-dimensional systems, 28

  \indexspace

  \item Picard, Charles E., 18
  \item Pletcher, Scott, 2
  \item Poisson identity, 12
  \item polynomial costs, 10, 11

  \indexspace

  \item quasi-linkage equilibrium, 3

  \indexspace

  \item recombination, 2, 7, 8, 57, 63, 68, 73, 104
  \item recombination measure, 54
  \item recombination operator, 54
  \item recombination tree, 55

  \indexspace

  \item segregating set, 8, 53--55
  \item selective cost, 3, 5, 6, 9--11, 15, 18, 25, 30, 73, 104
  \item shattering, 63, 66, 68
  \item somatic mutations, 1
  \item stable equilibrium, 27, 32, 43, 46, 49
  \item Stirling number, 115

  \indexspace

  \item thinning, 115
  \item total variation norm, 17, 40, 82
  \item Turelli, Michael, 3, 51, 53

  \indexspace

  \item vintage, 55, 60

  \indexspace

  \item walls of death, 4
  \item Wasserstein metric, 15, 16, 18, 27, 40, 73, 114
  \item Wasserstein, Leonid, 15
  \item Williams, George C., 1

\end{theindex}

\end{document}